%% file: Thesis_all.tex
\documentclass[12pt,oneside,final]{report}

\usepackage{fancyhdr}
\usepackage[centertags]{amsmath}
\usepackage{amsfonts}
\usepackage{amssymb}
\usepackage{amsthm}
\usepackage[super,comma,sort&compress]{natbib}
\usepackage{makeidx}
\usepackage{newlfont}
\usepackage{longtable}
\usepackage{epsfig}
\usepackage{graphicx}

\usepackage{hu-thesis} 
\usepackage{chapterbib}
\usepackage{xtocinc} 
\usepackage[active]{srcltx}  

\usepackage{chapterbib}
\newlength{\defbaselineskip}
\newcommand{\setlinespacing}[1]%
           {\setlength{\baselineskip}{#1 \defbaselineskip}}
\numberwithin{equation}{section}

\makeindex 

\begin{document}

\ifx\href\undefined\else\hypersetup{linktocpage=true}\fi

\nolistoftables
\nolistoffigures
\nobib
\dedicate{To Tanya and sons}
\nofront
\permissionfalse
\phd \copyrightyear{2003} \submitdate{December 2003}
\convocation{December}{2003}
\university{Harvard University}
\address{Cambridge, Massachusetts}
\dept{Department of Physics}
\title{Experimental X-Ray Studies of Liquid Surfaces}
\author{Oleg Grigorievich Shpyrko}
\supervisor{Peter S. Pershan}
\firstreader{Frans Spaepen}
\secondreader{David Weitz}

{
\typeout{:?0000} 
\beforepreface
\typeout{:?1111} 
}

\setcounter{page}{3}

{
\include{Abs}
}
\tableofcontents
\listoffigures
\listoftables
{ \typeout{Acknowledgements}
\include{Ack} }
\afterpreface
\def\baselinestretch{1}
\setlinespacing{1.66}
{
\typeout{Introduction}
\include{T0_all}
}

\include{T1_all}

\include{T2_all}

\include{hg_all}

\include{in_all}

\include{kna_all}

\include{k_all}

\include{water_all}

\include{sn_all}

\include{alloy_all}

\include{jop_all}

\include{hgau_all}

\include{biin_all}

\include{GaBi_prl_all}

\include{GaBiMono_all}

\include{GaBi_all}

\index{}



{\refstepcounter{chapter} \addcontentsline{toc}{chapter}{Index}}

\printindex

\end{document}

%% file: Abs.tex

%
%

\begin{abstract_new}

Over the past two decades \index{synchrotron!source} synchrotron
facilities dedicated to the generation of x-rays for study of
atoms, molecules and condensed matter have proliferated in Europe,
Asia and the United States. As a result of the special properties
of x-rays produced by these machines, there has been an enormous
growth in the experimental possibilities. In this work we will
demonstrate how these special properties can be used to carry out
hitherto impossible measurements of the $\AA$ngstr\"om level
structure of the free surfaces of liquids.

A large part of this thesis is devoted to surface-induced atomic
or  molecular layering phenomenon in liquids. While the atomic
structure of metallic and \index{liquid!dielectric} dielectric
liquids is similar in the bulk, the microscopic atomic structure
at the \index{liquid-vapor interface} liquid-vapor interface of
dielectric liquids is believed to be fundamentally different from
that of a metal: \index{liquid!metallic} metallic liquids exhibit
an oscillatory density profile extending over a few atomic lengths
into the bulk, as opposed to a more simple monotonic density
profile that is thought to be characteristic of
\index{liquid!dielectric} dielectric liquids formed from atoms or
small molecules. Unfortunately, this oscillatory density profile
is partially obscured by the surface roughness generated by
 thermally excited capillary waves. As a
result, extraction of the microscopic form of this local density
profile requires the special experimental and theoretical methods
that are the subject of this thesis. It will be shown that the
layering is present in all metallic liquids studied so far.
Although there have been x-ray reflectivity studies of other
liquids, the only measurement to probe the surface structure for a
non-metallic liquid made of small molecules is the one reported
here on the surface of water. In view of the fact that the surface
tension for liquid K is not very different from that of water
these measurements demonstrate that the appearance of layering in
metals is not simply a consequence of higher  surface tension. In
addition to studies of pure liquids, part of this thesis is
devoted to surface of binary metallic alloys. The phenomena we
investigated include \index{Gibbs!adsorption} Gibbs adsorption and
the \index{wetting!tetra-point} tetra-point
\index{wetting!transition} wetting transition.

\end{abstract_new}


%% file: Ack.tex

\prefacesection{Acknowledgements}
\def\baselinestretch{1.0}
\setlinespacing{1.15}

I would like to thank my thesis advisor, Professor \index{Pershan,
Peter} Peter Pershan, for many suggestions and constant support
and guidance during this research.

My research was done in collaboration with three postdoctoral
fellows who contributed their knowledge and experience as well as
provided guidance on many projects - \index{Tostmann, Holger}
Holger Tostmann, \index{Huber, Patrick} Patrick Huber and
\index{Grigoriev, Alexei} Alexei Grigoriev. All of us shared a lot
of sleepless synchrotron night shifts together.

I would like to thank my fellow physics graduate student
\index{Fukuto, Masafumi} Masa Fukuto for being like a "big
brother" to me during all these years. Special thanks go out to
other members of our research group for their support and
thoughtful discussions regarding our projects and physics in
general - \index{Penanen, Konstantin} Konstantin Penanen,
\index{Heilmann, Ralf} Ralf Heilmann, \index{Gang, Oleg} Oleg
Gang, \index{Yano, Yohko} Yohko Yano, \index{Steimer, Christoph}
Christoph Steimer and \index{Alvine, Kyle} Kyle Alvine.

I acknowledge the contributions from our collaborators
\index{Ocko, Benjamin} Ben Ocko and \index{Deutsch, Moshe} Moshe
Deutsch, as well as the beamline support at X22B and X25 beamlines
at NSLS in \index{Brookhaven} Brookhaven National Lab and CMC-CAT
and ChemMatCARS sectors at APS, \index{Argonne} Argonne National
Lab. Specifically, I would like to mention \index{Lin, Binhua}
Binhua Lin, \index{Graber, Timothy} Tim Graber, \index{Gebhardt,
Jeffrey} Jeff Gerbhardt, \index{Meron, Mati} Mati Meron,
\index{Gog, Thomas} Thomas Gog and \index{DiMasi, Elaine} Elaine
DiMasi.

Over these years I had received a great deal of encouragement from
my family - parents Nina and Grigory, my brother Igor and
stepfather Vlad. I would also like to thank my physics
schoolteacher \index{Rozenwayn, Alexey} Alexey Rozenwayn for
igniting my interest in the wonderful world of science.

\medskip
\noindent
Cambridge, Massachusetts \hfill Oleg Shpyrko\\
December, 2003


%% file: T0_all.tex

\chapter{Introduction}

For many centuries  metals have played a crucial role in history
of civilization - from primitive tools of the Iron Age to the
modern wonders of science and technology. With the sizes of
high-tech device fabrication technology rapidly approaching the
\index{nanoscale} nanometer length scale frontier, the properties
of the surface become of a great interest. As such the
understanding of surface phenomena and the role surfaces of metals
and alloys play in various processes such as mixing, alloying,
refining,  oxidation and  wetting becomes increasingly important.
The present thesis is therefore motivated by both the need to
understand practical technological issues as well as the
fundamental questions related to thermodynamics of the surfaces of
liquid metals and their  alloys.

\medskip

For solid surfaces a number of local experimental techniques, such
as atomic force   microscopy (AFM) and scanning tunneling
microscopy (STM) are widely available and can provide an
atomic-size resolution information. In the past couple of years it
was shown that these microscopy techniques could be applied to
liquid systems, namely studies of wetting and capillarity
\cite{Salmeron00, Salmeron01, Mugele02}. However, the spatial
resolution achieved in these studies was limited to
\index{nanoscale} nanometer length scales, which is not enough to
resolve atomic structural features. Optical methods, such as
\index{ellipsometry} ellipsometry \cite{Nattland96}, can be used
to obtain some structural data from the free surfaces of liquids,
however they cannot provide any information about features much
smaller than the wavelength of visible light. Techniques which use
X-rays, on the other hand, make it possible to studying structural
features of the surface which occur on the length scale of a few
$\AA$. Advances in \index{synchrotron!radiation} synchrotron
radiation allow for these techniques to be applied to study the
structural phenomena at the free surfaces of liquids. Over the
past 20 years these methods has been applied to the study of a
number of liquid systems, including measurements of the surface
structure of liquid metals \cite{Flom93,  Sluis83, Magnussen95,
Regan95, Regan96, Lei96, Regan97, Regan97b, Tostmann98, Zhao98,
DiMasi98, Tostmann99, Tostmann99b, DiMasi99, Yang99, Yang00,
DiMasi00, Tostmann00, DiMasi00b, DiMasi01, Huber02, Huber02b,
Shpyrko03, Shpyrko03b}, \index{superfluid helium} superfluid
 helium \cite{Lurio91, Lurio92, Lurio93, Penanen95,
Penanen00, Penanen00b}, \index{liquid!crystals} liquid crystals
\cite{Pershan87, Braslau85, Ocko86, Kellogg95, Braslau88, Ocko87,
Swislow91, Lucht01},  water \cite{Braslau85, Braslau88,
Shpyrko03c} and various other systems \cite{Swislow91, Schwartz92,
Schlossman91b, Schlossman91, Bommarito96, Fukuto97, Fukuto98,
Doerr99, McClain99, Fukuto99, Heilmann01, Heilmann01b, Prange01,
Mitrinovic01, Li01, Kraack02, Fukuto02, Gang02}.

\medskip
This work describes X-ray surface techniques that we have applied
to studying surface phenomena such as surface-induced layering in
metallic liquids, capillary wave excitations in both
\index{liquid!metallic} metallic and \index{liquid!dielectric}
dielectric liquids, as well as \index{wetting!critical} critical
wetting phenomena in
 binary metallic alloys.

\section{Overview}

This thesis is structured as follows:

\medskip
 Chapter 2 includes a discussion of the scientific background of
 various x-ray surface techniques used in
this work, as well as the practical limitations of these methods
as applied to studies of liquids in general and
\index{liquid!metallic} metallic liquids in particular.

\medskip

Chapter 3 contains a brief review of some of the previous
experiments which provide the groundwork for the measurements
described in this thesis, mainly the studies of surface-induced
layering in liquid  Hg \cite{Magnussen95}, Ga \cite{Regan95} and
In \cite{Tostmann99}.

\medskip
Chapter 4 covers the temperature-dependent studies of
surface-induced layering phenomena in liquid Hg. Liquid Hg is the
first metal for which surface layering has been experimentally
confirmed and this is the first experiment where we attempted to
study the temperature-dependent behavior of the layering
structure.

\medskip
 Chapter 5 reviews experimental studies of
liquid In in more details. This is the first study where we have
thoroughly examined the off-specular contributions from thermally
excited capillary wave excitations and the validity of the
capillary wave theory in application to metallic liquid surfaces.

\medskip
Chapters 6 and 7 include a discussion of x-ray specular
reflectivity and diffuse scattering measurements from the surface
of liquid \index{alkali metals} alkali metals. Chapter 6 deals
with experimental results from binary sodium-potassium alloy,
while Chapter 7 deals with studies of surface of pure potassium.
These studies resolve a long-standing fundamental question of
whether the surface-induced atomic layering is a result of
high-surface tension acting as an effective \index{hard wall} hard
wall potential. Unlike most high-surface tension metals studied so
far, the relatively low surface tensions of the alkali metals are
comparable to water and are not that much larger than some other
dielectric liquids. Furthermore since their electronic properties
are \index{nearly free electron} nearly-free electron like the
relation between their structure and their electronic properties
is more easily treated than for most other metals.

In view of the relatively low surface tension the surface
 roughness due to thermally excited capillary
waves partially obscures the surface layering in \index{alkali
metals} alkali metals. Although this is somewhat similar to the
\index{Debye-Waller} Debye-Waller effect that is common in
crystallography, the problem of thermal excitations on a
two-dimensional surface is subtler. The procedures for extracting
the surface structure from finite temperature measurements will
also be discussed in these Chapters.

\medskip
Chapter 8 contains a discussion of x-ray specular reflectivity and
diffuse scattering measurements from the surface of water. Water
is one of the most basic substances known to mankind, and it is
commonly used as a substrate in many structural measurements, such
as \index{self-assembled monolayers} self-assembled monolayers,
while at the same time the $\AA$ level molecular structure of
water surface is only partially known \cite{Du93}. The new
generation of \index{synchrotron!source} synchrotron sources
provides an opportunity to significantly improve our knowledge of
the surface of water. X-ray surface studies of liquid potassium
described in Chapter 7 demonstrated that by combining x-ray
\index{x-ray!specular reflectivity} specular reflectivity and
x-ray \index{x-ray!diffuse scattering} diffuse scattering it is
possible to obtain structural information on the surface layering
of liquids whose low surface tensions limit the reflectivity
measurements to values of $q_z$ that are less than $2
\pi/$layering~distance. The experiments described in this section
make use of the method described in Chapter 7 to confirm the
absence of molecular layering at the free surface of water.

\medskip
Chapter 9 deals with experimental studies of liquid Sn. Although
liquid Sn does exhibit a strong layering peak that is similar to
those observed in all of the other metals that have been studied,
there is a secondary maximum in the reflectivity at low angles
that was not observed in Ga\cite{Regan95}, In \cite{Tostmann99} or
K\cite{Shpyrko03}.
 A somewhat similar feature
was observed for liquid Hg \cite{Magnussen95}; however, in view of
the high vapor pressure for Hg the possibility that this feature
might have been due to an impurity was never excluded. In contrast
the fact that liquid Sn was studied at a third generation
synchrotron (APS) under UHV conditions allowed a combination of
resonance scattering and x-ray \index{x-ray!fluorescence}
fluorescence studies that absolutely exclude the possibility that
the effect is associated with impurities. This feature is
definitely an unexpected intrinsic property of the surface of
liquid Sn. The analysis to be presented in this chapter will
demonstrate that the data is consistent with the existence of a
high density atomic layer at the surface, which can be attributed
to spacing for the uppermost layer at the surface.

\medskip
 Chapter 10 and Chapter 11 reviews x-ray reflectivity and diffuse scattering results for a number of binary
 alloys: Chapter 10 concentrates on GaBi, GaIn and HgAu, while Chapter 11 covers KNa, BiIn and GaBi  alloys. In these
 chapters we address issues such as  \index{Gibbs!adsorption} Gibbs adsorption, surface segregation, formation of surface
 phases and surface wetting.

\medskip
Chapter 12 provides temperature dependent x-ray reflectivity
measurements for  mercury-gold alloy. Upon approaching the
\index{solubility limit} solubility limit of Au in Hg, a new
surface phase forms which is 1--2 atomic diameters thick and has a
density of about half that of bulk Hg. We present a surface
\index{phase diagram} phase diagram, summarizing the evolution of
this unexpected surface structure through comparatively small
changes in temperature and composition.

\medskip
Chapter 13 describes \index{x-ray!resonant scattering} resonant
x-ray reflectivity measurements from the surface of liquid
Bi$_{22}$In$_{78}$. In contrast to \index{Gibbs!adsorption} Gibbs
adsorption of a complete monolayer observed in all liquid alloys
studied to date we find only a modest surface Bi enhancement, with
35~at\% Bi in the first atomic layer. This suggests that surface
adsorption in Bi-In is dominated by attractive interactions that
increase the number of Bi-In neighbors at the surface.

\medskip
Chapters 14 through 16 contain a discussion of the results of an
experimental study of \index{wetting!tetra-point} tetra-point
wetting on the free surface of binary alloy mixture of
 gallium and bismuth. Almost all experimental and
theoretical studies describing \index{wetting!transition} wetting
transition published to date address dielectric liquids in which
the wetting is dominated by long-ranged \index{interactions!Van
der Waals} Van der Waals interactions \cite{Dietrich88}. The
single exception is wetting near the bulk liquid/gas critical
point for which the vanishingly small difference between the
density of the liquid and the gas cause the amplitude of  the
long-range Van der Waals force to also become vanishingly small.
In this case wetting should be dominated by \index{short-range
interactions} short-range forces. \cite{Ross99, Ross01}. In
contrast, wetting in \index{liquid!metallic}metallic liquids is
naturally governed by short-range \index{interactions!Coulomb}
Coulomb screened potentials.

 The measurements described in these chapters determined the microscopic
(i.e. $\AA$ngstr\"om level) compositional gradient of the
liquid/liquid interface between the wetting layer and the bulk
liquid \cite{Ebner77, Davis96}. For this binary system, the
wetting transition is pinned to a bulk \index{monotectic point}
monotectic temperature, creating a rare case of
\index{wetting!tetra-point} tetra point wetting, where four phases
coexist in the bulk: a Ga-rich liquid, a Bi-rich liquid, a solid
Bi, and a vapor\cite{Dietrich97, Serre98}.

In Chapter 16 the compositional gradient at the liquid/liquid
interface is interpreted in terms of a \index{mean-field square
gradient model} mean-field square gradient model that was extended
by inclusion of capillary waves as the dominant thermal
fluctuations. This theory makes use of the influence parameter,
$\kappa$, which governs the contribution of the concentration
gradient to the surface energy. With this in hand, the
liquid/liquid \index{interfacial energy} interfacial energy, which
is very difficult to measure directly, was determined. With
knowledge of the interfacial energy it was possible to separate
the thermal capillary contributions from the intrinsic width of
the liquid/liquid interface.

\medskip
\medskip

\section{Citations to Previously Published Work}

As is customary in physics department,the format of this thesis is
introductory chapters followed by reprints of previously published
papers. The references to the published materials are listed
below:
\medskip
  The overview presented in Chapter 3 is based on the following papers:
\begin{itemize}
  \item  H. Tostmann, E. DiMasi, P.S. Pershan, B.M. Ocko, O.G. Shpyrko and M.
Deutsch, "Structure of Liquid Metals and the Effect of Capillary
Waves: X-Ray Studies on Liquid Indium", \textit{Phys. Rev. B} 59,
783 (1999)
  \item E. DiMasi, H. Tostmann, O. G. Shpyrko, M. Deutsch, P.S. Pershan
and B.M. Ocko, "Surface Induced Order in Liquid Metals and Binary
Alloy", \textit{Journal of Physics: Condensed Matter} 12, 209
(2000)
  and
  \item E. DiMasi, O. M. Magnussen, B. M. Ocko, M. Deutsch, H.
Tostmann, O. G. Shpyrko, M. J. Regan and P. S. Pershan
"Temperature Dependent Surface Structure of Liquid Mercury"
\textit{Bull. Am. Phys. Soc.} 42, 310 (1997)
\end{itemize}

\medskip
Chapter 4 is based on
\begin{itemize}
 \item E. DiMasi, O. M. Magnussen, B. M. Ocko, M. Deutsch, H.
Tostmann, O. G. Shpyrko, M. J. Regan and P. S. Pershan
"Temperature Dependent Surface Structure of Liquid Mercury",
\textit{Bull. Am. Phys. Soc.} 42, 310 (1997)
\end{itemize}

\medskip
Chapter 5 presents results published in
\begin{itemize}
\item  H. Tostmann, E. DiMasi, P.S. Pershan, B.M. Ocko, O.G. Shpyrko and M.
Deutsch, "Structure of Liquid Metals and the Effect of Capillary
Waves: X-Ray Studies on Liquid Indium", \textit{Phys. Rev. B} 59,
783 (1999)
\end{itemize}

\medskip

Chapter 6 contains the results presented in a publication
\begin{itemize}
  \item  H. Tostmann, E. DiMasi, P.S. Pershan, B.M. Ocko, O.G. Shpyrko and M.
Deutsch, "Microscopic Surface Structure of Liquid Alkali Metals",
\textit{Phys. Rev. B} 61 (1999) 7284

\end{itemize}

\medskip

Chapter 7 is based on
\begin{itemize}
\item O.G. Shpyrko, P. Huber, P.S. Pershan, B.M.
Ocko, H. Tostmann, A. Grigoriev and M. Deutsch, "X-ray Study of
the Liquid Potassium Surface: Structure and Capillary Wave
Excitations", \textit{Phys. Rev. B} 67, 115405 (2003).
\end{itemize}

\medskip
Chapter 8 is based on
\begin{itemize}
\item O.G. Shpyrko, M. Fukuto, P.S. Pershan, B.M.
Ocko, I. Kuzmenko and M. Deutsch "Surface Layering of Liquids: The
Role of Surface Tension", \emph{Phys. Rev. B}  69 245423, (2004)
\end{itemize}

\medskip
Chapter 9 is derived from two papers:
\begin{itemize}
\item O.G. Shpyrko, A. Grigoriev , C. Steimer, P.S. Pershan, B. Lin, T. Graber, J. Gerbhardt, M. Meron,
 B.M. Ocko and M. Deutsch "Anomalous Layering at the Liquid Sn Surface",
 \emph{Phys. Rev. B}, 70, 224206 (2004)
\item A. Grigoriev , O.G. Shpyrko, C. Steimer, P.S. Pershan, B. Lin, T. Graber, J. Gerbhardt, M. Meron,
 B.M. Ocko and M. Deutsch "Surface Oxidation of Liquid Tin", \emph{Surf. Sci.} 575, 3, 223 (2005).
\end{itemize}

\medskip
Chapters 10 and 11 presents reviews from
\begin{itemize}
\item H. Tostmann, E. DiMasi, O. G. Shpyrko, P. S. Pershan, B. M. Ocko
and M. Deutsch, "Surface Phases in Binary Liquid Metal Alloys: an
X-ray Study", \textit{Ber. Bunsenges. Phys. Chem.} 102, 1136-1141
(1998)
\end{itemize}
 and
\begin{itemize}
\item E. DiMasi, H. Tostmann, O. G. Shpyrko, M. Deutsch, P.S. Pershan
and B.M. Ocko, "Surface Induced Order in Liquid Metals and Binary
Alloy", \textit{Journal of Physics: Condensed Matter} 12, 209
(2000)
\end{itemize}

\medskip
Chapter 12 presents results from
\begin{itemize}
\item E. DiMasi, H. Tostmann, O. G. Shpyrko, M. Deutsch, P.S. Pershan
and B.M. Ocko, "Resonant X-Ray Scattering from the Surface of a
Dilute Hg-Au Alloy", \textit{Materials Science V} Vol. 590. Eds.
Mini, Perry and Stock. Materials Research Society (2000).
\end{itemize}

\medskip
Chapter 13 contains measurements previously published in
\begin{itemize}
  \item E. DiMasi, H. Tostmann, O. G. Shpyrko, M. Deutsch, P.S. Pershan
and B.M. Ocko, "Pairing Interactions and Gibbs Adsorption at the
Liquid Bi-In Surface: a Resonant X-ray Reflectivity Study",
\textit{Phys. Rev. Lett.}  86, 1538 (2001).
\end{itemize}

\medskip
Chapters 14 through 16 include the measurements published in the
following four papers:
\begin{itemize}
  \item P. Huber, O.G. Shpyrko, P.S. Pershan, B.M. Ocko, E. DiMasi, M.
Deutsch "Short-Range Wetting at Liquid Gallium-Bismuth Alloy
Surfaces: X-ray Reflectivity Measurements and Square Gradient
  Theory", \emph{Phys. Rev. B} 68, 085409 (2003)
\end{itemize}

\begin{itemize}
  \item P. Huber, O.G. Shpyrko, P.S. Pershan, E. DiMasi, B.M. Ocko, H.
Tostmann and M. Deutsch "Wetting at the Free Surface of a Liquid
Gallium-Bismuth Alloy: An X-ray Reflectivity Study Close to the
Bulk Monotectic Point", \emph{Colloids \& Surfaces A.} 206, 515 (2002)
\end{itemize}

\begin{itemize}
\item H. Tostmann, E. DiMasi, O.G. Shpyrko, P.S. Pershan, B.M. Ocko and
M.Deutsch "Microscopic Structure of the Wetting Film at the
Surface of Liquid Ga-Bi Alloys", \emph{Phys. Rev. Lett.} 84 (2000) 4385

and

  \item P. Huber, O.G. Shpyrko, P.S. Pershan, B.M. Ocko, E. DiMasi, and M.
Deutsch "Tetra Point Wetting at the Free Surface of Liquid Ga-Bi",
Phys. Rev. Lett. 89, 035502 (2002).
\end{itemize}

\bibliographystyle{unsrt}


%% file: T1_all.tex
\chapter{Theoretical and Experimental Overview}

\section{X-rays}

The  \index{x-ray!spectrum} x-ray portion of the electromagnetic
spectrum is generally taken to encompass wavelengths from
$10^{-12}$ to $10^{-8}$ meters (0.01 to 100~$\AA$). Early in the
20th century \index{Laue} Laue and Knipping \cite{Laue12, Laue12a}
made use of the fact typical atomic and molecular scales are of
the order of 1 to 10~$\AA$ to demonstrate the first
 x-ray diffraction pattern from a single
crystal.  Ever since then x-rays have naturally become one of the
principal tools for studying atomic and molecular structures.

\section{Index of Reflection}
When \index{Roentgen, Wilhelm Conrad} Roentgen discovered X-rays
in 1895, he immediately searched for a means to focus them
\cite{Rontgen95, Rontgen96}. On the basis of slight refraction in
prisms he stated that the index of refraction of X-rays in
materials cannot be much more than n = 1.05 if it differed at all
from unity. Twenty years layer, \index{Einstein, Albert} Einstein
proposed \cite{Einstein18} that the refractive index for X-rays
was $n = 1 - \delta$ where $\delta \approx 10^{-6}$, allowing for
total \index{external reflection} external reflection at
\index{x-ray!grazing incidence diffraction} grazing incidence
angles. A simple classical model in which an electron of the
material is considered to be accelerated by the electromagnetic
field generated by x-rays shows that the complex \index{dielectric
contstant} dielectric constant $\varepsilon$ can be described as:

\begin{equation}
\label{eq:refindex1a} \epsilon = 1- \frac{4\pi \rho Z e^2 }{m
\omega^2}\left( 1+ \frac{i}{\mu\omega}\right)
\end{equation}
where $\rho$ is the electron density of the material, $\omega$ is
the frequency of the  electromagnetic field, $\mu$ is the
\index{x-ray!absorption length} absorption length, $e$ is the
electron charge, $m$ is the electron mass and Z is the atomic
number of the element. The results of a more accurate quantum
mechanical treatment are qualitatively similar:

\begin{equation}
\label{eq:refindex1} {n = \sqrt{\epsilon} = 1-{\delta}+{i\zeta}}
\end{equation}
where the real and imaginary parts of the \index{index of
refraction} index of refraction, $\delta$ and $\zeta$, account for
the and \index{x-ray!absorption} absorption of material,
respectively. They are defined as:

\begin{equation}
\delta=\frac{r_e}{2\pi}\lambda^2 \sum _k \frac{Z_k+f'_k}{V_m}
\label{eq:refindex3}
\end{equation}
and
\begin{equation}
\zeta=\frac{r_e}{2\pi}\lambda^2 \sum _k
\frac{f''_k}{V_m}=\frac{\lambda}{4\pi}\mu \label{eq:refindex4}
\end{equation}
where $r_e = 2.813 \cdot 10^{ - 5}~\AA$ is the classical radius of
the electron, $V_m$ is the volume of the  unit cell, $Z_k$ is the
number of electrons of atom $k$ in the unit cell, $f'$ and $f"$
are the \index{x-ray!anomalous scattering} anomalous scattering
coefficients~\footnote{For the relatively small angular range of
the x-ray reflectivity measurements described in this thesis the
angular dependence of the atomic \index{form factor} form factor
can be neglected.}, and $\lambda$ is the x-ray wavelength. If the
x-ray wavelength $\lambda$ is low in comparison with the principal
absorption energies, $f'_k \approx 0$  and $\delta \approx r_e
\lambda^2 \rho / (2 \pi)$.

\section{Critical Angle of Reflection}
In view of the fact that $\delta$ is positive, the refractive
index of a material is always smaller than unity \cite
{Compton23}.  X-rays incident on the material surface from the
vacuum will be refracted towards the surface. If the incident
angle (measured from the surface) is less than a
\index{critical!angle} \textit{critical angle} $\alpha_c$ that is
given by:
\begin{equation}
\cos(\alpha_c) = n = 1 - \delta \label{eq:refindex5}
\end{equation}
the x-rays will be totally reflected. Since $1 \gg \delta > 0$ the
expression for critical angle can be approximated as:
\cite{AlsNielsen01}

\begin{equation}
\alpha_c^2 = 2\delta = \frac {r_e \lambda^2} {\pi} \rho
\label{eq:refindex6}
\end{equation}

\section{X-Ray Surface Techniques: Overview}
Three major surface X-ray techniques are commonly used throughout
the measurements described in this thesis -
namely,\index{x-ray!specular reflectivity} X-ray specular
reflectivity, \index{x-ray!diffuse scattering} X-ray diffuse
scattering and \index{x-ray!grazing incidence diffraction} X-ray
grazing incidence diffraction. Each technique will be described in
detail below. In the case of binary systems each of these
techniques can be enhanced by tuning the x-ray energy through the
\index{x-ray!absorption edge} absorption edge of one, or another
of the elements. In this way the relative scattering amplitudes of
individual elements can be changed by factors that are sometimes
as large as a factor of two. In this thesis we have applied these
resonance effects to specular reflectivity measurements. In
addition, observation of characteristic x-ray
\index{x-ray!fluorescence} fluorescence as a function of
excitation energy can be used as a separate method for testing for
the presence of different elements.

\section{Reflectivity}

\subsection{Kinematics of Reflectivity Measurement} 
\begin{figure}[tp]
\begin{center}
\unitlength1cm
\begin{minipage}{12cm}
\epsfxsize=9cm \epsfysize=8.5cm \epsfbox{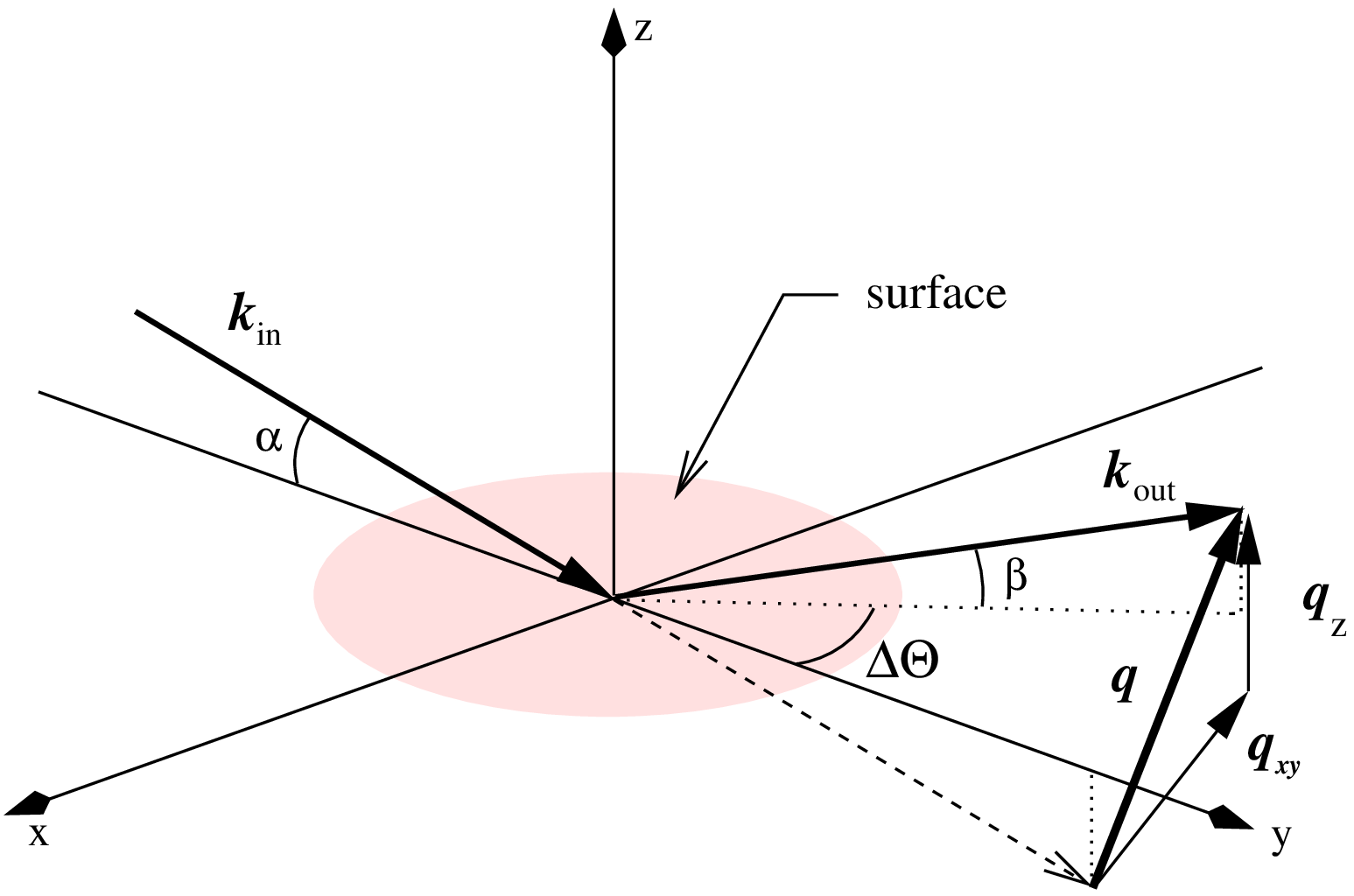}
\end{minipage}
\caption{\label{fig:kinema}Kinematics of the x-ray measurement.}
\end{center}
\end{figure}

In order to
introduce these x-ray techniques consider a simple case of a
\index{x-ray!monochromatic beam} monochromatic x-ray beam with
wavelength $\lambda$ reflected from a smooth, ideally flat and
abruptly terminated (that is, having step-function-like density
profile at the material-vapor interface) planar surface. Figure
\ref{fig:kinema} provides a general description of kinematics for
all of the x-ray surface techniques described in this work. For
any incident angle $\alpha$ there will be a specular reflection
observed at a detection angle $\beta=\alpha, \Delta\Theta=0$, both
incident and reflected beams located in the same
\index{specular!plane} (specular) plane perpendicular to the
surface plane. This type of reflection will be referred to as
\index{specular!reflection} \textit{specular}. Later in this
thesis we will discuss off specular scattering for which the
detector is located out of the plane of incidence ($\Delta\Theta
\neq 0$) and/or with $\beta\neq \alpha$.
\medskip
Conventional measurements of x-ray reflectivity from solid
surfaces commonly make use of what is known as $ \Theta-2\Theta$
instruments in which the sample is rotated by an angle $\Theta$
while the detector is rotated by twice that, or $2\Theta$ in order
to keep the incident angle $\alpha$ equal to the angle $\beta$.
This is not possible for liquid surfaces and the X-ray
reflectivity experiments described in this thesis require
simultaneous scanning both the incident angle $\alpha$ and output
angle $\beta$, while keeping $\beta=\alpha$, in the specular plane
geometrically defined as $\Delta \Theta=0$.

\medskip

\section{Ideal Surface:  Fresnel Reflectivity}

\index{Fresnel reflectivity} The intensity of the reflected
specular signal from an ideal flat surface can be calculated by
considering the usual boundary conditions for electromagnetic
waves. The result is what is known as \textit{Fresnel
relationship}, which gives the amplitude of specular reflection
for both s- and p-polarized waves as a function of incident angle
$\alpha$. The reflectivity, defined as a ratio of the reflected
beam intensity $I_{r}$ to the intensity of the direct beam $I_0$:
$R_F(\alpha)={I_{r}(\alpha)}/{I_{0}}$, is the modulus square of
this amplitude. For small angles the Fresnel reflectivity is
independent of the \index{x-ray!polarization} polarization
\cite{Tolan99}:

\begin{equation}
R_F(\alpha)={\biggl|
{{sin\alpha-\sqrt{{sin^2\alpha}-{sin^2\alpha_c}-2i\zeta}} \over
{sin\alpha+\sqrt{{sin^2\alpha}-{sin^2\alpha_c}-2i\zeta}}}
\biggr|}^2 \label{eq:FresnelA}
\end{equation}
Equation \ref{eq:FresnelA} can also be rewritten as a function of
the vertical component of \index{wavevector transfer} wavevector
transfer, $q_z$ (for $\alpha=\beta$ and $\Theta=0$, $\vec q = (0,
0, q_z)$)

\begin{equation}
R_F(q_z)= {\biggl|
{{q_z-\sqrt{{q_z}^2-{q_c}^2-32i{\pi}^2\zeta/{\lambda}^2}} \over
{q_z+\sqrt{{q_z}^2-{q_c}^2-32i{\pi}^2\zeta/{\lambda}^2}}}
\biggr|}^2 \label{eq:Fresnelqz}
\end{equation}
where $q_z$ is defined as:

\begin{equation}
q_z = {2\pi \over \lambda} (\sin\alpha+\sin\beta) \label{eq:qz}
\end{equation}
or, since $\alpha=\beta$

\begin{equation}
q_z = {4\pi \over \lambda} \sin\alpha
\label{eq:qz2}
\end{equation}
If \index{dispersion} dispersion effects are neglected, i.e. if
$f'_k \approx 0$, the \index{critical!wavevector} critical
wavevector can be written as

\begin{equation}
q_c = {4\pi\sin\alpha_c \over \lambda} \approx 4 \sqrt {\pi r_e
\rho} \label{eq:qc} ~~\footnote{Note that unlike $\alpha_c$, $q_c$
is independent of $\lambda$, which is a reason why reflectivity is
commonly expressed as a function of wavevector transfer component
$q_z$ rather than incident angle $\alpha$.}
\end{equation}
For large values of $q_z$, \index{Fresnel reflectivity} Fresnel
reflectivity asymptotically approaches
\begin{equation}
R(q_z) = {\biggl( {{q_c} \over {2q_z}} \biggr)}^4
\label{eq:Fresnelqz2}
\end{equation}

\section{Reflectivity from a Structured Surface}

\subsection{ Born Approximation vs. Dynamic Parratt Formalism}
The kinematical theory or first \index{Born approximation} Born
approximation greatly simplifies the expression for the intensity
of specular reflectivity for non-ideal surfaces that are either
rough or have structure. The basic assumption of the Born
approximation is that the amplitude of the transmitted wave is
taken equal to that of the incident wave \cite{Born64, Daillant99,
Tolan99}. This amounts to neglecting refraction effects. The
assumption is applicable for $q_z \gtrsim 5q_c$. In view of the
fact that most of the interesting features to be discussed in this
thesis occur for $q_z \gg q_c$, the majority of results will be
analyzed using the Born approximation. The studies of
\index{wetting!tetra-point} tetra-point wetting in the binary GaBi
alloy that will be discussed in Chapter 7 involve features at
values of $q_z \approx q_c$, and for these we will find it
necessary to make use of the interactive matrix formalism
developed by \index{Parratt formalism}
\textit{Parratt}\cite{Parratt54}, which takes into account the
refraction, or multiple scattering, that is neglected by Born
approximation. For the first six chapters of this work we will
limit ourselves with the simpler Born approximation result.

\subsection{Master Formula}
Within the first Born approximation, the x-ray reflectivity signal
is a direct measure of the average electron density of the sample
perpendicular to the surface normal $z$, $\rho(z)$. More
specifically, the ratio of the x-ray reflectivity to the Fresnel
reflectivity will deviate from unity by the absolute value of the
square of amplitude of a \index{surface!structure factor}
\textit{surface structure factor} $\Phi (q_z)$, that is defined as
the \index{Fourier transform} Fourier transform of the derivative
of the in-plane average of the electron density along the surface
normal \cite{Braslau88}:

\begin{equation}
\Phi (q_z) = \frac{1}{\rho_{\infty}} \int dz\frac{\langle d\rho(z)
\rangle }{dz} \exp(\imath q_z z) \label{eq:structure_ch1}
\end{equation}
where $\rho_{\infty}$ is taken to be the electron density of the
bulk. The reflectivity from a surface with structure is thus given
by
\begin{equation}
R(q_z)=R_F(q_z) \cdot {\biggl| \Phi (q_z) \biggr|} ^2
\label{eq:Fresnel3}
\end{equation}
The equations \ref{eq:structure_ch1} and \ref{eq:Fresnel3} are
known as a \index{Master formula} \textit{Master
formula}\cite{Pershan84}.

\section{Form Factor: Examples}

In this section we will illustrate the application of
Eq.~\ref{eq:structure_ch1} with a few heuristic examples of
density profiles that are relevant to the measurements that are
discussed later in this thesis.

\subsection{Ideally sharp flat surface}
An ideally flat surface in which the electron density transition
from the bulk to the vacuum is abrupt can be represented by a
density profile along the surface normal $\rho(z)$ which has a
shape of a \index{step function} step function:

\begin{equation}
 \frac {\langle \rho(z) \rangle}{\rho(\infty)} = \left\{ \begin{array}{ll}
         1 & \mbox{if $x \geqslant 0$};\\
         0 & \mbox{if $x < 0$}.
         \end{array} \right.
\label{eq:rho_ex1a}
\end{equation}
This corresponds to a $\delta$-function density derivative:
\[ \frac {d} {dz} \biggl( \frac {\langle \rho(z) \rangle}{\rho(\infty)}\biggr) = \delta(z) \]
and according to Eq.~\ref{eq:structure_ch1}, results in a
structure factor

\begin{equation}
\biggl| \Phi (q_z) \biggr| = 1
\label{eq:rho_ex1b}
\end{equation}
As should be expected, the predicted reflectivity from ideally
flat surfaces follows Fresnel law.

\subsection{Flat surface with a gradual interfacial profile}
\begin{figure}[htp]
\centering
\includegraphics[width=0.8\columnwidth]{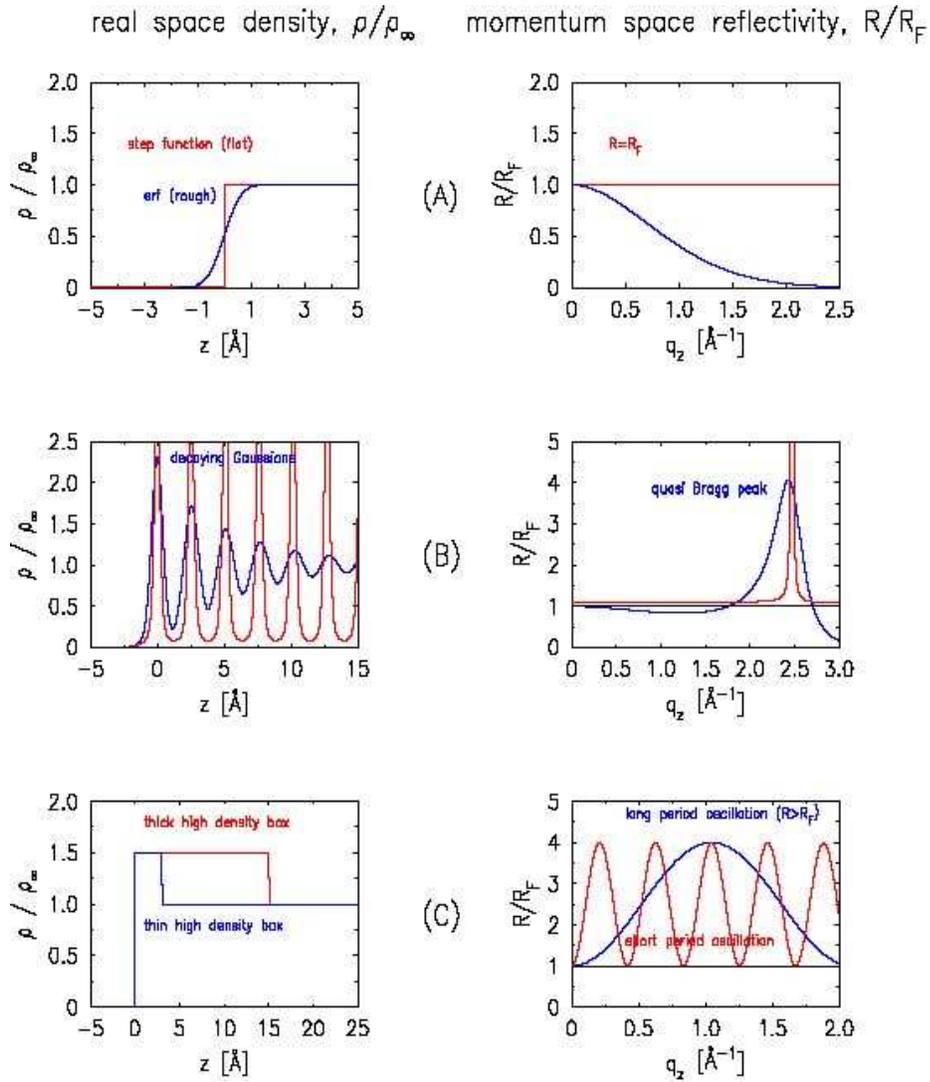}
\caption{\label{fig:model} Density profiles $\rho(z) /
\rho_\infty(z)$ and respective reflectivity curves
$R(q_z)/R_F(q_z)$ for a smooth and roughened surfaces (A),
crystalline and layering structure (B) and thin/thick high-density
layers (C)}
\end{figure}

For an interface in which density profile undergoes a monotonic
gradual transition from vacuum to bulk can be characterized by an
\index{error function} error-function of width $\sigma$:
\begin{equation}
\frac {\langle \rho(z) \rangle}{\rho(\infty)} = \text{erf} \biggl(
\frac {z} {\sigma \sqrt{2}} \biggr) \label{eq:rho_ex2a}
\end{equation}
The derivative of such an error function is the \index{gaussian}
gaussian of width $\sigma$,

\begin{equation}
\frac {d} {dz} \biggl( \frac {\langle \rho(z)
\rangle}{\rho(\infty)}\biggr) = \frac {1} {\sigma \sqrt{2\pi}}
e^{- \frac {x^2} {2 \sigma^2}} \label{eq:rho_ex2b}
\end{equation}
and according to Eq.~\ref{eq:structure_ch1},

\begin{equation}
 \Phi (q_z) = e^{- {{q_z}^2 \sigma^2} /{2}}
\label{eq:rho_ex2c}
\end{equation}
It follows that
\begin{equation}
R(q_z)=R_F(q_z) \cdot {\biggl| \Phi (q_z) \biggr|} ^2  = R_F(q_z) e^{-{q_z}^2 \sigma^2}
\label{eq:rho_ex2d}
\end{equation}
Figure \ref{fig:model}~(A) shows the two density profiles
described so far along with the corresponding forms for
$R(q_z)/R_F(q_z)$. In these plots the red line represents the
density profile and $R(q_z)/R_F(q_z)$ of for the ideal abrupt
surface. The blue lines illustrate the roughened density profile
and reflectivity for the gradual interface.

\subsection{Crystalline and Quasi-Crystalline Model}
A slightly more complicated case is the surface structure factor
 from a periodic lattice of a finite-size crystal that terminates at atomic planes.
If the \index{form factor} atomic form factors are neglected the
density profile perpendicular to the surface can be modeled by a
series of N $\delta$-functions ($N \gg 1$) located at $z = d$,
$2d$, $3d$, ... , $nd$, ... $Nd$, where d is the \index{lattice
spacing} lattice spacing parameter. This model can be
quantitatively described by a density function $\langle \rho(z)
\rangle$:

\begin{equation}
\frac {\langle\rho(z)\rangle}{\rho(\infty)} =  d \sum_{n=1}^N
\delta(z-nd) \label{eq:rho_ex3a}
\end{equation}
Applying this to Eq.~\ref{eq:structure_ch1} the structure factor
becomes
\begin{eqnarray}
\biggl| \Phi (q_z) \biggr| = q_z d \biggl| \int_{-\infty}^\infty dz \sum_{n=1}^N \delta(z-nd) \exp(\imath q_z z) \biggr| =   \nonumber \\
= q_z d \biggl| \sum_{n=1}^N e^{\imath q_z nd} \biggr| = q_z d
\frac{\sin(\frac{N q_z d}{2})}{\sin(\frac{q_z d}{2})}
\label{eq:rho_ex3b}
\end{eqnarray}
The function $q_z d \sin(N q_z d/2)/\sin(q_z d / 2)$ has sharp
peaks at $q_z d = 2 \pi$, $4 \pi$, $6 \pi$, etc. that are
associated with constructive interference from planes with
\index{lattice spacing} lattice spacing $d$ that are just the
\index{Bragg reflection} Bragg reflections from an array of
parallel planes. For the finite N the scattering amplitude of
these peaks increase and the widths decrease as the number of
atomic layers N is increased\footnote{Even for an infinitely large
N, the reflectivity corresponds to a finite number of layers
typically determined among other things by the finite value of
\index{x-ray!absorption length} absorption length and energy
bandwidth};

The crystalline density model is illustrated in Figure
\ref{fig:model}~(B) for $N=50$. The density profile (box on the
left) drawn with a red line describes a set of well-defined atomic
layers, and corresponds to the \index{quasi-Bragg peak} Bragg-like
peak in the momentum reflectivity space (box on the right). The
blue line in Figure \ref{fig:model}~(B) corresponds to a model for
surface induced atomic layering of the electron density that
decays with distance from the surface. Details for this model
which was first introduced to discuss the surface of liquid Hg
will be discussed later in this thesis. In view of the fact that
the layering is limited to just a few atomic layers at the
surface, the surface structure peak associated with this density
profile that is shown by the blue line in the box on the right is
broader and less pronounced, than that of the crystal model shown
by the red line. Such layering peak is commonly referred to as
\index{quasi-Bragg peak} \textit{quasi}-Bragg peak.

\subsection{High-density surface layer}
The final model to be discussed (Figure~\ref{fig:model}~(C)) is
constructed with a surface slab whose electron density is higher
than that of the bulk liquid. In this heuristic model atomic
layering is neglected. The simplest mathematical representation of
this is what is commonly referred to as a \index{box model}
\textit{two-box model}, a higher-density box of width $d$ and
density $\rho_{\infty}(1+A)$,  built on top of a semi-infinite
 box of lower density $\rho_{\infty}$ described by a step-function:

\begin{equation}
 \frac {\langle \rho(z) \rangle}{\rho(\infty)} = \left\{ \begin{array}{ll}
        0 & \mbox{if $x < 0$};\\
        1+A & \mbox{if $0 \leqslant x < d$};\\
         1 & \mbox{if $d \leqslant x$}.

         \end{array} \right.
\label{eq:rho_ex4a}
\end{equation}
The derivative to this density profile consists of two
$\delta$-functions, one located at $z=0$ with amplitude of (1+A),
and another one at $z=d$ with amplitude of $-A$:

\begin{equation}
\frac {d} {dz} \biggl( \frac {\langle \rho(z)
\rangle}{\rho(\infty)}\biggr) = (1+A)\delta(z) - A \delta(z-d)
\label{eq:rho_ex4b}
\end{equation}

\begin{equation}
\biggl| \Phi (q_z) \biggr| =\biggl| 1+A - A e^{\imath q_z d} \biggr|
\label{eq:rho_ex4c}
\end{equation}
The resulting \index{Fresnel reflectivity} Fresnel-normalized
reflectivity $R(q_z) / R_F(q_z)$ is an oscillatory function with a
period of $ 2 \pi / d$,  maximum value of 1+2A and a minimum value
of unity. These oscillations are commonly referred to in a
literature as \textit{Kiessig fringes} \cite{Kiessig31}.

This density model is illustrated in Figure \ref{fig:model}~(C)
for A=0.5. Two profiles are shown - one with a thinner high
density box and another one with a thicker density box. The
thicker layer produces more frequent oscillations, as is expected
since the period of oscillations $ 2 \pi / d$ is inversely
proportional to the thickness of the layer. In this model the
density oscillates between values of 1 and $(1+2A)^2=4$. For both
of these models the widths of the two interfaces have been
neglected. For real surfaces with finite widths there are two
effects. If the widths of the two interfaces are widely disparate,
the amplitude of the oscillations decays as $q_z$ exceeds the
reciprocal of the broader width and the reflectivity reaches a
plateau of either $A^2$ or $(1+A)^2$, depending on which of the
two interfaces has the narrower width. Eventually as $q_z$ exceeds
the narrower width the average reflectivity decreases. If both
widths are comparable the oscillations and average reflectivity
decay proportionally. As will be shown later, systems studied in
this thesis can be represented by a combination of the simple
density models discussed above.

\section{Thermal Capillary Wave Fluctuations}

\subsection{Solids vs. Liquids}
The model surfaces that have been discussed in Section 2.8 tacitly
assumed that the surface was perfectly flat.  For some crystal
surfaces this could be a reasonable approximation since the
constraints on atomic positions that are associated with the
crystalline structure keep the surface flat on atomic scale.
Although surface roughness that is associated with miscuts, atomic
steps and other similar defects gives rise to off specular diffuse
scattering that reduces the intensity of the specular peak,  this
scattering is often weak. In addition the reciprocal of the
\index{correlation length} correlation lengths that are associated
with these defects are usually much broader than the spectrometer
resolution. As a result for solid surfaces it is usually
relatively easy to separate the specular reflectivity peak from
the diffuse scattering. This is different in liquids for which the
gravity determined length scale for roughness associated with
thermally capillary waves is typically of the order of a
millimeter. Since the reciprocal of this length is smaller than
the \index{spectrometer resolution} spectrometer resolution, a
major fraction of the intensity that is measured at the specular
condition is due to diffuse scattering from thermal capillary
waves.

\subsection{Off-specular diffuse scattering}
\index{x-ray!diffuse scattering} The  manner in which surface
roughness gives rise to off-specular diffuse scattering is best
discussed in terms of \index{differential cross section}
differential cross section $d \sigma / d \Omega$. The geometry for
consideration of diffuse scattering is illustrated in Figure
\ref{fig:kinema}.The x-rays are incident at an angle $\alpha$ and
collected by the detector at an elevation angle $\beta$ and
azimuthal angle $\Delta \Theta$. The momentum transfer $\vec{q}$
can be decomposed into surface normal, $q_{z}$, and surface
parallel, $q_{xy}$, components given by:

\begin{equation}
    q_{z}=\frac{2\pi }{\lambda }(\sin \beta +\sin \alpha )
    \text{ \ \ \ \ and \ \ \ \ \ }
    q_{xy}=\frac{2\pi }{\lambda } \sqrt{\cos ^{2}\alpha +
    \cos ^{2}\beta -2\cos \alpha \cos \beta \cos \Delta \Theta } \: .
\label{eq:qzqxy}
\end{equation}
The general expression for the  \index{differential cross section}
differential cross section for x-ray scattering from a three
dimensional electron distribution $\rho(\vec{r})$ at a given
location $\vec{r}$
 can be expressed in terms of the electron  density correlation function
$\langle \rho(\vec{r}) \rho(\vec{r}\prime)\rangle =
\langle \rho(\vec{r}-\vec{r}\prime) \rho(0)\rangle$:\cite{Guinier63}

\begin{equation}
    \frac{d\sigma }{d\Omega }= V \left( \frac{e^{2}}{mc^{2}}\right)^2
\int d^3 (\vec{r} - \vec{r}\prime)  \langle \rho(\vec{r} - \vec{r}\prime) \rho(0)\rangle
\exp\left[\imath  \vec{q} \cdot (\vec{r} - \vec{r}\prime) \right]
\label{general}
\end{equation}
where $V$ is the illuminated volume. If the density distribution
is homogeneous within the $x$-$y$ plane but inhomogeneous normal
to the surface, the density correlation function depends on  the
relative distance $\vec{r}_{xy} - \vec{r}_{xy}\prime $ parallel to
the surface and the distances $z$ and $z\prime$ from the surface.
For an x-ray beam of a cross sectional area $A_0$ incident on the
surface at an angle $\alpha$ relative to the $x$-$y$ plane, the
illuminated area is $A_0/\sin(\alpha)$ and the differential cross
section can be written as a half-space integral:\cite{Braslau88}
\begin{equation}
\frac{d\sigma }{d\Omega }=  \frac{A_0}{\sin(\alpha)} \left(
\frac{e^{2}}{mc^{2}}\right)^2 \int_{z \geqslant 0} dz dz\prime d^2
\vec{r}_{xy} \langle \rho(\vec{r}_{xy},z) \rho(0,z\prime) \rangle
\exp\left[\imath q_z (z-z\prime) + \imath \vec{q}_{xy} \cdot
\vec{r}_{xy}  \right] . \label{surface}
\end{equation}
The problem of converting this expression to a form that
explicitly addresses surface capillary waves is that it is
difficult to separate sort distance surface fluctuations from
fluctuations within the bulk.  Mecke and \index{Dietrich,
Siegfried} Dietrich \cite{Mecke99} discuss this from a formal
theoretical perspective. From a somewhat more empirical point of
view it is reasonable to consider defining a local average such as

\begin{equation}
\langle \rho(\vec{r}_{xy},z) \rangle_\xi = \frac {1}{A_\xi}
\int_{A_\xi}  d^2 \vec{r'}_{xy} \rho(\vec{r}_{xy} -
\vec{r'}_{xy},z) \label{surface2}
\end{equation}
where $A_\xi$ is an area in the xy plane with dimensions of the
order of the bulk \index{correlation length} correlation length
$\xi$. For non-critical liquids most of the capillary excitations
are at wavelengths that are much larger than $\xi$ and $\langle
\rho(\vec{r}_{xy},z) \rangle_\xi$ is basically an average over all
of the atomic scale fluctuations. Far from the surface it is just
equal to the bulk electron density. A reasonable approximation in
the hydrodynamic limit of long wavelength capillary waves is to
assume that the dependence of $ \langle \rho(\vec{r}_{xy},z)
\rangle_\xi $ on the z-coordinate is just a single local function
of the distance from the position of the average surface,
$h(\vec{r}_{xy})$:

\begin{equation}
\langle \rho(\vec{r}_{xy},z) \rangle_\xi \approx \rho (z- h [
\vec{r}_{xy}])
\label{surface3}
\end{equation}
Making this assumption
\begin{equation}
\rho(\vec{r}_{xy},z) =  \langle \rho(\vec{r}_{xy},z) \rangle_\xi +
\delta \rho(\vec{r}_{xy},z) \label{surface4}
\end{equation}
where $\delta \rho(\vec{r}_{xy},z)$ represents atomic scale
fluctuations about the local average and

\begin{equation}
\langle \rho(\vec{r}_{xy},z) \rho(0 ,z') \rangle_\xi =
 \rho (z- h [ \vec{r}_{xy}]) \rho (z- h [0])
  + \langle \delta \rho(\vec{r}_{xy},z) \delta \rho(0,z') \rangle \label{surface5}
\end{equation}
The approximation that we make in analyzing the effects of
capillary fluctuations is to set

\begin{equation}
\langle \rho(\vec{r}_{xy},z) \rho(0,z') \rangle =  \langle
\rho(\vec{r}_{xy},z) \rho(0,z') \rangle_\xi \label{surface6}
\end{equation}
for all length scales with $|\vec{r}_{xy}| \gg \xi$. In fact, as
we will see below the integrated physical effects associated with
capillary waves only vary logarithmically with the smallest
distance, or largest capillary wavevector. As a result the error
in assuming that this is valid down to atomic length scales is
minor. With this substitution
\begin{eqnarray}
    \frac{d\sigma }{d\Omega }= \frac {A_0}{\sin \alpha} \left( \frac{e^{2}}{mc^{2}}\right)^2
\int dz dz' d\vec{r}_{xy} \langle \exp [iq_z(z-z')+i \vec{q}_{xy}
\cdot \vec{r}_{xy}]\rangle  \times \nonumber \\ \times
 \bigl\{ \rho (z- h [ \vec{r}_{xy}]) \rho (z- h [0])
+ \langle \delta \rho(\vec{r}_{xy},z) \delta \rho(0,z')
  \rangle \bigr\}
\label{surface7}
\end{eqnarray}
The second term in the integrand, which is only non-vanishing for
regions $|\vec{r}_{xy}| \leqslant \xi $, gives rise to both the
bulk diffuse scattering and surface scattering at momentum
transfers with large $q_{xy}$ \cite{Braslau88}. \index{Daillant,
Jean} Daillant et al. \cite{Fradin98, Fradin00} discussed some of
the problems that need to be addressed in order to try to separate
surface and bulk scattering in this large $q_{xy}$ region. In any
event, for discussion of diffuse scattering close to the specular
condition
 $q_{xy} \ll 1/\xi$ and this term is not
important and will be omitted. A change of variables yields
\begin{equation}
\frac{d\sigma }{d\Omega } \approx  \frac{1}{16\pi^2} \left(
\frac{q_c}{2} \right)^4 \frac{A_0}{\sin(\alpha)} \frac{| \Phi
(q_z)|^2}{q_z^2} \int_{|\vec{r}_{xy}| > \xi} d^2 \vec{r}_{xy}
\langle \exp \left\{ \imath q_z  [h(\vec{r}_{xy})-h(0)]  \right\}
\rangle \exp \left[ \imath \vec{q}_{xy} \cdot \vec{r}_{xy} \right]
\label{surface8}
\end{equation}
where the average appearing in Eq.~\ref{surface8} is the normal
statistical average of the thermal height-height fluctuations. The
quantity $q_c$ is the  \index{critical!angle} critical angle for
total external reflection of x-rays. The \index{surface!structure
factor} structure factor $\Phi(q_z)$ previously described in
Eq.~\ref{eq:structure_ch1} needs to be redefined with respect to
the height fluctuations at the surface, $h(\vec{r}_{xy})$:

\begin{equation}
\Phi (q_z) \equiv \int dz \frac{\partial \left<
\rho(\vec{r}_{xy},z) \right>_\xi} {\partial z}
\exp[iq_z(z-h(\vec{r}_{xy}))] = \int dz \frac{\partial
\rho(z-h(\vec{r}_{xy}))} {\partial z}
\exp[iq_z(z-h(\vec{r}_{xy}))] \label{eq:phi}
\end{equation}
For the case of thermally excited capillary waves on a liquid
surface, the height fluctuations can be characterized by their
statistical average \cite{Braslau88}
\begin{equation}
\langle [ h(\vec{r}_{xy}) - h (0) ]^2 \rangle =
\int_{|\vec{r}_{xy}|
> \xi} d^2 \vec{r}_{xy} \langle \exp \left\{ \imath q_z
[h(\vec{r}_{xy})-h(0)]  \right\} \rangle \exp \left[ \imath
\vec{q}_{xy} \cdot \vec{r}_{xy} \right] \label{Eq:braslau}
\end{equation}
\index{Sinha, Sunil} Sinha et al.\cite{Sinha88} have shown that
integration over these height fluctuations yields the following
dependence of the scattering on $q_{xy}$ and $q_z$:

\begin{equation}
\int_{|\vec{r}_{xy}| > \xi} d^2 \vec{r}_{xy} \exp \Biggl\{ \frac {
q_z^2}{2} \Biggl\langle [h(\vec{r}_{xy}) - h (0) ]^2 \Biggr\rangle
\Biggr\}   \exp \left[ \imath \vec{q}_{xy} \cdot \vec{r}_{xy}
\right] = \frac{C}{q_{xy}^{2-\eta}} \label{Eq:sinha}
\end{equation}
where
\begin{equation}
\eta = \frac{k_BT}{2\pi \gamma} q_z^2
\label{Eq:eta}
\end{equation}
and the value of $C$ can be determined with an approximate sum
rule \cite{Tostmann99} that is obtained by integration over all
 surface capillary modes having wavevectors smaller than the upper wavevector cutoff
$ q_{max} \equiv \pi/\xi $:
\begin{eqnarray}
C = \int_{|\vec{q}_{xy}| < \pi/\xi} d^2 \vec{q}_{xy} \int_{|
\vec{r}_{xy} | > \xi} d^2 \vec{r}_{xy} \Biggl\langle \exp \left\{
\imath q_z  [h(\vec{r}_{xy}) - h(0)]  \right\} \Biggr\rangle
\exp \left[ \imath \vec{q}_{xy} \cdot \vec{r}_{xy} \right] \nonumber \\
= \int_{|\vec{q}_{xy}| < \pi/\xi} d^2 \vec{q}_{xy} \int_{|
\vec{r}_{xy} | > \xi} d^2 \vec{r}_{xy}   \exp \left\{- \frac{
q_z^2}{2}
  \langle [h(\vec{r}_{xy}) - h(0)]^2 \rangle \right\}
\exp \left[ \imath \vec{q}_{xy} \cdot \vec{r}_{xy} \right] \nonumber \\
= 4\pi^2 \lim_{ \vec{r}_{xy} \rightarrow 0} \Biggl[ \exp \left\{ -
\frac{q_z^2}{2}  \langle [h(\vec{r}_{xy}) - h(0)]^2 \rangle
\right\}
  \Biggr]
= 4\pi^2
\end{eqnarray}
Therefore, the properly normalized \index{differential cross
section} differential cross section for scattering of x-rays  from
a liquid surface is:

\begin{equation}
    \frac{d\sigma }{d\Omega } = \frac{A_{0}}{\sin \alpha }
    \left( \frac{q_c}{2} \right) ^{4}\frac{k_B T}{16\pi ^2 \gamma }
    \left| \Phi (q_z)\right| ^2
    \frac{1}{q_{xy}^2} \left( \frac{q_{xy}}{q_{\mbox{\scriptsize max}}}
    \right) ^\eta \: .
\label{eq:final}
\end{equation}

\subsection{Specular Reflectivity and Local Density Profile}

The intensity measured at a specific scattering vector $\vec{q}$
is obtained by integrating Eq.~\ref{eq:final} over the solid angle
$d\Omega$ defined by the detector acceptance of the experiment:
\begin{equation}
    \frac{I}{I_0}= \int_{res} \frac{d\sigma }{d\Omega } = \frac{A_{0}}{\sin \alpha }
    \left( \frac{q_c}{2} \right) ^{4}\frac{k_B T}{16\pi ^2 \gamma }
    \left| \Phi (q_z)\right| ^2 \int_{res}
    \frac{1}{q_{xy}^2} \left( \frac{q_{xy}}{q_{\mbox{\scriptsize max}}}
    \right) ^\eta \: .
\label{eq:final2}
\end{equation}
For the specular reflection $\alpha=\beta$ and $\Theta =0$ and the
integral is centered at ${q}_{xy} = 0$. The projection of the
detector resolution onto the $x$-$y$ plane is typically
rectangular\cite{Braslau88} and the above mentioned integration
has to be done numerically. On the other hand a heuristically
useful approximate analytical formula can be obtained for the
specular reflectivity if the projection of detector resolution on
liquid surface is assumed to be a circle of radius $q_{res}$
\cite{Regan97b}. In that case, integral over the circular
\index{resolution function} resolution function defined as
$|q_{xy}|<q_{res}$ can be easily calculated in polar coordinates:

\begin{eqnarray}
\int_{res} \frac{1}{q_{xy}^2} \left( \frac{q_{xy}}{q_{\mbox{\scriptsize max}}}
    \right) ^\eta d^2 q_{xy} =
 \left( \frac{1}{q_{\mbox{\scriptsize max}}}
    \right) ^\eta  \int_{q_r} \int_{\phi} \frac{1}{q_{r}^{2-\eta}} q_r dq_{r} d\phi = \nonumber \\
= \frac{2 \pi} {{q_{\mbox{\scriptsize max}}}^\eta}  \int_{q_r} \frac{1}{q_{r}^{1-\eta}} dq_{r}d\phi =
  \frac{2 \pi} {{q_{\mbox{\scriptsize max}}}^\eta}
   \frac{q_r^{\eta}}{\eta} \biggl|_0^{q_{res}} =
 \frac{2 \pi}{\eta} \left( \frac {q_{res}}{q_{\mbox{\scriptsize max}}} \right) ^\eta
 \label{eq:circle0}
 \end{eqnarray}
 Substituting Eq.~\ref{eq:circle0}
into Eq.~\ref{eq:final} results in:
 \begin{equation}
    \frac{R(q_z)}{R_F(q_z)} = \left| \Phi (q_z) \right| ^2
    \left( \frac{q_{\mbox{\scriptsize res}}}{q_{\mbox{\scriptsize max}}}
    \right) ^\eta = \left| \Phi (q_z) \right| ^2
    \exp [ - \sigma _{\mbox{\scriptsize cw}}^2q_z^2 ]
\label{eq:circle}
\end{equation}
where $\Phi (q_z)$ is the local \index{surface!structure
factor}structure factor defined in Eq.~\ref{eq:phi},  $R_F (q_z)$
is the \index{Fresnel reflectivity} Fresnel reflectivity from
classical optics for a flat ($h(\vec{r}_{xy}) = h (0)$ for all
$h(\vec{r}_{xy})$) and structureless ($|\Phi (q_z)|^2 =1$)
surface, defined in Eq.~\ref{eq:Fresnelqz}. The quantity $\sigma
_{\mbox{\scriptsize cw}}$ denotes an  effective capillary wave
roughness:
\begin{equation}
\sigma _{\mbox{\scriptsize cw}}^2 = \frac{k_B T}{2\pi \gamma }
    \ln \left(
    \frac{q_{\mbox{\scriptsize res}}}{q_{\mbox{\scriptsize max}}
    }\right) \:  .
\label{eq:cw}
\end{equation}
that increases with increasing $q_z$.

The implication of this is that a true specular reflection can not
be properly defined for liquid surfaces. In fact it follows from
Eq.~\ref{eq:circle0} that the value of the signal that is measured
at the specular condition ($\alpha=\beta, \Delta\Theta=0$) depends
on the \index{resolution function} resolution
function\cite{Schwartz90}. One implication of the above analysis
is that for a liquid surface there is not the same type of true
specular reflectivity as would be obtained for a flat solid
surface. For example, for a flat solid surface

\begin{equation}
\lim_{
\vec{r}_{xy} \rightarrow 0} \langle [h(\vec{r}_{xy}) - h(0)]^2
\rangle \rightarrow 2 \langle h[0]^2 \rangle
\label{eq:squarex}
\end{equation}
and one term in the integral of the Eq.~\ref{Eq:sinha} will be
proportional to $\delta^2(q_{xy})$, which is just the specular
signal. The signal observed at the specular signal that is
dominated by this delta functions is not sensitive to the angular
acceptance of the resolution function. In contrast the signal
predicted from Eq.~\ref{eq:circle} from a liquid surface will
generally depend on the resolution function. For small $\eta$,
corresponding to small $q_z$ the dependence is weak and the
specular reflectivity from the liquid surface is not very
different from that of the signal due to the $\delta$-function of
the solid surface; however, as $q_z$ and $\eta$ increase and, as
demonstrated by Schwartz et al. \cite{Schwartz90} for the surface
of water, the magnitude of the specular signal is dependent on the
resolution. In fact, for $\eta \rightarrow 2$ the peak in
intensity at the specular condition vanishes and at these angles
there is no operational way to distinguish surface scattering from
scattering that arises from the bulk. In view of the fact that
$\eta \sim 1/\gamma$ this can be a problem for studies of
low-surface tension systems, such as \index{alkali metals} alkali
metals and dielectric liquids.

Historically the reflectivity from liquid surfaces was originally
discussed solely in terms of the Master formulae
\ref{eq:structure_ch1} and \ref{eq:Fresnel3}; however as shown by
this discussion the ratio of $R(q_z)$ to $R_F(q_z)$ depends on
both an intrinsic \index{surface!structure factor} structure
factor and a thermal term that is not quite identical to the
\index{Debye-Waller} Debye-Waller factor used for solid surfaces.
Nevertheless it is often convenient to define a quantity that we
will refer to as the thermal \index{surface!structure factor}
structure factor:

\begin{equation}
\langle \left| \Phi (q_z) \right|^2 \rangle_T=
\frac{R(q_z)}{R_F(q_z)} \label{eq:phiterm}
\end{equation}
that corresponds to a macroscopically averaged, or
\textit{thermal} density profile $\langle \rho(z) \rangle_T$ which
includes thermal fluctuations and is related directly to the
measured reflectivity $R(q_z)$:

\begin{equation}
\frac{1}{\rho_{\infty}} \int dz \frac{\langle \rho
(z)\rangle_T}{dz} \exp(\imath q_z z) = \frac{R(q_z)}{R_F(q_z)}
\label{eq:cw2a}
\end{equation}
This macroscopically averaged thermal density profile, $\langle
\rho (z) \rangle_T$, can be shown \cite{Regan95} to be the result
of the convolution of the intrinsic density profile defined
earlier in Eq.~\ref{surface3} for $h(x,y)=0$, with the associated
Gaussian distribution of thermal height fluctuations characterized
by capillary roughness $\sigma_{cw}$.

\begin{figure}[htp]
\begin{center}
\unitlength1cm
\begin{minipage}{18cm}
\epsfxsize=15cm  \epsfbox{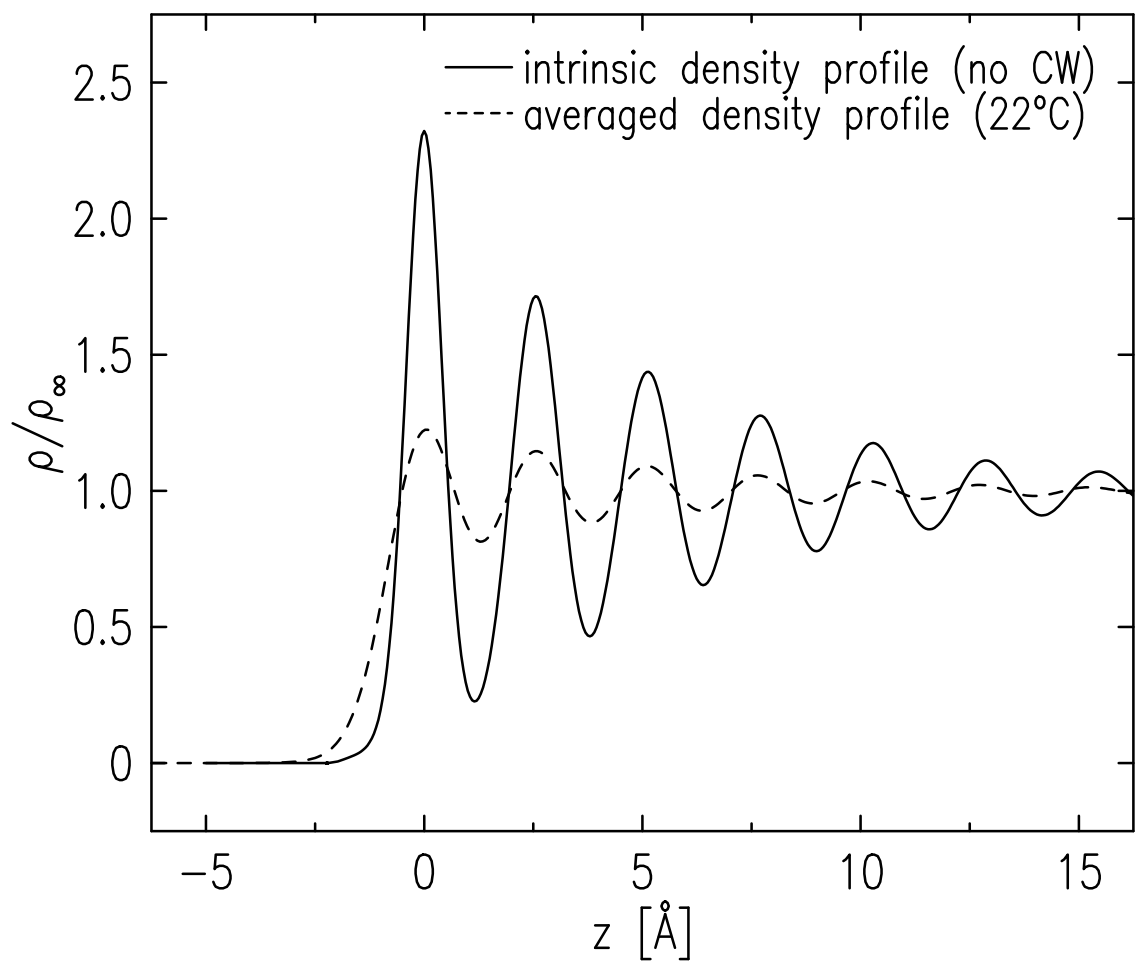}
\end{minipage}
\caption{\label{fig:intr_vs_capil} The difference between
intrinsic and thermal profile, an example taken from experimental
studies of liquid Gallium at room temperature.}
\end{center}
\end{figure}

Figure \ref{fig:intr_vs_capil} illustrates the difference between
the intrinsic density profile  and its thermally roughened density
profile equivalent for liquid gallium at room temperature: the
solid line shows the intrinsic density profile extracted from
experimental reflectivity data of liquid gallium by fitting it to
atomic layering density model and taking into account the thermal
fluctuations at the surface. The dashed line represents the
thermally averaged density profile that is obtained from the best
fit of Eq.~\ref{eq:cw2a} to the measured reflectivity at
T=22$^\circ$C.

As will be shown in Chapter 3, this intrinsic density profile for
gallium has been found to be independent of temperature and
resolution, while the thermal density profile which is essentially
a local density profile convolved with thermal capillary
excitations, does depend strongly on both the temperature and
resolution effects.

\subsection{Integration of the Resolution Function}

The problem of calculating the effect of the resolution on the
contributions of capillary excitations to the observed  signal is
to carry out an integration of the singular function
$1/{q_{xy}}^{2-\eta}$ over a two-dimensional space defined by the
detector resolution parameters. Although not impossible, the
procedures that must be followed to carry out a 2D numerical
integration over the region that encompasses a singularity slows
down the integrations considerably. We circumvented this problem
by analytically integrating the non-singular $ q_r^{2-\eta} q_r
dq_r$, where $q_r = \sqrt{q_x^2+q_y^2}$ and only then doing a
digital integral over the azimuthal angle $\phi$, taking $\phi =
\arctan (q_y / q_x)$. The resultant integral that must be done
numerically has the form:

\begin{equation}
\int_{res} \frac{1}{q_{xy}^2} \left( \frac{q_{xy}}{q_{\mbox{\scriptsize max}}}
    \right) ^\eta d^2 q_{xy} = \oint_{res} \frac{1}{q_{r}^2} \left( \frac{q_{r}}{q_{\mbox{\scriptsize max}}}
    \right) ^\eta q_{r} d q_{r} d \phi = \frac{1}{\eta} \oint_{\phi} \left( \frac{q_{res}(\phi)}{q_{\mbox{\scriptsize max}}}
    \right) ^\eta d \phi
\label{eq:square1}
 \end{equation}
where the piecewise continuous function $q_{res}(\phi)$ is defined
by the border of the projection of the resolution function on the
x-y plane of the liquid surface. With the singularity removed it
is straightforward to carry out the one-dimensional integration
over the variable $\phi$.

The practical situation corresponds to a rectangularly shaped
detector resolution that is defined by pairs of horizontal and
vertical slits. A pair of vertical slits with an opening of 2V and
horizontal slits with an opening of 2H would accept scattering
angles ranging from $\beta-\Delta\beta$ to $\beta+\Delta\beta$
vertically and $\Theta-\Delta\Theta$ to $\Theta+\Delta\Theta$
horizontally, where $\Delta\beta$ and $\Delta\Theta$ are defined
in low angle approximation ($\Delta\beta$, $\Delta\Theta \ll 1$)
as

\begin{equation}
 \begin{array}{ll}
\Delta\beta=\frac{V}{L} \\
\Delta\Theta=\frac{H}{L}
\end{array}
 \label{eq:rectres}
\end{equation}
The quantity L is the distance from the center of the sample to
the position of the detector slits.

According to \ref{eq:qzqxy}, the  projection of resolution
function in $q_{xy}$ space is the rectangle of the size $2Q_x$ by
$2Q_y$, where

\begin{equation}
 \begin{array}{ll}
Q_x = \frac{2\pi}{\lambda} \Delta\Theta \cos\beta=\frac{2\pi}{\lambda}\frac{H}{L} \cos\beta \\
Q_y = \frac{2\pi}{\lambda} \sin\beta \Delta\beta
=\frac{2\pi}{\lambda} \frac{V}{L}  \sin\beta
\end{array}
 \label{eq:rectres2}
\end{equation}
We assume that the detector accepts all radiation for which
$q_{xy}$ falls within this rectangular shape and rejects all that
falls outside. A representation of such rectangular resolution
function $\Xi(q_x, q_y)$ with the detector slits centered at an
arbitrary position ($q_{x0}, q_{y0}$) is:
\begin{figure}[htp]
\includegraphics[width=0.8\columnwidth]{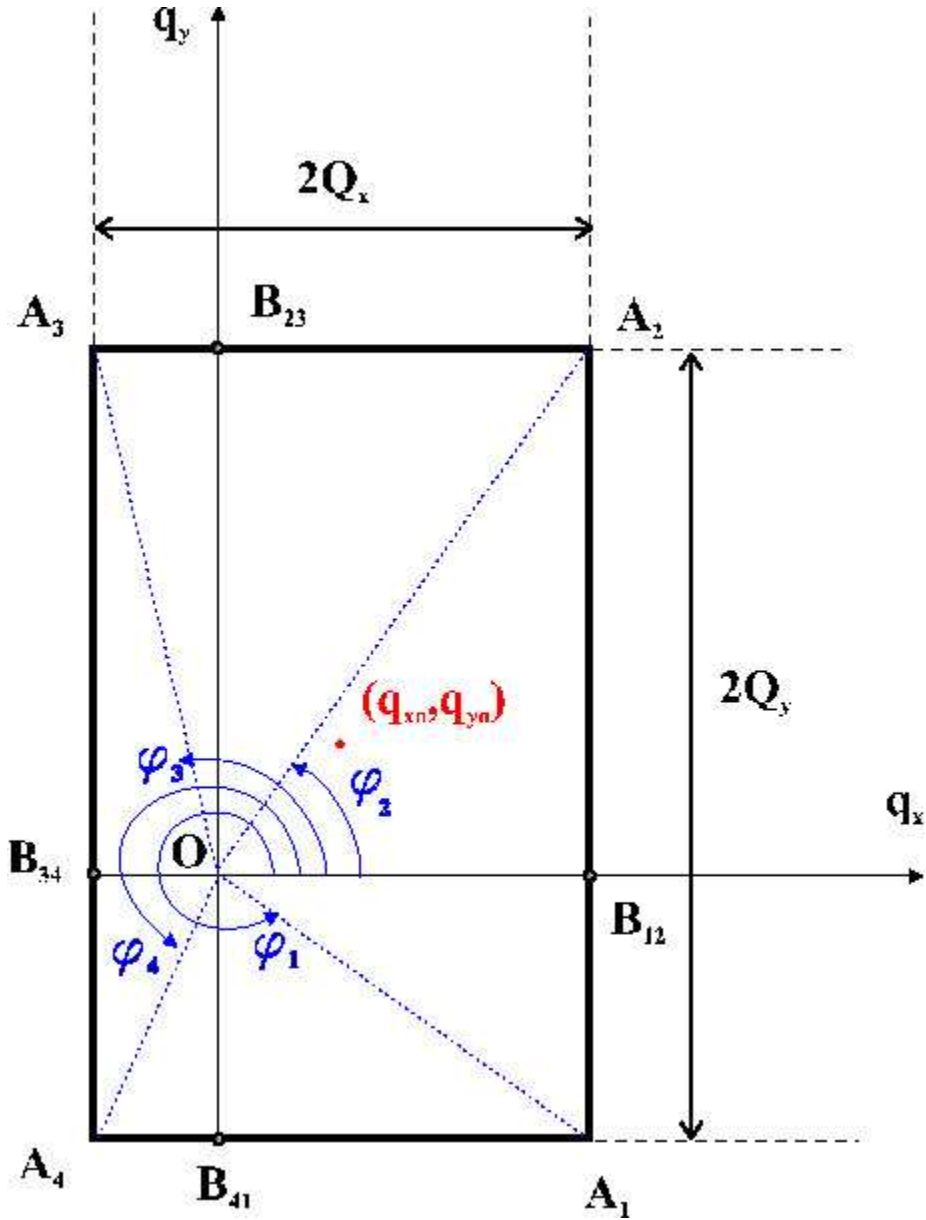}
\caption{\label{fig:rect1} The schematic describing the
integration of capillary wave contributions over the rectangular
resolution function in $q_{xy}$ space, for the case when the
singularity O is located \textit{inside} the resolution volume.}
\end{figure}
\begin{equation}
 \Xi(q_x, q_y) = \left\{ \begin{array}{ll}
        1 & \mbox{if $|q_x-q_{x0}| \leqslant Q_x$}\\
        0 & \mbox{if $ Q_x < |q_x-q_{x0}| $}\\
        1 & \mbox{if $|q_y-q_{y0}| \leqslant Q_y$}\\
        0 & \mbox{if $ Q_y < |q_y-q_{y0}|$}
         \end{array} \right.
\label{eq:rectres}
\end{equation}
For the spectrometer either centered at the specular, or close to
it $q_{res}(\phi)$ defines a rectangle that encloses the specular
position, $q_{xy}=0$.  For the spectrometer at an off specular
position the situation is slightly more complicated.

Consider the first case when $|q_{x0}|<Q_x$ and $|q_{y0}|<Q_y$,
which implies that the singularity is located inside the
resolution function. The diagram describing the resolution
function integration procedure in this case is shown in Figure
\ref{fig:rect1}. Point O corresponds to a singularity located at
$q_x=q_y=0$. The center of the rectangular resolution box is
$(q_{x0}, q_{y0})$. We designate the corners of the rectangle as
$A_1, A_2, A_3$ and $A_4$. The points $B_{12}, B_{23}, B_{34}$ and
$B_{41}$ are defined by the perpendicular projections $OB_{12},
OB_{23}, OB_{34}$ and $OB_{41}$. Additionally, we define the
angles that are formed by the lines $OA_1, OA_2, OA_3$ and $OA_4$
relative to the $q_x$ axis as $\phi_1, \phi_2, \phi_3$ and
$\phi_4$ respectively. From basic geometry,
\begin{equation}
\begin{array}{ll}
OB_{12}=Q_x+q_{x0}\\
OB_{23}=Q_y+q_{y0}\\
OB_{34}=Q_x-q_{x0}\\
OB_{41}=Q_y-q_{y0}
\end{array}
\label{eq:defres}
\end{equation}

\begin{equation}
\begin{array}{ll}
\phi_1 = 2\pi-\arctan (\frac{OB_{41}}{OB_{12}})=2\pi-\arctan(\frac{Q_y-q_{y0}}{Q_x+q_{x0}})\\
\phi_2 = \arctan (\frac{OB_{23}}{OB_{12}})= \arctan(\frac{Q_y+q_{y0}}{Q_x+q_{x0}})\\
\phi_3 = \pi-\arctan(\frac{OB_{23}}{OB_{34}})=\pi-\arctan(\frac{Q_y+q_{y0}}{Q_x-q_{x0}})\\
\phi_4
=\pi+\arctan(\frac{OB_{41}}{OB_{34}})=\pi+\arctan(\frac{Q_y-q_{y0}}{Q_x-q_{x0}})
\end{array}
\label{eq:defres2}
\end{equation}
Note that $\arctan(x)$ is defined for the region
$-\frac{\pi}{2}<x\leq\frac{\pi}{2}$. Then $q_{res}(\phi)$
described in \ref{eq:square1} can be written as:

\begin{equation}
q_{res}(\phi) = \left\{ \begin{array}{ll}
        \frac{OB_{12}}{\cos\phi} & \mbox{if $\phi_1 < \phi \leqslant \phi_2$}\\
        \frac{OB_{23}}{\sin\phi} & \mbox{if $\phi_2 < \phi \leqslant \phi_3$}\\
        -\frac{OB_{34}}{\cos\phi} & \mbox{if $\phi_3 < \phi \leqslant \phi_4$}\\
        -\frac{OB_{41}}{\sin\phi} & \mbox{if $\phi_4 < \phi \leqslant \phi_1$}
         \end{array} \right.
\label{eq:qres}
\end{equation}
The negative signs on the third and fourth line are necessary
because the $\cos(\phi)$ and $\sin(\phi)$ are negative for the
relevant angles. Equation \ref{eq:square1} can be rewritten as

\begin{equation}
\begin{array}{ll}
\oint_{res} \frac{1}{q_{xy}^2} \left( \frac{q_{xy}}{q_{\mbox{\scriptsize max}}}
    \right) ^\eta d^2 q_{xy} =  \frac{1}{\eta} \oint_{\phi} \left( \frac{q_{res}(\phi)}{q_{\mbox{\scriptsize max}}}
    \right) ^\eta d \phi = \\
      =({\eta  q_{\mbox{\scriptsize max}}^\eta})^{-1}  \bigg[
    \int_{\phi_1-2\pi}^{\phi_2} \left(\frac{Q_x+q_{x0}}{\cos\phi} \right)^\eta d\phi+
    \int_{\phi_2}^{\phi_3} \left(\frac{Q_y+q_{y0}}{\sin\phi} \right)^\eta d\phi+
    \int_{\phi_3}^{\phi_4} \left(\frac{Q_x-q_{x0}}{-\cos\phi} \right)^\eta d\phi+
    \int_{\phi_4}^{\phi_1} \left(\frac{Q_y+q_{y0}}{-\sin\phi}
\right)^\eta d\phi
    \bigg]
    \end{array}
\label{eq:rect_form}
 \end{equation}

\begin{figure}[htp]
\centering
\includegraphics[width=0.8\columnwidth]{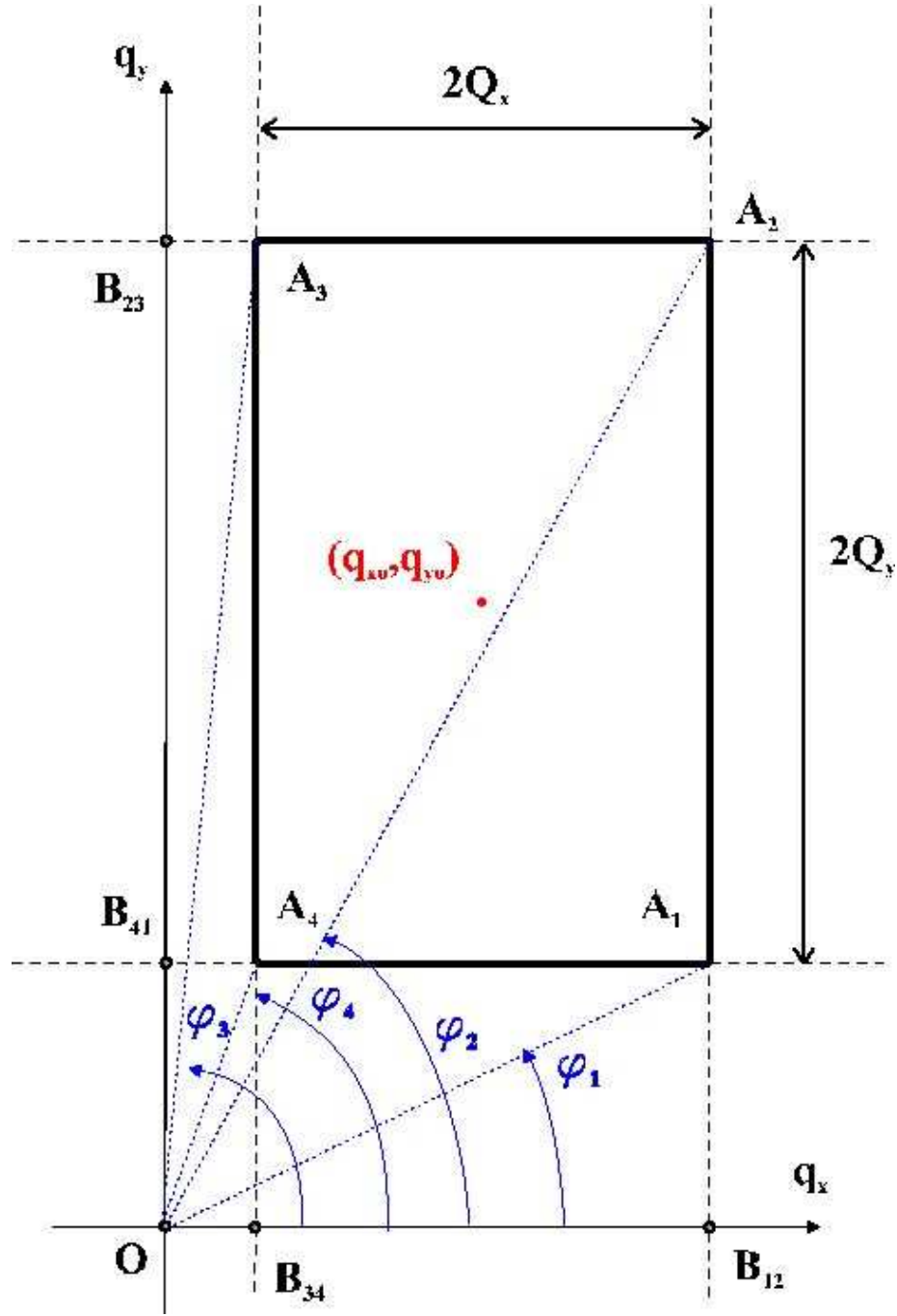}
\caption{\label{fig:rect2} A schematic describing the procedure
integration of capillary wave contributions over the rectangular
\index{resolution function} resolution function in $q_{xy}$-space
for the case when the singularity O is located \textit{outside} of
the resolution volume.}
\end{figure}
A similar treatment can be applied for the case when the
singularity is outside of the rectangular resolution box. One
example of this is illustrated schematically by Figure
\ref{fig:rect1}. In this case the projections $OB_{12}, OB_{23},
OB_{34}$ and $OB_{41}$, as well as corresponding angles $\phi_1,
\phi_2, \phi_3$ and $\phi_4$ can be defined in a similar way:

 \begin{equation}
\begin{array}{ll}
OB_{12}=q_{x0}+Q_x\\
OB_{23}=q_{y0}+Q_y\\
OB_{34}=q_{x0}-Q_x\\
OB_{41}=q_{y0}-Q_y
\end{array}
\label{eq:defres3}
\end{equation}
and
\begin{equation}
\begin{array}{ll}
\phi_1 = \arctan( \frac{OB_{41}}{OB_{12}}) =\arctan( \frac{q_{y0}-Q_y}{q_{x0}+Q_x})\\
\phi_2 = \arctan( \frac{OB_{23}}{OB_{12}})=\arctan(\frac{q_{y0}+Q_y}{q_{x0}+Q_x})\\
\phi_3 = \arctan(\frac{OB_{23}}{OB_{34}})=\arctan(\frac{q_{y0}+Q_y}{q_{x0}-Q_x})\\
\phi_4 =
\arctan(\frac{OB_{41}}{OB_{34}})=\arctan(\frac{q_{y0}-Q_y}{q_{x0}-Q_x})
\end{array}
\label{eq:defres4}
\end{equation}
For the geometry illustrated in Figure \ref{fig:rect2} the final
integral is:
\begin{equation}
\begin{array}{ll}
\oint_{res} \frac{1}{q_{xy}^2} \left(
\frac{q_{xy}}{q_{\mbox{\scriptsize max}}}
    \right) ^\eta d^2 q_{xy} =  \frac{1}{\eta} \oint_{\phi} \left( \frac{q_{res}(\phi)}{q_{\mbox{\scriptsize max}}}
    \right) ^\eta d \phi = \\
    =({\eta  q_{\mbox{\scriptsize max}}^\eta})^{-1}  \bigg[
    \int_{\phi_1}^{\phi_2} \left(\frac{q_{x0}+Q_x}{\cos\phi} \right)^\eta d\phi+
    \int_{\phi_2}^{\phi_3} \left(\frac{q_{y0}+Q_y}{\sin\phi} \right)^\eta d\phi-
    \int_{\phi_4}^{\phi_3} \left(\frac{q_{x0}-Q_x}{\cos\phi} \right)^\eta d\phi-
    \int_{\phi_1}^{\phi_4} \left(\frac{q_{y0}-Q_y}{\sin\phi} \right)^\eta d\phi
    \bigg]
    \end{array}
\label{eq:rect_form2}
 \end{equation}
Note that if the spectrometer is sufficiently far off of the
specular condition $q_{x0} \pm Q_x$ and/or $q_{y0}\pm Qy$ can be
negative. Nevertheless this form for the resolution integration is
valid.

\section{Experimental Details}

In this section we will discuss a number of practical
considerations in performing x-ray measurements on liquid
surfaces.

\subsection{Synchrotron Radiation Sources}
A study of surface layering requires a reflectivity measurement
out to the values of vertical component that are at least of the
order of wavevector transfer for which the layering peak maximum
is expected, $q_z^{peak} = \pi / a$, \cite{Magnussen95} where $a$
is the atomic radius. For liquid metals $a$ is typically on the
order of $1.0-1.5~\AA$, \cite{CRC85} making $q_z^{peak} \approx
2.0-3.0~\AA^{-1}$. If we neglect the reflectivity enhancement due
to the layering and consider just the \index{Fresnel reflectivity}
Fresnel reflectivity $R(q_z)\approx {( {{q_c} / {2q_z}}  )}^4$
(Eq.~\ref{eq:Fresnelqz2}). For liquid metals, typical value of
\index{critical!wavevector} critical wavevector $q_c$ is on the
order of $0.05~\AA^{-1}$. Therefore \index{Fresnel reflectivity}
Fresnel reflectivity at $q_z = q_z^{peak} \approx 2.5~ \AA^{-1}$
and $R_F(q_z^{peak})={( {{q_c} / {2q_z^{peak}}} )}^4 \approx
{10}^{-7}$. The effect of thermally excited capillary wave
fluctuations reflectivity is to reduce the reflectivity by an
additional \index{Debye-Waller} Debye-Waller-like factor: $DW =
\exp(-\sigma_{cw}^2 (q_z^{peak})^2)$. For a typical value of
$\sigma_{cw}=1.0~\AA$, $DW = 10^{-3}$, which reduces the
reflectivity by additional three orders of magnitude. Therefore,
in order to obtain reasonable signals while measuring the specular
reflectivity signal out to the layering peak, one needs an x-ray
source capable of providing on the order of ${10}^{10}$ photons
per second. In addition, since liquid metal samples in the
experimental chamber can not be too large (typically $<$ 2 to 4
cm) the x-ray beam must be relatively small and well-collimated.
This also requires \index{x-ray!monochromatic beam} monochromatic
radiation.

\begin{figure}[htp]
\includegraphics[width=0.8\columnwidth]{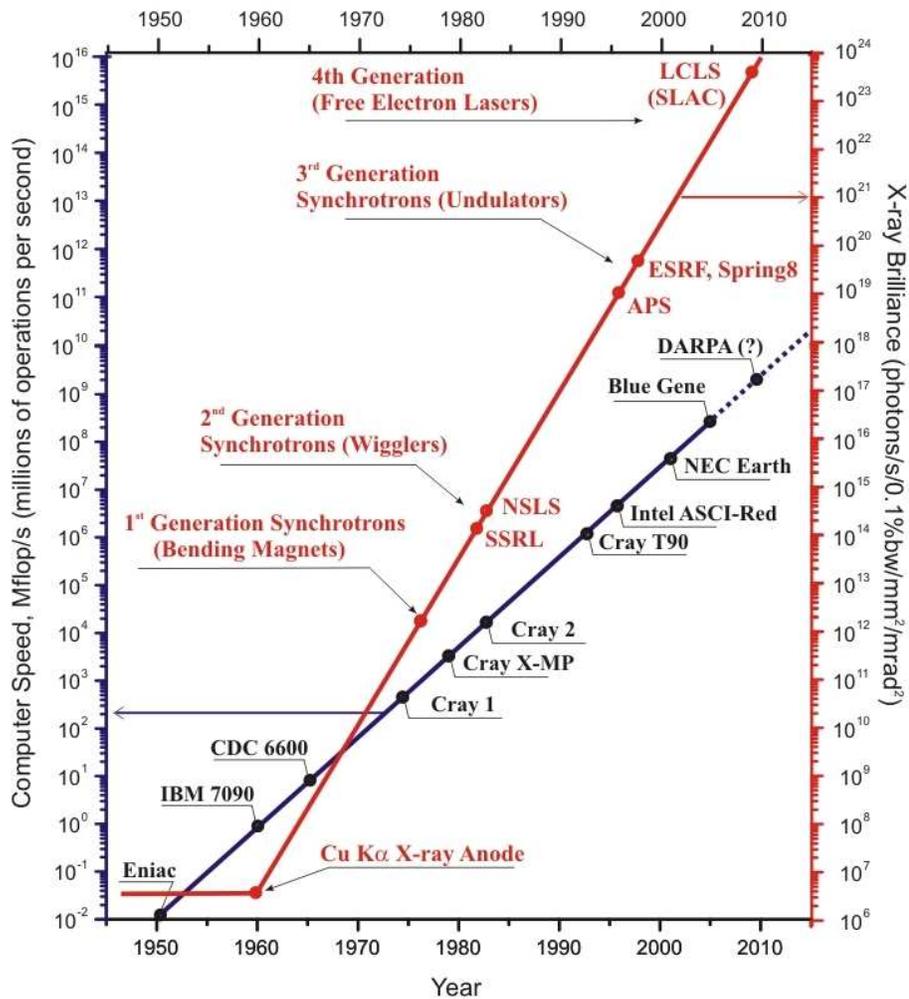}
\caption{\label{fig:IBMX} A comparison between
\index{synchrotron!radiation} synchrotron radiation development
(x-ray source brightness, expressed in photons/sec/0.1\%
bandwidth /${mm}^2$/${mrad}^2$ and the growth in leading-edge
computing speed (expressed in millions of operations/sec) which
occurred over the past 60 years}
\end{figure}

Advances in \index{synchrotron!radiation} synchrotron radiation
over the past twenty years made it possible to satisfy all of
these conditions. Figure \ref{fig:IBMX} shows the comparison
between the advances in x-ray source brightness and the growth of
computing power which occurred over the past half a century.

\subsection{Liquid Surface Spectrometer}

Measurement of the angular dependence of the reflectivity from a
solid sample can be achieved by simply rotating the sample with
respect to the incident beam. This is obviously unacceptable for
studying bulk liquids, since the surfaces must remain horizontal.
Therefore, studying liquids requires a specially designed
goniometer that is commonly called \textit{liquid surface
spectrometer} \footnote{A more appropriate name for this
instrument is  liquid surface \textit{reflectometer}, or perhaps
liquid surface \textit{diffractometer}, however, historically the
name 'liquid surface \textit{spectrometer}' has been established,
even though x-ray energy spectra is virtually never measured by
such 'spectrometer'}.

\begin{figure}[htb]
\begin{center}
\unitlength1cm
\begin{minipage}{22cm}
\epsfxsize=15cm   \epsfbox{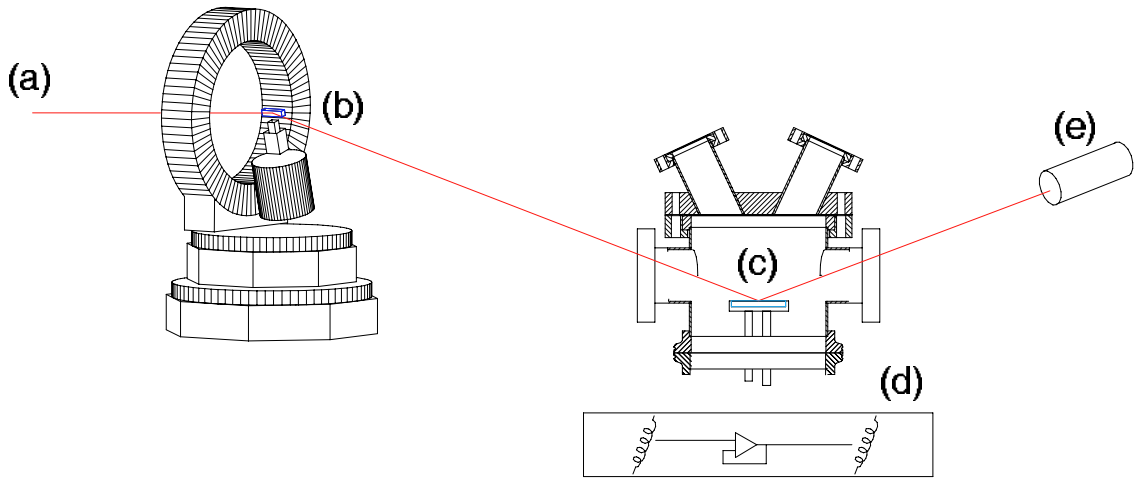}
\end{minipage}
\caption{\label{fig:liqspec} A schematic picture of liquid
spectrometer setup. Shown are: the incoming synchrotron beam (a),
the steering crystal (b), the liquid sample within the
\index{UHV!chamber} UHV chamber (c), temperature control and the
active \index{vibration isolation} vibration isolation system (d)
and the detector (e)}
\end{center}
\end{figure}

In Figure \ref{fig:liqspec} we present a sketch of the
instrumentation for realizing the kinematics that was shown in
Figure \ref{fig:kinema} for a liquid surface. A horizontal
monochromatic beam from the \index{synchrotron!source} synchrotron
source (a) is Bragg-reflected by a \index{sterring crystal}
\textit{steering crystal} (b) that can be rotated around the axis
parallel to the incident beam by an angle $\chi$ while keeping the
angle between the incident beam and the \index{reciprocal lattice
vector} reciprocal lattice vector $\overrightarrow{G}$ of the
crystal fixed. In this way the angle $\alpha$ that the
\index{Bragg reflection} Bragg-reflected beam makes with respect
to the horizontal plane of the sample changes such that

\begin{equation}
G \sin \chi = \frac{2\pi}{\lambda} \sin \alpha
\label{eq:Gvector}
\end{equation}
Not shown on the figure is the incident arm located between the
\index{sterring crystal} steering crystal and the sample,
containing the slits defining the size of the beam and the
normalization monitor. The size of the slit is smaller than the
size of the beam and is usually on the order of 10 microns for low
incident angles and opened up to 300-500 microns for higher angles
to increase the flux. The normalization monitor is used to measure
the intensity of the incident beam, which is then used to
normalize the scattered signal.
 In view of the fact that the beam follows a cone the incident arm
as well as the sample stage containing the sample (c) have to move
both vertically and horizontally in order to follow the beam.

The beam reflected from the sample is registered by a detector (e)
that must also move horizontally in order to follow the
reflection. If the distance from (b) to (c) is equal to the
distance from (c) to (e) the vertical position of the detector can
be held fixed.

\goodbreak

In fact, the actual mechanical implementation of what might appear
a rather simple instrument is quite sophisticated. Depending on
the particular setup, a typical liquid surface spectrometer
requires motion of more than twenty individual motors, some of
which require precision in the range of $1\mu$m - $100\mu$m for
linear motors and a few millidegrees for rotational motors. These
include motors responsible for movement of monochromator crystal,
sample height (which is defined by the incident angle $\alpha$),
detector height and detector rotation (which determines output
angle $\beta$, as well as various motors responsible for optical
delivery of the beam (mirrors and slits). In addition there  are
two sets of motorized slits before the sample and two on the
detector side, each of which has four separate motors. The motion
of most of the spectrometer motors is controlled by commands
issued through a computer interface, according to a liquid surface
geometry code \index{EPICS} \cite{epics, spec}. However, for the
geometry code to work properly, the exact positions of these
motors have to be individually aligned with respect to both the
x-ray beam and the positions of other motors. Furthermore several
tracking parameters, such as the distances that relate the linear
displacement of the sample and detector to the various angles,
have to be carefully determined before liquid surface spectrometer
can be used. Also, in the course of the measurement there is often
slight drifts in the position of the incident beam, the shape of
the liquid sample, etc. As a consequence key motor positions and
\index{tracking parameters} tracking parameters have to be
periodically checked during the experiment, and if necessary
adjusted \cite{alignment_xray, alignment_x22b}.

\subsection{Active Vibration Isolation System}

One of the experimental \index{vibration isolation} difficulties
in dealing with liquids is that even minor vibrational noise
disturb the sample, causing ripples on the surface that deflect
the reflected beam away from the detector. It is therefore crucial
to isolate the liquid from vibrations coming from either outside
of the experimental hutch, or from the movements of liquid
spectrometer itself. Suppression of external vibrations are
accomplished by a commercially available \index{Halcyonics GmbH}
(Halcyonics GmbH) \textit{active vibration isolation system} (d).
The motion of the table top containing the sample is detected by a
highly sensitive piezoelectric accelerometer consisting of a mass
resting on a piezoelectric disc. The acceleration signal obtained
by \index{accelerometer} accelerometer is processed by analog
control circuit and amplifier. A correction signal is then
supplied to a magnetic actuator that is located on the movable
table that supports the liquid sample. Active feedback systems
such as the MOD-1M provided by Halcyonics is capable of
suppressing external vibrations in all 6 degrees of freedom with
frequencies $>0.6 Hz$ while supporting up to 90kg of equipment
\cite{Gaul98}. Minimization of the effect of vibrations caused by
the movements of the spectrometer is further achieved with a
pre-determined \textit{sleep time}, normally on the order of 5 to
10 seconds, is used between the completion of the motor movement
and the beginning of signal detection count.

\subsection{Ultra High Vacuum Chamber}

A study of atomic structure at the surface would be impossible
unless the surface can be kept atomically clean. Most
\index{liquid!metallic} metallic liquids oxidize very easily, and
in order to maintain the surfaces free of oxide and other
contaminants, the sample has to be kept in the atmosphere with
sufficiently small partial pressures of oxygen, water vapor and
other non-inert gases. The way to achieve this is by using an
\index{UHV!chamber} \textit{Ultra High Vacuum} (UHV) system. A
schematic design of the UHV chambers used in liquid metal
experiments described in this thesis is shown in Figure
\ref{fig:chamber}. The chamber contains a sample (A) placed in Mo
holder (B) mounted on top of the heater (not shown), with
thermocouple and heater wires accessible through bottom flange M.
\index{x-ray!windows} Be windows (C) are capable of withholding a
pressure differential of more than $10^5$ Pa, yet at the same time
are 99$\%$ transparent to the x-rays. The vacuum inside the
chamber is produced by ion pump (D) and turbo pump (E) backed up
by a rotary station, connected to the turbo pump through a hose
(F), and is monitored by vacuum gauge and RGA system which can be
connected at flange (K). Ion sputtering system (G) is used to
clean the surface of the sample, and is connected to the Ar gas
line (H)~\footnote{During the x-ray measurements only the ion pump
is used and the turbo pump is turned off to reduce vibrations,
however, during sputtering ion pump is turned off and only turbo
pump is used}. The sample can be monitored through the glass
window on a view port window (I). A linear transfer mechanism (L)
provides the ability of wiping the sample and moving the
\index{thermal radiation shield}  radiation shield in and out with
the help of bellow system (J). The chamber components are
supported by a rigid frame (N). The photo (top inset of Figure
\ref{fig:chamber}) shows the smaller of the two chambers actively
used in the x-ray experiments. The design of the two chambers is
quite similar, the advantage of the larger chamber is a larger
volume and higher "ceiling" which allows for a greater range of
motion of linear transfer mechanism (which can be mounted both
vertically and horizontally), as well as providing for an
additional port, while the disadvantage is the increased weight
and size of the chamber. These UHV systems are capable of
sustaining the 1~atm pressure differential to the room, with the
partial pressures of oxygen and other major contaminants two
orders of magnitude lower.

\subsection{Liquid Metal Sample preparation}
\begin{figure}[htp]
\centering
\includegraphics[width=0.7\columnwidth]{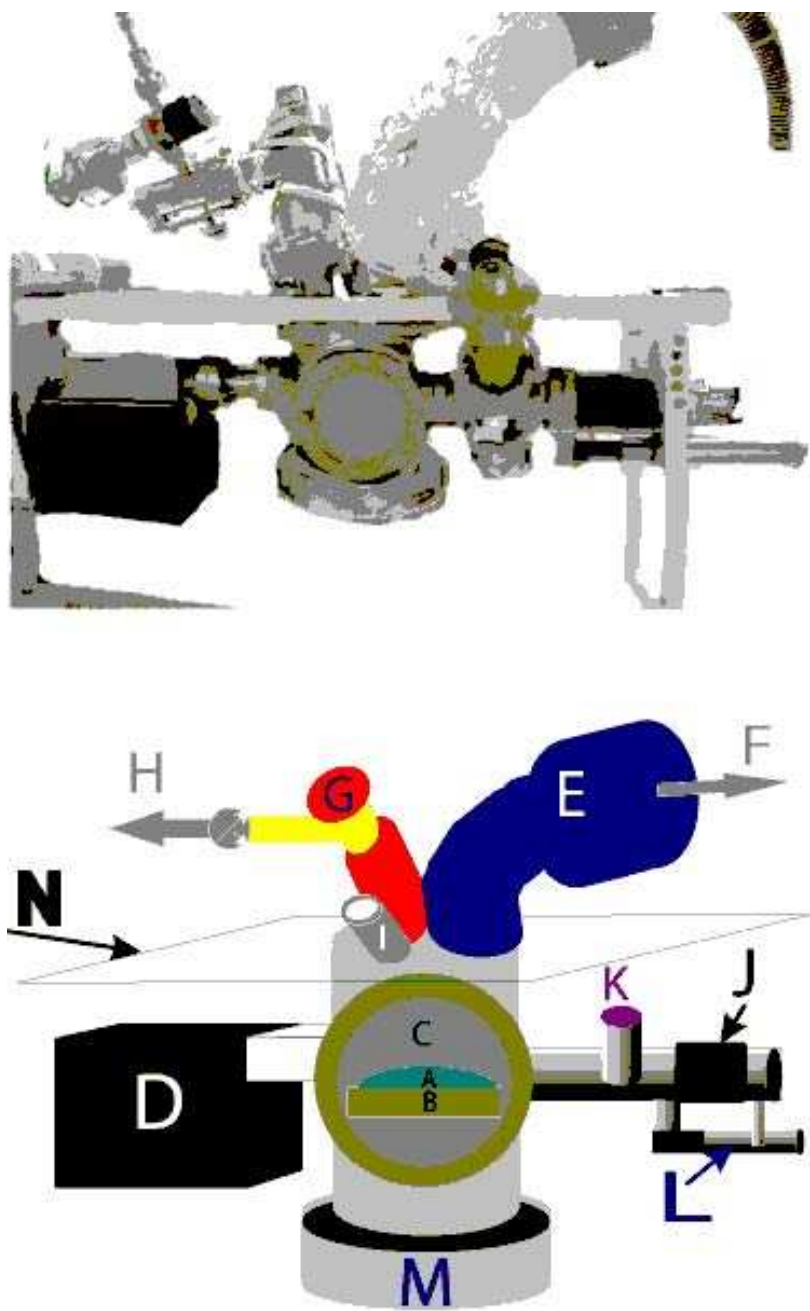}
\caption{\label{fig:chamber} A photo and schematic of a typical
UHV chamber setup: Liquid Sample (A) contained in sample holder
(B), output Be window (C) (input Be window is located on the other
side of the chamber), ion pump (D), turbo pump (E), hose leading
to rotary pump (F), ion sputtering gun (G), a hose leading to Ar
cylinder (H), \index{viewport} viewport for monitoring the sample
(I), a bellow (J), vacuum gauge and a port for RGA system (K),
linear transfer mechanism (L), bottom flange containing
thermocouple, heater and electrical contacts feedthroughs (M) and
chamber frame (N).}
\end{figure}

A typical liquid metal sample is produced from melting 50g-200g of
metal, of the highest purity that is commercially available
(typically on the order of 99.9999$\%$ purity). The metal can be
supplied in form of ingots, powder or a single briquette. The
metal is melted in glass hardware which has been chemically
cleaned after being soaked in an \index{ultrasonic bath}
ultrasonic bath (Branson Ultrasonic Cleaner, Model 1510). The
melting is done in a glove box filled with either an inert gas
atmosphere in case of a binary alloy, or in the
\index{UHV!chamber} UHV chamber itself, in case of a pure system.
Despite these precautionary measures, the surface of the metal
typically contains a macroscopic amount of oxide, which must be
mechanically removed by scraping the liquid surface with a clean
glass or Mo wiper. The wiper is typically a Mo foil mounted on the
linear transfer mechanism (J-L in Figure 2.8) that allows
manipulation  of the wiper inside the chamber. Wiping the sample
works well for removing macroscopic quantities of oxide, but when
the contamination is present in a form of just a few monolayers
cleaning is better accomplished by means of an \textit{ion
sputtering} process described below. The sample is then frozen and
is kept under inert gas atmosphere or under vacuum until the UHV
chamber is mounted on the sample stage of surface spectrometer
where it undergoes what is known as a $\textit{bakeout}$ process
during which the chamber and its components are gradually heated
up to 150-200 $^\circ$C, effectively degassing the chamber walls
of any substances absorbed there. During this procedure the system
is continuously pumped by 100 L/min \index{pump!turbo} turbo pump
(Varian, Model: DS-102) (the pump is connected to chamber through
the port indicated by part E  in Figure \ref{fig:chamber}). After
the completion of the bakeout, the chamber is cooled down to room
temperature and the pressure is lowered further by switching to
the \index{pump!ion} ion pump (part D in Figure
\ref{fig:chamber}), which is more efficient at low pressures.
During initial stages of pumping and bakeout the atmosphere inside
the chamber is periodically checked by a \index{residual gas
analyzer} residual gas analyzer (RGA) system (connected through
port K), capable of distinguishing individual chemical species
present in the UHV chamber by a method of mass spectroscopy.
Typically the bakeout procedure has to be performed at least twice
- the sample is usually prepared at Harvard, several weeks prior
to the experiment and transported to one of the
\index{synchrotron!source} synchrotron facility under vacuum. In
view of the fact that the \index{x-ray!windows} Be x-ray windows
are brittle, for this first bake out and for transportation to the
synchrotron they are replaced with thick (2 mm) glass windows. On
arrival at the \index{synchrotron!source} synchrotron the vacuum
is briefly broken and the glass windows are replaced by the Be
windows.  The window replacement can be performed within 20-30
minutes, but to further minimize exposure to air, prior to opening
the chamber it is backfilled with nitrogen atmosphere which
prevents oxygen and water from absorbing at the chamber walls.
After the windows are replaced and the chamber is evacuated, the
second bake out is done. In view of the preliminary
\index{bakeout} first bakeout and the short exposure to air, this
second bakeout can be typically completed in 10-15 hours, as
compared to 24 hours for the first.

\subsection{Surface Cleaning by Ion Sputtering}

Despite the efforts to minimize the exposure to contaminants
described in the previous section, the surface of the sample is
frequently covered by a thin native film of oxide that cannot be
removed by scraping. Moreover, the \index{bakeout} bakeout process
described previously always results in some of the contamination
desorbed from the walls being deposited on the surface of the
sample. To minimize this contamination the sample is protected
with a \index{thermal radiation shield} radiation shield during
bakeout. During ion sputtering the chamber is backfilled with an
inert gas atmosphere of $\approx 10^{-6}$ Torr, in our case Argon.
The ion sputtering gun system uses high voltage (2-5 kV)to first
ionize Ar atoms, then accelerate the resulting Ar$^+$ ions to form
a high-energy ion beam, which is directed onto the sample. When
the primary Ar$^+$ ions strike the sample their energy is
transferred to the target atoms of the sample, which is sufficient
to eject them from the surface. The sputtering Ar$^+$ beam can be
de-focused to cover the entire area of the sample, normally on the
order of a few cm$^2$, or focused on a particularly contaminated
part of the surface of the sample. Because the surfaces of liquid
metals are highly reflective, presence of even a single monolayer
of oxide can be easily detected by eye, making it possible to
monitor the sputtering progress through a viewport (I) shown on
Figure \ref{fig:chamber}. Presence of impurities can be also
checked with x-ray reflectivity or fluorescence measurements.

\subsection{Temperature control}
\begin{figure}[htp]
\centering
\includegraphics[width=0.5\columnwidth]{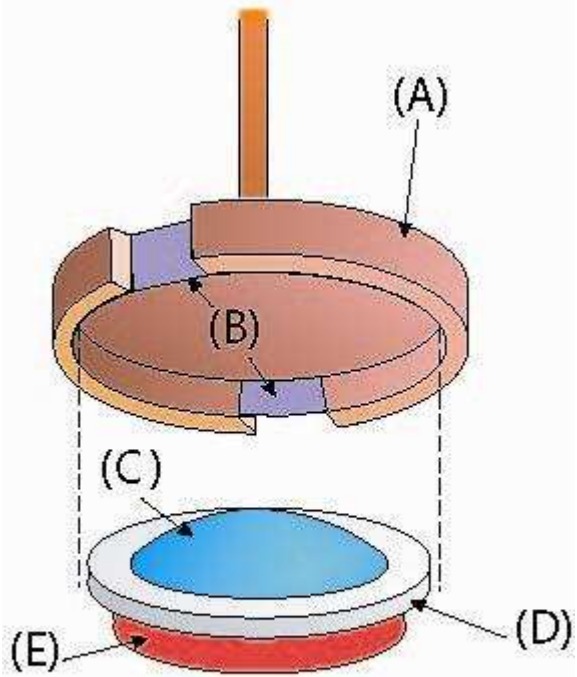}
\caption{\label{fig:shield} A schematic picture describing
radiation shield setup: radiation shield (A) with Be foil walls
(B), liquid sample (C) contained in a Mo sample pan (D),
positioned directly on top of a \index{Boralectric resistive
heater} Boralectric resistive heater (E). For the measurement the
shield is lowered so that the sample is completely surrounded by
the shield and its' walls}
\end{figure}

The temperature of the sample is controlled by a heater mounted
directly below the sample holder, that is capable of achieving
temperatures well in excess of 400 $^\circ$C. A Eurotherm
temperature controller with a feedback loop maintains temperature
drifts well under $0.1^\circ$C. In order to reduce temperature
gradients that arise from radiation losses to the relatively cool
chamber walls, a special radiation shield could be installed to
maintain the sample in a thermally isotropic environment. The
shield construction is shown in Figure \ref{fig:shield}. The
shield is mounted on a linear transfer mechanism (part L of Figure
\ref{fig:chamber}) which allows for the shield to be moved by 5-7
cm during sputtering or wiping. Care must be taken to avoid
touching the sample as it would inevitably introduce oxide,
impurities or formation of \index{amalgam} amalgam at the surface.
The radius of the cover is designed to be $\approx$ 5mm larger
than that of the sample. The radius of the typical sample holder
is 25-30mm, and the height of the sample is 3-5~mm. The height of
the shield is 15~mm. The shield is made out of copper to improve
thermal equilibration time. For studies involving high-temperature
measurements the shield temperature can be controlled by a
separate heater. In experiments which do not require a high degree
of temperature control the radiation shield construction can be
replaced by a 0.5mm thick sheet of \index{molybdenum} Mo foil,
approximately 4cm $\times$ 4cm in size which can be mounted on the
same \index{linear transfer mechanism} linear transfer mechanism.

\subsection{Curvature Effects}
The liquid metals that we have studied do not seem to wet
non-reactive substrates such as Si, or metals such as Mo and W. As
a result the finite contact angles that they make with the sample
pan always lead to samples with finite curvature. To some extent
this effect could have been mitigated if the diameter of the
\index{liquid!metallic} metallic liquid was sufficiently large;
however, the fact that the sample must be mounted in a UHV chamber
places restrictions on sample size. In this section we will
describe a method for measuring the \index{curvature!macroscopic}
macroscopic surface curvature profile.

\begin{figure}[htp]
\begin{center}
\unitlength1cm
\begin{minipage}{12cm}
\epsfxsize=8cm \epsfysize=8cm \epsfbox{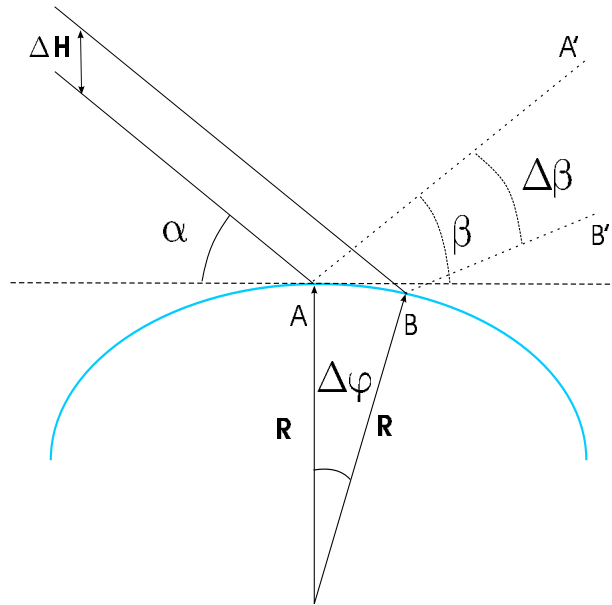}
\end{minipage}
\caption{\label{fig:curvature} Illustration of reflection geometry
from curved sample.}
\end{center}
\end{figure}

Consider a well-collimated x-ray beam of vertical size $\Delta H$
incident on a surface of a curved liquid sample at angle $\alpha$,
as shown on Figure \ref{fig:curvature}. If the local radius of
curvature $R$ is sufficiently large, and if the curvature is
sufficiently uniform,  the resulting \index{x-ray!footprint}
footprint (distance between points A and B on the surface) on the
sample can be expressed as:
\begin{equation}
AB=\frac{\Delta H}{\tan \alpha}
\label{eq:curvature1}
\end{equation}
The angle $\delta \phi$ between surface normals at A and B
therefore can be written as:
\begin{equation}
\delta \phi = \frac{AB}{R} = \frac{\Delta H}{R \tan \alpha}
\label{eq:curvature2}
\end{equation}
and the rays reflected from A and B - AA' and BB' will diverge by
an angle $\Delta \beta = 2 \delta \phi$ or
\begin{equation}
\Delta \beta =  \frac{2 \Delta H}{R \tan \alpha}
\label{eq:curvature2}
\end{equation}
as a result the height $\Delta H'$ of the beam observed a distance
L away from the sample is

\begin{equation}
\Delta H' = L \Delta \beta =  \frac{2 L \Delta H}{R \tan \alpha}
+ \Delta H
\label{eq:curvature3}
\end{equation}
The \index{curvature!local} local radius of curvature (average
curvature over the illuminated part of the sample) can be
determined from a measurement of $\Delta H'$ for a fixed incident
angle $\alpha$. By repeating the measurement of $\Delta H'$ while
moving the beam across the sample (by scanning the sample height,
for example) we can obtain information on the local curvature at
different points on the sample surface. In addition, from the
variation in the average scattering angle "beta" one can obtain a
measure of the global curvature.

In view of the fact that the local curvature measurement is
affected by beam divergence, non-uniform beam profile and diffuse
scattering from the sample, local curvature measurements typically
produce higher numbers than global curvature values.

A series of such scans is called a \index{walking scan}
$\textit{walking scan}$ because the beam walks across the surface
of the sample. To illustrate consider what happens when the sample
height is increased by a value $\Delta sh$. If the sample surface
was ideally flat, then the position of the reflected beam observed
at the detector position would change by double that value, $2
\Delta sh$. On the other hand, because of the sample curvature,
the average reflected angle $\beta$ will be changed by a value
$\Delta \beta$ given by an expression similar to
\ref{eq:curvature2}:

\begin{equation}
\Delta \beta =  \frac{2 \Delta sh}{R \tan \alpha}
\label{eq:curvature4}
\end{equation}
Note that in this case the sample height displacement $\Delta sh$
replaces the height of the beam $\Delta H$, but otherwise the
geometry remains the same.

The shift in the vertical position of reflected beam, or 'detector
height', $DH$, due to sample height displacement $\Delta sh$ can
be expressed as:

\begin{equation}
\Delta DH = 2 \Delta sh + L \Delta \beta = 2 \Delta sh + \frac{2 L \Delta sh}{R \tan \alpha}
\label{eq:curvature5}
\end{equation}
\begin{figure}[htp]
\begin{center}
\unitlength1cm
\begin{minipage}{15cm}
\epsfxsize=13cm \epsfysize=12cm   \epsfbox{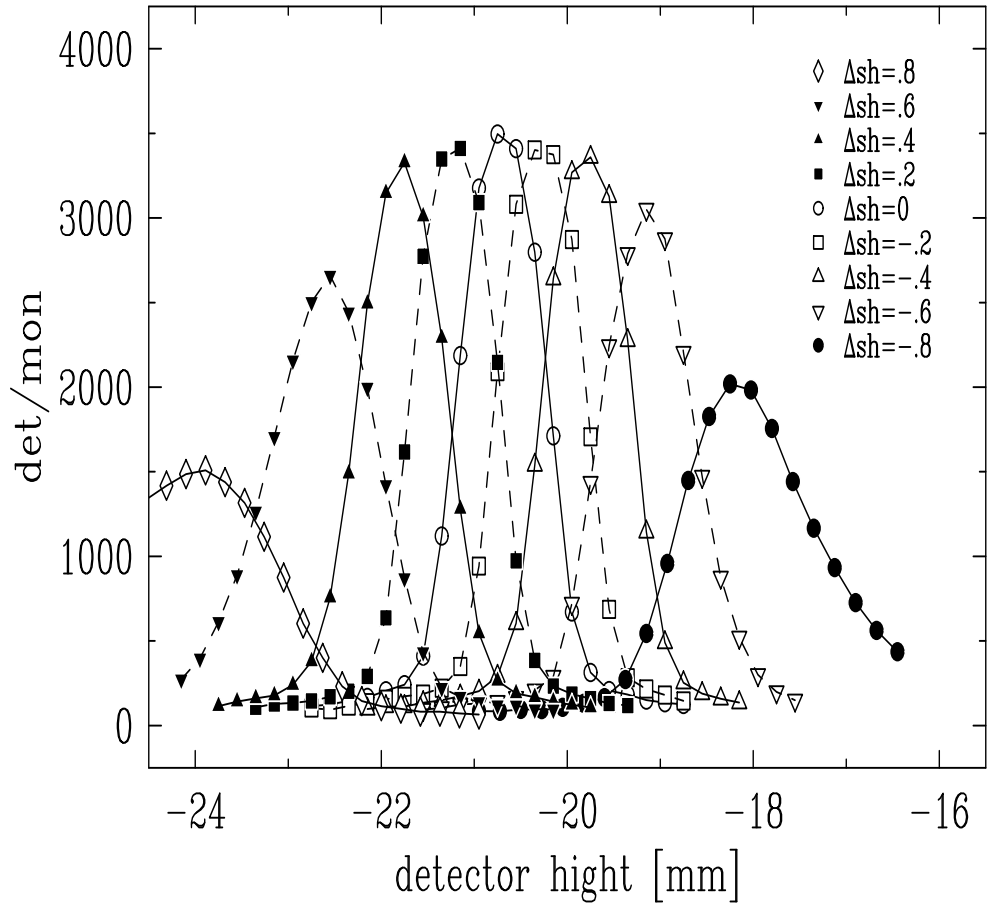}
\end{minipage}
\caption{\label{fig:walk1} Graphical representation of a "walking
scan" procedure: a series of detector height scans for various
sample height position, taken on a sample of liquid GaBi alloy.
Sample height position was changed in increments of 0.2mm, see
legend for more details.}
\end{center}
\end{figure}

Figure \ref{fig:walk1} provides an illustration of a typical
walking scan. The sample height was incrementally changed by 0.2mm
in both directions from the nominal position of spectrometer
geometry ($\Delta sh = 0$), and the detector height was taken at
each position of sample height. The scans were performed for
$\alpha=74$ mrad with detector L=402mm away from the sample. The
vertical opening of the detector slit was 1mm. As can be seen
qualitatively from this figure, the detector height scans centered
around $\Delta sh = 0$ are the narrowest, have the highest
intensity, and smallest spacing between the adjacent scan. As the
beam travels further from the sample's center on either side, the
scans become broader, peak intensity decreases while the distances
between the neighboring scans increases. This implies that the
radius of curvature is largest around nominal position $\Delta sh
= 0$ and decreases to the sides, as is expected. This particular
sample is completely convex, but depending on the size and the
wetting angles the sample can become concave in the middle.

A more quantitative approach consists of plotting the values
$\Delta DH - 2 \Delta sh$ as a function of $\Delta sh$, since
according to Eq.~\ref{eq:curvature5}

\begin{equation}
\Delta DH - 2 \Delta sh = \frac{2 L \Delta sh}{R \tan \alpha}
\label{eq:curvature6}
\end{equation}
the sample curvature $R^{-1}$ is therefore proportional to the
derivative of such plot:
\begin{figure}[htp]
\begin{center}
\unitlength1cm
\begin{minipage}{15cm}
\epsfxsize=12cm   \epsfbox{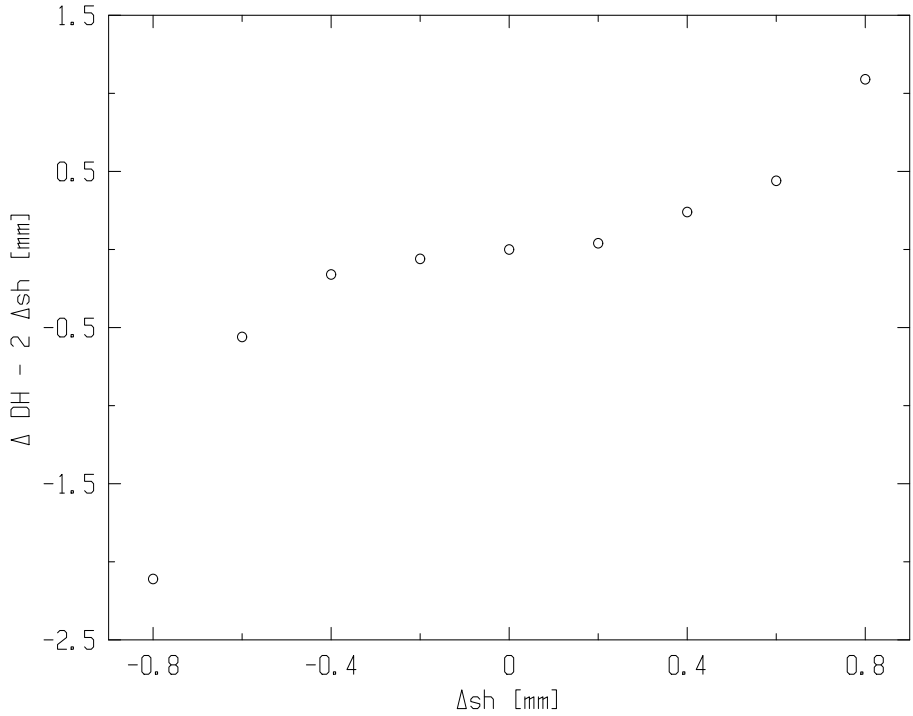}
\end{minipage}
\caption{\label{fig:walk2}  The displacements of the center
positions of detective height scans from the walking scan shown in
Figure \ref{fig:walk1}, adjusted for twice the amount of the
sample height displacement to account for a geometrical change in
position of the beam for an ideally flat reflecting surface, as a
function of sample height position. The derivative of this plot is
the effective measure of the \index{curvature!local} local surface
curvature, $R^{-1}$. }
\end{center}
\end{figure}

\begin{equation}
\frac{1}{R} = \frac{\tan \alpha}{2 L} \frac{\delta(\Delta DH-2
\Delta sh)}{\delta \Delta sh} \label{eq:curvature7}
\end{equation}
Plugging in the values for incident angle $\alpha=74$ mrad and the
distance from the sample L=402mm, the slope of the plot of $
{\delta(\Delta DH-2 \Delta sh)}/{\delta \Delta sh}=0.45$ at
$\Delta sh \approx 0$ corresponds to a value of $R^{-1}\approx 9
\cdot 10^{-5} m^{-1}$ or $R \approx 11$ meters. On the other hand,
the slope corresponding to $\Delta sh = -0.8$mm, as well as the
respective sample curvature at this position on the sample, are an
order of magnitude higher.




\bibliographystyle{unsrt}


%% file: T2_all.tex
\chapter{Surface-Induced Atomic Layering}

\section{Theoretical predictions}

\begin{figure}[htp]
\centering

\includegraphics[width=12cm]{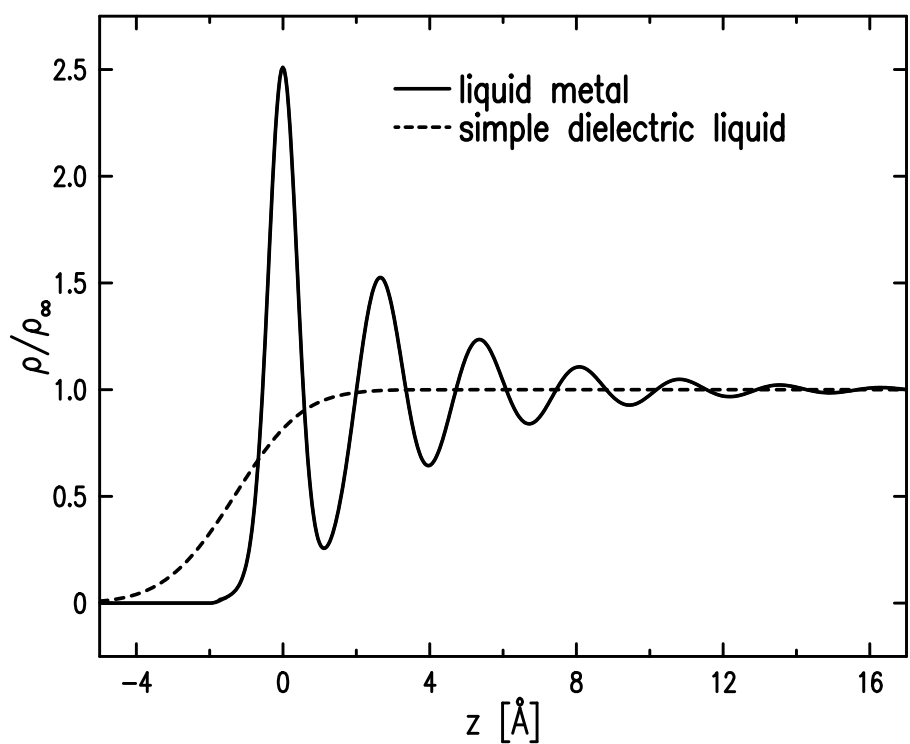}
\caption{\label{fig:LMvsDK} Surface-normal average of the density
at the surfaces of liquid metal (solid line) and dielectric liquid
(broken line).}
\end{figure}

The experimental studies described in Chapters 4  through 7 of
this thesis were motivated by the theoretical prediction by
Rice~\cite{Rice74} of the existence of a well-defined atomic
layering structure at the \index{surface!layering}
\index{liquid-vapor interface}liquid/vapor interface of a liquid
metal. The effect is the most prominent at the first surface
layer, with each subsequent layer becoming less and less ordered
such that deep in the bulk of the liquid all memory of the surface
is lost. By this prediction the average \index{local electron
density} local electron density introduced in Eq.~\ref{surface3},
$\langle \rho (z)\rangle_\xi $, along the surface normal can be
represented by a decaying oscillating function such as in Figure
\ref{fig:LMvsDK}. On the other hand for a
\index{liquid!dielectric} dielectric liquid consisting of
relatively small molecules, such as H$_2$O, CCl$_3$, C$_6$H$_2$
etc., the $\langle \rho (z)\rangle_\xi $ is believed to be a
simple monotonic rise from zero on the vapor side to bulk value on
the liquid side.

\subsection{Velasco's Theory}
Recent theoretical and computer simulations by \index{Velasco,
Enrique} Velasco et al. \cite{Velasco02} offer a somewhat
different interpretation of the atomic layering phenomena than by
Rice. Their principal argument is that the layering phenomena
found in atomic liquids up until now is of more universal nature,
and that the many-body character of metallic interactions may not
necessarily play a leading role in forming layering. Velasco et
al. claim that one should expect surface-induced atomic layering
in any liquid for which the ratio of the melting temperature $T_m$
to the critical temperature $T_c$, $T_m / T_c$ is sufficiently
small. Another way to rephrase this hypothesis: the surface
layering will appear in every liquid at low enough temperature, if
it is not first preempted by solidification, below which the
liquid becomes metastable and the free liquid-vapor interface
cannot be observed at equilibrium. In fact this is similar to the
criteria for which \index{liquid!crystals} liquid crystals will
form if they are not precluded by crystallization. This
explanation of layering phenomenon, therefore, does not rely on
nonadditive particle interactions, as does Rice's theory, but
depends solely on thermodynamic stability of the liquid-vapor
interface. The potential shortfall of the theory offered by
Velasco is that the potentials they use to describe the layering
phenomena are not realistic, in the sense that they may not be
capable of explaining some of the other thermodynamic properties
of liquids.

\section{Candidates for X-ray Study of Atomic Layering}

In this section we will consider the conditions which might be
used to define good candidates for x-ray reflectivity study of
surface-induced atomic layering. To study atomic surface layering
phenomena one has to start with a surface that is atomically
clean. While a combination of chemical procedures and ion
sputtering may produce samples that produce an atomically clean
surface, this condition must also be maintained throughout the
length of the experiment. For most metallic systems the principal
surface contamination is the surface oxide that will form in the
presence of even small amounts of oxygen or water in the system.
Oxide layers that a visible to the eye will usually form at oxygen
pressures as low as $10^{-8}$ Torr. The only way to achieve and
maintain a clean surface for these metals is to study them under
\index{UHV!conditions}  ultra high vacuum (UHV) conditions, which
implies that the candidate for such study must have a low enough
vapor pressure for UHV techniques to be applied successfully.  The
\index{alkali metals} alkali metals are exceptions to this since
the surface oxides seem to dissolve in the bulk. \cite{Addison84,
Borgstedt87} Similarly liquid Hg can be kept relatively clean
under slightly reducing atmospheres.

The second problem is to have sufficient intensity to measure the
reflectivity at values of $q_z$ that are of the order of the value
at which the \index{surface!layering} surface layering structure
peak is expected, $q_{peak}\approx 2\pi/a$ where $a$ is the atomic
or molecular radius. In view of the fact that the bare Fresnel
reflectivity falls off rapidly with increasing $q_z$, i.e.
$R_F(q_z) \approx (q_c/2q_z)^4$ where $q_c$ is the
\index{critical!wavevector} critical wavevector, there is some
advantage to studying materials with larger atomic numbers for
which both $q_c$ is larger and $a$ is smaller. Thermal capillary
excitations add to the problem since, as discussed in Sect. 2.9,
the capillary reduced roughness causes the reflectivity to fall
exponentially with $q_z^2$. The single parameter that best
summarizes this effect is $\eta = (k_B T /2 \pi\gamma) q_z ^2$
(Eq.~\ref{Eq:eta}). Thus another advantage in material property is
a low melting temperature and high surface tension.

The relative effect of capillary fluctuations on the reflectivity
from different materials with comparable \index{surface!structure
factor}surface structure factors can be most easily estimated by
considering the effect of the capillary waves on just the
\index{Fresnel reflectivity} Fresnel reflectivity at the value of
$q_z=q_{peak}=2\pi/a$ for which the layering peak is expected,
where $a$ is the atomic size: $R(q_z)/|\Phi(q_z)|^2=R_F(q_z)
\exp(-\sigma_{cw}^2 q_z^2)$. Although the square of the surface
\index{surface!structure factor}structure factor
$|\Phi(q_{peak})|^2$ will enhance $R(q_{peak})$ by a factor $\sim
100$, this factor is about the same for all of the pure metals
that have been studied thus far. The effect on the Fresnel
reflectivity at $q_z=q_{peak}$ can be approximated by

\begin{equation}
\frac{R(q_{peak})}{|\Phi(q_peak)|^2}
 \sim \bigg(\frac{q_c}{q_{peak}}\bigg)^4
\exp{\bigg[- \frac{k_B T_m q_{peak}^2}{{2 \pi }{\gamma}} \ln
\left(
    \frac{q_{\mbox{\scriptsize res}}}{q_{\mbox{\scriptsize max}}
    }\right)\bigg]} \sim \rho_e^2
a^4 \exp{\bigg[- \frac{T_m} {\gamma a^2}~constant\bigg]}
\label{Rpeak}
\end{equation}

Table \ref{tab:Metals} contains a list of relevant material
parameters for a number of pure elements. The elements are listed
in the decreasing order of $R(q_{peak})/|\Phi(q_{peak})|^2$, the
estimated reflectivity at $q_{peak}$ obtained by multiplying
\index{Fresnel reflectivity} Fresnel Reflectivity $R_F$ by
\index{Debye-Waller} Debye-Waller capillary wave factor. Not
surprisingly, the first three elements for which the
\index{surface!layering} atomic layering has been experimentally
observed - mercury, gallium and indium (in that order) can be
found at the top of the list. In the rest of this chapter we will
review some of the experimental data obtained during these
measurements and its interpretation.


\begin{table}
\begin{tabular}{crrrrrrrr}
\hline
   {\bf Z} & {\bf Name} & {\bf $T_m$,} & {\bf $\gamma$,} & {\bf $P_{vap}$,} & {\bf $q_{peak}$,} & {\bf $\sigma_{cw}$,} & {\bf $\eta(q_{peak})$} & {\bf $R(q_{peak})/|\Phi(q_{peak})|^2$} \\
     &   & {$^\circ$  K}  &   {mN/m} &  {Torr} & { $\AA^{-1}$} & {$\AA$}  &  &   \\
     \hline
      \hline

        80 &    Mercury &     234.28 &        498 &     2.e-06 &        2.3 &       0.72 & 0.53&   3.9e-09 \\
        31 &    Gallium &     302.93 &        718 &     $<$1e-11 &        2.6 &       0.68 & 0.61&   4.0e-10 \\
        49 &     Indium &     429.32 &        556 &     $<$1e-11 &        2.2 &       0.92 & 0.80 &   3.8e-10 \\
        48 &    Cadmium &      594.1 &        570 &     1.e-01 &        2.1 &       1.07 &  1.02 &  1.7e-10 \\
        50 &        Tin &     505.12 &        560 &     $<$1e-11 &        2.2 &       1.00 & 0.99 &   1.2e-10 \\
        81 &   Thallium &      576.6 &        464 &     3.e-08 &        2.1 &       1.17 & 1.22 &   1.1e-10 \\
        82 &       Lead &     600.65 &        458 &     3.e-09 &        2.1 &       1.20 & 1.23 &   1.0e-10 \\
        79 &       Gold &     1337.6 &       1338 &     2.e-05 &        2.5 &       1.05 & 1.34 &   7.5e-11 \\
        83 &    Bismuth &      544.5 &        378 &     2.e-10 &        2.1 &       1.26 & 1.43 &   3.0e-11 \\
        30 &       Zinc &     692.73 &        782 &     1.e-01 &        2.6 &       0.99 & 1.33 &   1.2e-11 \\
        55 &     Cesium &      301.6 &         70 &     2.e-06 &        1.4 &       2.18 & 1.73 &   1.5e-12 \\
        29 &     Copper &     1356.6 &       1300 &     3.e-04 &        2.8 &       1.07 & 1.83 &   1.1e-12 \\
        37 &   Rubidium &      312.2 &         85 &     1.e-06 &        1.5 &       2.01 & 1.71 &   8.4e-13 \\
        47 &     Silver &     1235.1 &        828 &     2.e-03 &        2.5 &       1.28 & 2.06 &   6.7e-13 \\
        19 &  Potassium &      336.8 &        110 &     1.e-06 &        1.5 &       1.83 & 1.59 &   4.9e-13 \\
        56 &     Barium &       1002 &        277 &     1.e-01 &        1.6 &       1.99 & 2.10 &   3.4e-13 \\
        11 &     Sodium &        371 &        191 &     1.e-07 &        1.9 &       1.46 & 1.54 &   3.3e-13 \\
        12 &  Magnesium &        922 &        559 &     2.e+00 &        2.3 &       1.35 & 1.85 &   1.1e-13 \\
         3 &    Lithium &      453.7 &        398 &     1.e-10 &        2.3 &       1.12 & 1.36 &   9.1e-14 \\
        20 &    Calcium &       1112 &        361 &            &        1.8 &       1.84 & 2.27 &   2.3e-14 \\
        51 &   Antimony &      903.9 &        367 &            &        2.3 &       1.64 & 2.79 &   1.2e-14 \\
        32 &  Germanium &     1210.6 &        621 &     8.e-07 &        2.6 &       1.46 & 2.98 &   2.0e-15 \\
        14 &    Silicon &       1683 &        865 &     3.e-04 &        2.7 &       1.46 & 3.20  &   1.5e-16 \\
        52 &  Tellurium &      722.7 &        180 &     1.e-01 &        2.5 &       2.10 & 5.71&   2.6e-21 \\
        34 &   Selenium &        490 &        106 &     2.e-03 &        2.6 &       2.25 & 6.76 &   6.8e-24 \\
        18 &      Argon &       83.8 &         13 &            &        4.1 &       2.65 & 23.64 &   2.1e-62 \\
        10 &       Neon &      24.56 &          5 &     4.e+00 &        7.1 &       2.28 & 52.05 & 6.6e-125 \\

\end{tabular}
\caption{\label{tab:Metals}Comparison of candidates for
Surface-induced layering study. \cite{CRC85, Podesta02, Iida93,
Kaye86, Emsley98} }
\end{table}

\section{Distorted Crystal Density Model}

In view of the fact that the  phase information is lost in a
reflectivity measurement, it is impossible to extract $\langle
\rho (z) \rangle _\xi$ from a single reflectivity measurement
\cite{Tolan99} by simply performing a reverse Fourier transform of
the x-ray reflectivity data (see Eq.~\ref{eq:phi}). So-called
\index{model independent methods} model independent methods
\cite{Tolan99, Chou97, Zheng01} have been developed that purport
the ability to extract average local density profile defined in
Eq.~\ref{surface3}, $\langle \rho (z)\rangle _\xi$; however, as
has been shown by \index{Pershan, Peter} Pershan \cite{Pershan00}
there is a certain degree of ambiguity that can not be avoided. As
a result our solution is to assume a model for the density
profile, which is then Fourier transformed and fitted to the
experimentally determined structure factor.

\subsection{Model Description} The simplest approximation for the oscillating electron density
profile describing the surface-induced layering phenomena in
single component liquid metals can be constructed by considering a
physical model which consists of a series of well-defined atomic
2D layers ordered parallel to the surface, so-called
\index{distorted crystal model} \textit{distorted crystal model}
\cite{Magnussen96}. The layers are separated from each other by a
distance $d$. The width of each subsequent layer is gradually
increased, causing an overlap between layers as one goes deeper
into the bulk, until eventually the liquid approaches a common
bulk-like liquid behavior with no sign of layering.
Mathematically, the electron density can be described by a
semi-infinite sum of equally spaced \index{gaussian} Gaussians,
with identical total integrated density: \footnote{This neglects
the changes in scattering \index{form factor} form factor
amplitude, $f(q)$, which are typically small for small $q$.}
\begin{equation}
     \frac{\langle{\rho} (z)\rangle} {\rho _\infty} = \sum _{n=0} ^\infty
    \frac{d/\sigma _n}{\sqrt{2\pi }}
    \exp \left[ -(z-nd-z')^{2}/2\sigma _n ^2 \right]
\label{eq:rho}
\end{equation}

 As
will be shown below, if the widths have the form $\sigma _n ^2 =
n\bar{\sigma}^2 + \sigma _{0}^{2}$, and $\bar{\sigma}$ and $\sigma
_0$ are constants, the sum of the \index{Fourier transform}
Fourier transforms will have a simple analytic form. The model
density approaches the bulk density $\rho_{\infty}$ for $\sigma_n
\gg d$.

\subsection{An Alternative Density Model}
An alternative model for the layering profile can be built from
superposition of an \index{error function} error-function and an
exponentially decaying sinusoidal:

\begin{equation}
     \frac {\langle{\rho} (z)\rangle}{\rho _\infty} =
      \left[ 1 + A \; \exp\left({-\frac{z-z'}{\lambda}}\right) \sin\left(\frac{2\pi
      (z-z_0-z')}{d}\right)\right]
      erf\left(\frac {z}{\sigma_0}\right)
\label{eq:rho_alt}
\end{equation}
Although the profiles produced by this model are very similar to
those of the \index{distorted crystal model} distorted crystal
model, there are some disadvantages. For example, if an empirical
model is to be used it should have the fewest number of arbitrary
parameters. This model contains five \index{fitting!parameters}
parameters ($A, \lambda, d, z_0$ and $\sigma_0$), as opposed to
just three in the case of a distorted crystal model ($\sigma_0,
\bar{\sigma}, d$). Another disadvantage of this model is that in
contrast to the distorted crystal model, it cannot be clearly cast
in terms of atomic distributions. However, the best fit electron
density functions from both of these empirical models are
essentially identical \cite{Regan95} and the only arguments to
favor one or the other are really subjective.
\medskip
\begin{table}
\begin{tabular}{rrrrr}
\hline
 && Table (a) && \\
  \hline \hline
{\bf Variable} & {\bf Name} & {\bf Units} & {\bf Typical Value} & {\bf Fit?}  \\
\hline \hline
       vdh & vertical slit size &         mm &      0.1-4 &         no \\
\hline
       hdh & horizontal slit size &         mm &      0.1-4 &         no \\
\hline
         L & sample-to-detector &         mm &        400 &         no \\
\hline
$\Delta \Theta$ & background offset &     degree &    0.1-0.3 &         no \\
\hline
$\alpha$ & incident angle &     degree &     0.1-10 &         no \\
\hline
$\lambda$ & wavelength &     $\AA$ &    0.6-1.6 &         no \\
\hline
 &&    & & \\
 && Table (b) && \\
  \hline \hline
{\bf Variable} & {\bf Name} & {\bf Units} & {\bf Typical Value} & {\bf Fit?}  \\
\hline \hline
   $Q_c$  & Critical Wavevector & $\AA^{-1}$ &       0.05 &         no  \\
\hline
$Q_{max}$ &    $Q_{max}$ & $\AA^{-1}$ &    1.0-3.0 &         no \\
\hline
$\gamma$ & surface tension &  dyn/cm &     70-700 &  sometimes \\
\hline
        T  & temperature &          K &    250-500 &         no \\
\hline

 &&   & & \\
 && Table (c) && \\
  \hline \hline
  {\bf Variable} & {\bf Name} & {\bf Units} & {\bf Typical Value} & {\bf Fit?}  \\
\hline \hline
$\sigma_0$ & avg atomic width &     $\AA$ &    0.3-1.0 &        yes \\
\hline
$\overline{\sigma}$ & broadening width &     $\AA$ &    0.3-3.0 &        yes \\
\hline
         d & interatomic spacing &     $\AA$ &    2.0-3.0 &        yes \\
\hline
\end{tabular}
\caption{\label{tab:fitpar}A listing of typical fitting parameters
related to atomic  layering model described here. Table (a)
contains parameters describing the spectrometer geometry, table
(b) lists parameters relevant to the studied material and table
(c) includes parameters describing the density profile}
\end{table}
\medskip

\subsection{Fitting Functions}
\index{fitting!function} In order to extract a real space density
profile from the experimental data it is necessary to construct a
computer code that relates the reflectivity to the model. There
are three things that must be included in this code. The first is
a numerical expression that will accurately represent the surface
\index{surface!structure factor} structure factor $|\Phi(q_z)|$
associated with the \index{electron density model} electron
density model (Eq.~\ref{eq:phi}). The second is a numerical
function that takes account of the integral over the singular
function that describes the capillary effects. The method that was
used to achieve this was discussed in Sect. 2.9.4. The third, and
final ingredient is to subtract from scattering at the specular
condition the intensity of the theoretically predicted diffuse
scattering at the position where the off-specular background
scattering is measured. This code is then fit to the measured data
using the \index{C-PLOT} C-PLOT non-linear least square fitting
procedures with varying numbers of self adjusting parameters.

The parameters used by these functions can be divided into three
general categories. The first corresponds to the constraints
imposed by the \index{spectrometer} spectrometer. These include
details such as the x-ray wavelength, the dimensions of the
detector slits, the lengths of the various arms of the
spectrometer. These are necessary in order to implement the
digital integration over the spectrometer resolution. Some of
these parameters are listed in Table \ref{tab:fitpar} (a). The
second has to do with the thermal capillary excitations that were
discussed in Section 2.9. These include the surface tension
$\gamma$ and the temperature $T$ of the material, as well as the
\index{critical!wavevector}  critical wavevector $q_c$ and are
listed in  Table \ref{tab:fitpar} (b). Finally there are the
parameters in Table \ref{tab:fitpar} (c) that are needed to model
the intrinsic electron density profile. The parameters in Table
\ref{tab:fitpar} (a) and (b) are known and the measurements are
analyzed by \index{fitting!function} fitting the \index{distorted
crystal model} distorted crystal model to the data by varying only
the parameters in Table \ref{tab:fitpar} (c).

In view of the fact that at least approximate values for the
surface tension for most systems are known the first step is to
use these values for obtain a first best fit of the structure
factor $\Phi(q_z)$ to the reflectivity. This form for $\Phi(q_z)$
is then used to more precisely define the value of the surface
tension that was used to account for the capillary effects by
fitting the diffuse scattering data. If necessary, the
reflectivity data fitting may have to be adjusted with the revised
surface tension, as the entire process is reiterated until a
self-consistent density model and surface tension values
accurately describing both reflectivity and diffuse scattering
data are found. This procedure is slightly more complicated for
alloys since the density profiles depend on the relative
concentration of the two species. Additionally, the simple
\index{distorted crystal model} distorted crystal model is often
not sufficient to describe layering in binary alloys, the
structure of which can be complicated by surface segregation.
These effects will be discussed in detail when the data for the
alloys are presented.

\section{X-Ray Measurements of Liquid Mercury, Gallium and Indium}

\subsection{X-Ray Reflectivity Comparison}

\begin{figure}[htp]
\begin{center}
\unitlength1cm
\begin{minipage}{12cm}
\epsfxsize=12cm \epsfysize=16cm \epsfbox{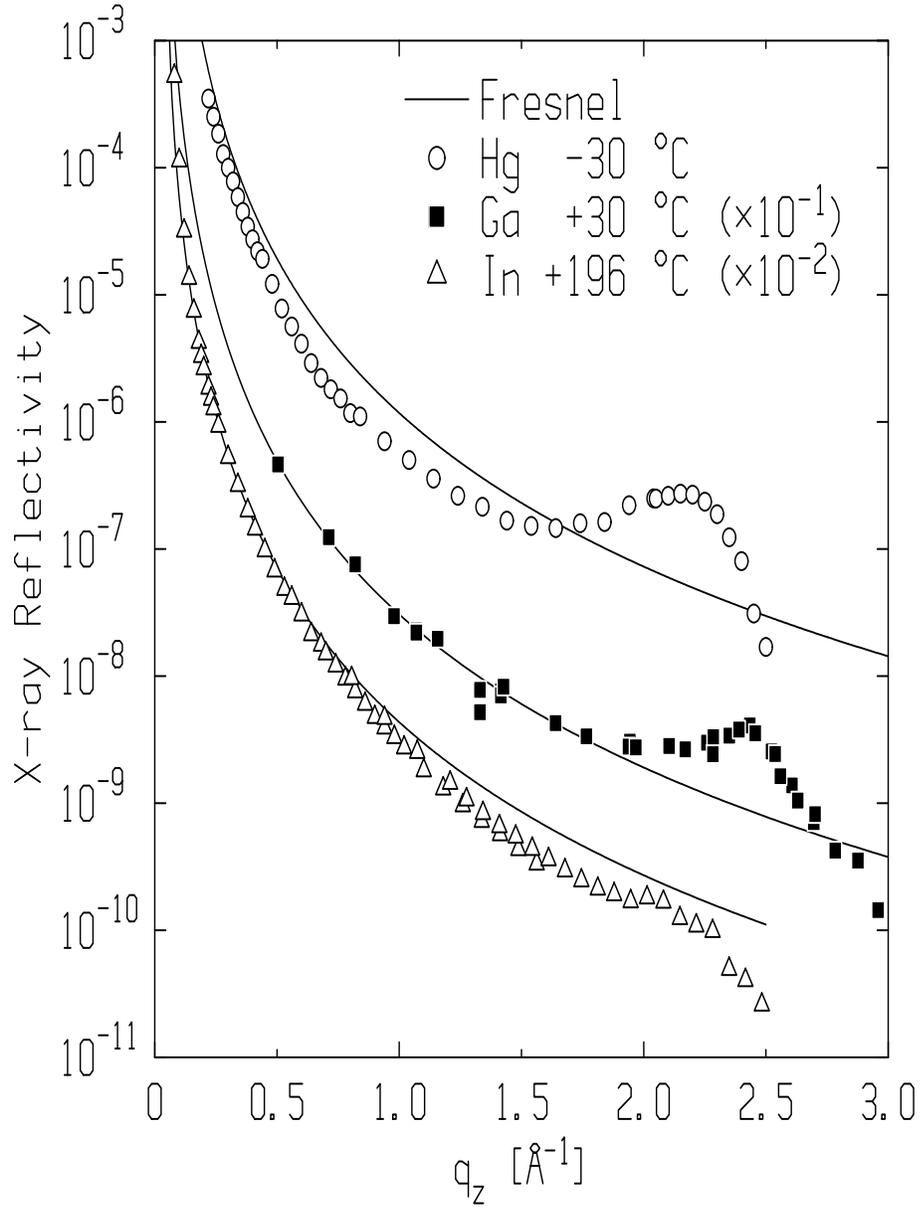}
\end{minipage}
\caption{\label{fig:HGGAIN} The plot showing x-ray reflectivity
data taken from liquid Hg (circles), Ga (squares) and In
(triangles), along with theoretically predicted Fresnel curves
(lines). }
\end{center}
\end{figure}

\begin{figure}[htp]
\begin{center}
\unitlength1cm
\begin{minipage}{12cm}
\epsfxsize=12cm \epsfysize=16cm \epsfbox{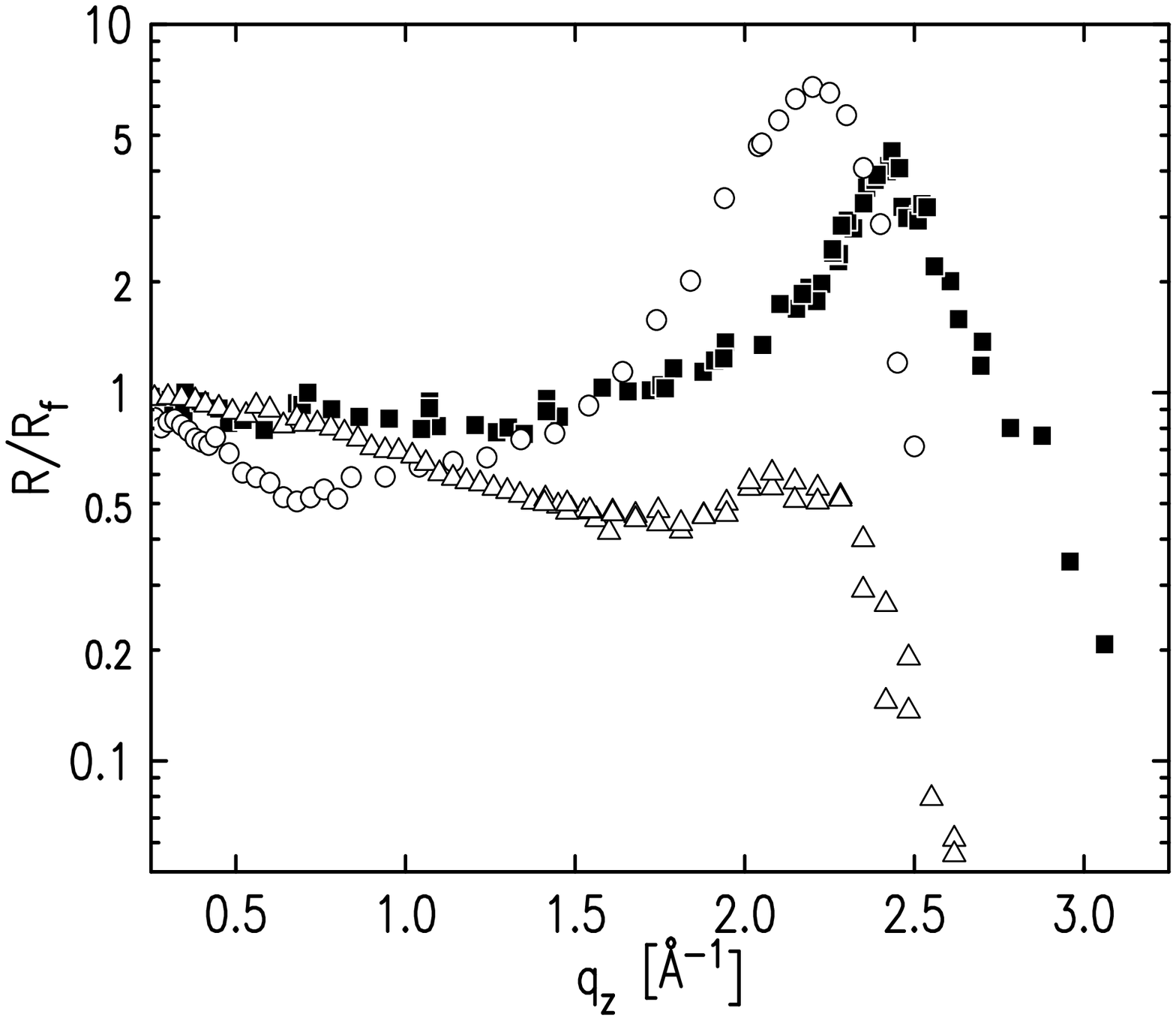}
\end{minipage}
\caption{\label{fig:HGGAIN2} The plot showing Fresnel-normalized
x-ray reflectivity data taken from liquid Hg (circles), Ga
(squares) and In (triangles).}
\end{center}
\end{figure}

The raw reflectivity data from Hg \cite{Magnussen95}, Ga
\cite{Regan95} and In \cite{Tostmann99} is shown in Figure
\ref{fig:HGGAIN}. For clarity of representation, the data sets for
indium and mercury have been offset by multipliers of 10 and 0.1,
respectively. The lines represent \index{Fresnel reflectivity}
Fresnel reflectivity curves, without taking into account capillary
roughening or any surface structure or capillary contributions.
Figure \ref{fig:HGGAIN2} shows the reflectivity data for these
three metals normalized to the respective Fresnel reflectivities.

Although the prominence of local maxima at the position of the
layering peaks decreases in moving from Hg through Ga to In, this
does not reflect the true strength of the intrinsic, or local
layering that was discussed in Section 2.9.4.  As can be seen in
Table 3.1, when the values of $q_{peak}$ are taken into account,
this ordering of the three metals happens to coincide with the
order of the increasing melting temperatures ($-35^{\circ}$C for
Hg, $25^{\circ}$C for Ga and $170^{\circ}$C for In), and
consequently the value of $\eta(q_{peak})$ at the melting
temperature also increases in the order of Hg, Ga and In.  In
fact, when the thermal effects are removed the intrinsic layering
is comparable for all three metals.

The effects of thermal capillary waves on the reflectivity can be
most directly seen by measuring the temperature dependence of the
reflectivity for any one metal. This has now been done for the two
metals with the lowest melting temperatures (Ga \cite{Regan95},
and Hg \cite{Magnussen95}) since it is for these metals that the
practical range of temperatures is largest. In the next section we
will illustrate this effect for Ga. Although we will not show
them, the results for Hg are similar.

\begin{figure}[htp]
\begin{center}
\unitlength1cm
\begin{minipage}{18cm}
\epsfxsize=12cm   \epsfbox{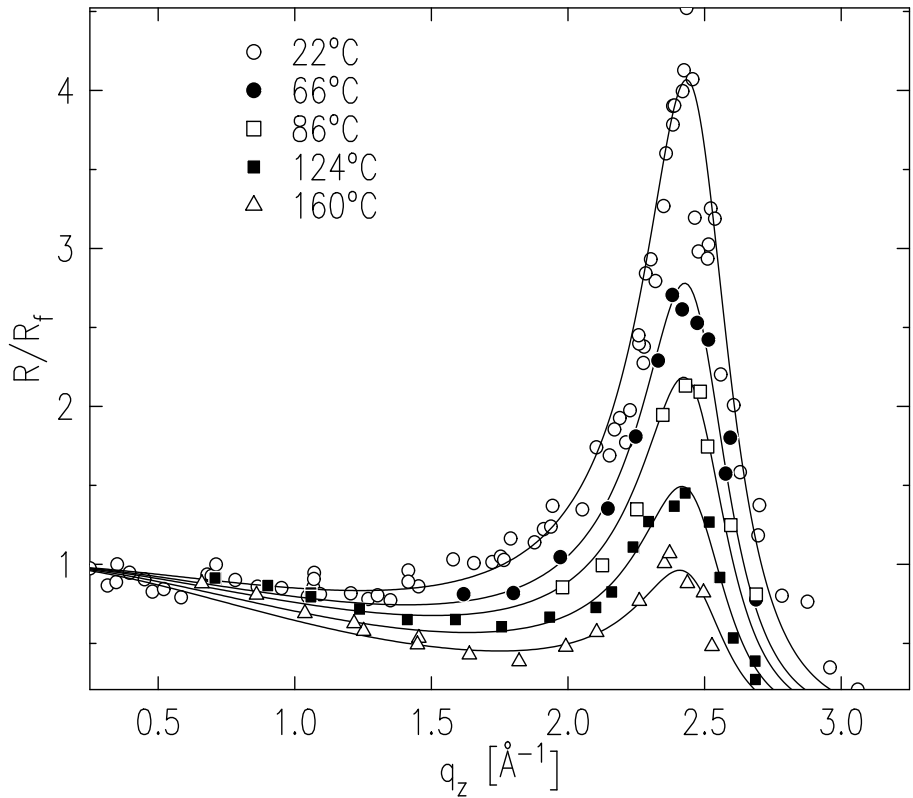}
\end{minipage}
\caption{\label{fig:Ga_temp} Fresnel-normalized x-ray reflectivity
data measured for liquid Ga in the temperature range of
$22^{\circ}$C to $160^{\circ}$C. }
\end{center}
\end{figure}

\subsection{Temperature-Dependent Reflectivity: Liquid Ga}
In Figure \ref{fig:Ga_temp} we show \index{Fresnel reflectivity}
Fresnel-normalized reflectivity data sets measured for liquid Ga
at $22^{\circ}$C, $66^{\circ}$C, $86^{\circ}$C, $124^{\circ}$C and
$160^{\circ}$C. The layering peak is most prominent for the lowest
temperature and gradually decreases as the temperature increases.

\begin{figure}[hp]
\begin{center}
\unitlength1cm
\begin{minipage}{20cm}
\epsfxsize=15cm   \epsfbox{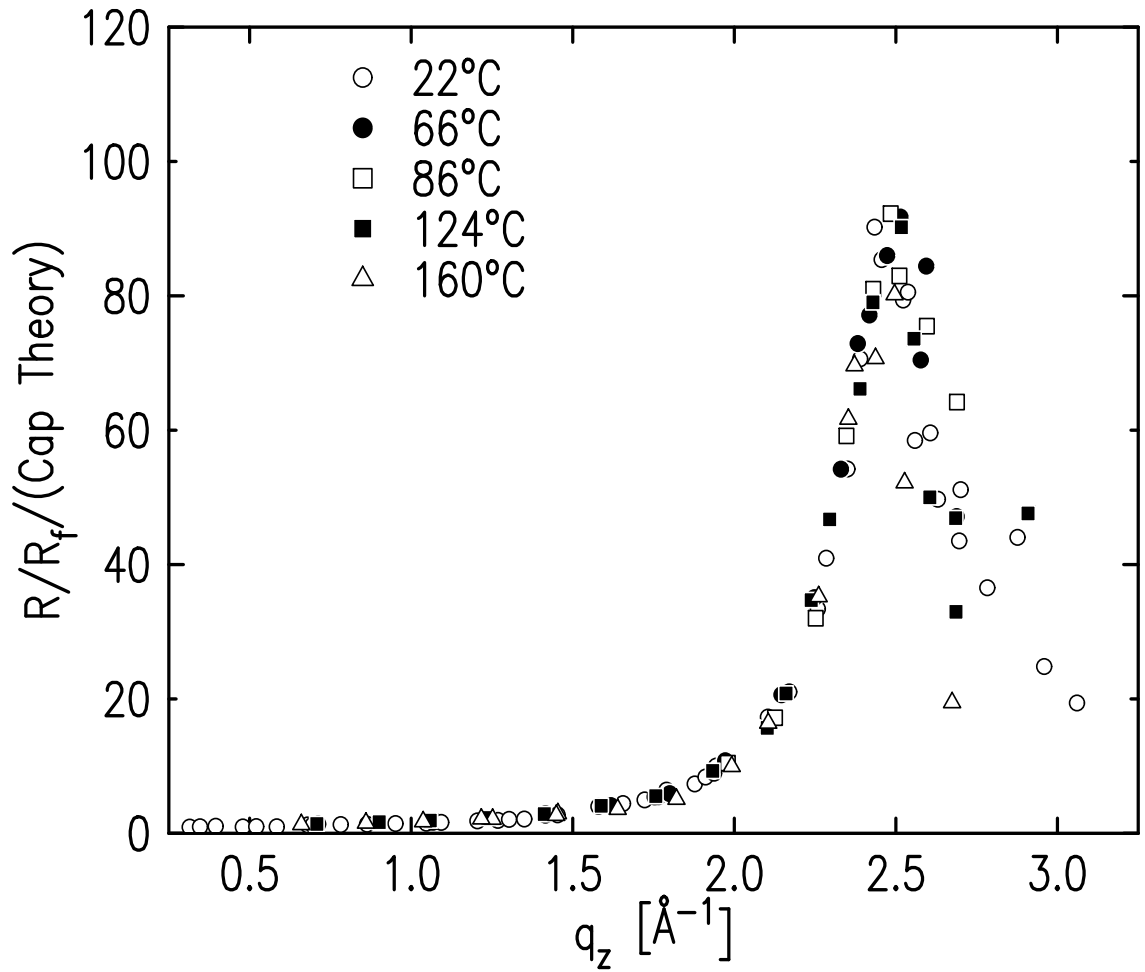}
\end{minipage}
\caption{\label{fig:Ga_tnorm} The temperature-dependent
reflectivity datasets pictured in Figure \ref{fig:Ga_temp}
normalized to the thermal capillary wave fluctuations. }
\end{center}
\end{figure}

In Figure \ref{fig:Ga_tnorm} we show the intrinsic surface
\index{surface!structure factor} structure factor that is obtained
by dividing the data by the predicted effect of capillary
excitations (Eq.~\ref{eq:final}) using the published surface
tension value of 718 mN/m. As can be seen, with the predicted
capillary effects removed the intrinsic structure factor is
independent of temperature.

\subsection{X-Ray Diffuse Scattering Data: Liquid In}
In this thesis a \index{x-ray!diffuse scattering} \textit{diffuse
scattering scan} implies a measurement of signal as a function of
$\beta$ at $\Delta \Theta=0$ while keeping incident angle $\alpha$
fixed. The observed scattering therefore follows a trajectory in
the ($q_x, q_z$) plane with $dq_z / dq_x =\tan \beta$ and $q_y$=0.
The signal measured in the specular reflectivity and diffuse
scattering geometries are, however, related since the signal
measured at $\beta=\alpha$ in the diffuse measurement is the
specular signal.

\begin{figure}[htp]
\begin{center}
\unitlength1cm
\begin{minipage}{16cm}
\epsfxsize=12cm  \epsfysize=16cm  \epsfbox{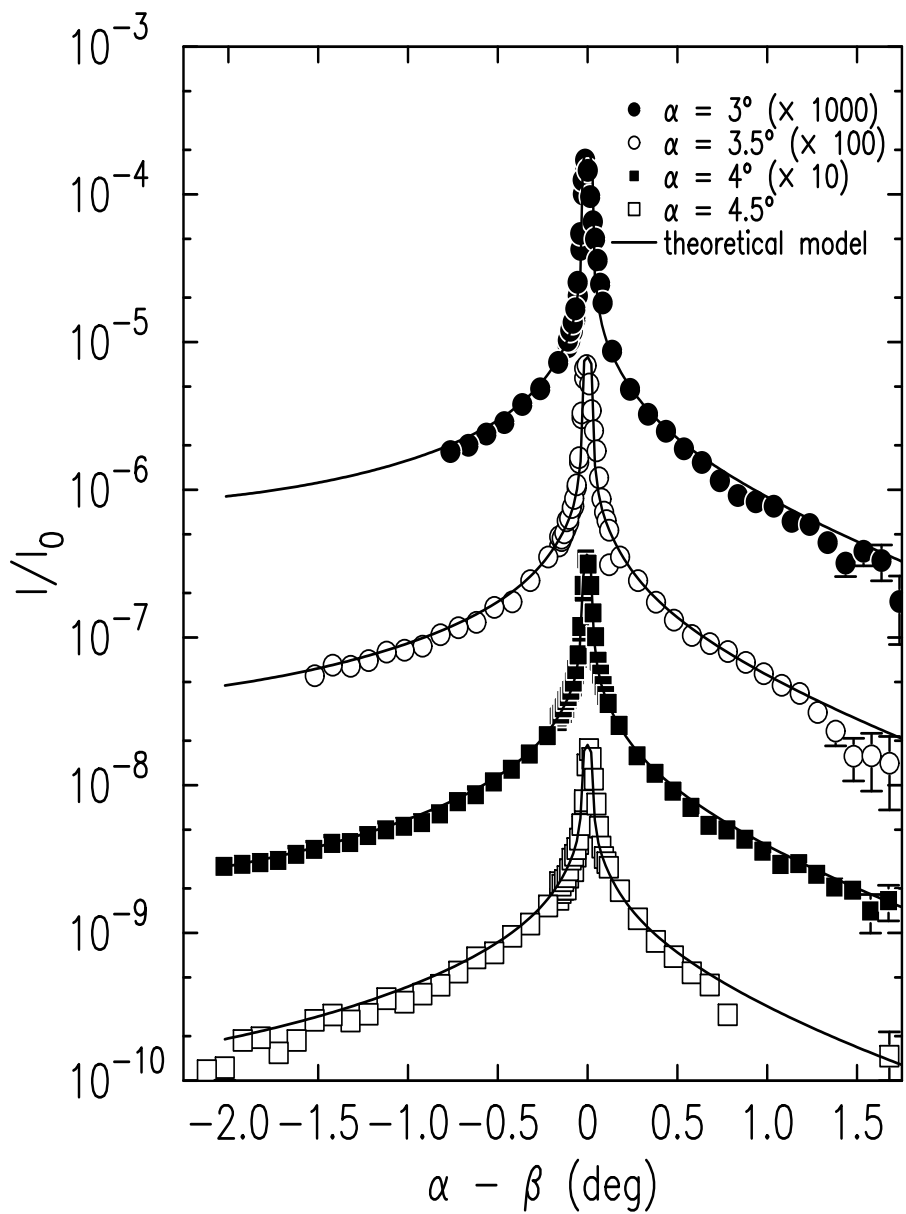}
\end{minipage}
\caption{\label{fig:Indiffuse} Diffuse scattering data for liquid
indium (symbols), along with the theoretical simulations which
follow from capillary wave theory (lines), taken at 3.0, 3.5, 4.0
and 4.5 degrees, which correspond to $\eta$ values of 0.049,
0.066, 0.087 and 0.110 respectively.}
\end{center}
\end{figure}

The results of  diffuse scattering scans that illustrate
confirmation of the capillary wave prediction for liquid In are
shown in Figure \ref{fig:Indiffuse}, which shows a set of diffuse
scattering scans taken at various values of incident angle
$\alpha$ (symbols), along with theoretical predictions of the
capillary wave theory. For all experimental measurements, the
signal originating at the surface was separated from the bulk
diffuse background by subtracting intensities measured with the
detector moved approximately $0.4^\circ$ transverse to the
reflection plane.
 The solid lines are the
intensities that are theoretically predicted by numerically
integrating the capillary prediction (Eq.~\ref{eq:final}) over the
known \index{resolution function} resolution function, using the
method described in Section 2.9.3, while also taking into account
the background subtraction procedure.

The excellent agreement between the experimental data and
theoretical predictions demonstrated in Figure \ref{fig:Indiffuse}
is a strong indication that one can rely on elements of capillary
wave theory in order to successfully deconvolve the surface
\index{surface!structure factor} structure factor information
related to the intrinsic atomic structure of the liquid surfaces.
This method of deconvolving thermal contribution is going to be
discussed in great details in Chapter 4 of this work.



\bibliographystyle{unsrt}

%% file: hg_all.tex
\chapter{Temperature Dependent Surface Structure of Liquid
Mercury}

\section{Abstract}
In this chapter we present x-ray reflectivity measurements of
liquid mercury between $-36^{\circ }$C and $+25^{\circ }$C. The
surface structure can be described by a layered density profile
convolved with a thermal roughness $\sigma _T$. The layering has a
spacing of 2.72~\AA\ and an exponential decay length of 5.0~\AA .
Surprisingly, $\sigma _T$ is found to increase considerably faster
with temperature than the $\sqrt{T}$ behavior predicted by
capillary wave theory, in contrast with previous measurements on
Ga and \index{liquid!dielectric} dielectric liquids.

\section{Introduction}

The effect of temperature on the surface structure of a liquid is
quite different from that of a solid surface. In the liquid,
thermal surface waves are excited at all wavelengths from the
particle spacing to a long wavelength gravitational cutoff, and
produce a surface roughness on the order of the particle spacing.
For non-metallic liquids, these capillary waves broaden the
liquid--vapor interface, which is a monotonically decreasing
density profile with a width of
    several~\AA .\cite{brasl88}

Metallic liquids exhibit a more complex surface structure in which
the atoms are stratified parallel to the liquid--vapor interface
in layers that persist into the bulk for a few atomic diameters.
This layering arises from the strongly density dependent non-local
interionic
    potential\cite{rice}
and is unique to metallic liquids. Nevertheless, the effect of
temperature is still expected to be principally in the form of
capillary waves, which roughen the surface-normal profile and
diminish the layering amplitude. X-ray reflectivity measurements
have confirmed the existence of surface layering in liquid
    Hg,\cite{magnu95a}
    Ga,\cite{regan95a}
    In,\cite{indium}
and several
    alloys.\cite{alloys98,Lei}
Models based on capillary waves as the only mechanism for surface
roughening were found to describe the temperature dependent
layering of liquid
    Ga\cite{regan96a}
as well as \index{x-ray!diffuse scattering} diffuse scattering
measured from the liquid In
    surface.\cite{indium}

In previous comparisons of the x-ray reflectivity of liquid Hg and
    Ga,\cite{compare}
two important differences were identified. Although reflectivities
for both metals exhibit \index{quasi-Bragg peak} quasi--Bragg
peaks indicative of surface layering, the Hg data have a minimum
at low momentum transfer ($q_z\approx0.6$~\AA$^{-1}$) not found in
Ga
    (see Fig.~2 in Ref.~9).
To describe this minimum, the model for the surface structure had
to be more complex than that of Ga. One successful model
incorporated a low density region persisting a few~\AA\ into the
vapor side of the interface. This feature was taken to be
intrinsic to Hg. In addition, the room temperature layering peak
appeared to be broader for Hg than for Ga, leading to the
conclusion that in Hg, surface layering decayed over a much
shorter length scale.

Subsequent measurements yielded variations in this low-$q_z$
minimum, prompting us to address the possible effect of impurities
at the surface. An important difference between experiments on Hg
and Ga is that due to the low vapor pressure, Ga can be measured
\index{UHV!chamber}  under ultra high vacuum (UHV) conditions and
cleaned in situ by argon sputtering, which is not possible for Hg.
In this work, we compare Hg measured in a reducing H$_2$
    atmosphere\cite{H2}
to measurements taken in a    UHV compatible chamber having a very
low oxygen partial pressure.

The present temperature dependent study has an important advantage
over previous room temperature measurements.  Near the melting
point, the layering peak has a large amplitude and is easily
distinguished from the region of the low-$q_z$ minimum. This
allows a robust determination of the layering decay length that is
not affected by the details of the models in the near-surface
region. This is not the case at room temperature, where the
shallow peak and pronounced minimum yield more ambiguous fits.  We
conclude in the present work that Ga and Hg have comparable
layering decay lengths. This underscores the value of temperature
dependent measurements for comparisons of liquid metals' surface
structure.

\section{Experimental Details}

X-ray reflectivity
    measurements\cite{brasl88}
were carried out using the liquid surface \index{spectrometer}
spectrometer at beamline X22B at the \index{synchrotron!source}
\index{Brookhaven} National Synchrotron Light Source, at an x-ray
wavelength of 1.24~\AA\ and resolution of
    $0.035$~\AA$^{-1}$.\cite{resolution}
The background intensity, due mainly to scattering from the bulk
liquid, was subtracted from the specular signal by rocking the
detector out of the reflection plane.

\section{Sample Preparation}
Liquid mercury was prepared in two different sample chambers. In
the first, the Hg
    sample\cite{bethaddr}
was contained in a glass reservior with a teflon stopcock and
glass filling capillary, mounted on a high vacuum chamber
comprised of a cylindrical thin-walled Be can mating to stainless
steel flanges with viton o-ring seals.  After several cycles of
evacuating the chamber to $10^{-4}$~Torr and backfilling with
900~Torr of dry H$_2$ gas, the liquid Hg was poured from the
reservoir into a ceramic sample pan. Surfaces appeared clean to
the eye and yielded reproducible reflectivity measurements for as
long as four days. When the sample was deliberately exposed to
air, the reflectivity exhibited deep oscillations characteristic
of rapid oxide
    formation.\cite{Hgox}

A second Hg reservoir, capillary and valve made from stainless
steel was mounted on top of a UHV chamber. Following a bakeout,
\index{bakeout} the sample was poured, with the Hg vapor pressure
of about $10^{-2}$~Torr then  determining the total pressure in
the chamber.  Small patches of oxide were initially observed, but
disappeared from the surface within a few hours. Hg oxide is
expected to decompose due to the low oxygen partial pressure  in
this pseudo UHV
    environment.\cite{riceHg}
In addition, we observe a continuous evaporation and
recondensation of Hg, constantly renewing the Hg
    surface.\cite{renew}
The resulting Hg surface was stable, as determined by x-ray
reflectivity, for eleven days.

In both chambers, a copper cold finger extended from the bottom of
the sample pan into a liquid nitrogen dewar. Temperature was
monitored near the sample pan, and calibrated in separate
experiments with a sensor immersed in the Hg. Corrected sample
temperatures are reported with relative errors of $\pm 1^\circ$C.
For the UHV chamber, the gradient between sample and thermometer
was greater, and there is an additional systematic error which may
be as large as $8^\circ$C. Both chambers rested on an active
\index{vibration isolation}  vibration isolation unit.

\section{X-Ray Reflectivity Measurements}
\begin{figure}[tbp]
\centering
\includegraphics[width=0.8\columnwidth]{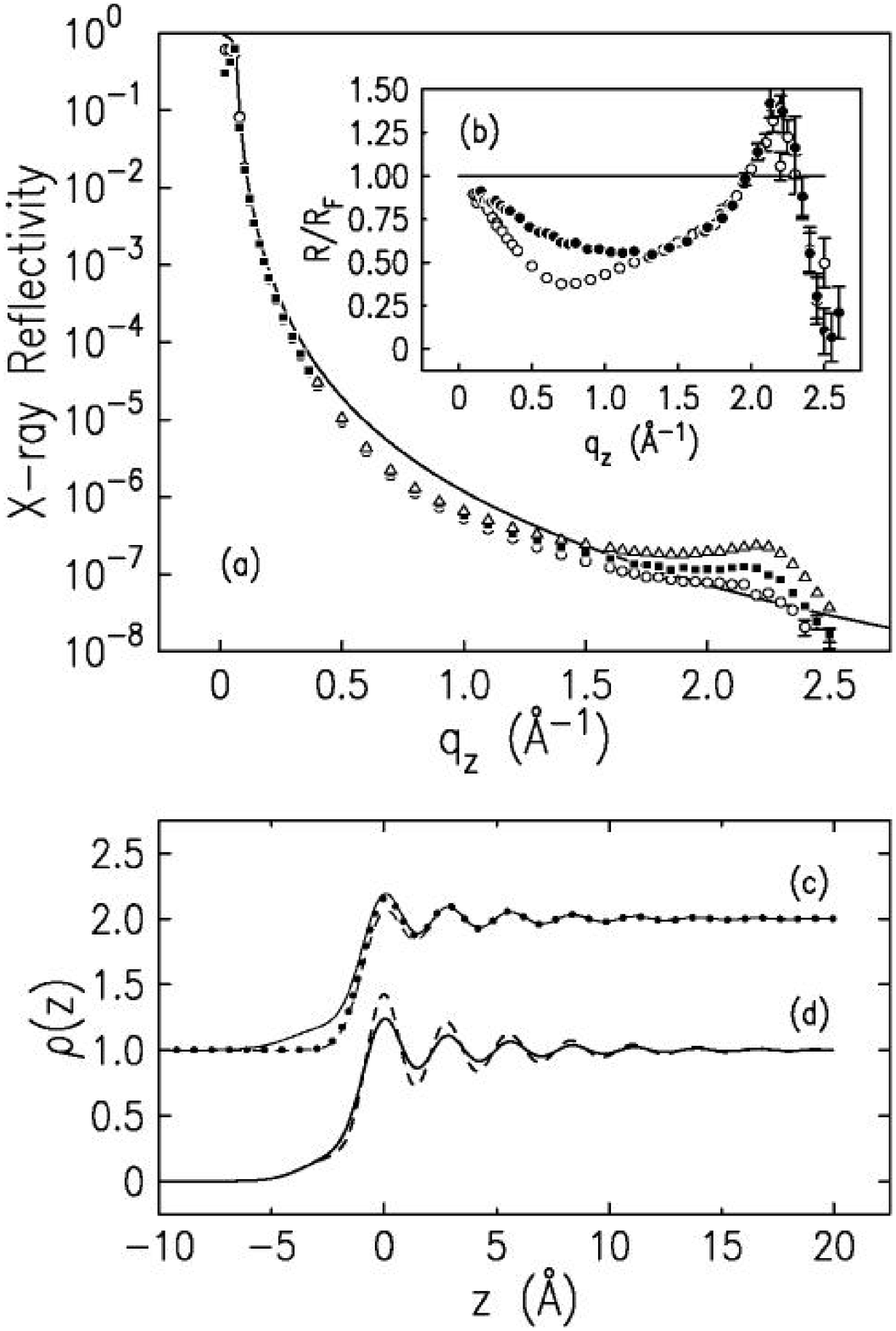}
 \caption{(a) X-ray reflectivity of liquid
Hg. $\triangle$: $-35^\circ$C; \protect\rule{1.4ex}{1.4ex}:
$0^\circ$C; $\circ$: $+23^\circ$C; (---): calculated Fresnel
reflectivity $R_F$. (b) Room temperature reflectivity normalized
by $R_F$ (---) for two Hg samples. $\circ$: glass chamber with
H$_2$ gas. $\bullet$: UHV chamber. (c) Calculated local layering
profiles having $d=2.72$~\AA , $\overline{\sigma}=0.46$~\AA , and
$\sigma _T=1.0$~\AA . ($\cdots$): no surface modification. (---):
adlayer model. (- - -): surface depletion model. (d) Best fit
profiles for vacuum data. (---): Room temperature ($\sigma _T =
1.0$~\AA ). (- - -): $T=-36^\circ$C ($\sigma _T = 0.8$~\AA ). }
\label{hg_fig:refl}
\end{figure}

Our x-ray reflectivity measurements reveal that the surface
induced layering is strongly temperature dependent.
    Fig.~\ref{hg_fig:refl}(a)
shows the reflectivity as a function of $q_z$\/ measured in the
glass chamber for three temperatures between $-35^{\circ }$C and
room temperature. The \index{quasi-Bragg peak} quasi--Bragg peak
observed at $q_z\approx 2.2$~\AA $^{-1}$ indicates that the liquid
metal is stratified near the surface. The peak position, amplitude
and width give the length scales for the surface induced layering,
which extends several \AA\ into the bulk liquid. The other main
feature of the data is the minimum in the reflectivity at $q_z
\approx 0.6$--1.2~\AA $^{-1}$, the depth of which differs between
samples. This minimum is much more pronounced under the high
pressure H$_2$ environment (900~Torr) than under the Hg vapor
pressure of $10^{-2}$~Torr, as shown for two sets of room
temperature measurements in
    Fig.~\ref{hg_fig:refl}(b).
These low-$q_z$ data indicate that the details of the electron
density within 5~\AA\ of the surface are affected by the sample
environment.

\section{Electron Density Model}
We have quantified the surface layering by constructing a model
for the surface-normal density profile, parameterized as reported
    previously,\cite{magnu95a,regan95a,regan96a}
and fitting the reflectivity calculated from the model to the
experimental data. The surface-normal density profile is
constructed from slabs of electron density parallel to the
liquid--vapor interface. The slabs represent planes of atoms
separated by a distance $d$, which are disordered in-plane, and
which have mean-squared surface-normal displacements increasing
with depth. The electron density profile is composed of a sum of
Gaussian terms:
\[
\rho (z)=\rho _\infty\sum_{n=0}^\infty \frac{d/\sigma
_n}{\sqrt{2\pi }}\exp \left[ -(z-d)^2/2\sigma _n^2\right] \;,
\]
where $\rho _\infty$ is the electron density of bulk Hg, and we
represent the mean-squared displacements in the $n$th layer by
$\sigma _n^2=n\overline{\sigma }^2+\sigma _T^2$, where $\sigma _T$
characterizes the roughness of the layer nearest the surface. Far
into the bulk, $n\overline{\sigma }^2$ becomes large, and no
layering is present, so that $\overline{\sigma}$ quantifies a
layering decay length. With this form for $\rho (z)$, the
reflectivity, proportional to the Fresnel reflectivity $R_F$ of an
abruptly terminated homogeneous electron density, can be
calculated analytically and yields
\begin{eqnarray*}
    \frac{R(q_z)}{R_F} & = &
        \left| \frac 1{\rho _\infty} \int
        \limits_{-\infty }^{+\infty }(\partial \rho /\partial z)\exp
        (iq_zz)\,dz\right| ^2 \\
     & = & F^2\,(q_zd)^2\exp (-\sigma _T^2q_z^2)
    \times \\
     & & \left[
    {1-2\exp (-q_z^2\overline{\sigma }^2/2)\cos (q_zd)+
    \exp(-q_z^2\overline{\sigma }^2)}
    \right] ^{-1}\;,
\end{eqnarray*}
where the reduced scattering \index{form factor} form factor
$F=(f(q_z)+f^{\prime })/(Z+f^{\prime })$ has been inserted, as
presented
    previously.\cite{note}

The parameters $\sigma _T$, $\overline{\sigma }$, and $d$ control
the height, width and position of the layering peak, and
characterize a simple local layering profile
    (Fig.~\ref{hg_fig:refl}(c), dotted line).
Although this model is sufficient for liquid
    Ga\cite{regan95a,regan96a}
and     In,\cite{indium} it does not yield a good fit to the the
low-$q_z$ reflectivity of liquid Hg. To account for the low-$q_z$
reflectivity we have considered profiles which incorporate density
terms in addition to the layered model, modifying the first Hg
layer, the first several layers, and/or the region several \AA\
towards the vapor side of the interface. These additional features
are independent of temperature. The modification having the best
fit for all data is a single additive Gaussian term $f_A
\rho_\infty (d / \sigma _A \sqrt{2 \pi}) \exp [-(z-z_A) / 2\sigma
_A ^2]$, positioned a few \AA\ into the vapor region, with a width
$\sigma _A \approx 1.5$~\AA\ and a density $f_A$ relative to the
bulk varying from 0.1--0.3 between samples.
    (Fig.~\ref{hg_fig:refl}(c), solid line).
However, this model is difficult to distinguish mathematically
from one in which the density of the first Hg layer is decreased.
In particular, the data with the most exaggerated low-$q_z$ minima
can be fit by models with a broadened, depleted first Hg layer and
an expanded spacing between the first two layers
    (Fig.~\ref{hg_fig:refl}(c), dashed line).
Models having alternating long and short layer spacings, as might
result from atomic pairing, do yield destructive interference at
low $q_z$ but do not fit the data well.

Deviations from the simple oscillatory profile are smaller for
samples poured in a \index{UHV!chamber} UHV chamber at 10$^{-7}$
Torr following a \index{bakeout} bakeout than for measurements in
a glass chamber evacuated to only 10$^{-4}$ Torr. To explain this
difference, we attribute the modified density profile to the
presence of impurities. In recent self-consistent quantum monte
carlo simulations of the Hg-vapor interface, no vapor-side tail or
appreciable modification of the surface layer was observed in the
computed density
    profile.\cite{Chek98}
As discussed above, we believe the presence of macroscopic oxide
is unlikely, although oxygen, water or other contaminants
introduced with the Hg sample may be present in small amounts.
These may form a passivated, low-density adlayer or incorporate
themselves into the first Hg layer to alter its density.

\section{Discussion}
\begin{figure}[tbp]
\centering
\includegraphics[width=0.8\columnwidth]{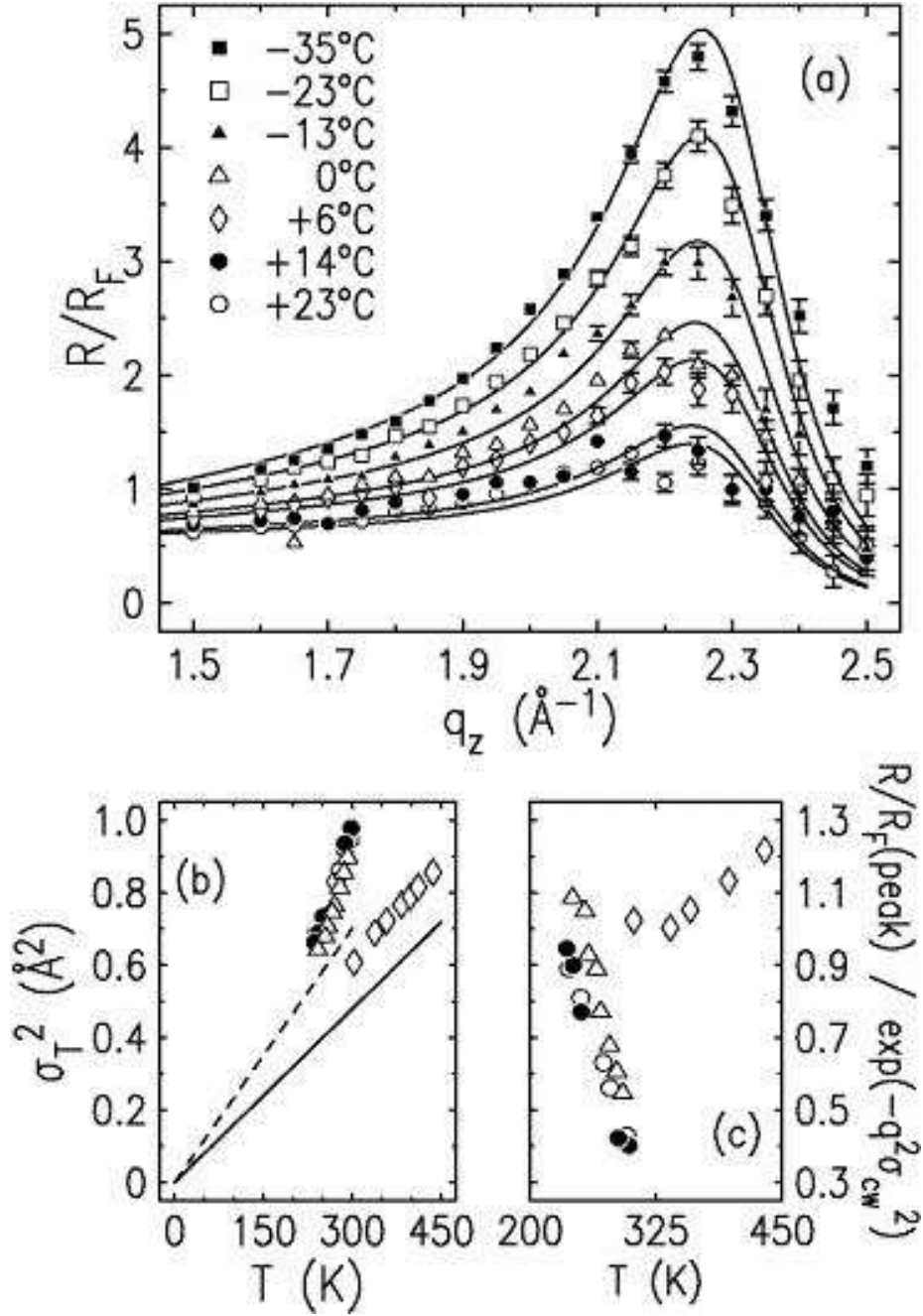}
\caption{(a) $R/R_F$ for different temperatures, from sample
measured in H$_2$ atmosphere. Solid lines are fit curves for which
$\sigma _T$ has been varied. (b) Temperature dependence of $\sigma
_T ^2$. $\triangle$: Measurement in vacuum). $\circ$: Measurement
in H$_2$ gas). $\bullet$: Second sample measured in H$_2$ gas.
$\diamond$: Ga data from Ref.~8. Lines: Capillary wave form
$\sigma _{cw}^2(T)$ for Hg (- - -) and Ga (---). (c) Temperature
dependence of $(R/R_F) / \exp(-q^2 \sigma _{cw}^2)$ at the
layering peak, normalized at the melting point, for Ga
($\diamond$) and Hg (symbols as in (b)). } \label{hg_fig:T}
\end{figure}

Although the available x-ray data do not allow a unique
determination of the near-surface region, we obtain reliable
information on the temperature dependence of the surface-normal
density profile. All Hg data are described by local layering
parameters $d = 2.72 \pm 0.02$~\AA\ and $\overline{\sigma}= 0.46
\pm 0.05$~\AA . These values are obtained for both H$_2$ and
vacuum measurements, and are independent of the model surface
modification. The temperature dependence of the reflectivity can
be quantified solely by the variation of $\sigma _T$, as shown in
    Fig.~\ref{hg_fig:T}(a)
for measurements in H$_2$ atmosphere. This parameter controls the
amplitude of the layering peak and is directly related to the
amplitude of the density oscillations at the surface, as shown in
    Fig.~\ref{hg_fig:refl}(d) for vacuum measurements,
varying from $0.8$~\AA\ near the melting point to about $1.0$~\AA\
at room temperature. The other parameters, if allowed to vary for
each temperature independently, show no systematic dependence on
$T$.

In particular, this means that the layering decay length of $5.0
\pm 0.5$~\AA\ for Hg is independent of temperature, as was
previously shown for
    Ga.\cite{regan96a}
This is somewhat surprising, since it is not obvious that layering
should persist over essentially the same length scale in the face
of thermal roughening which effectively broadens the interfacial
width. Such a model, in which the local surface-normal density
profile is independent of the surface height fluctuations, is
intrinsic to the analyses long used for calculating the scattering
cross section from rough
    surfaces.\cite{sinha88,sanyal91}
Previous observations on liquid metals demonstrated that such an
interpretation appears to be
    justified.\cite{indium,regan96a}
In fact, it has been thought that the liquid metal surface may
provide the best justification of this approach, since the
measured reflectivity peak is a dramatic structural feature that
is both straightforward to characterize and strongly affected by
surface height fluctuations that vary considerably with
temperature.

In the capillary wave model, thermally excited surface waves
produce mean-squared displacements for which the restoring force
is due to the surface tension $\gamma$:
$$
    \sigma _{\mbox{\scriptsize cw}}^2(T)=
    \frac{k_BT}{2\pi \gamma }\ln \left( \frac{k_{\max }}{k_{\min }
    }\right) \;,
$$
where $k_{\max } = 1.2$~\AA$^{-1}$ and $k_{\min } = 7.4\times
10^{-3}$~\AA$^{-1}$ are wavevector cutoffs determined by the
atomic size and the instrumental
    resolution, respectively.\cite{brasl88}
This model has been successfully applied to dielectric
    liquids\cite{brasl88,masa97}
and to liquid
    Ga.\cite{regan96a}
    Fig.~\ref{hg_fig:T}(b)
compares $\sigma _{\mbox{\scriptsize cw}}^2$ calculated as a
function of temperature for Hg and Ga (dashed and solid lines),
for appropriate values of the surface
    tension,\cite{iida93}
bulk interatomic
    spacing,\cite{iida93}
and instrumental resolution, to the surface roughness $\sigma _T$
extracted from reflectivity measurements of Hg and Ga (symbols).
The temperature dependence of $\sigma _T$ differs fundamentally
between Hg and Ga. For Ga, the slope of the experimental points
matches that of the capillary wave model, offset only by a
constant term $\sigma _0 \approx 0.3$~\AA , attributed to an
intrinsic
    roughness.\cite{regan96a}
This allows the quantity $\sigma _T$ for Ga (denoted in Ref.~8 by
$\sigma_c$) to be written as $\sigma _T ^2 = \sigma _0 ^2 + \sigma
_{\mbox{\scriptsize cw}} ^2$. Liquid Ga, therefore, has a
temperature independent local surface layering profile, roughened
by capillary waves.

By contrast, $\sigma _T^2(T)$ for Hg has a much steeper slope than
predicted by capillary wave theory, exhibiting an additional
temperature dependent roughness. This result is independent of the
local layering model, as shown in
    Fig.~\ref{hg_fig:T}(c),
where we plot $(R/R_F)/\exp(-q^2 \sigma _{\mbox{\scriptsize cw}}
^2)$ for Hg and Ga at their layering peaks vs.\ $T$, normalized at
the melting point. The Ga data increase only slightly with $T$,
within their scatter of about 15\%. The Hg data exhibit a steep
slope far exceeding the $\approx 15$\% scatter. The temperature
dependence of the Hg--vapor interface roughness has also been
extracted from recent self-consistent quantum monte carlo
    simulations.\cite{Chek98}
While similar linear behavior of $\sigma _T ^2 (T)$ is observed,
direct comparison of the slope is complicated by the differences
in capillary wave cutoffs and surface tension between the
experimental and simulation samples.

The temperature dependence of $\gamma $, even considering the
possible effects of impurities, can be ruled out as a cause. For
pure Hg,
    $d\gamma /dT\approx -0.2\times 10^{-3}$~N/mK,\cite{iida93}
an order of magnitude too small to explain our observations.
Optical measurements on Hg surfaces with varying degrees of
cleanliness have found $d\gamma /dT$ in the range of $-0.7\times
10^{-3}$ to $+0.3\times 10^{-3}$~N/mK depending on the method of
sample
    preparation.\cite{kolev97}
The observation that oxide or contamination can produce $d\gamma
/dT>0$ for Hg is notable, but this would cause a reduction of
$\sigma _T$ with increasing $T$ rather than the increase observed,
and so can not account for our results.

\section{Summary}
We have measured specular x-ray reflectivity from liquid mercury
between $-36^{\circ }$C and $+25^{\circ }$C. The longitudinal
density profile exhibits density oscillations due to strongly
temperature dependent surface induced layering, as well as a
sample dependent low-density feature near the liquid--vapor
interface, independent of $T$. We conclude that the surface
structure of liquid Hg has a temperature dependent component in
addition to capillary wave roughening. This may take the form of
temperature dependent height fluctuations which increase the
mean-squared atomic displacements beyond the effect of capillary
waves, or of a temperature dependent local density profile. The
specular reflectivity measurements presented here can not
distinguish between these two cases. Measurements of
\index{x-ray!diffuse scattering} diffuse scattering from the Hg
surface may provide further information. These results
re-emphasize the fact that roughness intrinsic to the
surface-normal density profile in general can not be unambiguously
separated from height fluctuations, despite previous treatments of
scattering from the liquid surface.

%% file: in_all.tex
\chapter{X-ray Studies on Liquid Indium}

\section{Abstract}

In this chapter we describe x-ray reflectivity (XR) and small
angle off-specular \index{x-ray!diffuse scattering} diffuse
scattering (DS) measurements from the surface of liquid Indium
close to its melting point of $156^\circ$C. From the XR
measurements we extract the surface structure factor convolved
with fluctuations in the height of the liquid surface. We present
a model to describe DS that takes into account the surface
\index{surface!structure factor} structure factor, thermally
excited capillary waves and the experimental resolution. The
experimentally determined DS follows this model with no adjustable
parameters, allowing the surface \index{surface!structure factor}
structure factor to be deconvolved from the thermally excited
height fluctuations. The resulting \index{local electron density}
local electron density profile displays exponentially decaying
surface induced layering similar to that previously reported for
Ga and Hg. We compare the details of the local electron density
profiles of liquid In, which is a \index{nearly free electron}
nearly free electron metal, and liquid Ga, which is considerably
more covalent and shows directional bonding in the melt. The
oscillatory density profiles have comparable amplitudes in both
metals, but surface layering decays over a length scale of $3.5\pm
0.6$~\AA\ for In and $5.5\pm 0.4$~\AA\ for Ga. Upon controlled
exposure to oxygen, no oxide monolayer is formed on the liquid In
surface, unlike the passivating film formed on liquid Gallium.

\section{Introduction}

The structure of the free surface of a liquid metal is
fundamentally different from that of a \index{liquid!dielectric}
dielectric liquid such as water or Argon. For dielectric liquids,
the interatomic or intermolecular potential is long-ranged and the
nature of the interactions does not change significantly across
the liquid--vapor interface. Theoretical modeling and
\index{molecular Dynamics simulations} molecular dynamics studies
result in an interfacial density profile that varies monotonically
from the bulk liquid density to the bulk vapor
density.\cite{abra82} This can be contrasted with results of
computer simulations of dielectric liquid noble gases in contact
with their solid phase.\cite{chap77} There, the hard wall provided
by the solid phase is shown to induce surface layering in the
liquid with a well defined lamellar structure.

In liquid metals, the potential energy function changes
drastically from a screened short-ranged
\index{interactions!Coulomb} Coulomb potential in the bulk liquid
phase where the conduction electrons are delocalized
 to a long-ranged \index{interactions!Van der Waals} van der~Waals type potential in the
vapor phase where all electrons are localized.\cite{rice74} In
this case, density functional calculations\cite {evan81,iwam92}
and computer simulations\cite{harr87} indicate an oscillatory
surface-normal density profile. This surface induced layering can
be attributed to the fact that strong \index{interactions!Coulomb}
Coulomb interactions between the abruptly truncated conduction
electrons at the surface and the ion cores impose a constraint
ordering the surface in a well defined lamellar structure. This
constraint can be thought of as an effective hard wall ionic
potential at the surface where the conduction  electron density
abruptly terminates.

The existence of surface layering in liquid metals has only recently been
verified unambiguously by experiment, first for liquid Hg\cite{magn95} and
then for liquid Ga.\cite{regan95} Comparison of the temperature
dependence of this layering, however, reveals qualitative differences
between
the two metals.\cite{regan96a,dimasi98} This observation
indicates that details of the interactions such as  the degree of
covalency in the metal may affect the surface structure.

For crystalline metals that can be described by  weak
pseudo--potentials, electron--ion scattering is relatively weak
and the non localized or itinerant electrons can be treated as if
they were essentially free.\cite{ashcroft76} \index{alkali metals}
Alkali metals with their nearly parabolic energy bands are the
prime example of this class of \index{nearly free electron} nearly
free electron (NFE) like metals. In analogy with this, the term
NFE metals applies to liquid metals whose conduction electrons
have similar itinerant character and are only weakly perturbated
by a small ionic pseudopotential.\cite{ziman61} NFE liquid metals
can be contrasted with those whose conduction electrons are
partially localized in covalent
    bonds.\cite{selloni95}
Evidence for such covalent character in liquid metals may be
obtained from measurements of the optical properties as well as
from the bulk \index{liquid!structure factor}liquid structure
    factor.\cite{march90}

In fact, both liquid Ga and Hg show substantial deviations from
NFE behavior. In liquid Ga, a shoulder in the bulk
\index{liquid!structure factor} liquid structure factor indicates
the presence of directional bonding in the
melt.\cite{nart72,bell89,gong93,cicc94} Liquid Hg displays a
pronounced asymmetry in the first peak of the bulk
\index{liquid!structure factor} liquid structure factor, and the
optical constants of liquid Hg deviate markedly from the
predictions of the Drude theory for free
    electrons,\cite{croz72}
which, for example,  adequately describes the optical constants of
the \index{alkali metals} alkali
    metals.\cite{mayer63,heyer86}
In order to understand which aspects of surface layering may be
affected by deviations from NFE character, it is necessary to
investigate liquid metals with less tendency towards covalent
bonding. The \index{alkali metals} alkali metals would be ideal
candidates in this regard. Unfortunately, the very low surface
tension of liquid alkali metals will result in a large thermally
induced surface roughness, making it extremely difficult to
observe the surface layering experimentally. This is exacerbated
by the fact that \index{alkali metals} alkali metals are very
reactive in air but cannot be investigated under
\index{UHV!conditions} Ultra High Vacuum (UHV) conditions due to
their high vapor pressure.
Liquid Indium, which has a high surface tension,
reasonably low melting point and low vapor pressure does not suffer
from these shortcomings. Moreover, it is considered to be NFE like for the
following reasons. First, the  first peak in the bulk structure factor
of liquid In is highly symmetric, indicating the absence of
significant orientational bonding in the
    melt.\cite{orto66}
Second, quantitative analysis of the shape of the
\index{liquid!structure factor} bulk liquid structure factor
yields a twelve-fold coordination with an interatomic distance
that decreases with increasing temperature.\cite{ocke66} This is
precisely what is expected for  a liquid that can be described by
an ideal hard sphere model. Third, the measured optical properties
of liquid In agree with the Drude free electron theory over a
large range of wavevectors.\cite{schu57} Even more interesting,
the optical constants of mixtures of liquid Hg and In differ
appreciably from free electron behavior for Hg rich compositions,
but gradually approach free electron behavior with increasing In
content.\cite {hodg67} A main objective of this study is to
compare the surface structure of the more NFE like liquid metal In
to that of the more covalent liquid Ga.

This fundamental interest in the relationship between the surface structure
and electronic properties of liquid metals bears directly on the topic of
surface reactions, which is of considerable practical importance.
 This is especially true
for liquids, where the atoms are mobile, and reactions at the surface can
directly affect alloying, phase formation and other properties of the bulk.
For this reason, surfaces  and reactions at
surfaces play a critical role in process and extractive
metallurgy.\cite{rich74} We previously demonstrated that on exposure to
oxygen the surface of liquid Ga becomes coated with a uniform 5~\AA\ thick
passivating oxide film that protects the underlying bulk phase from further
oxidation.\cite{regan97} We will show below that the oxidation behavior of
liquid In is fundamentally different from that of Ga.

\section{Experimental Details}

The experiments reported here were primarily performed at the
beamline X25 at the National \index{synchrotron!source}
Synchrotron Light Source \index{Brookhaven} (NSLS). Following a
toroidal focussing mirror, a Ge (220) crystal was used to select
the x-ray wavelength $ \lambda = 0.653$~\AA\ and to tilt the beam
downward onto the horizontal liquid metal surface. The geometry of
this liquid \index{spectrometer} spectrometer has been described
    elsewhere.\cite{regan97b}
Additional data were acquired at beamline X22B at the
\index{Brookhaven} NSLS, using a similar scattering geometry and
an x-ray wavelength of 1.24~\AA . The size of the x-ray beam
striking the liquid metal surface was determined by
1~mm~(horizontal)~$\times $~0.1~mm~(vertical) slits downstream of
the monochromator and the measured signal was normalized to the
incoming beam intensity.

The experimental resolution, determined  primarily by slits in
front of the detector, was varied for different experimental
configurations. Specular reflectivity was primarily measured using
a conventional NaI scintillation detector.  Typical slit settings
for measuring this specular reflectivity at X25 were
4~mm~(h)~$\times $~4~mm~(v). At 600~mm from the sample this yields
a $q_z$  resolution of 0.06~\AA $^{-1}$. For comparable resolution
at the longer wavelength at X22B, a vertical detector slit of 8~mm
was used. Small angle off-specular surface diffuse scattering was
measured using a  position sensitive
    detector
(PSD) aligned within the reflection plane.\cite{PSD} For a given
incident angle, the PSD detects both the specular reflection and
off-specular diffuse intensity in the reflection plane up to
$\approx 0.5^{\circ }$  away from the specular peak. We used a
coarse vertical resolution for the PSD in the multi channel mode
(64 channels on 20\,mm). Away from the specular peak, where the
intensity varies slowly with position, data from neighboring
channels were averaged together for improved statistics. Counting
all photons hitting the detector, and disregarding the positional
information, the PSD can be used as a conventional counter. In
this ``single channel'' mode the slit settings were 20~mm
vertically and 3~mm horizontally, corresponding to a $q_z$
resolution of 0.32~\AA $^{-1}$. For all experimental arrangements,
the signal originating at the surface was separated from the bulk
diffuse background by subtracting intensities measured with the
detector moved approximately $\pm $0.4$^{\circ }$ (corresponding
to $\approx 0.67\,{\AA}^{-1}$) transverse to the reflection
plane.\cite{regan97b}

The linearity
of the PSD was checked by comparison of diffuse scans
taken with the PSD  to diffuse data taken  by moving
the scintillation detector within the reflection plane, and by measuring the specular
reflectivity with the stationary  PSD in a single-channel mode. As we will discuss
below, once scaled for the different  resolution, the different types of
scans perfectly overlap each other.

 Indium of 99.9995\% purity was contained in a
 Molybdenum pan within an Ultra High Vacuum
(UHV) chamber to maintain a clean surface.\cite{XPS} Details of
the UHV chamber design can be found elsewhere.\cite{regan97b} The
sample was heated  above the melting point of liquid In to $170\pm
5~^{\circ }$C. After a three day \index{bakeout} bakeout, the
pressure was $8\times 10^{-10}$~Torr, with a partial oxygen
pressure $< 8\times 10^{-11}$, for which the formation time  for
an oxide monolayer is calculated to be several days.\cite{roth90}
Although patches of oxide were present on the In ingot when it was
introduced into the chamber, this  oxide was removed by Argon ion
sputtering (2.5~keV, 20~$\mu $A ion current) at an Ar partial
pressure of about $5\times 10^{-5}$~Torr. The impact of the ion
beam induced
 mass flow along the surface that
transported the oxide patches towards the ion beam, thus cleaning
the entire liquid--vacuum interface and not only the area hit
directly by the Ar$^{+}$
    ions.\cite{fine81}
Data presented here were always
taken within one to two hours of sputter cleaning the sample.
Since
data acquired without sputtering but within a few hours reproduced each
other, and because of the unique growth properties of oxide on liquid In
which  will  be presented below, we are confident that the measured
scattering is intrinsic to a clean liquid Indium surface.

A major improvement over experiments described previously is the
incorporation of active \index{vibration isolation} vibration
isolation. Previous measurements eliminated mechanically induced
surface waves by utilizing the viscous drag at the bottom of thin
samples. These samples had curved surfaces, requiring a time
consuming technique to measure the
    reflectivity\cite{regan97b} and
making it very difficult to measure at \index{x-ray!grazing
incidence diffraction} grazing incidence, important for the
detection of \index{in-plane ordering} in-plane ordering. In this
work, the rigid UHV chamber and ion pump assembly \index{pump!ion}
were mounted on an active vibration
    isolation unit,\cite{JRS}
similar to that used previously in our studies of liquid
    Hg.\cite{magn95}
With this system we were able to use a sample pan 5~mm deep and
60~mm in diameter, resulting in flat samples where the angular
deviation of the surface normal from the vertical across the
surface was negligible compared to the critical angle for total
\index{external reflection} external reflection of x-rays.

\section{Theory}

\begin{figure}[tbp]
\unitlength1cm \epsfig{file=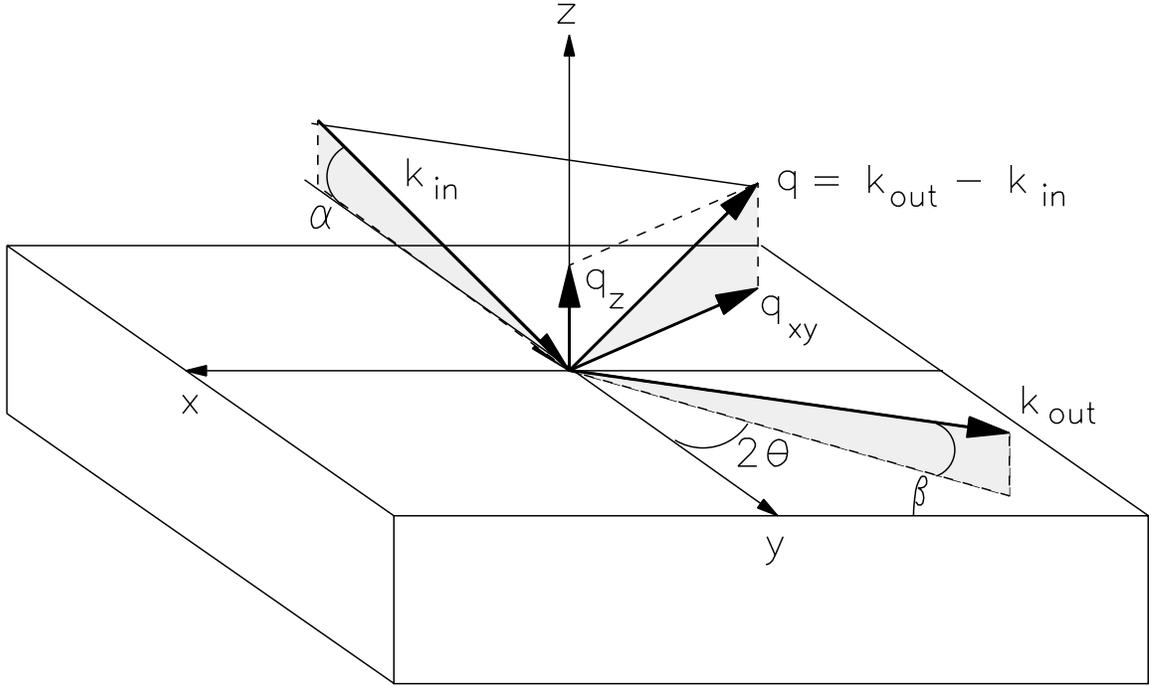, angle=90,
width=1.0\columnwidth} \caption{ Sketch of the geometry of x-ray
scattering from the liquid surface with $\alpha$ and $\beta$
denoting incoming and outgoing angle, the incoming and outgoing
wavevector $\vec{k}_{in}$ and $\vec{k}_{out}$ respectively and the
azimuthal angle $2\theta$. The momentum transfer $\vec{q}$ has an
in-plane component $q_{xy}$ and a surface-normal component $q_z$.
} \label{in_fig:geom}
\end{figure}

The scattering geometry is given in Figure~\ref{in_fig:geom},
defining the liquid surface to be lying in the $x$-$y$ plane, with
x-rays incident at an angle $\alpha$ and collected by the detector
at an elevation angle $\beta$ and azimuthal angle $2\theta$.
The momentum transfer $\vec{q}$ can be decomposed
into surface normal, $q_{z}$, and surface parallel, $q_{xy}$, components
given by:
\begin{equation}
    q_{z}=\frac{2\pi }{\lambda }(\sin \beta +\sin \alpha )
    \text{ \ \ \ \ and \ \ \ \ \ }
    q_{xy}=\frac{2\pi }{\lambda } \sqrt{\cos ^{2}\alpha +
    \cos ^{2}\beta -2\cos \alpha \cos \beta \cos 2\theta } \: .
\end{equation}
In the following, we develop a formula for the \index{differential
cross section} differential cross section $d\sigma/d\Omega$ for
x-ray scattering from a rough and structured liquid surface. The
general expression for the  \index{differential cross section}
differential cross section for x-ray scattering from a three
dimensional electron distribution $\rho(\vec{r})$ at a given
location $\vec{r}$
 can be expressed in terms of the electron  density correlation function
$\langle \rho(\vec{r}) \rho(\vec{r}\prime)\rangle =
\langle \rho(\vec{r}-\vec{r}\prime) \rho(0)\rangle$:\cite{guinier}
\begin{equation}
    \frac{d\sigma }{d\Omega }= V \left( \frac{e^{2}}{mc^{2}}\right)^2
\int d^3 (\vec{r} - \vec{r}\prime)  \langle \rho(\vec{r} -
\vec{r}\prime) \rho(0)\rangle \exp\left[\imath  \vec{q} \cdot
(\vec{r} - \vec{r}\prime) \right] \label{in_general}
\end{equation}
where $V$ is the illuminated volume.
If the  density distribution  is homogeneous within the $x$-$y$ plane but inhomogeneous
normal to the surface, the density correlation function depends on  the different
positions $\vec{r}_{xy} - \vec{r}_{xy}\prime $
on the surface
and the distances $z$ and $z\prime$ from the surface.
For an x-ray beam of a cross sectional
area $A_0$ incident on the surface at an angle $\alpha$ relative to the
$x$-$y$ plane, the illuminated area is $A_0/\sin(\alpha)$
and the differential cross section can be written:\cite{bras88}
\begin{equation}
\frac{d\sigma }{d\Omega }=  \frac{A_0}{\sin(\alpha)} \left(
\frac{e^{2}}{mc^{2}}\right)^2 \int dz dz\prime d^2 \vec{r}_{xy}
\langle \rho(\vec{r}_{xy},z) \rho(0,z\prime) \rangle
\exp\left[\imath q_z (z-z\prime) + \imath \vec{q}_{xy} \cdot
\vec{r}_{xy}  \right] . \label{in_surface}
\end{equation}
For a homogeneous liquid that is sufficiently far away from any critical
region, the only long wavelength excitations that give rise to significant scattering
are thermally excited surface capillary waves. In this case,  there is some
length $\xi$ such that for $|\vec{r}_{xy}| > \xi$ the density density
correlation function can be expressed in terms of a  density profile
$\tilde{\rho} [z - h( \vec{r}_{xy} )]$ defined relative to the local position
of the liquid surface $h ( \vec{r}_{xy} )$
and  capillary wave induced relative variations
of the height, $h(\vec{r}_{xy}) - h (\vec{r}_{xy}\prime) $.\cite{comment1}
\begin{equation}
\label{in_height} \langle \rho(\vec{r}_{xy},z) \rho(0,z\prime)
\rangle  \stackrel{ |\vec{r}_{xy}| > \xi}{\longrightarrow}
\tilde{\rho} (z - [h(\vec{r}_{xy} ) - h (0)] ) \tilde{\rho}
(z\prime)
\end{equation}
Defining $\delta \rho(\vec{r}_{xy},z) = \rho(\vec{r}_{xy},z) -  \tilde{\rho}(z-h(\vec{r}_{xy}))$,
the differential cross section from a liquid surface can  now be written:
\begin{eqnarray}
\frac{d\sigma }{d\Omega }=  \frac{A_0}{\sin(\alpha)} \left( \frac{e^{2}}{mc^{2}}\right)^2
\int dz dz\prime d^2\vec{r}_{xy}
\exp\left[\imath q_z (z-z\prime) + \imath \vec{q}_{xy} \cdot \vec{r}_{xy}\right] \nonumber \\
\times \left\{ \tilde{\rho} (z - [h(\vec{r}_{xy} ) - h (0)] )
\tilde{\rho} (z\prime) + \langle \delta \rho(\vec{r}_{xy},z)
\delta \rho(0,z\prime)\rangle \right\} . \label{in_sepacross}
\end{eqnarray}
The second term in the integrand is only non-vanishing for regions
$|\vec{r}_{xy}| \leq \xi $. This term gives rise to both the bulk
 diffuse scattering and surface
scattering at momentum transfers having large $q_{xy}$
components.\cite{bras88} The effect of this term can be separated
experimentally from the specular reflection and may be omitted in
the following discussion of scattering close to the specular
condition. A change of variables yields
\begin{equation}
\frac{d\sigma }{d\Omega } \approx  \frac{1}{16\pi^2}
\left( \frac{q_c}{2} \right)^4 \frac{A_0}{\sin(\alpha)}
\frac{| \Phi (q_z)|^2}{q_z^2}
\int_{|\vec{r}_{xy}| > \xi} d^2 \vec{r}_{xy}  \langle
\exp \left\{ \imath q_z  h(\vec{r}_{xy})  \right\} \rangle
\exp \left[ \imath \vec{q}_{xy} \cdot \vec{r}_{xy} \right]
\end{equation}
where $q_c$ is the \index{critical!angle} critical angle for total
\index{external reflection} external reflection of x-rays and the
surface structure factor
\begin{equation}
\Phi (q_z) = \frac{1}{\rho_{\infty}} \int dz \frac{d\tilde{\rho}
(z) }{dz} \exp(\imath q_z z) = \frac{-\imath q_z}{\rho_{\infty}}
\int dz  \tilde{\rho} (z)  \exp(\imath q_z z)
\label{in_eq:structure}
\end{equation}
is the Fourier transform of the  local or  intrinsic density profile
$ \tilde{\rho} (z) $,
 independent of $\vec{r}_{xy}$.
In this equation, $\rho_{\infty}$ is the bulk electron density.

For the case of thermally excited capillary waves on a liquid
surface, the height fluctuations can be characterized by their
statistical average $\langle | h(\vec{r}_{xy}) |^2
\rangle$.\cite{bras88} \index{Sinha, Sunil} Sinha et al. have
shown that integration over these height fluctuations yields the
following dependence of the scattering on $q_{xy}$ and
$q_z$:\cite{sinha88}
\begin{equation}
\int_{|\vec{r}_{xy}| > \xi} d^2 \vec{r}_{xy}  \langle \exp \left\{
\imath q_z  h(\vec{r}_{xy})  \right\} \rangle \exp \left[ \imath
\vec{q}_{xy} \cdot \vec{r}_{xy} \right] =
\frac{C}{q_{xy}^{2-\eta}} \label{in_sinha}
\end{equation}
with $C$ to be determined and
$$
\eta = \frac{k_BT}{2\pi \gamma} q_z^2 .
$$
The value of $C$ can be determined
by integration over all
 surface capillary modes
having wavevectors smaller than the upper wavevector cutoff
$ q_{max} \equiv \pi/\xi $:
\begin{eqnarray}
C = \int_{|\vec{q}_{xy}| < \pi/\xi} d^2 \vec{q}_{xy}
\int_{| \vec{r}_{xy} | > \xi} d^2 \vec{r}_{xy}
\langle \exp \left\{ \imath q_z  h(\vec{r}_{xy})  \right\} \rangle
\exp \left[ \imath \vec{q}_{xy} \cdot \vec{r}_{xy} \right] \nonumber \\
\approx 4\pi^2 \lim_{ \vec{r}_{xy} \rightarrow 0}
 \langle \exp \left\{ \imath q_z h(\vec{r}_{xy})  \right\} \rangle = 4\pi^2
\end{eqnarray}
Therefore, the properly normalized \index{differential cross
section} differential cross section for scattering of x-rays  from
a liquid surface is:

\begin{equation}
    \frac{d\sigma }{d\Omega } = \frac{A_{0}}{\sin \alpha }
    \left( \frac{q_c}{2} \right) ^{4}\frac{k_B T}{16\pi ^2 \gamma }
    \left| \Phi (q_z)\right| ^2
    \frac{1}{q_{xy}^2} \left( \frac{q_{xy}}{q_{\mbox{\scriptsize max}}}
    \right) ^\eta \: .
\label{in_eq:final}
\end{equation}

The intensity measured at a specific scattering vector $\vec{q}$
is obtained by integrating Eq.~\ref{in_eq:final} over the solid
angle $d\Omega$ defined by the detector acceptance of the
experiment. For the specular reflection $q_{xy} = 0$
($\alpha=\beta$ and $2\theta =0$), the integral is centered at
${q}_{xy} = 0$. The projection of the detector resolution onto the
$x$-$y$ plane is rectangular\cite{bras88} and the above mentioned
integration has to be done numerically. However, an analytical
formula can be obtained for the specular reflectivity if the
detector resolution is assumed to be a circle of radius $q_{res}$,
 independent of $q_z$\cite{regan97b}:
\begin{equation}
    \frac{R(q_z)}{R_f(q_z)} = \left| \Phi (q_z) \right| ^2
    \left( \frac{q_{\mbox{\scriptsize res}}}{q_{\mbox{\scriptsize max}}}
    \right) ^\eta = \left| \Phi (q_z) \right| ^2
    \exp [ - \sigma _{\mbox{\scriptsize cw}}^2q_z^2 ]
\label{in_eq:circle}
\end{equation}
where
$$
R_f (q_z) = \left| \frac{q_z - \sqrt{q_z^2 -q_c^2}}{q_z +
\sqrt{q_z^2 -q_c^2}} \right|^2 \approx \left(\frac{q_c}{2
q_z}\right)^4 \ \  \ \ { for}\ \ q_z \stackrel{>}{\sim} 5q_c
$$
is the Fresnel reflectivity from classical optics for a flat ($h(\vec{r}_{xy}) = h (0)$
for all $h(\vec{r}_{xy})$)
and structureless ($|\Phi (q_z)|^2 =1$) surface. The quantity
$\sigma _{\mbox{\scriptsize cw}}$
denotes the capillary wave roughness:
\begin{equation}
\sigma _{\mbox{\scriptsize cw}}^2 = \frac{k_B T}{2\pi \gamma }
    \ln \left(
    \frac{q_{\mbox{\scriptsize res}}}{q_{\mbox{\scriptsize max}}
    }\right) \:  .
\label{in_eq:cw}
\end{equation}
In the experiments presented here, the measured specular
reflectivity was analyzed by numerical integration of the
theoretical model over the rectangular resolution defined by the
detector slits. In practice, the specular reflection is measured
with coarse resolution and the results based on
Eq.~\ref{in_eq:circle} fall within 5\% of the exact values
obtained from integrating over Eq.~\ref{in_eq:final}. On the other
hand, the off specular diffuse scattering is measured with much
finer resolution and near the specular peak, $\vec{q}_{xy} = 0$,
it has to be analyzed by numerical integration of
Eq.~\ref{in_eq:final} over the slit defined resolution. In
addition, numerical integration of Eq.~\ref{in_eq:final} is
necessary to obtain the correct ratio of specular reflectivity to
off-specular   diffuse scattering.

In Eq.(\ref{in_eq:structure}) we have introduced a local structure
factor which is the Fourier transform of the local or intrinsic
density profile. An alternative representation is to define a
macroscopically averaged density profile $\langle \rho(z)
\rangle_T$ such that
$$
\frac{1}{\rho_{\infty}}
\int dz \frac{\langle \rho (z)\rangle_T}{dz} \exp(\imath q_z z) = \Phi (q_z) \exp(-1/2 \sigma_{cw}^2 q_z^2)
$$
This macroscopically averaged density profile, $\langle \rho (z) \rangle_T$, can be shown to be equal to
the convolution of the local density profile, $ \tilde{\rho} (z) $, with the associated
Gaussian distribution of height fluctuations characterized by
$$
\sigma_{cw}^2 = \langle \left| h(\vec{r}_{xy}) \right|^2 \rangle = \frac{k_BT}{2\pi\gamma}
\ln \left(\frac{1}{|\vec{r}_{xy}| q_{max}}\right) .
$$
Unfortunately, it is difficult to extract a precisely defined $\langle \rho (z) \rangle_T$ from a single
reflectivity measurement because of the above mentioned problem of the variation of the resolution with
$q_z$.

A second difficulty with reflectivity  measurements, even when $| \Phi (q_z)|^2$ can be separated
from the thermal effects, is that the  direct inversion of $| \Phi (q_z)|^2$ in order to obtain
 $ \tilde{\rho} (z) $  is not possible because the phase information necessary to perform the
Fourier transform in Eq.~\ref{in_eq:structure} is lost in any
intensity measurement. The common practice to overcome this
problem is to assume a model for the density profile, which is
then Fourier transformed and fitted to the experimentally
determined structure factor. In this study, the liquid metal is
modelled by layers of different electron densities parallel to the
surface. The layers represent planes of atoms, separated by a
distance $d$, and having increasing width the further they are
below the surface. Mathematically, the electron density is
described  by  a semi-infinite sum of Gaussians, normalized to the
bulk density $\rho _{\infty }$:
\begin{equation}
     \tilde{\rho} (z) =\rho _\infty \sum _{n=0} ^\infty
    \frac{d/\sigma _n}{\sqrt{2\pi }}
    \exp \left[ -(z-nd)^{2}/2\sigma _n ^2 \right]
    \otimes  F_{\mbox{\scriptsize In}}(z).
\label{in_eq:rho}
\end{equation}
Here $\otimes $ denotes convolution, $F_{\mbox{\scriptsize
In}}(z)$ is the atomic x-ray scattering \index{form factor} form
factor of In, and $\sigma _n ^2 = n\bar{\sigma}^2 + \sigma
_{0}^{2}$, where $\bar{\sigma}$ and $\sigma _0$ are constants.
This form for $\sigma _n$ produces a quadratic increase in the
Gaussian width with distance $z$ below the surface, so that the
parameter $\bar{\sigma}$ is related to the decay length for
surface layering and the model density approaches the bulk density
$\rho_{\infty}$ for $\sigma_n \gg d$. An advantage of using
Gaussian functions to model $ \tilde{\rho} (z) $ is that an
expression for $\langle \rho (z) \rangle_T$  is obtained from
Eq.~\ref{in_eq:rho} by replacing $\sigma_n$ with $\sigma_n^T$ and
defining $(\sigma_n^T)^2 = \sigma_{cw}^2 + \sigma_n^2$. Note that
the effective width of the individual layers becomes
$(\sigma_n^T)^2$ which explicitly demonstrates the additivity of
the different contributions to  the surface roughness from
capillary modes (for which ${q}_{xy} < q_{max}$) and short
wavelength modes (for which ${q}_{xy}  > q_{max}$) that are
incorporated into the model for $ \tilde{\rho} (z) $.

\section{Experimental Results}

\begin{figure}[tbp]
\unitlength1cm \epsfig{file=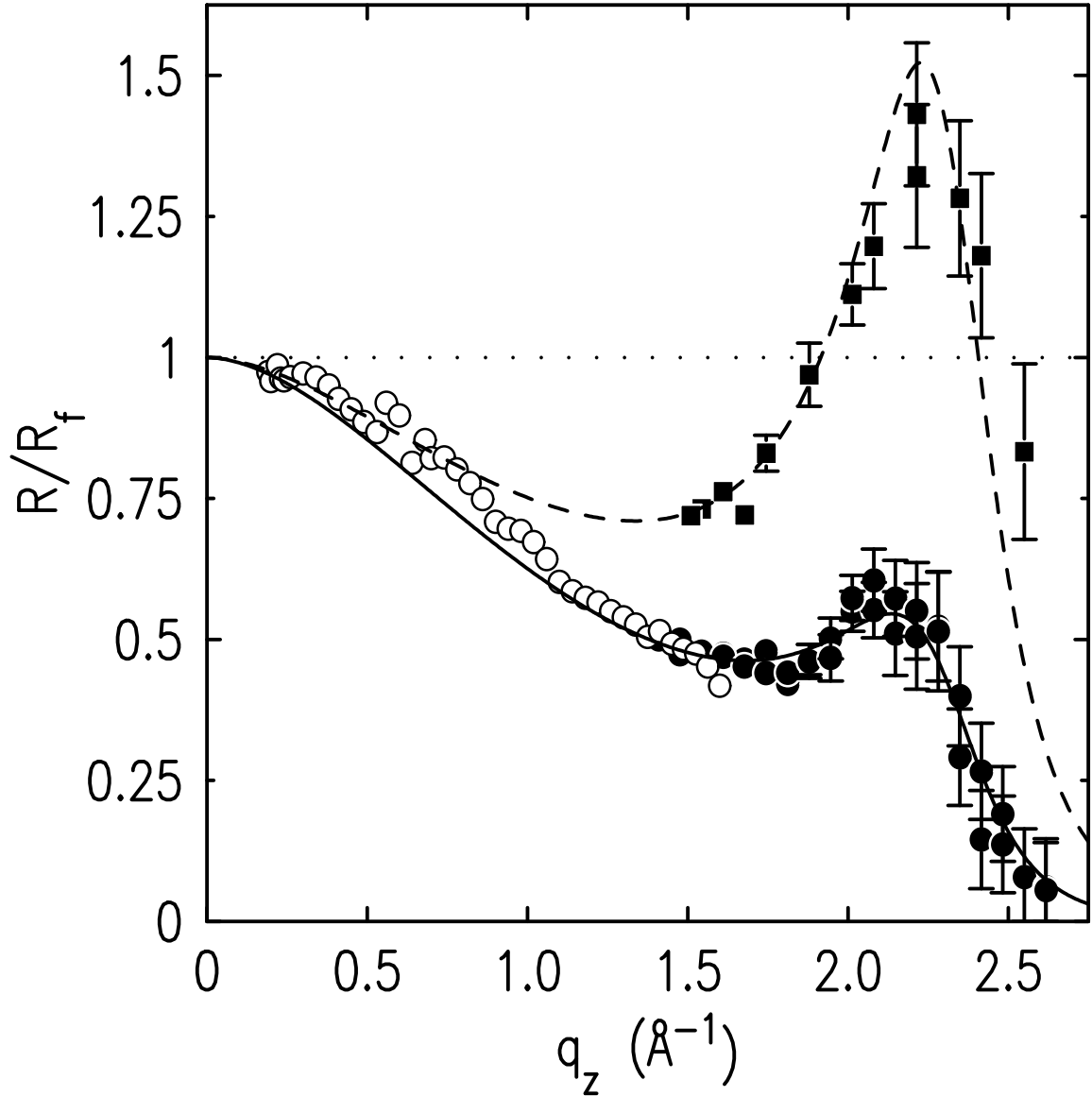, angle=90,
width=1.0\columnwidth} \caption{Specular x-ray reflectivity for
liquid Indium at 170~$^\circ$C taken with vertical detector
resolutions of 0.06~\AA$^{-1}$ (scintillation detector; open
circles: X22B, closed circles: X25)
 and 0.32~\AA$^{-1}$ (PSD in single channel mode; squares: X25).
The reflectivity is normalized to the Fresnel reflectivity $R_f$
of a flat In surface. Solid lines are fits as described in the
text. Data from X22B was not incorporated in any fits.}
\label{in_fig:refl}
\end{figure}

The specular reflectivity measured from liquid In at
$170\pm 5$~$^\circ$C
demonstrates the presence of surface layering.
    Figure~\ref{in_fig:refl} shows the
reflectivity as a function of $q_z$, normalized to the Fresnel
reflectivity of a sharply terminated In surface with a
\index{critical!angle} critical angle $q_c=0.052$~\AA $^{-1}$. The
circles show the data taken with the scintillation detector, with
a $q_z$ resolution of 0.06~\AA $^{-1}$. Open circles represent the
small angle data taken at X22B ($\lambda$=1.24~{\AA}) whereas the
closed circle data have been taken at X25 ($\lambda$=0.65~{\AA}).

 Filled
squares represent data taken with the PSD in a single channel
mode, with the vertical resolution coarser by a factor of 5. The
large increase in signal with decreasing resolution immediately
demonstrates the significance of the diffuse scattering. The
reflectivity at small and intermediate wavevectors falls below the
Fresnel value calculated for an ideally flat and abruptly
terminated surface. This is due to the roughness of the liquid In
surface which scatters photons out of the specular condition. The
most prominent feature in these data is the broad peak centered
near 2.2~\AA $^{-1}$. This peak is due to constructive
interference of x-rays  from layers ordered parallel  to the
surface. This very distinctive structural feature of liquid metal
surfaces has been observed earlier for
    Hg\cite{magn95} and
    Ga.\cite{regan95}
The two data sets show the same features, although as a
consequence of integrating the diffuse scattering over coarser resolution,
the intensity is greater for data taken with the PSD.

The lines in
    Figure~\ref{in_fig:refl}
illustrate fits of the density profile model discussed above to these data.
Three surface layering parameters, $d$, $\overline{\sigma}$, and $\sigma _0$,
determine the form of $\Phi (q_{z})$, which is then used in
    Eq.~\ref{in_eq:final}
along with $T= 170^\circ $C and the experimentally determined
value for the surface tension of liquid In at that temperature, $
\gamma = 0.556$~N/m.\cite{iida93} As discussed above, we assign
the short wavelength cutoff for capillary waves, $q_{max} \approx
\pi/\xi$ , to be $\sim 1 {{\AA}^{-1}}$ where $\xi$ was taken to be
the nearest neighbor atomic distance in the bulk
melt.\cite{comment2} The integral over
    Eq.~\ref{in_eq:final}
is then performed numerically over the known resolution volume.
Surface layering parameters
obtained from the scintillation detector data are
$d=2.69\pm 0.05$~\AA ,
$\overline{\sigma} = 0.54 \pm 0.06$~\AA , and
$\sigma _0 = 0.35 \pm 0.04$~\AA\ (solid line in
    Figure~\ref{in_fig:refl}).
Essentially the same parameters (with larger error due to poorer
statistics) result from a fit of the model to
the PSD data (broken line in
     Figure~\ref{in_fig:refl}).

\begin{figure}[tbp]
\centering
\includegraphics[width=1.0\columnwidth]{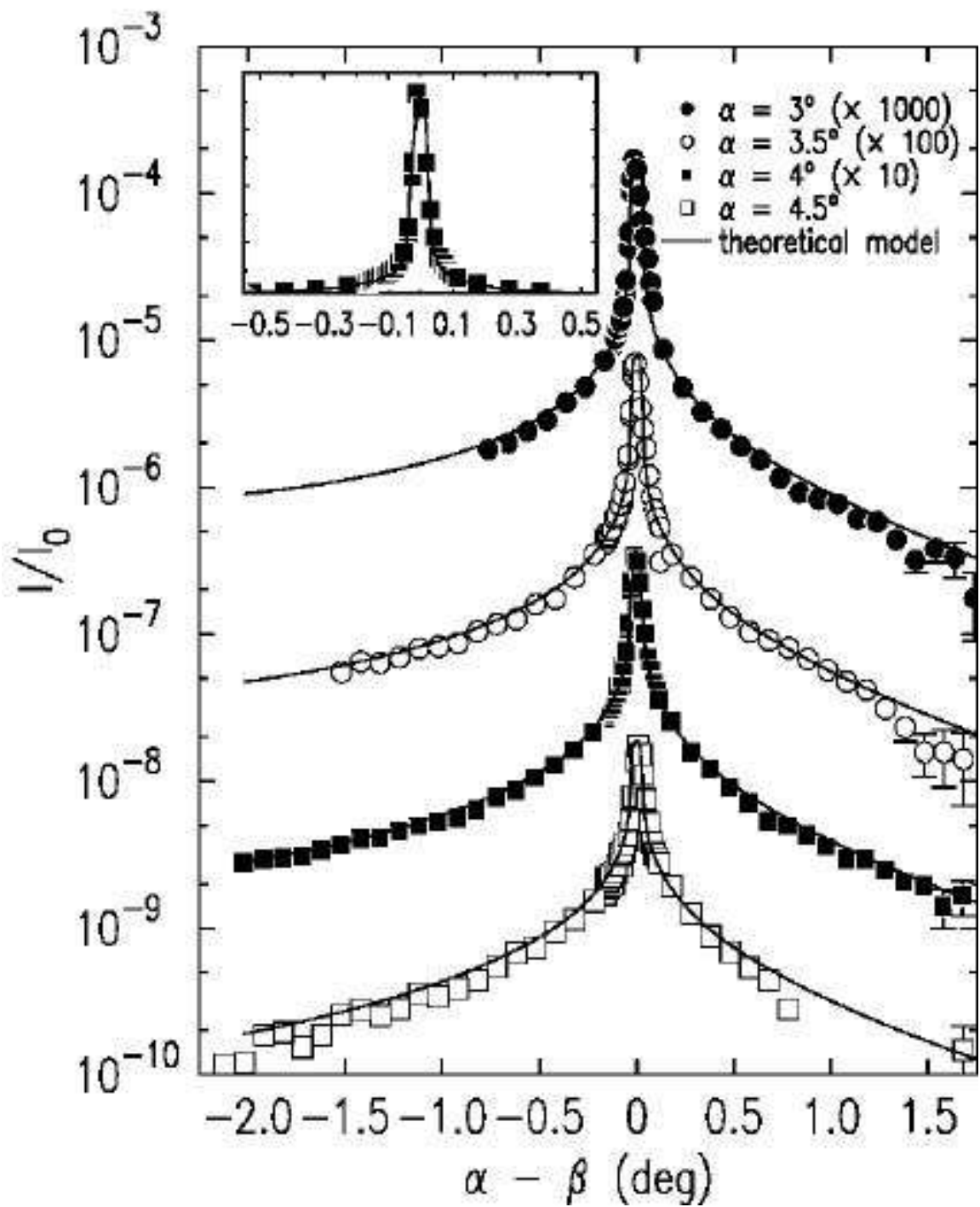}

\caption{Diffuse scattering as a function
of scattering angle $\protect\beta$ for different fixed incoming
angles $\protect\alpha$. Solid lines: Diffuse scattering
calculated  from the experimentally determined structure factor
with  no further adjustable parameters. Inset: linear plot
emphasizing the fit near the specular peak for $\protect\alpha =
4.5^{\circ}$%
} \label{in_fig:diff}
\end{figure}

Although the resolution dependence of the reflectivity of liquid
metals provides a test of the thermal capillary wave theory
prediction, a more rigorous test is the measurement of the
spectral density of the off specular \index{x-ray!diffuse
scattering} diffuse scattering. We have measured diffuse intensity
over a $\beta $ range straddling the specular condition.
Intensities normalized to the direct beam are  shown in
    Figure~\ref{in_fig:diff}
for several choices of $\alpha $. Data at extreme values  of
$(\beta -\alpha) $ are limited by the intense background due to
bulk scattering at large $\beta$ and by the scattering geometry
at small $\beta$. The asymmetry of the wings centered around the specular
ridge ($\beta -\alpha =0)$ arises from the $\beta $ dependence
of both the exponent $\eta $ and the surface structure factor $\Phi (q_z)$.

The solid lines are obtained by calculating the intensity from
    Eq.~\ref{in_eq:final},
integrating numerically over the \index{resolution function}
resolution function and subtracting a similar calculation for the
background scan taken out of the reflection plane. The calculation
incorporates the surface structure factor $\Phi (q_z)$ determined
from the reflectivity measurements along with the other fixed
quantities ($\gamma $, $k_B T$, and $q_{\mbox{\scriptsize max}}$),
without further adjustable parameters. We find excellent agreement
between this model and the experimental data for the entire range
studied, including the wings
    (Figure~\ref{in_fig:diff}) and the specular region (inset of
    Figure~\ref{in_fig:diff}).
The implications of this agreement will be discussed
in the next section.

To study the oxidation properties of liquid Indium
we exposed the In surface to controlled amounts of oxygen through a
bakeable UHV leak valve.
During oxidation, macroscopic oxide clumps, large enough to
be observed by eye, formed  at the edges of the sample, while the
center remained clean. X-ray reflectivity measured during oxidation showed
no changes until the floating oxide regions grew large enough to reach the
area illuminated by the x-ray beam, at which point the macroscopically rough surface
scattered  the reflected signal away from the specular condition.
This result is in sharp contrast to the
formation of the  highly uniform, 5~\AA\  thick, passivating oxide layer
we previously observed to form on liquid Ga under the influence
of the same amount of oxygen.\cite{regan97}

\section{Discussion}

\begin{figure}[tbp]
\epsfig{file=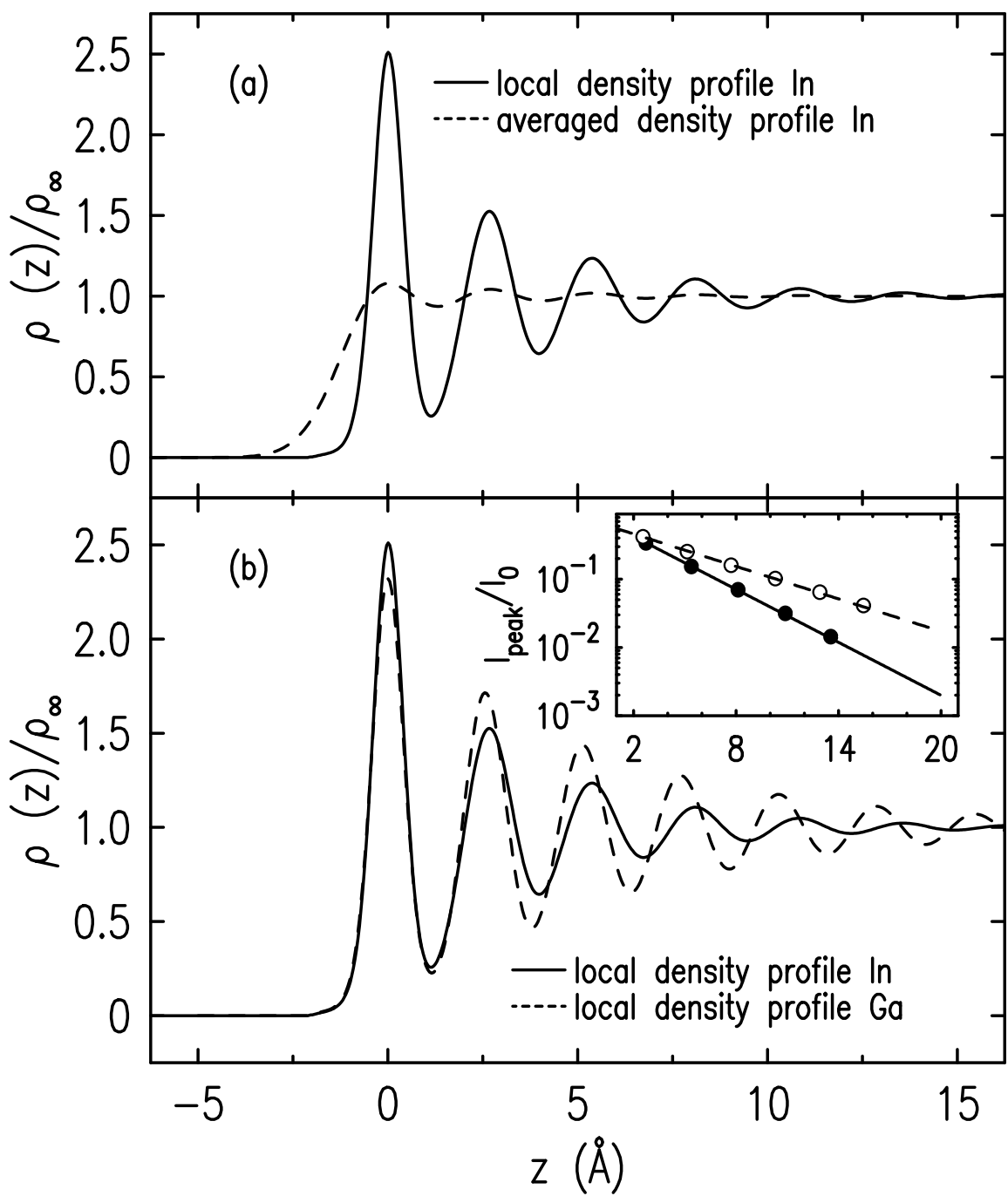, angle=90, width=1.0\columnwidth}
\caption{ (a) Comparison of the local real space density profile
for liquid Indium (---) with the thermally averaged density
profile for the same metal (- - -). The averaged density profile
is directly accessible by experiment and has been measured at
170$^{\circ}$\,C. (b)  Real space local density profile for liquid
Gallium (- - -) and
liquid Indium (---) after the removal of thermal broadening. Densities $%
\protect\tilde{\rho} (z)$ are normalized to the bulk densities
$\protect\rho_{\infty}$ of liquid Gallium and Indium. Inset: Decay
of the amplitude of the surface--normal density profile for liquid
Ga (open circles) and liquid In (filled circles). The lines
represent the fit of this decay of the surface layering to the
form $\exp(-z/l)$ for liquid Ga (---) and In (- - -). }
\label{in_fig:rho}
\end{figure}

The macroscopic density profile extracted from the reflectivity
measurements is resolution dependent since the density
oscillations are smeared out by thermally activated capillary
waves as described in the Theory section. This smearing depends on
the temperature and surface tension and hence also varies for
different liquids. The  macroscopic density profile directly
extracted from the experiment is the  averaged density profile
$\langle \rho (z)\rangle_T$  shown in previous
publications.\cite{magn95,regan95,regan96a,regan97b} In order to
compare the intrinsic layering properties in different liquid
metals, it is necessary to remove these thermal effects that vary
from metal to metal and experiment to experiment and to obtain the
intrinsic or  local density profile $ \tilde{\rho} (z)$. Previous
temperature dependent reflectivity measurements  of liquid Ga
around the specular position have shown that the resolution
dependent roughness is well described by the capillary wave (CW)
prediction.\cite{regan96a} Similar conclusions were reached from
$T$-dependent measurements of liquid paraffins.\cite{ocko94} Here,
we employed a different approach  to verify that  the CW
prediction holds for liquid In, performing  diffuse scattering
measurements at a single temperature at different angles of
incidence. The magnitude and the angular dependence of the diffuse
scattering is  in full compliance with the predictions of the
capillary wave model without
 using any adjustable parameters. This is demonstrated by the perfect agreement
of the diffuse scattering profiles with the theoretical curves in
    Figure~\ref{in_fig:diff}. This allows us to directly calculate the local
density profile $ \tilde{\rho} (z) $ by simply setting $\eta =0$.
A comparison between the local density profile $ \tilde{\rho} (z)$
and the temperature averaged density profile $\langle \rho (z)
\rangle_T$  is made in Figure~\ref{in_fig:rho}(a).

In this
single-$T$ approach the effects of the intrinsic, $T=0$, roughness
$\sigma _0$ and the cutoff $q_{\mbox{\scriptsize max}}$ cannot
be unambiguously
    separated,\cite{gelf90}
as was possible for the $T$-dependent measurements in
    \index{alkanes} alkanes\cite{ocko94}
and     Ga.\cite{regan96a}
However, this does not affect our main conclusion that the surface
roughness, as probed by the DS, is entirely due to thermally activated
capillary waves.

The local density profile of liquid In is compared with that of Ga
in Figure~\ref{in_fig:rho}(b). Profiles derived from the extremal
parameters of fits to $R(q_z)$ and from extremal values for
temperature, surface tension and resolution function are
indistinguishable on the scale of the figure. The amplitude of the
first density peak for In is comparable to that of Ga. This
reflects the fact that in the region nearest the surface, the
density oscillations are determined primarily by $\sigma _0$,
which has similar values for the two metals. Further towards the
bulk, layering is seen to decay faster for In than for Ga. We
quantify this observation by the excellent fit of  the peak
amplitude of the electron density profile to the form $\exp (-z/l
)$. The decay lengths $l =3.5$~{\AA}$\pm 0.6$ for In and $l
=5.5$~{\AA}$\pm 0.4$ for Ga differ by an amount well outside  of
the  experimental error. This is illustrated in the inset of
Figure~\ref{in_fig:rho}(b), a semilogarithmic plot of the maxima
of the density oscillations as a function of distance from the
first surface layer. The lines represent the fit of the peak
amplitude to the exponential form and the slope represents the
decay length.

\begin{figure}[tbp]
\centering
\includegraphics[width=1.0\columnwidth]{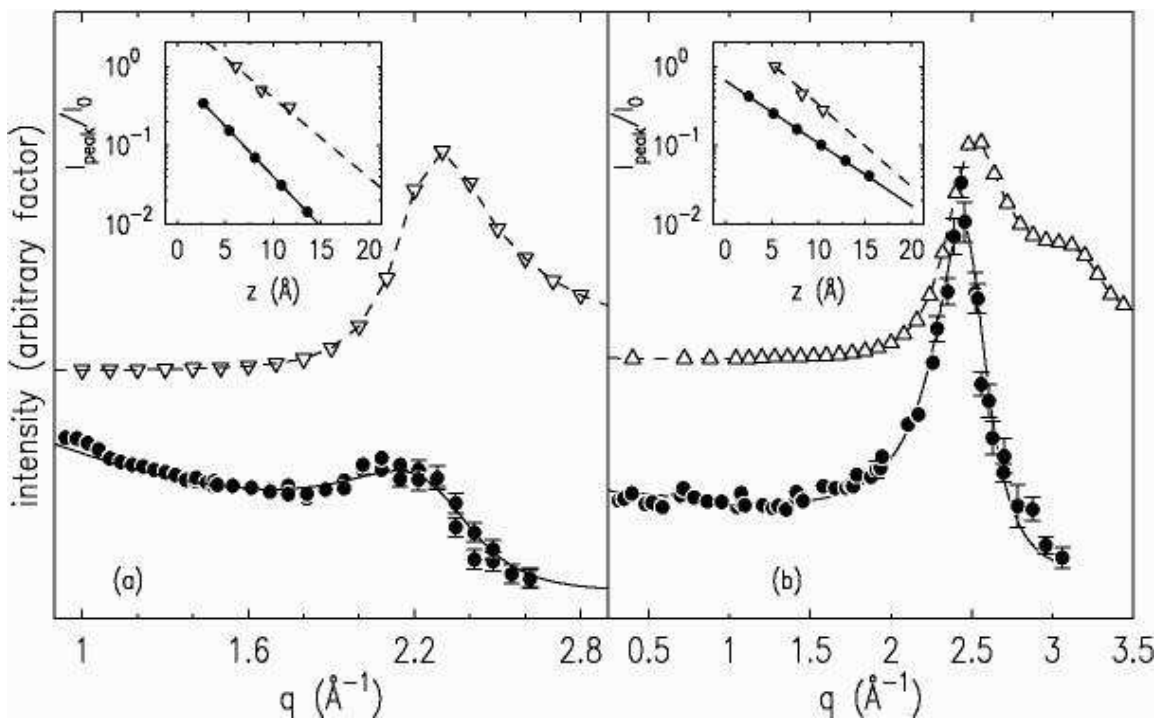}
\caption{(a) Bulk ($\triangle$) and surface ($\bullet$) structure
factors for liquid In. The bulk structure factor data are taken
from Orton et al.\protect\cite{orto66} The solid line is a fit of
the model explained in the text to the experimentally determined
surface structure factor. The broken line is a guide for the eye.
The inset compares the decay of the longitudinal surface density
oscillations (filled circles; solid line represents fit to an
exponential) to the decay of the bulk pair correlation function
(open triangles; broken line represents fit to exponential). The
coordinate $z$ of the x-axis represents the distance from the
surface in the case of the decaying surface layering and the
radial distance from a reference atom in the bulk liquid in the
case of the decaying bulk pair correlation.
 The data for the bulk pair correlation
function are taken from Waseda.\protect\cite{waseda80} (b) Bulk
($\triangle$) and surface ($\bullet$) structure factors for liquid
Ga. The bulk structure factor data are taken from Narten et
al.\protect\cite{nart72} Solid lines are fits to reflectivity
data, broken lines are guides for the eye. The inset compares the
decay of the longitudinal surface density oscillations (filled
circles; solid line represents fit to exponential) to the decay of
the bulk pair correlation function (open triangles; broken line
represents fit to exponential). The data for the bulk pair
correlation function are taken from Waseda.\protect\cite{waseda80}
}
 \label{in_fig:bulk}
\end{figure}

It is also instructive to compare the surface structure factor
measured in the reflectivity experiments to the bulk structure
factor measured by standard x-ray or \index{neutron scattering}
neutron diffraction (Figure~\ref{in_fig:bulk}(a): Indium;
Figure~\ref{in_fig:bulk}(b): Gallium). The widths of the peaks
shown are  inversely proportional to the decay lengths of the
corresponding correlation functions: the pair correlation function
$g(r)$\cite{waseda80} of the bulk and the layered density profile
$ \tilde{\rho} (z)  $  of the surface. Plotting the amplitudes  of
$g(r)$ and $ \tilde{\rho} (z)  $ on a semi-log scale (inset of
Figure~\ref{in_fig:bulk}) shows that they lie on straight lines.
This indicates that the correlations decay exponentially, and the
decay lengths are obtained as the  negative inverses of the
slopes. The inset plots show clearly that while the surface and
bulk decay lengths are about the same for In, in Gallium the
surface induced layering decays much more slowly than the bulk
pair correlation function. This is consistent  with the NFE nature
of In. For Ga, by contrast, the tendency towards covalency may
disrupt the hard sphere packing and enable surface correlations
that differ from those in the bulk.

Another piece of evidence suggesting that directional bonding in
the melt might have an effect on surface induced layering stems
from the ratio between the surface layer spacing and the bulk
nearest neighbor distance. This nearest neighbor distance is taken
from the analysis of the pair correlation function which is the
Fourier transform of the structure factor depicted in
Figure~\ref{in_fig:bulk}. Due to truncation problems and related
subtleties with the  data analysis, the  bulk nearest neighbor
distance cannot be directly inferred from the maximum in the bulk
liquid structure factor and we use the nearest neighbor distance
given by Iida and Guthrie.\cite{iida93} For liquid In, this ratio
of surface layer spacing to bulk nearest neighbor distance is
2.69~{\AA}/3.14~{\AA} = 0.86, close to the value of $\sqrt{2/3}
\approx 0.82$ for the hard sphere packing that would be expected
for a NFE liquid metal.
 For liquid Ga, this ratio
is larger, 2.56~{\AA}/2.78~{\AA} = 0.92. In this case, formation
of directional bonds, eventually leading to $ Ga_2$ dimers, may
prevent close packing.

Tomagnini et al.  recently considered a possible relationship
between surface induced layering in LM and the stability of
crystal facets at metal surfaces. Although
\index{surface!premelting} premelting of crystalline interfaces is
quite common in non-metallic crystals, most metals have at least
one close packed face which does not premelt, remaining solid up
to the melting temperature\cite{toma98}. /index{Tomagnini}
Tomagnini et al. use \index{molecular Dynamics simulations}
Molecular Dynamics simulations to demonstrate that when the period
of the surface induced layering at the liquid surface, which they
assumed to be $2\pi/q_0$  where $q_0$ is the position for the peak
in the bulk liquid structure factor, is commensurate with the
distance between lattice planes along the normal to the crystal
facet, this particular facet is stabilized and resistant to
\index{surface!premelting} premelting. It would be interesting to
correlate the premelting properties of Ga and In crystals with our
results for the period of the surface layering and in particular,
to examine if the effect of directional bonding that affects the
surface structure in the liquid affects premelting of the solid as
well.

We see here that for both Ga and In the period of the layering at
the liquid surface is significantly longer than $2\pi/q_0$.
Further, for In the period is reasonably close to the value that
would have been expected between the close packed layers if the
crystal structure of NFE In was such a crystal. In fact the
crystal structure of In is tetragonal, space group I4/mmm.

Our \index{x-ray!diffuse scattering} diffuse scattering
measurements show that the roughness of the liquid In surface can
be attributed entirely to thermally excited capillary waves. This
indicates that there are no other detectable inhomogeneities on
the surface, such as might be caused by microscopic oxide patches.
This observation is in concert with independent experiments on
organic Langmuir monolayers on water. If the monolayer is
homogeneous\cite{masa,gour97}, the DS can be described by
Eq.(\ref{in_eq:final}) with no adjustable parameters and no excess
scattering is observed --- just as  is the case for metallic
liquid In. On the other hand, if the organic monolayer is
compressed beyond its elasticity limit and becomes inhomogeneous,
Eq.(\ref{in_eq:final}) no longer describes the experimentally
determined DS, and excess scattering is observed  that must
originate from sources other than thermally activated capillary
waves.\cite{masa} This demonstrates the viability of diffuse
scattering as a tool for  studying surface inhomogeneities. This
is of particular interest for investigations on liquid alloys
where --- depending on the type of alloy --- a rich surface
behaviour is expected, ranging from concentration fluctuations for
alloys displaying surface segregation to critical fluctuations for
alloys with a critical consolute
    point.\cite{tost98}

Controlled oxidation of the liquid metal surfaces reveals a
further striking difference between Ga and In. Although exposure
of liquid Ga to air or to a large amount of oxygen leads to the
formation of a macroscopically thick  and rough oxide film,
exposure to controlled amounts of oxygen under
\index{UHV!conditions} UHV conditions produces a well defined
uniform 5~\AA\ thick oxide
    film.\cite{regan97}
This film protects the underlying bulk phase to a certain
extent from further corrosion, as further dosage at low oxygen pressures
($<2 \times 10^{-4}$~Torr)
was found to have no effect on the oxide thickness or
coverage fraction. This pasivating of the surface  is rather unusual for solid metals, Al
being prominent among the few examples. Liquid In, unlike Ga, was not found
to form a passivating oxide film. Instead, macroscopic clumps of oxide formed
which did not wet the clean In surface.
This corrosion mechanism is rather common for solid metal surfaces,
the oxidation of Fe being the best known example. It is not clear to what
extent this fundamental difference in the mechanism for corrosion is
related to differences in the  structure  of the liquid surface and to what extent to the chemical
affinity for oxidation of the respective liquid metal.

\section{Summary}

We have measured the x-ray reflectivity and small angle off
specular surface diffuse scattering from liquid In at $170^\circ
$C. Our results can be quantitatively explained by the convolution
of thermally excited  surface waves and a temperature independent
surface structure factor, corresponding to theoretically predicted
surface layering. The absence of excess   diffuse scattering
beyond that due to thermally excited capillary waves demonstrates
that the liquid--vapor interface is homogeneous in the
surface-parallel direction. The intrinsic layering profile of
liquid In, obtained by removing the capillary wave roughening, is
compared to that previously reported for liquid Ga. For Ga,
surface layering persists  farther  into the bulk than  is the
case for In. This may be attributed to directional bonding in the
Ga bond stabilizing surface induced layering over a larger
distance than  is the case for liquid In. Further evidence
suggesting a correlation between the degree of covalency in the
melt and the surface structure stems from the observation that the
compression of the surface layer spacing relative to the bulk
nearest neighbor distance is close to the behavior expected for an
ideal hard sphere liquid for In but considerably different for Ga.
Controlled oxidation of liquid In results in the formation of a
macroscopic  rough oxide, unlike the passivating microscopic
uniform oxide film that forms on liquid Ga.



%% file: kna_all.tex
\chapter{Surface Structure of KNa alloy}

\section{Abstract}

In this chapter we describe an x-ray scattering study of the
microscopic structure of the surface of a liquid alkali metal. The
bulk \index{liquid!structure factor} liquid structure factor of
the \index{eutectic}  eutectic K$_{67}$Na$_{33}$ alloy is
characteristic of an ideal mixture, and so shares the properties
of an elemental liquid alkali metal. Analysis of off-specular
\index{x-ray!diffuse scattering} diffuse scattering and specular
x-ray reflectivity shows that the surface roughness of the K-Na
alloy follows simple capillary wave behavior with a surface
structure factor indicative of surface induced layering.
Comparison of the low-angle tail of the K$_{67}$Na$_{33}$ surface
\index{surface!structure factor} structure factor with the one
measured for liquid Ga and In previously suggests that layering is
less pronounced in \index{alkali metals} alkali metals. Controlled
exposure of the liquid to H$_2$ and O$_2$ gas does not affect the
surface structure, indicating that oxide and hydride are not
stable at the liquid surface under these experimental conditions.

 \section{Introduction}

The structure of the free surface of a liquid metal (LM) is
fundamentally different from that of a \index{liquid!dielectric}
dielectric liquid, due to the strong coupling between conduction
electrons and ion cores.\cite{harr87} At the liquid-vapor
interface of a LM, the \index{interactions!Coulomb} Coulomb
interaction between the free electron \index{Fermi!gas}  Fermi gas
and the classical gas of charged ions acts like an effective
\index{hard wall} hard wall and forces the ions into ordered
layers parallel to the surface. The existence of surface-induced
layering in  LM  has  been verified unambiguously by experiment
for liquid Hg\cite{mercury}, Ga\cite{gallium} and In.\cite{indium}
Layering is also present in liquid binary alloys, even though it
may be suppressed by surface segregation\cite{gaprb} or the
formation of surface phases.\cite{bunsen,srn}

Even though surface layering appears to occur in LM in general,
comparison of the detailed surface structure
reveals qualitative differences between different
LM.\cite{indium,mercurytemp}
In fact, very little is understood about how the
surface induced layering
predicted for an ideal LM
is affected by the details of the more complicated electronic  structure
exhibited by the polyvalent metals studied so far.
One important goal therefore is to study the surface structure
of those LM which show the most ideal  electronic structure
in the bulk phase, characterized by itinerant conduction electrons
which are only weakly perturbed by a small
ionic pseudopotential. These LM are
referred to as nearly-free electron or NFE-type liquid
    metals.\cite{ziman61}
In a previous study we have compared the surface structure of
liquid In, which is considered to be a NFE-type
liquid metal, with the surface structure of liquid Ga, which displays
a considerable degree of covalency in the
    bulk.\cite{indium}
However, even though liquid In has less tendency towards
covalent bonding than Ga,
it is trivalent, and the most simple models do not adequately
describe its electronic structure or metallic
    properties.\cite{ashcroft}
The most ideal NFE-type liquid metals are the monovalent alkali
metals (AM) which  have a half-filled $s$-band and an almost
spherical \index{Fermi!surface}  Fermi
    surface.\cite{march}
The metallic properties of the AM are explained by the most
simple solid state models such as the Drude
    theory.\cite{ashcroft}
The fact that the conduction band is primarily formed
from  $s$-electron states
and  that the electron-ion interaction can be modeled by a
weak pseudopotential have led to numerous
theoretical studies on the electronic structure.
In fact, as far as the surface structure
of LM is concerned, most of the theoretical studies and
computer simulations have been applied to liquid Na, K and
    Cs.\cite{harr87,alkalitheory}

A particularly interesting question concerns the possible correlation
between surface tension and surface structure. In general, LM  have
surface tensions that are at least an order of magnitude
higher than the surface tensions of all other types of
    liquids.\cite{srn}
Liquid \index{alkali metals} alkali metals are the only exception
to this pattern. For example, Ga and Cs melt at about the same
temperature
 but the surface
tension of Ga is greater than that of Cs by about an order of magnitude.
If the high surface tension of
liquid metals is related to surface induced layering, it would
be reasonable to expect that liquid AM should display weak
or no layering at the surface. This is contradicted by
computer simulations showing pronounced layering for liquid
    Cs.\cite{harr87}

\begin{figure}[tbp]
\epsfig{file=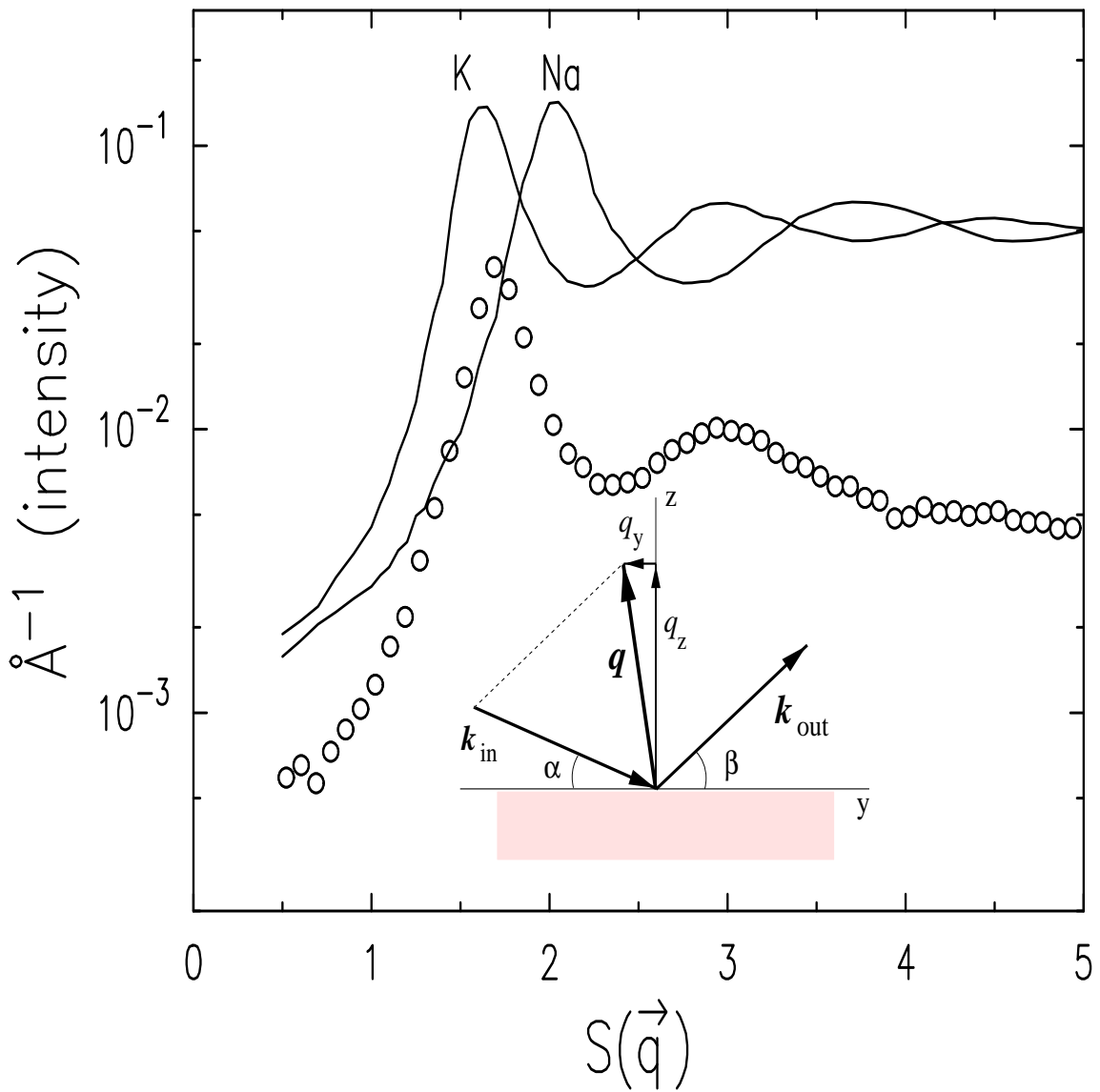, angle=0, width=1.0\columnwidth}
\caption{ Bulk liquid structure factor measured
 for liquid K$_{67}$Na$_{33}$ ($\circ$)
compared to the  bulk structure factors of pure liquid K and Na
from Ref.~14 (---). Inset: geometry for x-ray reflectivity and
off-specular diffuse scattering as described in the text.
}\label{kna_fig:phase}
\end{figure}

The alkali alloy  K$_{67}$Na$_{33}$ is more suitable
experimentally for such a study than are the elemental
liquid AM.
Among the elemental alkali metals,  Li deviates from NFE-type
    behavior,\cite{ashcroft,addison}
while the high  equilibrium vapor pressures
of the other elemental AM prevents the use of an
 ultra-high vacuum (UHV) environment over a long
period of time without noticeable evaporation of material.
Experiments under  UHV conditions are mandated by the high
reactivity of liquid AM towards water and oxygen.\cite{addison}
When liquid K and Na are alloyed in the \index{eutectic}  eutectic
composition (67~at\%K), however, a vapor pressure well below
$10^{-10}$ Torr is obtained. The low melting point of this alloy
($-12.6^\circ$C) is an additional advantage, since the thermal
surface roughness is considerably reduced.\cite{indium} This
binary alloy appears to be well  suited for emulation of  the
properties of an  elemental liquid AM. The total structure factor
$S(q)$ of bulk liquid K$_{67}$Na$_{33}$, shown in
Figure~\ref{kna_fig:phase}, is characteristic of random mixing,
without subpeak or asymmetry of the first peak.\cite{waseda} The
first peak of $S(q)$ for the alloy lies between those of pure  K
and Na, as expected for an ideal
    mixture.\cite{waseda}
More importantly, the electronic
structure of the homovalent K-Na alloys is as NFE-like as the
electronic structure of the pure
    components.\cite{shimoji}
Even if there were some homocoordination or surface enrichment of
K driven by  the  lower surface tension of K as compared to Na, it
would not be discernible in our structural measurements, since
x-ray scattering is sensitive to the distribution of the electron
density, which is virtually identical for K and Na (0.25~e/\AA$^3$
and 0.27~e/\AA$^3$ respectively).

 \section{Sample Preparation}
The eutectic alloy was prepared by mixing liquid K and Na in an
inert atmosphere  glove box ($<$~2~ppm $  H_2O$, $<$1~ppm $ O_2$).
The alloy  was transferred into a stainless steel reservoir and
sealed with a Teflon O-ring. The design of the reservoir is
similar to the one used in recent Hg
experiments.\cite{mercurytemp} The reservoir is connected to a
stainless steel UHV valve which is attached to a UHV flange. This
assembly was connected to the UHV chamber, which was baked out at
150$^{\circ}$~C until a vacuum in the $10^{-10}$~Torr range
resulted. After reaching this pressure, the Mo sample pan (20~mm
diameter, 1~mm depth) in the UHV chamber was sputtered clean with
Ar$^+$ ions. Sputtering  removes the native Mo-oxide layer and
facilitates the wetting of the sample pan by the AM.  After
sputtering, the UHV valve to the reservoir was opened and the
liquid alloy metered into the sample pan. As we will show below,
our measurements under these conditions are consistent with a
surface free of oxide on an atomic level.

 \section{Experimental geometry}
Experiments were  carried out using the liquid surface
\index{spectrometer} spectrometer at beamline X25  at the National
Synchrotron \index{synchrotron!source} Light Source, operating
with an x-ray wavelength $\lambda=$0.65~{\AA} and a resolution of
$\Delta q_z \approx 0.05$~{\AA}$^{-1}$. In the x-ray reflectivity
(XR) geometry, incident angle $\alpha$  and reflected angle $\beta
= \alpha$ are varied simultaneously with $\beta$ detected within
the plane of incidence. The background intensity, due mainly to
scattering from the bulk liquid, is subtracted from the specular
signal by displacing the detector out of the reflection plane by
slightly more than one resolution width.\cite{indium,gaprb,masa}
The specular reflectivity from the surface, $R(q_z)$, is a
function of the normal component, $q_z =
(4\pi/\lambda)\sin\alpha$, of the momentum transfer
$\vec{k}_{out}-\vec{k}_{in}=$ (0,0,$q_z$). $R (q_z)$ yields
information about surface roughness and surface-normal structure
and may be approximated as\cite{indium}
\begin{equation}
    R(q_z) =  R_f(q_z)
        \left| \Phi (q_z) \right| ^2
\left(\frac{q_{res}}{q_{max}}\right)^{\eta} \label{kna_eq:circle}
\end{equation}
where $R_f (q_z)$ is the Fresnel reflectivity of a flat,
laterally homogeneous surface.
The surface structure factor,  $\Phi (q_z)$,
is the Fourier transform of the gradient of the intrinsic
surface-normal  density profile.  The power-law term
with
$\eta=(k_BT/2\pi\gamma) q_z^2$  accounts
for roughening of the intrinsic density profile
by capillary waves (CW).
Scattering of x-ray photons by CW  outside of the detector
resolution $q_{res}$ result in a loss in observed
intensity. The short-wavelength cutoff $q_{max}$
is determined by the particle size $a$ with
$q_{max} \approx \pi/a$.
The observed reflectivity is reduced when either
 the temperature $T$  is increased
or the surface tension $\gamma$ is reduced.

For off-specular \index{x-ray!diffuse scattering} diffuse
scattering (DS), the incoming angle $\alpha$ is kept constant
while  $\beta$ is varied, straddling the specular condition (see
inset of Figure~\ref{kna_fig:phase}). The bulk diffuse background
has to be subtracted from the measured overall intensity. An
acurate description of the theoretical diffuse scattering requires
inclusion of the contribution of the surface to the bulk
background scattering.\cite{indium,masa} The lineshape of the DS
scans is determined principally by the power-law term in
Eq.~\ref{kna_eq:circle}.  Analysis of DS gives access to in-plane
correlations of the surface and surface inhomogeneities that might
be present.\cite{masa} Simultaneous analysis of XR and DS allows
for a consistent determination of surface roughness and structure.
As has been pointed out recently,\cite{petersrn} it is only for
$\eta < 2$ that the reflected intensity displays the cusp-like
singularity centered at the specular condition $\alpha = \beta$.
For $\eta \geq 2$, it is not possible to distinguish surface
scattering from bulk diffuse scattering since both are dominated
by correlations of the same length scale. The value of $q_z$ at
which $\eta$ reaches the limit of 2  at room temperature is about
1.7~\AA$^{-1}$ for K$_{67}$Na$_{33}$.

\section{X-Ray Diffuse Scattering Measurements}
\begin{figure}[tbp]
\epsfig{file=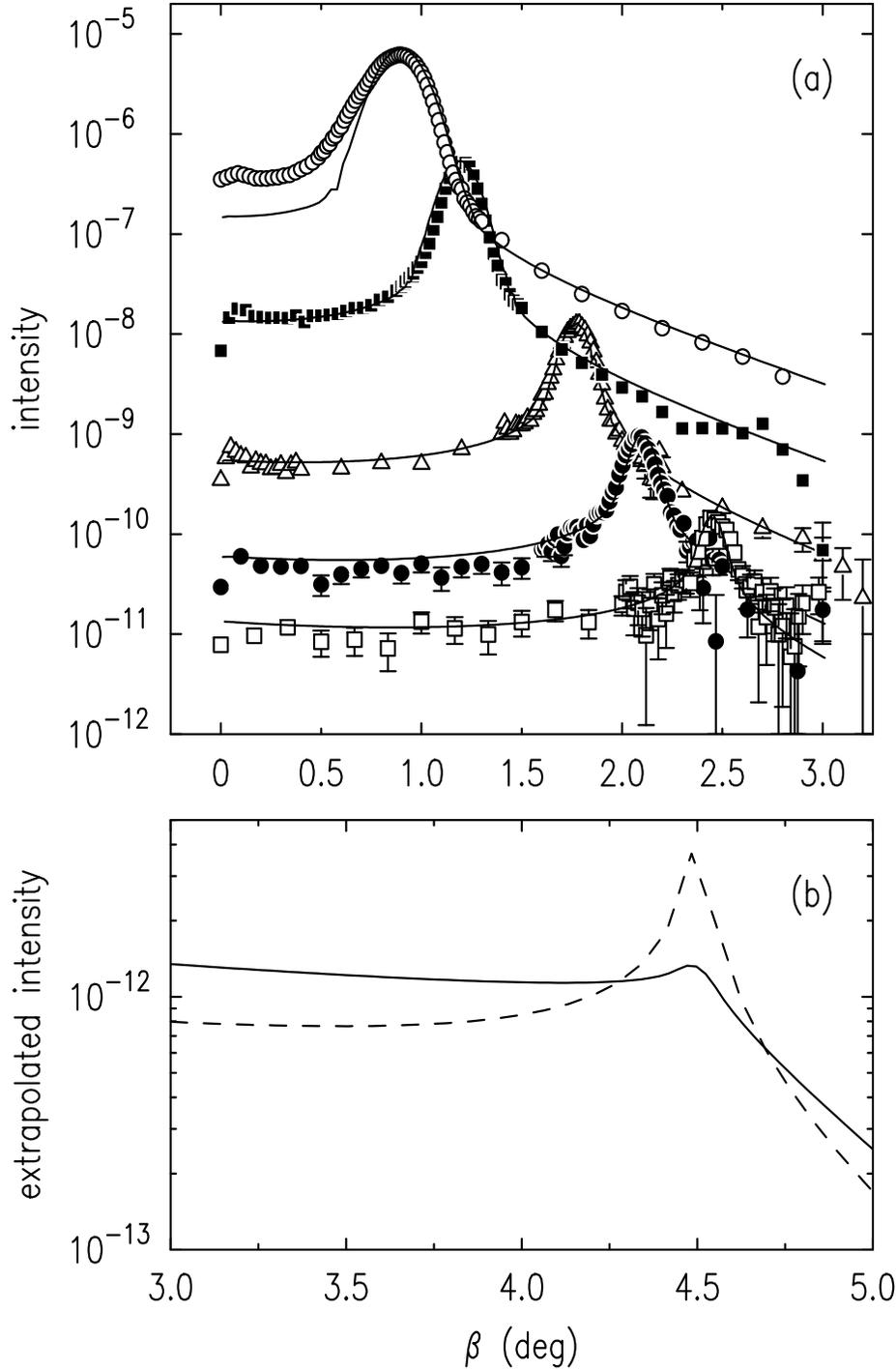, angle=0, width=0.8\columnwidth}
\caption{ (a) Off-specular diffuse scattering from the surface of
liquid $  K_{67}Na_{33}$ for different incoming angles $\alpha$:
(\protect$\circ$): 0.9$ ^{\circ}$ (multiplied by 25 for clarity),
(\protect\rule{1.5mm}{1.5mm}): 1.2$  ^{\circ}$ ($\times$ 5),
 (\protect$\triangle$): 1.77$  ^{\circ}$,
(\protect$\bullet$): 2.1$  ^{\circ}$ ($\times$ .2),
(\protect$\Box$): 2.45$  ^{\circ}$ ($\times$ .1). (b) Extrapolated
diffuse scattering (assuming no surface layering) for $\alpha =$
4.5~$  ^{\circ}$: (---) resolution of the present experiment
($\Delta q_z \approx$ 0.03~{\AA}$^{-1}$). ( - - - ) assuming the
resolution attainable at a low divergence x-ray source: ($\Delta
q_z \approx 3 \times 10^{-4}$~{\AA}$^{-1}$). (intensity $\times$
100). } \label{kna_fig:diffuse}
\end{figure}

Diffuse scattering measurements from the surface of $
K_{67}Na_{33}$ taken at different incoming angles $\alpha$
(Figure~\ref{kna_fig:diffuse}(a)) show that the liquid alloy has a
uniform surface roughened by capillary waves. The solid lines
through the data are calculated from CW theory (power-law term in
Eq.~\ref{kna_eq:circle}) with the surface roughness  entirely due
to thermally activated surface waves. This model has no adjustable
parameters, and the surface tension of 0.110$\pm0.003$~N/m is
consistent with macroscopic measurements.\cite{tension} The
agreement between the data and CW theory is excellent. The only
limitation is in the low-angle region ($\alpha=0.9^\circ$ and
$\beta < \alpha$), where the \index{x-ray!footprint} footprint of
the incoming beam is larger than the flat part of the sample.
Apart from this deviation, the sample shows no excess scattering
which would be expected from an inhomogeneous
surface.\cite{indium,masa} The absence of excess diffuse
scattering indicates that  the alloy surface is free of
microscopic patches or islands of possible contaminants such as
oxide or hydride.

 \section{X-Ray Reflectivity Measurements}
\begin{figure}[tbp]
\epsfig{file=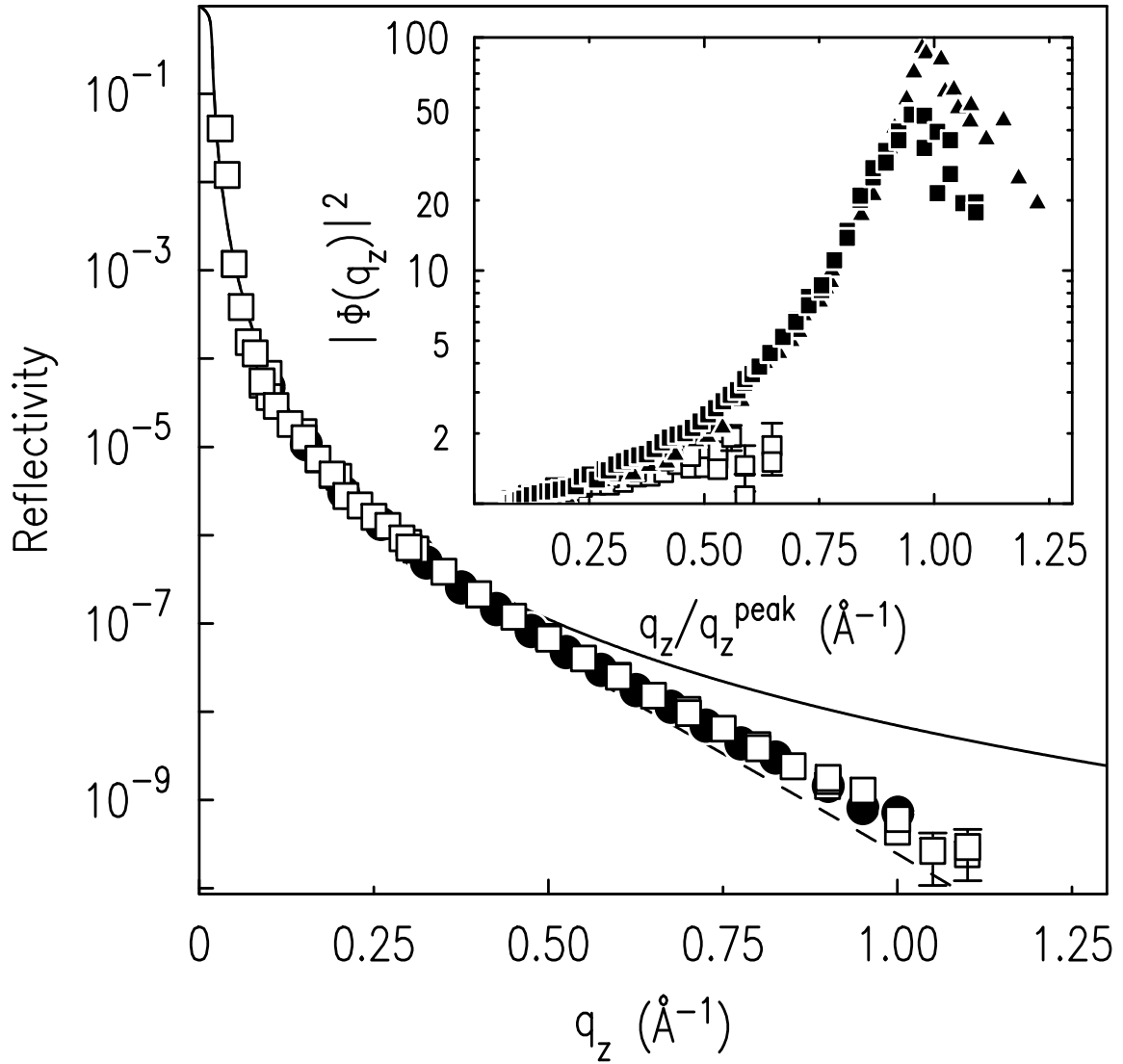, angle=0, width=1.0\columnwidth}
\caption{X-ray reflectivity from a clean liquid $  K_{67}Na_{33}$
surface at room temperature ($\Box$) and $  K_{67}Na_{33}$
following an exposure to 3000~L of oxygen ($\bullet$) Also
included for reference: (---) calculated Fresnel reflectivity
 from an ideal $  K_{67}Na_{33}$ surface; (- - -) calculated
reflectivity $R_{cw}$ from a $  K_{67}Na_{33}$ surface roughened
by capillary waves at room temperature. Inset: Surface structure
factor $|\Phi(q_z)|^2$ of liquid $  K_{67}Na_{33}$ at room
temperature ($\Box$). For comparison the surface structure factors
for liquid Ga (closed triangles) and In (closed boxes) are
included. $q_z$ is normalized to the position of the first peak in
the bulk $S(q)$} \label{kna_fig:reflectivity}
\end{figure}

As a result of this excellent agreement between CW theory and DS,
the reflectivity from the alloy surface shown in
Figure~\ref{kna_fig:reflectivity} (open squares) can be compared
to the calculated
 reflectivity  from a K$_{67}$Na$_{33}$ surface
which is roughened by CW but displays no structure (i.e.
$\Phi(q_z)=1$). The dashed line in
Figure~\ref{kna_fig:reflectivity} displays this theoretical
 prediction and it is obvious
that the data rise above it, consistent with the low-angle
behavior of surface induced layering observed
previously.\cite{gallium,indium} exist. This point is emphasized
further in the inset where the measured $R(q_z)$ is divided by
$R_f$ and the power-law CW term in Eq.~\ref{kna_eq:circle} to
obtain a direct measure of the surface structure factor
$\Phi(q_z)$.  As can be seen, the value of $\Phi(q_z)$ rises above
unity with increasing $q_z$ as expected when constructive
interference of x-rays due to surface  layering is present. The
low angle tail of the surface structure factor of
K$_{67}$Na$_{33}$   is now compared with that of liquid Ga and In
reported previously .\cite{gallium,indium} To normalize $q_z$ to
the atomic size, $q_z$ has been divided by $q_z^{max}$ of the
first maximum of $S(q)$. For K$_{67}$Na$_{33}$, $q_z^{max}\approx
1.7$~{\AA}$^{-1}$ whereas for Ga and In $q_z^{max}\approx
2.4$~{\AA}$^{-1}$. It is evident that the surface structure factor
indicative of surface-normal layering  does not increase as
quickly for  K$_{67}$Na$_{33}$ as for  Ga or In. Assuming that the
same layering model that successfully describes the surface of
Ga\cite{gallium} and In\cite{indium} also applies to the KNa alloy
the measurement indicates that either the surface layering of the
alloy is significantly weaker than that of both Ga or In, or the
length over which the layering decays into the bulk is
signficantly shorter. One can speculate that there might be a
correlation between these layering length scales and the surface
tension. This would mean that weak or quickly decaying layering is
expected for LM with exceptionally low surface tension. Clearly,
data extending well beyond $q_z = 1$~\AA$^{-1}$ are needed to
support this conclusion.

 \section{Capillary Wave Contributions}
Several factors contribute to the difficulty of extending
reflectivity measurements to values of $q_z$ large enough to cover
the expected position of the layering peak. The reflected
intensity falls over almost ten orders of magnitude, while the
background from the bulk structure factor increases, as this value
of $q_z$ is approached. We must also consider that the exponent
$\eta$ approaches its limiting value of 2 for $q_z \rightarrow
1.7$~{\AA}$^{-1}$ so that the specular  cusp becomes broadened and
ultimately lost in the diffuse background. Because of these
effects, simply increasing the photon flux is not sufficient to
improve the measurement substantially. A better approach is to
significantly improve the resolution. This is illustrated in
Figure~\ref{kna_fig:diffuse}(b), where the calculated DS for
$\alpha = 4.5$~degrees (corresponding to $q_z = 1.5$~{\AA}$^{-1}$)
is shown for the current experimental resolution (solid line) and
for a resolution sharper by a factor of 200 (dashed line).
Although the intensity is much reduced in the latter case, the
peak is much better defined. Reflectivity measurements conducted
at  low divergence \index{synchrotron!source} synchrotron x-ray
sources that are presently becoming available may make such
extensions of the $q_z$ range possible.

 \section{Oxidation Studies}
To explore the effect of possible contaminants on the surface
structure, we deliberately exposed the liquid alloy surface to
controlled amounts of pure oxygen and hydrogen gas, measured in
units of monolayer surface coverage (Langmuir).\cite{galliumoxide}
The liquid alkali surface has been exposed to up to 3000~L oxygen
and up to 6000~L hydrogen. As an example,$R(q_z)$ at the maximum
exposure of O$_2$ is shown as filled circles in
Figure~\ref{kna_fig:reflectivity}. In no case was a measurable
effect on $R(q_z)$ observed, indicating no change in the
surface-normal structure. This insensitivity of the liquid
K$_{67}$Na$_{33}$ surface to small amounts of oxygen is
 in striking contrast to similar experiments on
liquid Ga\cite{galliumoxide} and Hg\cite{lam} where the formation
of a thin homogeneous oxide film has been observed upon exposure
of the liquid surface to as little as 200~L of oxygen. This
difference is due to the fact that the low surface tension of
liquid AM precludes segregation of oxide from the bulk. By
contrast, high surface tension LM such as Ga or Hg lower their
high overall surface energy by segregation of a contaminant  film.
These results are also in agreement with photoelectron
spectroscopy experiments requiring a similarly clean liquid AM
surface.\cite{oelhafen}

 \section{Summary}
In summary, we have shown that it is possible to investigate the
surface structure of an atomically clean liquid AM under UHV
conditions.  Exposing the sample to up to 6000~L of H$_2$ or O$_2$
has no effect on the atomic surface structure since the
exceptionally low surface tension of AM precludes contaminant
segregation. From DS we extract the surface roughness, which
agrees well with capillary wave theory, and with the macroscopic
surface tension reported in the literature. The $q_z$-dependence
of the
 surface structure factor determined from the XR
is consistent with  constructive interference of x-rays due to
surface-normal layering. This surface-induced layering in liquid $
K_{67}Na_{33}$ appears to be  much weaker than that found in high
surface tension LM such as In or Ga.

%% file: k_all.tex

\chapter{Layering in Alkali Metals: Liquid Potassium}

\section{Abstract}

\noindent In this chapter we present x-ray reflectivity and
\index{x-ray!diffuse scattering} diffuse scattering measurements
from the liquid surface of pure potassium. They strongly suggest
the existence of atomic layering at the free surface of a pure
liquid metal with low surface tension. Prior to this study,
layering was observed only for metals like Ga, In and Hg, the
surface tensions of which are 5-7 fold higher than that of
potassium, and hence closer to inducing an ideal \index{hard wall}
"hard wall" boundary condition. The experimental result requires
quantitative analysis of the contribution to the surface
scattering from thermally excited capillary waves. Our
measurements confirm the predicted form for the
\index{differential cross section} differential cross section for
diffuse scattering, $d\sigma /d\Omega  \sim 1/q_{xy}^{2-\eta}$
where $\eta = k_BT q_z^2/2\pi \gamma $, over a range of $\eta$ and
$q_{xy}$ that is larger than any previous measurement. The partial
measure of the surface structure factor that we obtained agrees
with computer simulations and theoretical predictions.

\section{Introduction}

In contrast to measurements of \index{liquid!dielectric}
dielectric liquids, which show a monotonically varying density
profile at the liquid-vapor interface, liquid metals exhibit a
phenomenon known as surface-induced layering \cite{Rice74,
Evans81}. This had been thought to occur because the
\index{interactions!Coulomb} Coulomb interactions between two
constituents of the liquid metal, the free electron
\index{Fermi!gas} Fermi gas \cite{March90} and the classical gas
of positively charged ions, suppress local surface fluctuations
that would otherwise conceal the layered atomic ordering that
occurs at \index{hard wall} hard walls \cite{Iwamatsu92}. On the
other hand recent molecular simulations by \index{Chac{\'o}n,
Enrique} Chacon et al. \cite{Velasco02} argue that surface
layering should occur in any liquid for which the ratio of melting
temperature $T_m$ and critical temperature $T_c$, $T_m/T_c$, is
sufficiently small. Until this measurement the only experimental
observations of surface-induced atomic layering have been
accomplished in a number of high surface tension metals - pure
liquid metallic systems such as Hg \cite{Magnussen95}, Ga
\cite{Regan95}, In \cite{Tostmann99} as well as in binary alloys
\cite{Tostmann98, Dimasi00}. In fact, for Ga, Hg and K  $T_m/T_c
\approx 0.13-0.15$ and these observations are consistent with
\index{Chac{\'o}n, Enrique} Chacon et al. \cite{Velasco02}. From
another point of view, all but one of the metallic systems studied
so far exhibit relatively high values of surface tension, a few
hundred mN/m, leaving unanswered the fundamental question of
whether the surface-induced layering is caused by the high surface
tension, which suppresses the long wavelength capillary waves, or
whether it is caused by the intrinsic metallic properties of the
studied systems. In the later case surface-induced layering should
also be found in low-surface tension metals. Monovalent
\index{alkali metals} alkali metals have a surface tension
comparable to that of water (72 mN/m) and their study should,
therefore, shed light on this issue. Moreover, monovalent
\index{alkali metals} alkali metals are the best available
examples of an ideal \index{nearly free electron}
nearly-free-electron metal \cite{Ashcroft76}, therefore, they are
particularly well suited for studying surface induced layering in
an ideal, simple liquid metal. The same properties which make
alkali metals ideal for studies of layering have also been the
motivation for the numerous computer simulations of liquid Na, K
and Cs \cite{Rice74, Evans81, Harris87, Ashcroft76, Hasegawa82,
Chacon84, Gomez94, Goodisman83}.

Until recently, experimental surface studies of the pure liquid
alkalis were thought to be impossible because of two fundamental
experimental problems. First, their high evaporation rates which
lead to high values of vapor pressure \cite{Iida93}, and high
reactivity \cite{Borgstedt87} to water and oxygen implied that
their surfaces could not be maintained atomically clean for
extended periods of time even under UHV conditions. Therefore, the
first study of alkali metal systems performed by us
\cite{Tostmann99} employed a binary KNa alloy that preserves most
of the characteristic features typical of a \index{nearly free
electron} nearly-free-electron metal, but for which the vapor
pressure and melting temperature were both lower than those of the
pure metals. Contrary to expectations, no measurable changes were
detected in that study in the x-ray reflectivity of the alloy's
surface over periods of many hours or after intentional exposure
to large doses of oxygen. This lack of surface contamination is
believed to be due to the high solubility of the oxides of these
metals in the bulk liquid \cite{Borgstedt87, Addison84}, and to
the low surface tension of liquid \index{alkali metals} alkali
metals which does not promote segregation of the oxide at the
surface. The fact that the surface of the KNa alloy was found to
remain atomically clean for extended periods of time suggests that
the same might be true for the pure metals. The second
experimental problem is that the large capillary-wave-induced
roughness due to the low surface tension renders the x-ray
reflectivity peak weak and inseparable from its' diffuse wings.
Thus, a clear layering peak in the measured reflectivity curve,
such as found in the high-surface-tension liquid metal studies
\cite{Magnussen95, Regan95, Tostmann99, Tostmann98, Dimasi00}, can
not be observed here. We demonstrate here that this problem can be
resolved by a subtle application of   diffuse scattering and
reflectivity measurements. A significant aspect of this chapter
pertains, therefore, to the methods of measurement and analysis
that are necessary in order to separate the effects due to the
thermal capillary waves from those of the surface-induced
layering. Using these methods, the present study clearly
demonstrates the existence of surface induced layering in liquid
K, with properties similar to those of the high-surface-tension
liquid metals studied previously \cite{Magnussen95, Regan95,
Tostmann99, Tostmann98, Dimasi00}.

\section{Experimental Details}

The sample preparation procedure and the experimental setup were
similar to our previous experiment on the KNa alloy
\cite{Tostmann99}. The sample was prepared in an Argon-filled
glove box ( $ < $ 2ppm $H_2$0, $ < $ 1 ppm $O_2$). The K with
purity of 99.999 $\%$, which was contained in a glass ampoule, was
melted and then transferred into a sealed stainless steel
reservoir inside the glove box using a glass syringe. The
reservoir was sealed with a Teflon o-ring and then mounted onto a
UHV chamber.

\begin{figure}[htp]

\begin{center}
\unitlength1cm
\begin{minipage}{9cm}
\epsfxsize=9cm  \epsfbox{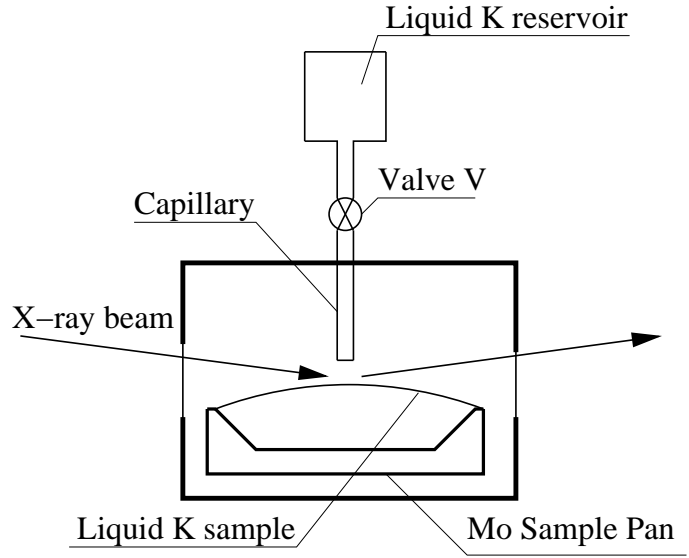}
\end{minipage}
\caption{ \label{fig:K_setup} Schematic description of the
experimental setup. }
\end{center}
\end{figure}

After a 24-hour \index{bakeout} bakeout of the entire chamber to
150 $^{\circ}$\,C, a base pressure of $10^{-9}$\,Torr has been
established inside the UHV chamber. The K in the reservoir was
then heated above the melting temperature $T_m$=63$^{\circ}$\,C
and dropped through a capillary into a Mo sample pan mounted
inside the UHV chamber by opening the valve V shown on Figure
\ref{fig:K_setup}. Previous to filling the pan with liquid K, the
pan was cleaned by sputtering with $Ar^{+}$ ions for several hours
in order to remove the native Mo oxide. This latter procedure
significantly improves the wetting of the sample pan by liquid K
and therefore helps to achieve a larger area of macroscopically
flat liquid surface than it has been possible for high surface
tension metals \cite{Regan95}. The sample pan was kept at a
constant temperature of 70 $^{\circ}$\,C. Pure K has a vapor
pressure of about $10^{-6}$\,Torr at this temperature, which
renders the requested UHV conditions almost impossible.  We
circumvented this problem by a differential \index{pump!turbo}
\index{pump!ion}  pumping setup, where we used a permanently
attached ion pump. By constantly pumping at the sample chamber, we
were able to achieve a base pressure of about $10^{-9}$\,Torr in
the chamber, three orders of magnitude lower than vapor pressure
of potassium.

\begin{figure}[htp]

\begin{center}
\unitlength1cm
\begin{minipage}{8.5cm}
\epsfxsize=8.5cm \epsfysize=8cm \epsfbox{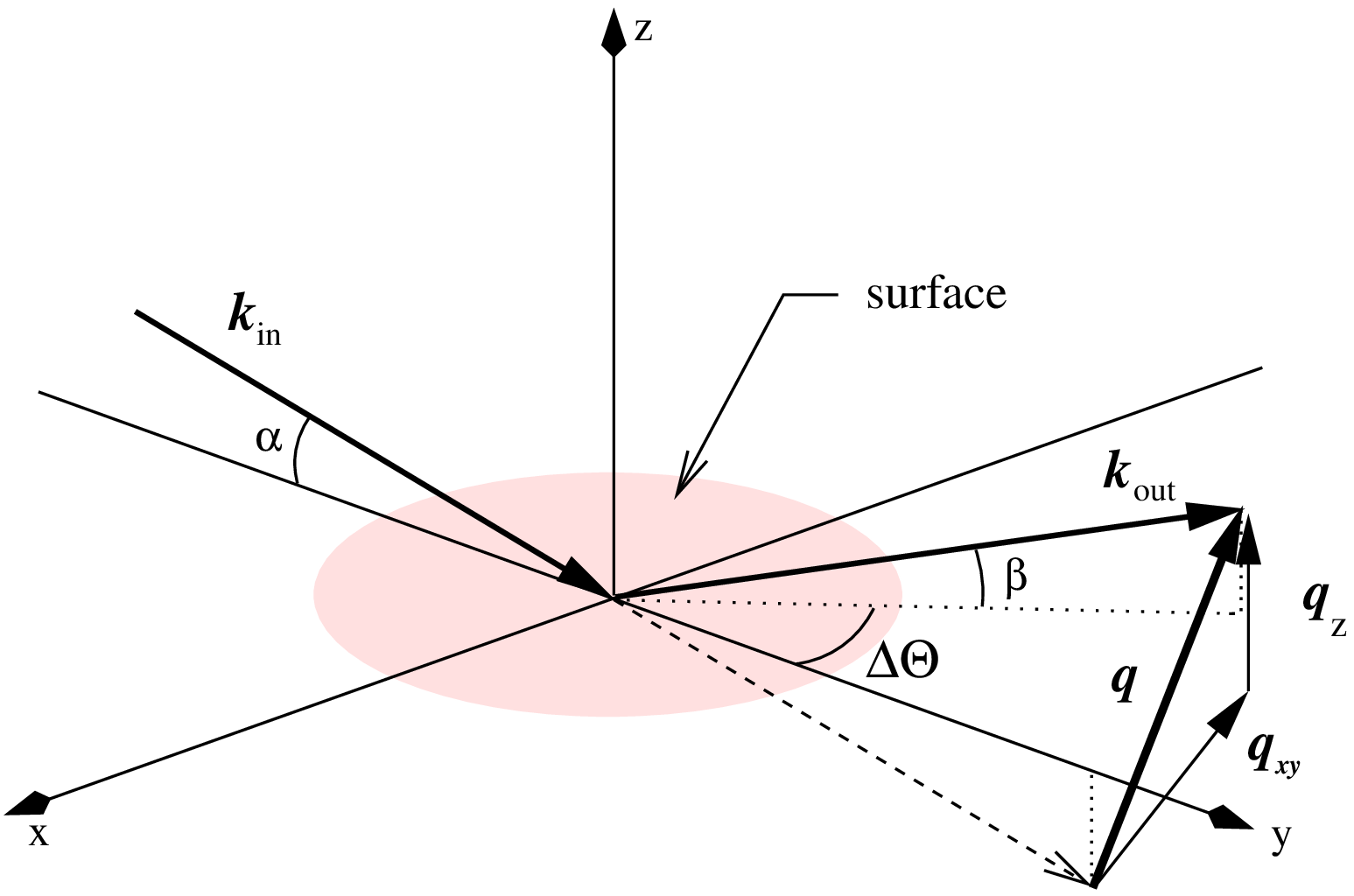}
\end{minipage}
\caption{\label{fig:K_Kinematics} Kinematics of the x-ray
measurement. }
\end{center}
\end{figure}

Measurements were carried out using the liquid surface
reflectometer at CMC-CAT beamline at the Advanced Photon Source,
\index{Argonne} Argonne National Laboratory, operating at an x-ray
wavelength of 1.531~${\AA}^{-1}$. The kinematics of the
measurement is illustrated in Figure \ref{fig:K_Kinematics}. Two
different surface-sensitive x-ray techniques were applied -
specular x-ray reflectivity and x-ray diffuse scattering. In the
specular x-ray reflectivity geometry, the incident angle $\alpha$
and the reflection angle $\beta$=$\alpha$ are varied
simultaneously while keeping $\Delta\Theta$ = 0 constant. The
signal is measured as a function of the surface-normal momentum
transfer $q_z = 2 \pi / \lambda \left( \sin \alpha + \sin \beta
\right)$ with the in-plane momentum transfer $q_{xy}= 0$.

In the x-ray diffuse scattering geometry, the detection angle
$\beta$ is varied within the plane of incidence $\Delta\Theta$ =
0, while the incident angle $\alpha$ stays fixed. The observed
scattering therefore follows a trajectory in the ($q_x, q_z$)
plane with $dq_z / dq_x =\tan \beta$ and $q_y$=0. The signals
measured in the two geometries are, however, related since the
signal measured at $\beta=\alpha$ in the diffuse measurement is
the specular signal.

The differential cross-section of the diffuse scattering signal is
given by \cite{Pershan00}:

\begin{equation}
\frac{d\sigma }{d\Omega } = \frac{A_{0}}{\sin \alpha } \left(
\frac{q_c}{2} \right) ^{4}\frac{k_B T}{16\pi ^2 \gamma } \mid \Phi
(q_z) \mid ^2 \frac{1}{q_{xy}^{2-\eta}} \left(
\frac{1}{q_{\mbox{\scriptsize max}}} \right) ^\eta \:
\label{eq:K_master}
\end{equation}
where $q_c$ \cite{AlsNielsen01} is the \index{critical!angle}
critical angle for total \index{external reflection} external
reflection of x-rays, $\gamma$ is the surface tension,
 $q_{max} \approx \pi/\xi$
is the value of the upper cutoff for capillary wave contributions
\cite{Pershan00}. The value for $\xi$ is empirically taken to be
the nearest neighbor atomic distance in the bulk
\cite{AlsNielsen01}. The scattering lineshape exponent $\eta$ is
given by \cite{Pershan00}

\begin{equation}
\eta = \frac{k_BT}{2\pi \gamma} q_z^2 \label{eq:k_eta}
\end{equation}
The principal goal of the present surface x-ray scattering
measurements is to determine the surface structure factor
\cite{Regan95}

\begin{equation}
\Phi (q_z) = \frac{1}{\rho_{\infty}} \int dz\frac{\left< d\rho(z)
\right> }{dz} \exp(\imath q_z z)
\label{eq:K_structure}
\end{equation}
which is given by the Fourier transform of the ($x,y$)-averaged
intrinsic density profile $ \left< \rho(z) \right> $ along the
surface normal $z$ in the absence of capillary waves. In this
equation, $\rho_{\infty}$ is the bulk electron density. The fact
that the peak of the \index{differential cross section}
differential cross section at the specular condition is algebraic,
rather than a $\delta$-function complicates the determination of
$\Phi(q_z)$ from the measured intensity. This can be done,
however, when both specular reflectivity and \index{x-ray!diffuse
scattering} diffuse scattering measurements are available.

\subsection*{Practical Considerations}

It follows from Eq.~\ref{eq:K_master} that the nominal specular
reflectivity signal, observed at $\alpha=\beta$ and
$\Delta\Theta=0$ corresponds to the integral of the $\left( 1 /
q_{xy} \right) ^{2-\eta}$ over the area $\Delta q_{xy}^{res}\left(
q_z \right) = \Delta q_{x}^{res} \times \Delta q_{y}^{res}$ which
is the projection of the instrumental \index{resolution function}
resolution function on the horizontal plane of the liquid surface.
The $q_z$-dependence of $\Delta q_{xy}^{res}\left( q_z \right)$ is
due to the increase of the projected width of the detector
resolution $\mid \Delta q_y^{res} \mid=(2 \pi / \lambda) \sin
\beta \Delta \beta $ on the plane of the surface as $\beta$
increases. The integration results in the reflectivity scaling as
$\left(\Delta q_{y}^{res} / q_{max} \right)^\eta / \eta $ and
since $\eta$ scales as $q_z^2$, the net effect is a
\index{Debye-Waller} Debye-Waller-like decrease in the
reflectivity as $q_z$ increases. For example, if one defines a
mean square surface roughness by taking an average over a length
scale $ \approx (\Delta q_{y}^{res})^{-1}$, this roughness can be
written as \cite{Schwartz90}
\begin{equation}
\sigma^2 (q_z) = (k_b T / 2 \pi \gamma) \ln (q_{max}/\Delta
q_y^{res}) \label{eq:K_sigma}
\end{equation}

From this one can show that the reflectivity is proportional to a
\index{Debye-Waller} Debye-Waller factor $\exp [-q_z^2 \sigma^2
(q_z)]$. The important point to note is that for liquid surfaces
the signal measured at the specular condition $q_{xy}=0$ depends
on the resolution of the reflectometer.

The validity of the simple capillary wave model in the high
$q$-range remains controversial and untested.  Although the model
assumes that the energy for surface excitations can be described
by the capillary form $\approx \gamma /q_{xy}^2 $ with a constant
surface tension, $\gamma$, over the entire range of length scales
from the macroscopic to a microscopic length of the order of $\xi
= \pi/q_{max}$, where $\xi$ is normally taken to be of the order
of the interatomic/molecular spacing $a$, this is certainly not
strictly true for lengths that are shorter than a length that we
might designate as $\xi_{cutoff} = \pi/q_{cutoff} > a$.
Nevertheless, the basic physics is not changed if we replace
$q_{max}$ by the smaller $q_{cutoff}$ since the differences
engineered by this are really just a matter of accounting, in the
sense that it determines the relative contributions between the
capillary wave and structure factor terms given in
Eq.~\ref{eq:K_master}. Nevertheless this is not a serious
objection to the present treatment, since the capillary wave model
is certainly accurate for values of $q_{xy}$ covered by the
present measurements. Even if the model were to fail for values of
$q_{xy}
> q_{cutoff} $, the shorter wavelength fluctuations  could be
incorporated into the definition of a $q_{cutoff}$-dependent
structure factor $\Phi(q_z)$. On the other hand, the fact that the
intensity of off-specular diffuse scattering predicted by
Eq.~\ref{eq:K_master} has been shown to be in satisfactory
quantitative agreement with measurements for a number of liquids
over a wide range of $q_y$ and $q_z$ implies that $q_{cutoff}$ is
reasonably close to $q_{max} = \pi / a$ and the extracted
structure factor $\Phi(q_z)$ is a reasonable measure of the local
profile.

There are three practical considerations involved in these
measurements.

The first is that it is necessary to insure that the measurement
corresponding to the integral over the resolution does not include
a diffuse scattering contribution from sources other than the
surface. For flat surfaces, for which the specular reflectivity
can be described by a $\delta$-function at $q_{xy}$=0, this can be
accomplished by subtracting the scattering measured at some finite
value of $q_{xy}$ from the signal measured at $q_{xy}$=0. For
liquid surfaces this separation is complicated by the fact that
the signal measured at finite $q_{xy}$ includes scattering from
the $1/q_{xy}^{2-\eta}$ tails of the cross-section. We circumvent
this problem in the present study by comparing the difference
between measurements at $q_{xy}$ =0 and at some $q_{xy}\neq 0$
with the same difference as predicted by the theoretical
differential cross section (Eq.~\ref{eq:K_master}). Thus, we
present all data in this study as differences between a signal
measured at $\lbrace \alpha, \beta, 0 \rbrace$ and at $\lbrace
\alpha,\beta, \Delta \Theta_{offset}$ = 0.2 degrees$\rbrace$. For
specular reflectivity $\alpha=\beta$; however, for off-specular
diffuse measurements $\beta \not= \alpha$.   diffuse scattering
from all sources other than the surface are essentially constant
over this range of $\Delta\Theta_{offset}$. This is because these
other sources of scattering, such as diffuse scattering from
either the bulk liquid or from the vapor above the surface, only
depend on the absolute value of the total momentum transfer $\mid
q \mid$ and this is essentially unchanged for small offsets in
$\Delta\Theta_{offset}$.

The second practical consequence of the $1 / q_{xy}^{2-\eta}$ form
is that as $\eta$ increases it becomes increasingly difficult to
distinguish the singular specular peak from the off-specular
power-law wings. At $\eta \geq 2$ this distinction is no longer
possible \cite{Pershan00}, even with infinitely sharp resolution.
Unfortunately, for liquid K, $\eta$ reaches the limit of 2 at
$q_z$=1.74~${\AA}^{-1}$, which is comparable to
$q_z$=1.6~${\AA}^{-1}$, where the \index{quasi-Bragg peak}
quasi-Bragg surface layering peak is expected to be observed.
However, in the following paragraph we show that as a result of
resolution effects, the limit at which the specular cusp can no
longer be distinguished from the background in practice is even
lower.

The impact of the finite resolution can be easily seen by
considering a simple example of a rectangular resolution function
of infinite length along the x axis but with a very small width
along the y axis. For small values of $\alpha$ this is often an
excellent approximation to the real resolution functions
\cite{Mitrinovic01}. With this approximation, the integration over
the $q_x$ component yields a $1/ {q_y}^{1-\eta}$ dependence for
the diffuse scattering, placing the value of $\eta$ where the
specular ridge disappears to be $\eta$=1 instead of $\eta$=2 . For
K, this $\eta$=1 limit is reached at $q_z$=1.2~${\AA}^{-1}$. The
detector slit dimensions which provide an optimal compromise
between intensity and resolution were discussed by one of us
\cite{Pershan00} and are $\left( width \right) \approx \left(
height \right) \times \sin \alpha$, where $\alpha$ is the incident
angle.

The third practical problem in extending x-ray reflectivity
measurements to large $q_z$ is the fast fall-off of the reflected
intensity. Much more serious is the \index{Debye-Waller}
Debye-Waller-like factor that was discussed above. Since the
surface tension $\gamma$ for K (110 mN/m at 70 $^{\circ}$\,C) is
much lower than that of a metal like Ga (770 mN/m) \cite{Iida93},
the factor of $\exp [-\sigma^2 q_z^2]$ with $\sigma^2 \sim
1/\gamma$ is orders of magnitude smaller than that of Ga at the
same values of $q_z$. Even at high-flux \index{synchrotron!source}
synchrotron facilities such as the Advanced Photon Source the low
count rates of the reflectivity signal observed at $q_z \approx
1.0-1.1~{\AA}^{-1}$ are among the dominant limiting factors of the
$q_z$ range accessible in this experiment.

\section{Measurements}
\subsection{Advantages of Linear Detector}
To partially compensate for the low intensity the off-specular
diffuse scattering measurements were carried out using a linear
position sensitive detector (PSD), rather than a point detector.

\begin{figure}[htp]
\begin{center}
\unitlength1cm
\begin{minipage}{12cm}
\epsfxsize=12cm \epsfysize=15cm\epsfbox{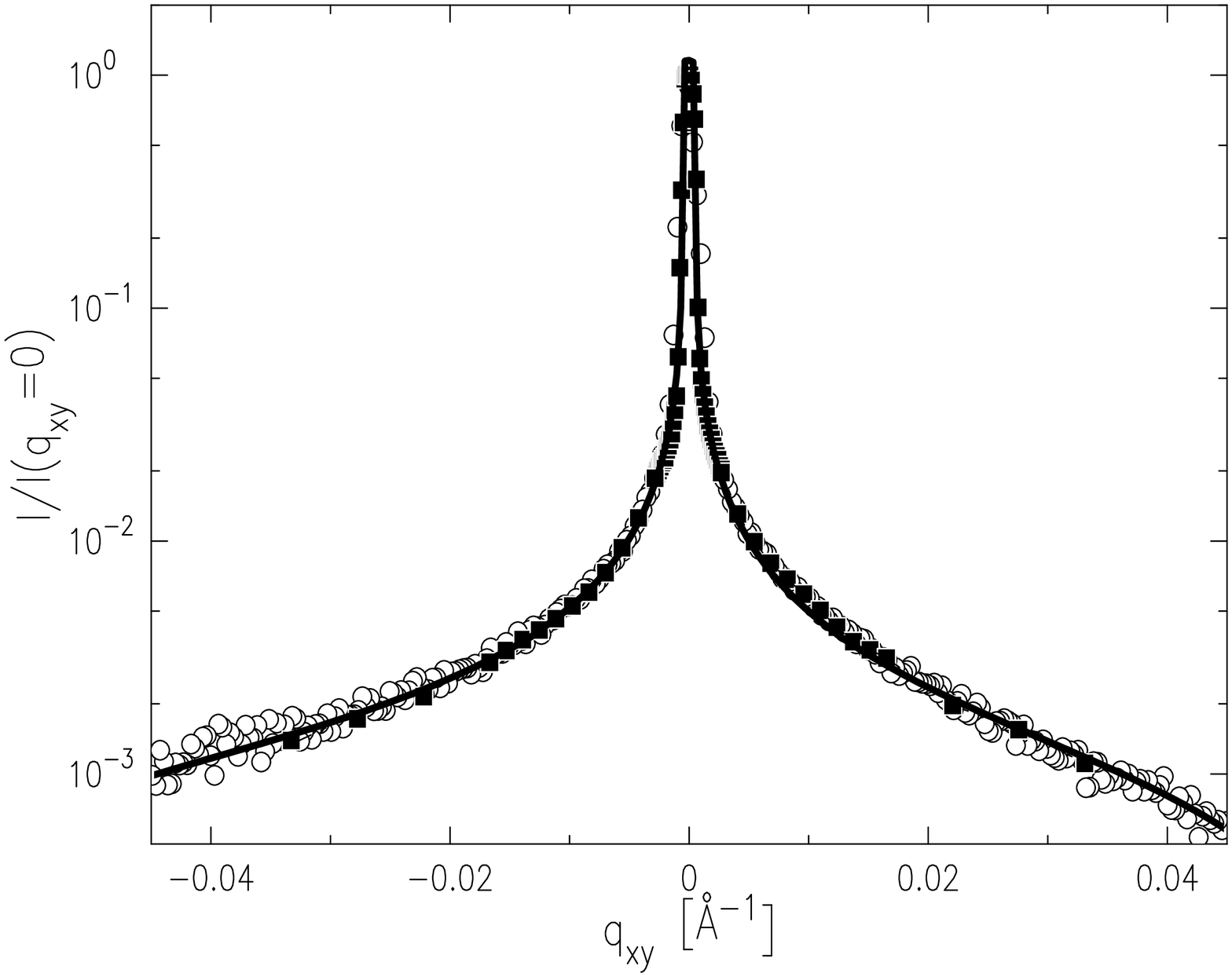}
\end{minipage}
\caption{ \label{fig:K_PSDcomp}Comparison of an X-ray diffuse
scattering scans taken with point detector - filled squares, and
position sensitive detector (PSD) - open circles. The specular
condition corresponds to $q_z$=0.3~${\AA}^{-1}$. The resolution of
the bicron detector (0.22mm vertically and 2.5mm horizontally) was
chosen to match that of a PSD. Also shown is a line which
represents theoretical modeling using the capillary-wave theory
for given resolution, temperature ($70^{\circ}C$) and surface
tension (110 mN/m). }
\end{center}
\end{figure}

In Figure \ref{fig:K_PSDcomp} we show a diffuse scattering profile
as a function of $q_{xy}$. The specular condition corresponds to
$q_z = 0.3~{\AA}^{-1}$. The profile exhibits a sharp peak whose
width is dominated by the instrumental resolution, however the
long tails are an intrinsic property of the capillary wave surface
roughness. We note that the data shown represents the intensity
difference for data obtained at $\Delta\Theta = 0$ and {\bf
$\Delta\Theta = \pm \Delta\Theta_{offset}$} with either a position
sensitive detector (PSD) or with a point detector. An advantage of
the PSD configuration is that it permits the entire profile to be
acquired simultaneously for a wide range of the output angles
$\beta$ for a fixed $\alpha$. We carefully checked for possible
instrumental errors associated with the PSD (saturation or
non-linear effects) by comparing data obtained with both the PSD
and a point detector. As shown, in Figure \ref{fig:K_PSDcomp},
data obtained with both configurations are indistinguishable.
Further, this comparison allowed an accurate determination of the
PSD spatial resolution, about 0.22 mm vertically.  To provide for
a reasonable comparison with the point detector, the vertical slit
was set to match the resolution of the PSD. The solid line in
Figure \ref{fig:K_PSDcomp} represents the theoretical prediction
for the different intensity, obtained by numerically integrating
$d \sigma / d \Omega$ in Eq.~\ref{eq:K_master} over the known
resolution. The known values of the incident angle $\alpha$,
temperature T, surface tension $\gamma$, x-ray energy and detector
resolution, were used without any adjustable parameters. The
essentially perfect agreement between both experimental data sets
and the theoretical simulations strongly supports the theoretical
model for the effect of the thermally induced surface excitation
on the diffuse x-ray scattering from the surface.

\subsection{Diffuse Scattering Data: Eta Effect}

\begin{figure}[htp]

\unitlength1cm
\begin{minipage}{20cm}
\epsfxsize=14cm   \epsfbox{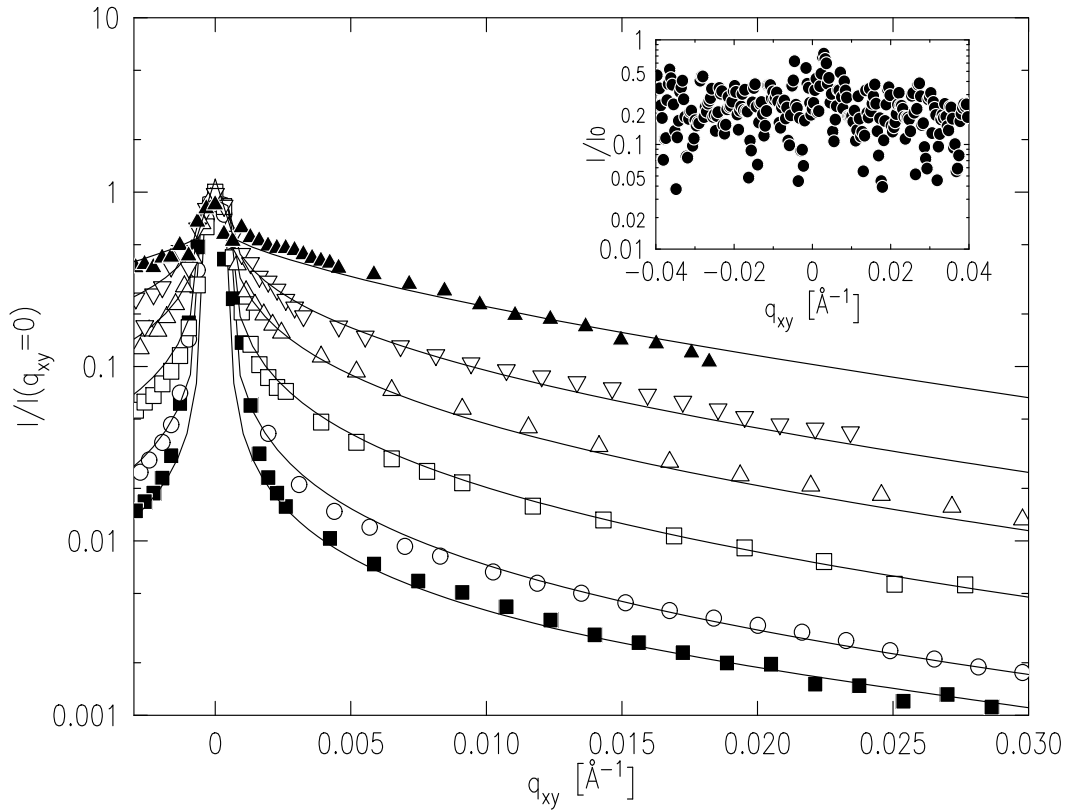}
\end{minipage}
\caption{\label{fig:K_diffuse}X-Ray diffuse scattering scans taken
with PSD, as a function of wavevector transfer component $q_{xy}$.
Specular condition corresponds to values of $q_z$ (bottom to top):
0.3, 0.4, 0.6, 0.8, 1.0 and 1.1~${\AA}^{-1}$. For comparison, the
scans are normalized to unity at $q_{xy}$=0. Lines represent
theoretically simulated scans for given values of $q_z$. The inset
shows a similar scan taken at 1.2~${\AA}^{-1}$, for which the
specular peak is marginally observable.}

\end{figure}

Further diffuse scattering data was collected with the PSD, which
proved to be especially useful at higher values of $q_z$, where
the weaker scattering makes it increasingly more difficult to
separate the specular signal from the power-law wings as $\eta$
approached 1. Figure \ref{fig:K_diffuse} shows a set of diffuse
scattering scans measured with a PSD and the corresponding
calculated theoretical predictions, for several $q_z$ values (0.3,
0.4, 0.6, 0.8, 1.0 and 1.1~${\AA}^{-1}$). Each curve represents a
difference between a diffuse measurement in the plane of incidence
($\Delta\Theta = 0$) and an average of two measurements taken by
displacing the detector out of the plane of incidence by an angle
of $\Delta \Theta_{offset} = \pm$ 0.2 degrees. Theoretical
simulations include the same subtraction operation as well. All
data sets are normalized to unity at $q_{xy}$=0 for easier
comparison. As $q_z$ gets larger, the ratio of specular signal to
the diffuse wings decreases, and the specular peak eventually
disappears under the off-specular diffuse wings. In the curve
measured at $q_z = 1.2~{\AA}^{-1})$, shown in the inset to Figure
\ref{fig:K_diffuse}, the specular peak is all but disappeared. The
solid lines that represent numerical integration of the
capillary-wave theory (Eq.~\ref{eq:K_master}) show a remarkably
good agreement with experimental data. The only adjustable
quantity in these theoretical curves is the form of $\Phi(q_z)$.
As discussed below, one common form was used for all data sets.

This agreement provides additional support for the absence of
surface contamination, since the presence of any kind of
inhomogeneous film at the surface would manifest itself in extra
x-ray diffuse signal which could not be explained by capillary
wave theory alone.

\subsection{Resolution Effects}

\begin{figure}[tp]

\begin{center}
\unitlength1cm
\begin{minipage}{12cm}
\epsfxsize=12cm \epsfysize=15cm \epsfbox{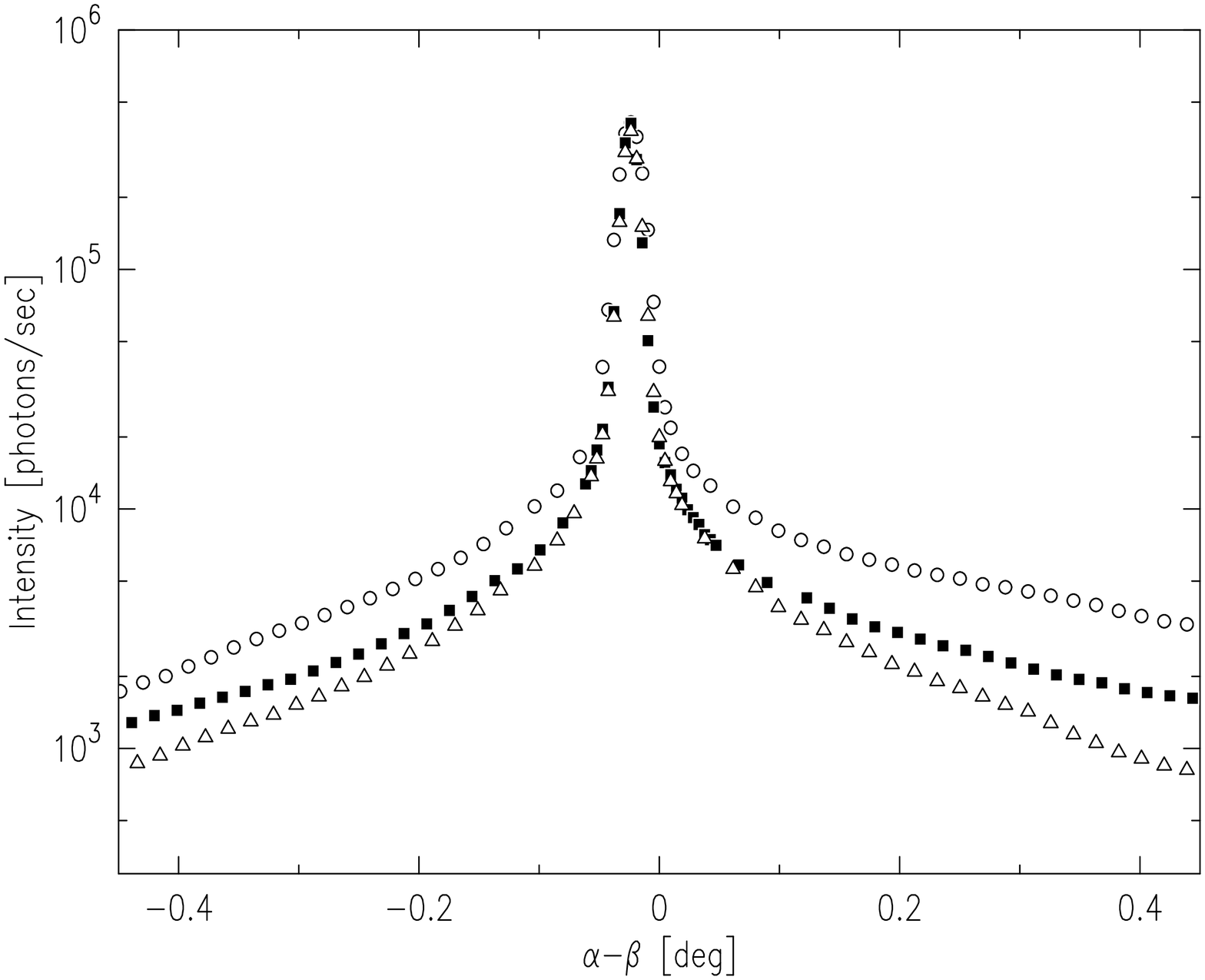}
\end{minipage}
\caption{\label{fig:K_res} Effects of varying the horizontal
resolution function on the diffuse scattering data. Diffuse scans
taken with PSD, peak position corresponding to
$q_z$=0.4~${\AA}^{-1}$, as a function of angular deviation from
specular condition ($\alpha-\beta$) with fixed vertical resolution
(.22mm), while horizontal resolution was changed (bottom to top:
2mm, 4mm, 8mm). }

\end{center}
\end{figure}

Figure \ref{fig:K_res} shows the effect of varying the shape of
the detector resolution on the diffuse scattering data. Changing
the horizontal size of the slit in front of the PSD has little
effect on either the measured intensity of the specular peak,
because of the intrinsic property of the singularity, or on the
diffuse scattering at small $q_{xy}$, because it is sharply peaked
around $q_{xy} = 0$. However, far away from the specular
condition, the signal is directly proportional to the area of the
resolution function, and scales with the size of the slits.

\subsection{Specular Reflectivity Data}
The excellent agreement between the experimental measurements and the
theoretical simulations strongly support our conclusion that by the
methods above one can account properly and accurately for both the
capillary-wave-induced diffuse scattering in, and the resolution effects on,
the measured data. We have, therefore, employed these results to analyze the
specular reflectivity data, to find out whether surface-induced layering
exists at the surface of K.
\begin{figure}[htp]

\begin{center}
\unitlength1cm
\begin{minipage}{12cm}
\epsfxsize=12cm   \epsfbox{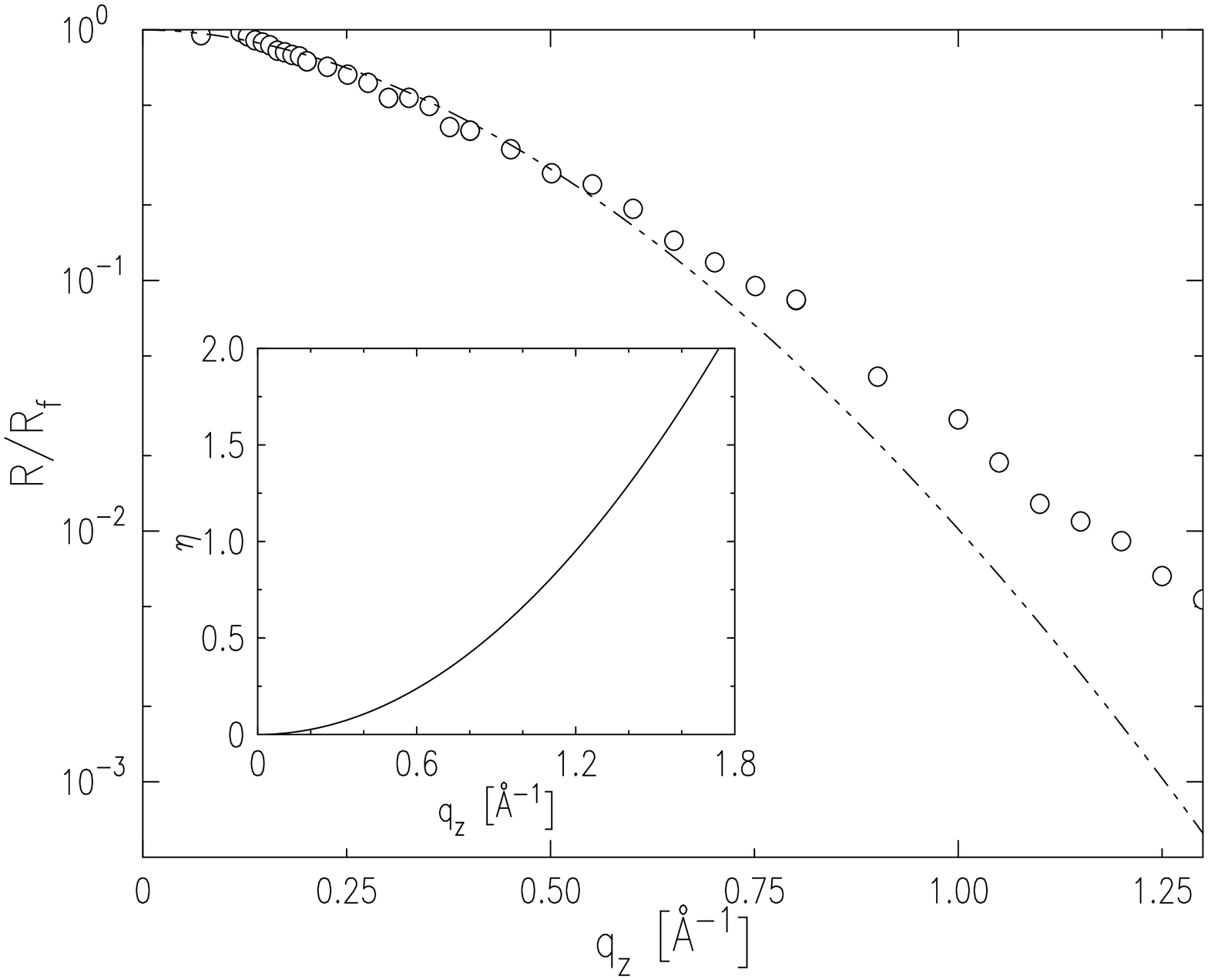}
\end{minipage}
\caption{\label{fig:K_RRF} Fresnel-normalized reflectivity signal
$R/R_f(q_z)$ for liquid K (circles) compared to capillary wave
predictions (dashed line). The inset shows the capillary wave
factor $\eta$ for liquid K as a function of $q_z$. }
\end{center}
\end{figure}

Specular reflectivity measurements were made by scanning $\alpha=\beta$
at both $\Delta\Theta=0$ and $\Delta\Theta = \pm \Delta\Theta_{offset}$. The $\beta$-resolution,
$\Delta \beta =$ 0.2 degrees,
was set by fixing the height of the detector slit at 2.5 mm,
located 714 mm away from the center of the sample.
Likewise, the horizontal resolution was defined by a 2.5 mm
wide slit located at the same position.
The ratio of the difference between the specular signal
measured at $\Delta\Theta=0$ and at $\Delta\Theta = \pm \Delta\Theta_{offset}$
 to the theoretical Fresnel reflectivity from the K liquid-vapor
interface is shown in Figure \ref{fig:K_RRF}. As expected, there
is a \index{Debye-Waller} Debye-Waller type effect that causes the
data to fall below unity with increasing $q_z$. However, the
measured curve (points) decreases slower than the theoretical
prediction (dashed line) for the \index{Debye-Waller}
Debye-Waller-like factor in Eq.~\ref{eq:K_master} convolved with
the resolution function.

\subsection{Surface Structure Factor Data}

\begin{figure}[htp]
\begin{center}
\unitlength1cm
\begin{minipage}{12cm}
\epsfxsize=12cm    \epsfbox{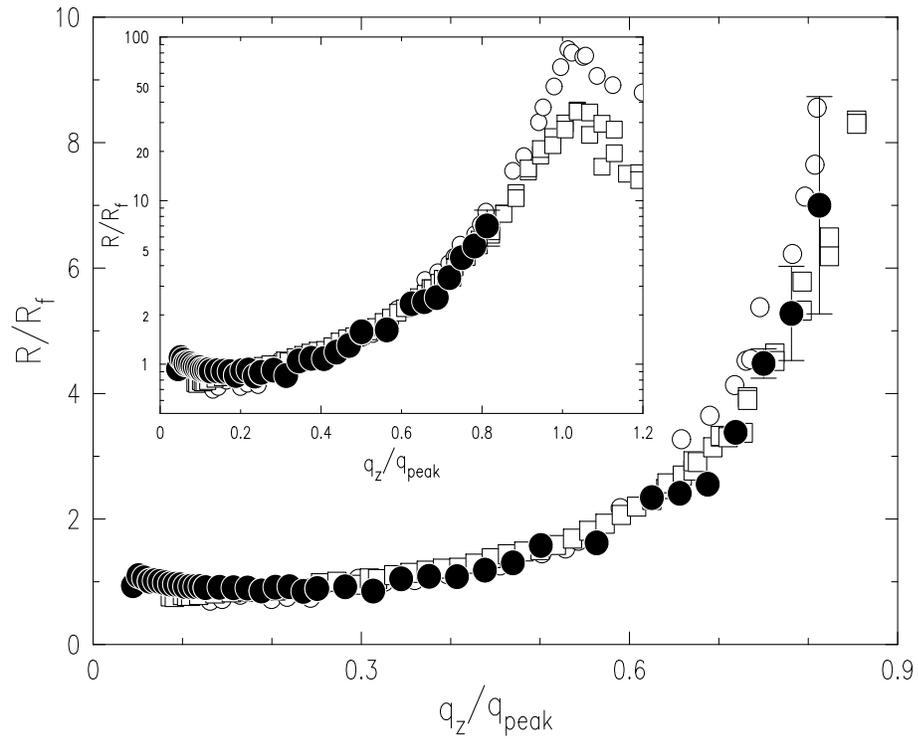}
\end{minipage}
\caption{\label{fig:K_Phi} Surface structure factor $\Phi(q_z)$
for liquid Ga (open circles), liquid In (open squares) and liquid
K (filled circles) obtained from X-ray reflectivity data by
deconvolving resolution, Fresnel reflectivity and capillary wave
contributions, shown here as a function of $q_z/q_{peak}$, here
$q_{peak}$ is the value of $q_z$ at which the layering peak is
observed (2.4~${\AA}^{-1}$ for Ga and 2.2~${\AA}^{-1}$ for In) or
is expected to be observed (1.6~${\AA}^{-1}$ for K). The inset
shows the same three sets of data extended to a greater range. }
\end{center}
\end{figure}

Figure  \ref{fig:K_Phi} shows the \index{form factor} form factor
$\Phi(q_z)$, obtained by dividing the data in Figure
\ref{fig:K_RRF} by the integral of the Debye-Waller factor (see
Eq.~\ref{eq:K_master}) over the resolution function. In order to
compare it with the surface structure factors measured previously
for other liquid metals, we have normalized $q_z$ by the value of
$q_{peak}$ - the value of $q_z$ at which the layering peak is
observed or is expected to be observed. In fact we are comparing
the electron density structure factors of the different metals.
Alternatively we might have compared  the atomic densities;
however, none of the atomic form factors vary by more than a few
percent over the measured range of $q_z$. The surface structure
factor of the pure K is found here to rise clearly above unity and
shows a similar functional behavior to that of the pure Ga and In,
metals for which the surface-induced layering has been
well-documented \cite{Regan95, Tostmann99}. The inset shows the
same plot comparing the surface structure factors extracted from
the specular reflectivity measurements of pure Ga, pure In and
pure K over an extended $q_{z}/q_{peak}$ range. Even though
experimental and intrinsic limitations (such as the "$\eta$
effect" discussed above) prevents us from resolving the layering
peak in its' entirety, it seems clear that liquid K exhibits
surface layering similar to that observed in other metals. These
experimental results are supported by computer simulations
predicting a surface-induced layering in pure potassium
\cite{Chekmarev98}.

\section{Summary and Discussion}

In this work we have expressed the x-ray scattering from the
liquid surface, both specular and diffuse, in terms of the product
of a thermal fluctuation term and a surface structure term
(Eq.~\ref{eq:K_structure}). This is analogous to the usual
treatment of \index{Bragg reflection} Bragg scattering from
three-dimensional crystals for which the intensity is expressed as
a product of a structure factor and a phonon-induced Debye-Waller
factor. For the liquid surface we have assumed that the thermal
fluctuations could be described by capillary waves over all length
scales extending from a distance $\xi \approx \pi/q_{max}$ that is
of the order of the atomic or molecular size, to a macroscopic
distance that is determined by the reciprocal of the reflectometer
resolution. The gravitationally determined capillary length  is
ignored since it is many orders of magnitude longer than the
resolution-determined length.  In fact the capillary wave model is
strictly true only for long wavelengths and the assumption that it
can be invoked for lengths as small as $\xi$ is only justified
empirically \cite{Magnussen95, Regan95, Tostmann99, Tostmann98,
Dimasi00}. However, the good agreement between the measured
 diffuse scattering and that
predicted by Eq.~\ref{eq:K_master}, found for a number of liquids
including the present one provides a rather general justification
for adopting this limit.

In the present study we generalized the methods that were
previously used for liquid surfaces by simultaneously carrying out
detailed measurements along and transverse to the specular axis.
As in these earlier studies we have been able to verify the $1 /
q_{xy}^{2-\eta}$ behavior of the capillary wave form of the
surface fluctuations to values of $\eta = 0.82$ and to obtain a
measure of the surface \index{surface!structure factor} structure
factor. Aside from recent work by Mitrinovic, Williams and
\index{Schlossman, Mark} Schlossman \cite{Mitrinovic01} which were
carried out to $\eta = 0.62$, the largest values of $\eta$
attained in any of the previous studies on liquids were less than
half of our value. In view of the fact that the measured value of
$\Phi(q_z)$ depends on proper isolation of the capillary diffuse
scattering our confirmation of the capillary model for diffuse
scattering from  K is  essential.

The most significant limitation to the present measurement came
about as a result of the fact that due to the thermally excited
capillary wave \index{Debye-Waller} Debye-Waller-like effect, the
largest value of $q_z$ at which the specular peak can be observed
is smaller than $1.6~{\AA}^{-1}$ at which  $\Phi(q_z)$ is expected
to have a peak. This limitation can be expressed in terms of the
capillary model of Eq.~\ref{eq:K_master} since as $\eta
\rightarrow 2$ the specular peak no longer exists. Another way to
see the same thing is that the scattering at the peak of
$\Phi(q_z)$ will be difficult to observe when the product
$\sigma^2(q_{peak})q_{peak}^2 \gg 1$. In essence, this condition
occurs when the effective surface roughness approaches that of the
surface layer spacing. Although the precise value of
$\sigma(q_{peak})$ depends on the resolution, for the resolutions
used to study K, $\sigma^2(q_{peak})q_{peak}^2 = 13.8$.
Nevertheless, in this chapter we demonstrated  by careful
measurement of the diffuse scattering with a number of different
resolution configurations that it was possible to quantitatively
extract values of  $\Phi(q_z)$ for K over a significant range of
$q_z$. These measurements establish that over the measured range
the $q_z$-normalized surface structure factor for pure liquid K is
nearly identical to that found in Ga and In, with a well defined
rise above unity as the peak in $q_z$ is approached from below.
This strongly suggests that the layering in K is identical to that
of other liquid metals, as implied by \index{Chac{\'o}n, Enrique}
Chacon et al. \cite{Velasco02}. Finally, we note that the present
method could also be applied to non-metallic liquids having
similar surface tensions \cite{Chacon01, Soler01, Chapela77}. The
unresolved and intriguing question of whether the local surface
layering in metallic systems is different from that of
non-metallic liquids when the surface tensions involved are
similar will have to await future studies.

\bibliographystyle{unsrt}

%% file: water_all.tex

\chapter{Search for Layering in Non-metallic Liquids: Water}

\section{Abstract}

Recent measurements show that the free surfaces of liquid metals
and alloys are always layered, regardless of composition and
surface tension, a result supported by three decades of
simulations and theory. Recent theoretical work claims, however,
that at low enough temperatures the free surfaces of \textit{all}
liquids should become layered, unless preempted by bulk freezing.
Using x-ray reflectivity and \index{x-ray!diffuse scattering}
diffuse scattering measurements we show that there is no
observable surface-induced layering in water at T=298 K, thus
highlighting a fundamental difference between
\index{liquid!dielectric} dielectric and
\index{liquid!metallic}metallic liquids. The implications of this
result for the question in the title are discussed.

\section{Introduction}
\begin{figure}[tbp]
\centering
\includegraphics[width=1.0\columnwidth]{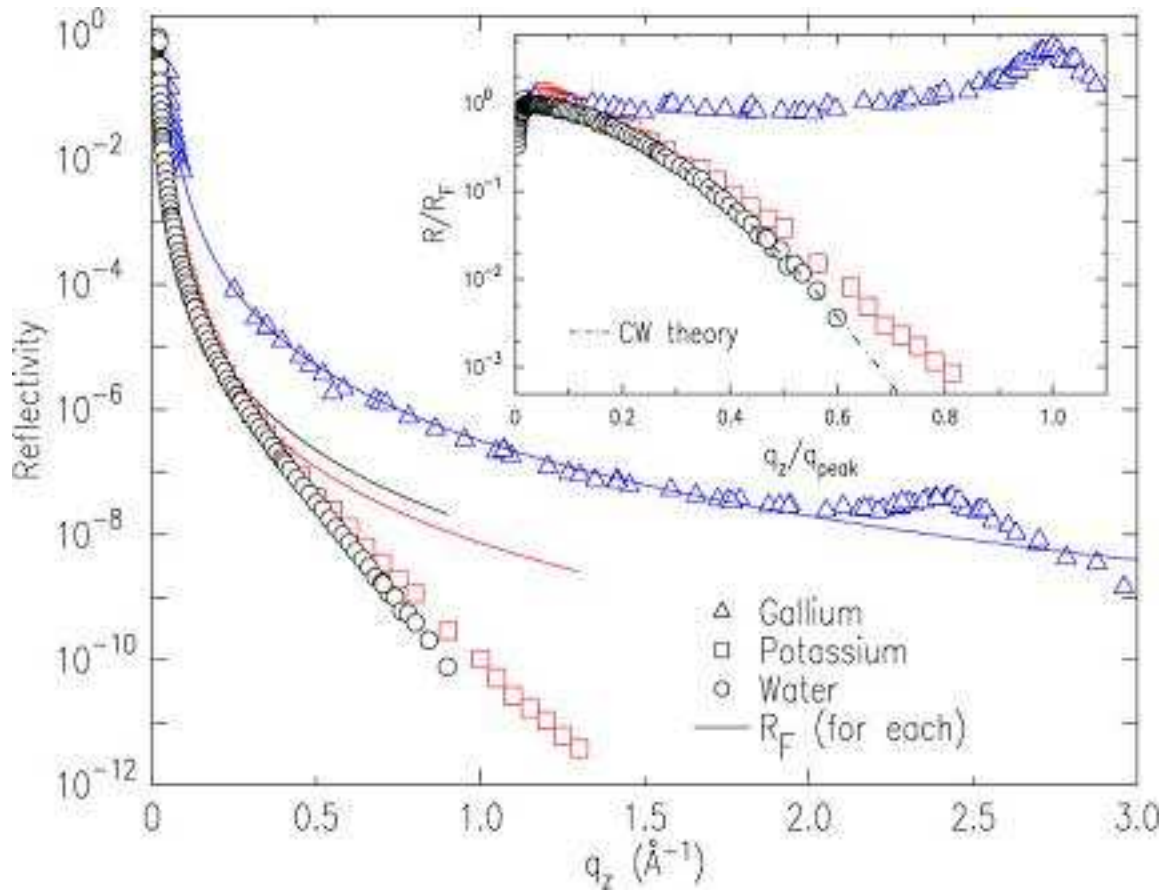}
\caption{\label{fig1} X-ray reflectivity from free liquid surfaces
of the indicated samples. Points - measured [$R(q_z)$], lines -
calculated for an ideally flat and steplike interface
[$R_F(q_z)$]. The inset shows the ratio of the two. }
\end{figure}

The free surface of liquid metals and alloys were demonstrated
experimentally over the last few years to be layered, i.e. to
exhibit an atomic-scale oscillatory surface-normal density
profile.\cite{Magnussen95, Regan95,Tostmann00,Yang99} This is
manifested by the appearance of a \index{quasi-Bragg peak}
Bragg-like peak in the x-ray reflectivity (XR) curve, $R(q_z)$, as
shown in Fig. \ref{fig1} for Ga.\cite{Regan95} The wave-vector
transfer position of the peak, $q_{peak}$ is related to the
layering period $d$ by $q_{peak}=2\pi/d$. The layered interface is
in a marked contrast with the theoretical description of the
liquid-vapor interface of a simple liquid. This theory, prevailing
for over a century, depicts the density profile as a monotonic
increase from the low density of the vapor and the high density of
the bulk liquid.\cite{Buff65,Rowlinson82} This view was supported
by XR measurements on many non metallic liquids measured over the
last two decades, including water,\cite{Braslau85} \index{alkanes}
alkanes, \cite{Ocko94} and quantum liquids, \cite{Lurio92} which
showed no Bragg-like peaks. However, as discussed below, the
measurements in all of these studies were restricted to the small
q range, i.e. $q_z \ll \pi/$atomic size, and would not have
detected \index{surface!layering}  surface layering if it existed.

Early simulations on non-metallic liquids demonstrate that atomic
layering is ubiquitous near a hard flat surface
 and this has been observed for liquid Gallium.\cite{Huisman97}
On the other hand, for the liquid  vapor interface it is tempting
to think that the large surface tension $\gamma$ of liquid metals
such as Hg ($\gamma\approx500$ mN/m), Ga ($\gamma\approx750$
mN/m), and In ($\gamma\approx550$ mN/m) might be the explanation
for the SL observed at their surface. This assumption is partially
mitigated by the observation of SL at the free surface of liquid K
where $\gamma\approx100$ mN/m only. Here we report x-ray
scattering results showing that the free surface of water, which
has nearly the same surface tension as K, does not exhibit SL
features in the reflectivity profiles, thereby suggesting that
surface tension by itself does not explain SL. This conclusion
rests on the validity of the capillary wave theory (discussed
below).

\index{Rice, Stuart} Rice \textit{et al}. \cite{Rice74} first
predicted SL in liquid metals three decades ago. They argued that
the layered interface structure for liquid metals is a consequence
of the strong dependence of the effective ion-potential energy on
the steeply varying electron density across the liquid/vapor
interface. At the low density vapor phase the electrically neutral
atoms interact through a \index{interactions!Van der Waals} van
der Waals interaction only. In the metallic, higher density,
liquid phase the electrons are delocalized, and the much more
complex interactions involve an interplay between a quantum
\index{Fermi!gas}  Fermi fluid of free electrons and a classical
liquid of charged ion cores. Rice concludes that this substantive
change in the effective-ion-potential stabilizes the short range
surface fluctuations with the result that the atoms near the
surface form a layered structure. \cite{Rice74} Calculations
employing the glue model of metallic cohesion support these
conclusions. \cite{Celestini97} By contrast, Soler \textit{et
al.}\cite{Soler01} claim that SL is solely due to the formation of
a dense layer at the surface. This layer is not restricted to
\index{liquid!metallic} metallic liquids, but may form also in
nonmetallic liquids due to nonisotropic interactions, such as
remnant covalent bonding in liquid Si\cite{Soler01} and the highly
directional interactions in \index{liquid!crystals} liquid
crystals,\cite{Pershan82} or in non-uniform Lennard-Jones fluids
with unbalanced attractive forces.\cite{Katsov02} Additionally,
the SL has been found in colloidal systems, \cite{Madsen01} which
suggests that layering phenomenon may be more fundamental than
previously thought. More recently, based on extensive simulations,
\index{Chac{\'o}n, Enrique} Chac{\'o}n \textit{et
al.}\cite{Chacon01} argued that SL does not require the many-body,
delocalize-electron interactions of a liquid metal at all. Rather,
according to them, SL is a universal property of all liquids at
low enough temperatures, $T \lesssim T_c/a$, whenever not
preempted by bulk solidification. Here $T_c$ is the critical
temperature of the liquid, and $a \approx 4-5$.

\section{Effect of Capillary Waves}
The dichotomy between these two views could be resolved, in
principle, by XR measurements  on selected materials
\cite{Deutsch98} to see which exhibit, or not, a layering peak at
some $q_{peak}$. Unfortunately, practical considerations limit the
number of elemental liquids that can be studied to a relatively
few. One of the major problems is that the measurable $q_z$ range
is more often than not limited to values much less than $q_{peak}$
by the strong off-specular diffuse scattering  caused by thermal
capillary waves. The effect of the capillary waves is to induce a
surface roughness $\sigma \sim \sqrt{T/\gamma}$, where $\gamma$ is
the surface tension. The consequence of this, which is shown in
Fig.~\ref{fig1}, for three liquids at room temperature, is to
reduce the reflectivity $R(q_z)$ below that of the theoretical
\index{Fresnel reflectivity} Fresnel reflectivity from an
interface with an idealized flat, steplike surface-normal density
profile. For low-$\gamma$ liquids such as water ($\sim$ 70 mN/m)
and K ($\sim$100 mN/m) the reduction is significant. For Ga, where
$\gamma \approx 750$ mN/m the effect is almost negligible at room
temperature. Even that small reduction is almost completely offset
at room temperature by the SL effect that peaks at $q_z \approx
2.5~${\AA}$^{-1}$. At higher temperatures, however, the effect is
quite prominent. \cite{Regan95}

Although the rapid falloff in $R/R_F$ of water prevents its
measurements out to $q_z\approx q_{peak}=2.0~\AA^{-1}$, our recent
studies of liquid K \cite{Shpyrko03} demonstrated that if surface
layering is present its signature can still be observed clearly at
$q_z<<q_{peak}$ even for low $\gamma\lesssim 100$\ mN/m liquids.
This is accomplished by carefully accounting for the effects of
capillary waves, based on diffuse x-ray scattering (DS)
measurements. We present here an x-ray study of the surface
structure of water over the most extended $q_z$-range published to
date, and including \index{x-ray!diffuse scattering} diffuse
scattering \cite{Braslau85, Schwartz90}. For the present
measurements the intrinsic \index{surface!structure factor}
surface structure factor of water can be extracted directly from
the raw $R(q_z)$ without resorting to any structural model for the
interface. Comparison between the water surface structure factor,
for which there is no evidence of SL, and that of K and Ga
suggests that surface tension is not the dominant cause of the SL
observed in liquid metals.

\section{Experimental Details}
X-ray measurements were carried out on the CMC-CAT liquid surface
diffractometer, APS, \index{Argonne} Argonne National Laboratory,
at a wavelength of $\lambda=1.531~{\AA}$. The purified water
sample was contained in a Langmuir trough\cite{Schwartz92} mounted
on the diffractometer. The surface was periodically swept with a
teflon barrier, monitoring $\gamma$ with a film balance, to ensure
a clean surface.\cite{Gaines66}

\begin{figure}[tbp]
\epsfig{file=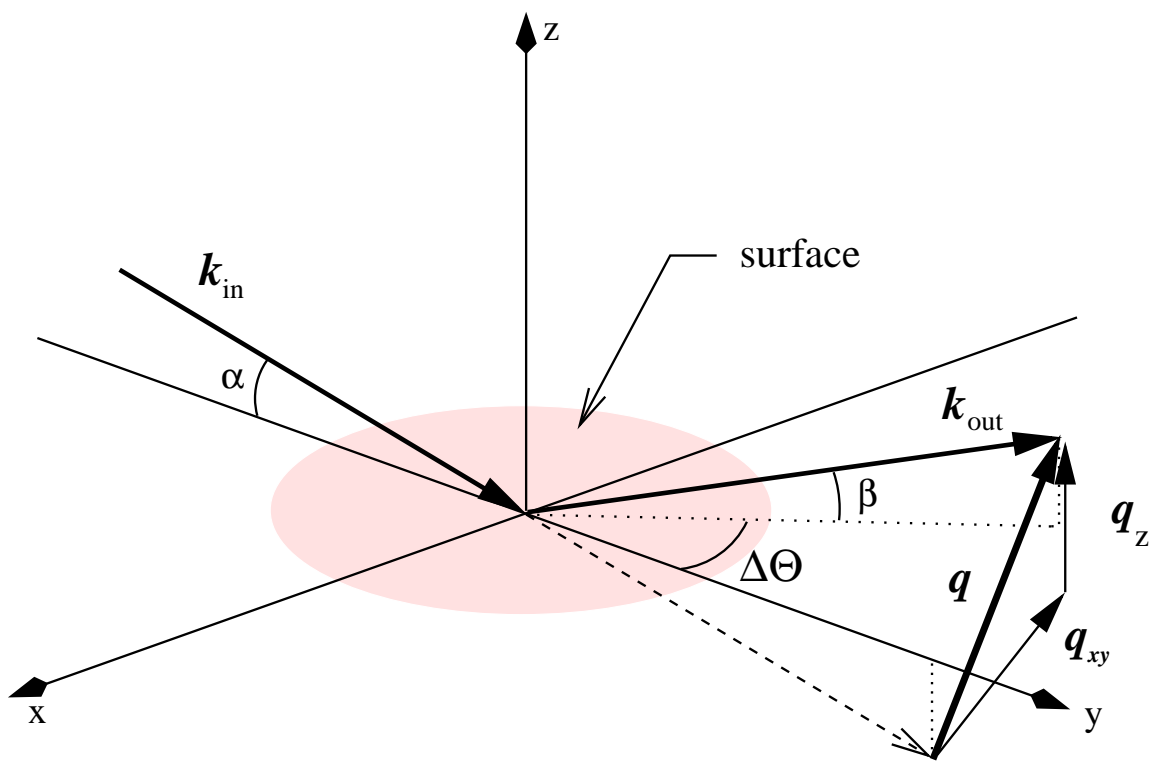, angle=0, width=1.0\columnwidth}
\caption{\label{fig2} Schematic description of the experimental
setup. }
\end{figure}
Both XR \cite{Deutsch98} and DS \cite{Pershan00} are well
documented techniques. Experimental geometry setup employed in the
experiments described in this work is shown in Fig.~\ref{fig2}.
For x-rays striking the surface at a grazing angle $\alpha$ and
detected at an output angle $\beta$ \textit{in the specular
plane}, the surface-normal ($z$) and in-plane surface-parallel
($y$) components of the wavevector transfer are $q_z = (2 \pi /
\lambda) \left( \sin \alpha + \sin \beta \right)$ and $q_{y} = (2
\pi / \lambda) \left( \cos \alpha - \cos \beta \right)$. In a XR
measurement both angles are varied, keeping $\alpha = \beta $, so
that $q_{y}=0$, and the reflected signal, measured vs $q_z$ and
divided by the incident intensity $I_0$, yields $R(q_z)$. In a DS
measurement, $\beta$ is varied for a fixed $\alpha$, and hence
both $q_z$ and $q_y$ vary.

Within the \index{Born approximation} Born
approximation,\cite{Deutsch98} $R(q_z)/R_F(q_z)=|\Phi (q_z)|^2
W(\eta,q_z)$, where $\Phi (q_z) = (\rho_{\infty})^{-1} \int
[\textrm{d}\left< \rho(z) \right>/\textrm{d}z] \exp(\imath q_z z)
\textrm{d}z$\ is the conventional \index{surface!structure factor}
structure factor of the liquid-vapor interface, $\left< \rho(z)
\right>$\ is the \textit{intrinsic} (i.e. in the absence of
capillary wave smearing) surface-parallel-averaged electron
density profile along the surface-normal $z$ direction, and
$W(\eta,q_z)$ accounts for the smearing of the intrinsic density
profile by capillary waves. The aim of our XR measurement is to
determine $\Phi (q_z)$ to observe a possible layering peak. Since
both $\Phi (q_z)$ and $W(\eta,q_z)$ depend on $q_z$, extracting
$\Phi (q_z)$ directly from the measured $R(q_z)$ is possible as
$|\Phi(q_z)|^2=R(q_z)/W(\eta,q_z)$ only if $W(\eta,q_z)$ is known
independently. For this we use the DS data.

\section{X-ray Diffuse Scattering Data}
\begin{figure}[tbp]
\centering
\includegraphics[width=1.0\columnwidth]{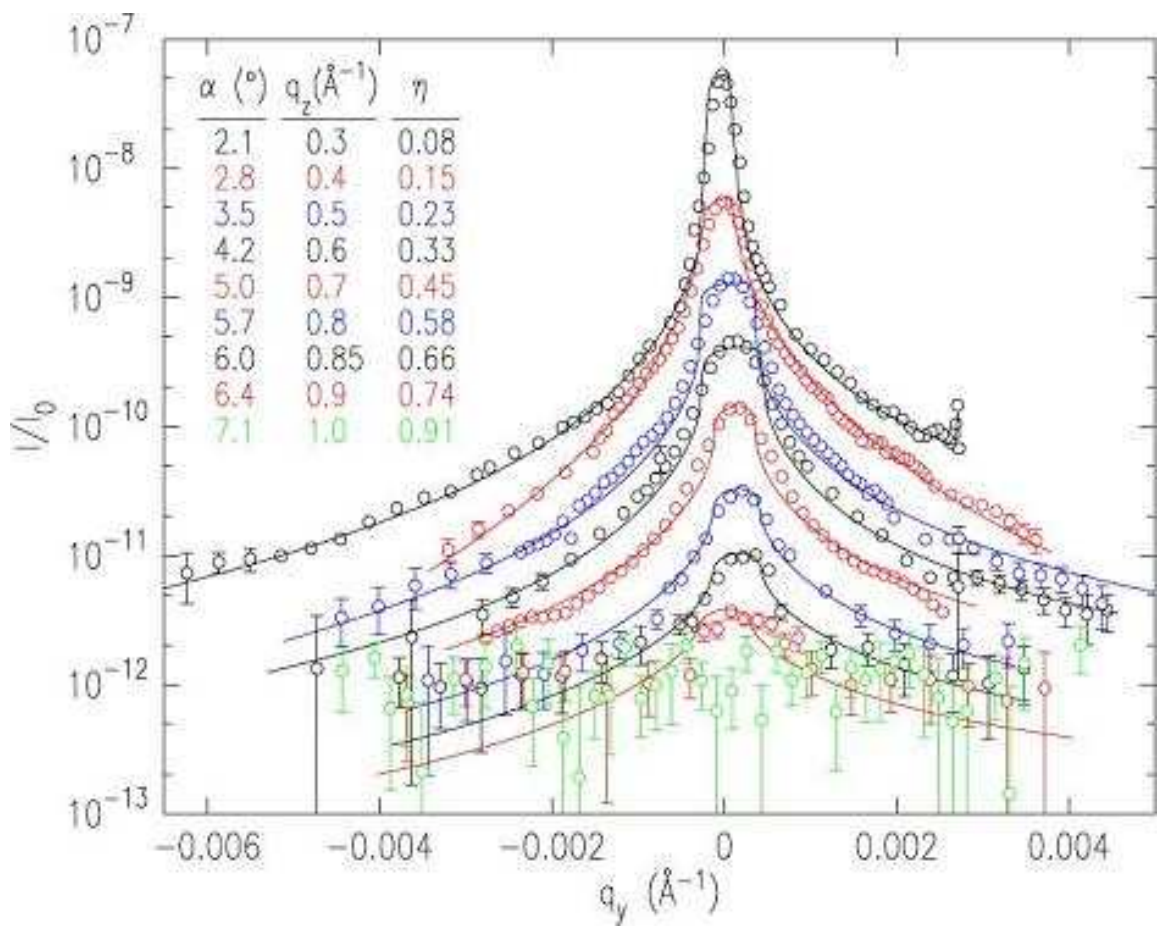}
\caption{\label{fig3} Comparison of measured \index{x-ray!diffuse
scattering} diffuse scattering
 with capillary wave theory predictions for the
angles of incidence $\alpha$ (top to bottom): 2.1$^{\circ}$,
2.8$^{\circ}$, 3.5$^{\circ}$, 4.2$^{\circ}$, 5.0$^{\circ}$,
5.7$^{\circ}$, 6.0$^{\circ}$, and 6.4$^{\circ}$. The last data set
corresponding to 7.1$^{\circ}$ shown in the inset no longer
exhibits a distinguishable specular peak. The $q_z$ values
corresponding to the specular condition $q_y=0$~\AA$^{-1}$ are
0.3, 0.4, 0.5, 0.7, 0.8, 0.85, 0.9, and 1.0~$\AA^{-1}$,
respectively. }
\end{figure}

The measured DS, shown in Fig.~\ref{fig3}, is given by theory as
$I_{DS}=\int[\textrm{d}\sigma(q_x-q'_x,
q_y-q'_y,q_z-q'_z)/\textrm{d}\Omega]\textrm{d}\omega(q'_x,q'_y,q'_z)$,
where $\textrm{d}\omega$ is the angular resolution of the
diffractometer and
\begin{equation}
\frac{d\sigma }{d\Omega } = \frac{A_{0}}{8\pi\sin \alpha } q_z ^2
R_F(q_z) \mid \Phi (q_z) \mid ^2 \frac{\eta}{q_{xy}^{2-\eta}}
\left( \frac{1}{q_{ max}} \right) ^\eta \: \label{eq:master}
\end{equation}
is the scattering cross-section. Here $q_c$ is the
\index{critical!angle}  critical angle for total \index{external
reflection} external reflection of x rays, $q_{max} \approx
\pi/\xi$ is the upper cutoff for capillary wave contributions,
with $\xi$ of order of the atomic diameter, and $ \eta
=(k_BT)/(2\pi \gamma)q_z^2 $.

Two subtle complications are encountered in analyzing the DS data.
First, nonsurface DS contributions (scattering from the bulk,
sample chamber \index{x-ray!windows} windows, etc.) can
significantly distort the shape of the DS scans. These background
contributions are measured by offsetting the detector by
$\Delta\Theta = \pm 0.3^\circ$ from the plane of incidence, and
are already subtracted from the data shown in Fig.~\ref{fig3}.
Second, in the fixed-$\alpha$ DS scans, carried out by scanning
$\beta$, $q_z$, and thus $|\Phi(q_z)|^2$, also vary. Fortunately,
$\delta q_z \approx \delta q_y /\beta$ and for the range of $q_y$
displayed in Fig.~\ref{fig3} and typical $q_z$ values, the changes
are small enough to be neglected, e.g. a maximal $\delta q_z
\approx 0.06~{\AA}^{-1}$\  at the largest $\beta$\ for
$\alpha=3.5^{\circ}$. Thus, it is a good approximation for each of
the DS scans to treat $\Phi(q_z)$ as a fixed function of $\alpha$.
$\mid\Phi(q_z )\mid^2$ is then obtained by dividing the measured
DS and XR curves by $W(\eta,q_z)=\int[A_{0}/(8\pi\sin \alpha)] q_z
^2 R_F(q_z) (\eta/q_{y}^{2-\eta})q^{-\eta}_{max}
\textrm{d}\omega$. \cite{Shpyrko03}

The theoretical curves calculated using Eq.~\ref{eq:master} with
$T=298~K$ and $\gamma=72~$mN/m are shown as lines in Fig.
\ref{fig3}. As $q_z$ increases, so do both $\eta \propto q_z^2$
and the intensity of the off-specular power-law wings relative to
that of the specular peak at $q_y=0$. The curve at
$q_z=1~{\AA}^{-1}$ demonstrates the capillary-wave-imposed limit
where the specular signal at $q_y=0$, which contains the surface
structure information, becomes indistinguishable from the DS
signal at $\pm q_y > 0$. In principle, this limit arises from the
fact that for $\eta \geq 2$ the singularity at $q_{y}=0$ in
$d\sigma/d\Omega$ vanishes and there is no longer any criterion by
which the surface scattering can be differentiated from other
sources of diffuse scattering. In practice, the fact that the
projection of the resolution function on the horizontal x-y plane
is very much wider transverse to the plane of incidence than
within the plane of incidence reduces this limit to a value closer
to $\eta \approx 1$. \cite{Shpyrko03} Figure~\ref{fig3} exhibits
excellent agreement between the theoretical DS curves calculated
from Eq. \ref{eq:master} with the measured DS over several decades
in intensity and one decade in $\eta$, \textit{without any
adjustable \index{fitting!parameters} parameters}. This confirms
the applicability of the capillary wave theory for the surface of
water over the $q_z$ range studied here, $0 \leq q_z \leq
0.9~{\AA}^{-1}$. Measurements of \index{x-ray!diffuse scattering}
diffuse scattering for small $q_z$ (i.e. small $\eta$) to values
of the surface parallel component of the wave vector transfer
$q_x$ of the order of $\pi/$atomic size have been done by
\index{Daillant, Jean} Daillant \textit{et. al} \cite{Fradin00,
Mora03} by moving the detector out of the plane of incidence for
\index{x-ray!grazing incidence diffraction} grazing incident
angles. At such large wave vectors the observed scattering must be
interpreted as the superposition of scattering due to surface
capillary waves and bulk diffuse scattering from the liquid below
the surface. The separation of these two contributions is rather
subtle and was only accomplished through nontrivial calculations
of the noncapillary terms. Eventually \index{Daillant, Jean}
Daillant \textit{et. al} concluded that as the value of $q_x$
approaches the atomic scale (i.e., $\pi/$atomic size) the surface
tension varies with $q_x$.\cite{Mecke99} In principle, this
\index{dispersion} dispersion should affect the sum rule that is
the origin of the $1/q_{xy}^\eta$ in the \index{differential cross
section} differential cross section in Eq.(\ref{eq:master}). On
the other hand, considering that the dispersion is a relatively
small effect that will only change the argument of a logarithmic
term, its effect on the present analysis can be neglected.

\section{X-ray Reflectivity Data}
\begin{figure}[tbp]
\centering
\includegraphics[width=1.0\columnwidth]{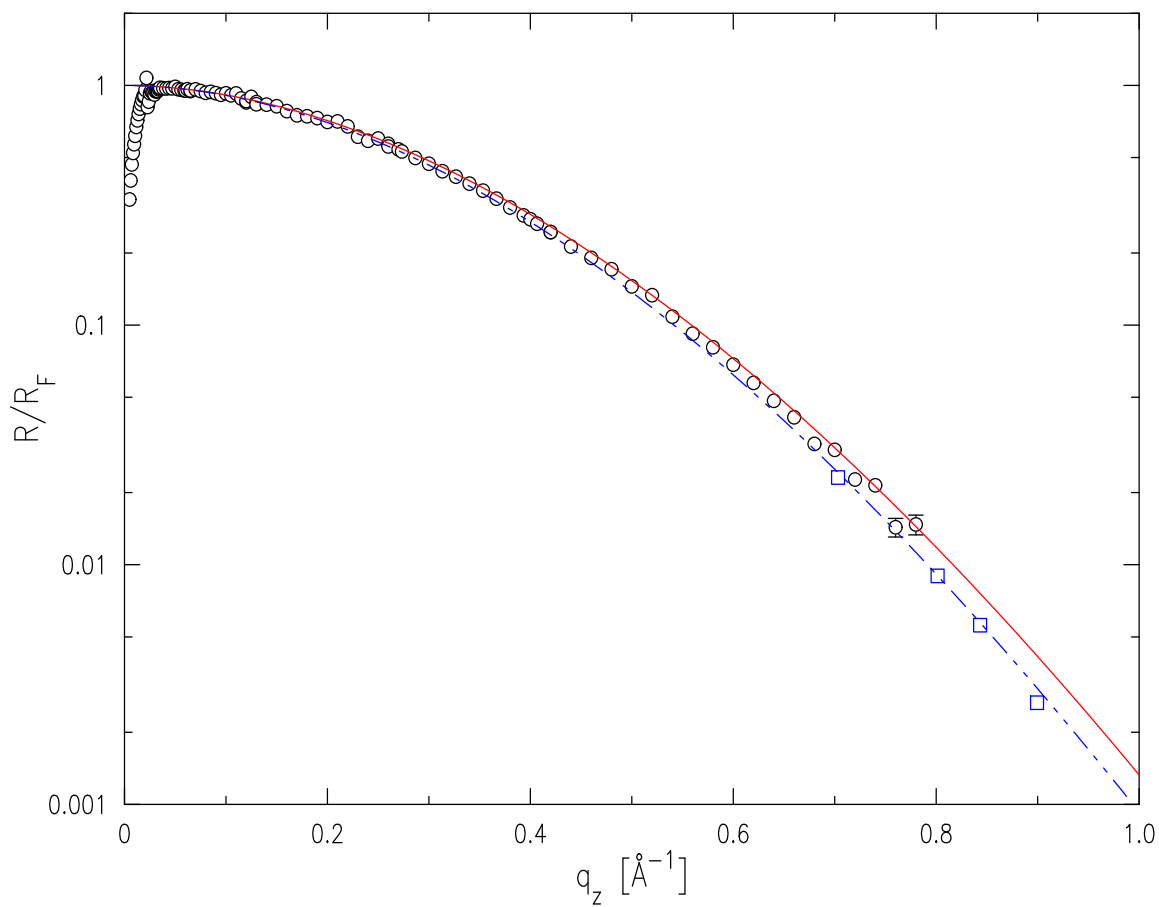}
\caption{\label{fig3a} Fresnel-normalized X-ray reflectivity from
free liquid surfaces water, taken for two different values of
resolution with bicron detector (circles) and linear detector
(squares). Blue and red lines show theoretical expected values for
the two measurements, respectively, assuming no surface structure.
}
\end{figure}

Fig.~\ref{fig3a} shows a specular reflectivity data taken for
liquid water, normalized by Fresnel function. Low angle data was
taken with a point bicron detector, but at high angles we used a
position sensitive, wire detector in order to effectively resolve
specular peak along with diffuse wings. Bicron detector has a
higher efficiency than linear detector and is therefore preferable
at low angles where $\eta$-effect is negligible. Due to slightly
different \index{resolution function} resolution functions of the
two detectors, the x-ray reflectivity data does not match
precisely. Shown with blue and red lines are the theoretical
predictions for the x-ray reflectivity signal assuming capillary
wave contributions according to Eq.(\ref{eq:master}. Red line
corresponds to resolution function used for data taken with bicron
detector, and blue (dashed) line used resolution function of wire
detector. Since the data doesn't deviate significantly from these
theoretical curves, one can conclude that the surface of water
does not exhibit a strong structure factor observed in metallic
liquids.

\begin{figure}[tbp]
\centering
\includegraphics[width=1.0\columnwidth]{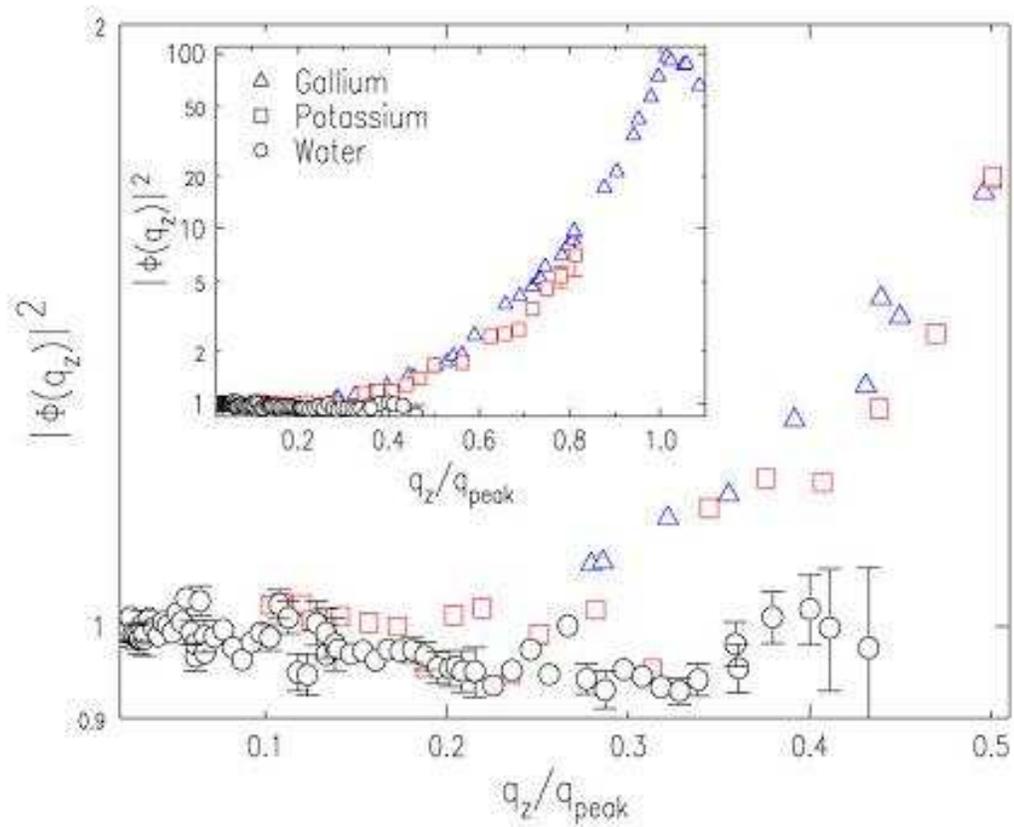}
\caption{\label{fig4} Comparison of the structure factor squared
$\mid \Phi(q_z) \mid ^2$ for water (circles), liquid potassium
(squares), and liquid gallium (triangles). The wavevector $q_z$ is
normalized to the expected position of the layering peak
$q_{peak}$ of each sample. The inset shows the data on an extended
scale. For discussion see text.}
\end{figure}

$|\Phi(q_z)|^2$ is then obtained directly from the measured XR
curve, as $R(q_z)/W(\eta,q_z)$ as discussed above. It is shown in
Fig.~\ref{fig4} (circles) along with previously measured results
for K (squares) and Ga (triangles).

The rise of the Ga $|\Phi(q_z)|^2$ to $\sim$100 at $q_{peak}$ due
to layering can be clearly seen on the inset of Fig.~\ref{fig4}.
For K, the capillary-wave-imposed limit only allows obtaining
$|\Phi(q_z)|^2$ for $q_z \lesssim 0.8 q_{peak}$. Nevertheless, the
value of $|\Phi(q_z)|$ for K starts to deviate from unity for
values of $q_z/q_{peak} \approx 0.3$. Furthermore, over the range
for which it can be measured it is basically identical to the
\index{surface!structure factor} structure factor of Ga. This is a
clear indication that the surface of liquid K has essentially the
same SL as that of Ga, which is also nearly identical to that of
the other liquid metals that have been studied to date, e.g.,
In\cite{Tostmann99} and Sn\cite{Shpyrko03b}.  For water, however,
no deviation of $|\Phi(q_z)|^2$ from unity is observed even at the
highest measurable $q_z/q_{peak}\approx0.5$. This suggests that
surface-induced layering does not occur at the surface of water.
The different behavior, in spite of the similar $\gamma$ of water
and K, leads to the conclusion that the \index{surface!layering}
surface layering in K, and by implication in other liquid metals,
is not merely a consequence of its surface tension.

The absence of layering in water, and its presence in potassium,
seems at first sight to corroborate the claim of \index{Rice,
Stuart} Rice \textit{et al.} \cite{Rice74} that layering is a
property arising from the metallic interaction of the liquid. On
the other hand, Chac{\'o}n \textit{et al.}\cite{Chacon01}, who
maintain that surface-induce layering is a general property of all
liquids, regardless of their interactions, predict that layering
should occur only at temperatures $T/T_c \lesssim 0.2$, where
$T_c$ is the critical temperature of the liquid. Although
supercooling is often possible, the practical limit for most
reflectivity measurements is the melting temperature $T_m$. Thus
the smallest $T/T_c$ for any liquid is on the order of $T_m/T_c$.
For liquid metals $T_m/T_c\approx 0.15$ (K), 0.13 (Hg), 0.07 (In),
0.066 (Sn), and 0.043 (Ga). Since these values are $ < 0.2$, by
Chac{\'o}n's criteria surface layering is expected, and indeed
demonstrated experimentally to occur, in all of
them.\cite{Magnussen95,Regan95,Tostmann99, Shpyrko03, Shpyrko03b}
By contrast, for water, where $T_m/T_c=0.42 \gg 0.2$, Chac{\'o}n's
criteria predict that the appearance of surface layering is
preempted by bulk freezing, and thus no SL should occur at room
temperature, as indeed found here.

\section{Summary}
In summary, we have shown that the surface of water does not
exhibit SL even though its surface tension is not significantly
different from that of liquid K for which SL was observed. Thus
the atomic layering of liquid K cannot be the result of surface
tension alone. Although this seems to support Rice's argument that
the metallic phase is essential for SL, we note that SL is
exhibited by both \index{liquid!crystals} liquid
crystals\cite{Pershan82} and other intermediate-size organic
molecules.\cite{Dutta99} Consequently, there can be other criteria
for SL, in addition to those proposed by Rice. One possibility,
proposed by Chac{\'o}n \textit{et. al}, is that SL should be
ubiquitous for all liquids that can be cooled to temperatures of
the order of $0.2T_c$. Unfortunately, this is difficult to explore
experimentally since suitable liquids are rather scarce. For
example, liquid noble gases, the archetypical
\index{interactions!Van der Waals} van der Waals liquids, have all
$T_m/T_c > 0.55$. For liquid helium the ratio is lower but in view
of both the presence of the superfluid transition and the low
scattering cross section this is a difficult system to study.
X-ray reflectivity measurements at 1~K, i.e., $T_m/T_c \sim 0.2$,
did not exhibit evidence for SL\cite{Lurio92}. Similarly, the
polar liquids oxygen and fluorine have $T_m/T_c$ of only 0.38 and
0.37. There are, however, some low melting organic van der Waals
liquids, e.g. propane and 1-butene, that have $0.2 < T_m/T_c <
0.25$, and may be suitable for addressing this issue. It would be
also very interesting to measure the surface structure of
materials such as Se ($T_m/T_c=0.28$), Te (0.3), Sc (0.28), and Cd
(0.22), for which $T_m/T_c >0.2$. If it were possible to probe
surface layering in these high $T_m/T_c$ materials such
experiments could test Chac{\'o}n's argument. Unfortunately, these
materials have relatively high vapor pressure and that precludes
the use of UHV methods for insuring atomically clean surfaces. The
surfaces of some alkalis and mercury can be kept clean without
\index{UHV!methods} UHV methods. For most of the experimentally
accessible metals this is not true.


%% file: sn_all.tex

\chapter{Surface of Liquid Sn: Anomalous layering}

\section{Abstract}

\noindent X-ray reflectivity measurements on the free surface of
liquid Sn are presented in this chapter. They exhibit the
high-angle peak, indicative of surface-induced layering, also
found for other pure liquid metals (Hg, Ga and In). However, a
low-angle peak, not hitherto observed for any pure liquid metal,
is also found, indicating the presence of a high-density surface
layer. \index{x-ray!fluorescence} Fluorescence and
\index{x-ray!resonant scattering} resonant reflectivity
measurements rule out the assignment of this layer to
surface-segregation of impurities. The reflectivity is modelled
well by a 10$\%$ contraction of the spacing between the first and
second atomic surface layers, relative to that of subsequent
layers. Possible reasons for this are discussed.

\section{Introduction}

\index{Rice, Stuart} Rice and coworkers\cite{Rice74} predicted
that the atoms at the free surface of a liquid metal should be
stratified to a depth of a few atomic diameters. This layering
phenomenon has been experimentally  confirmed two decades later by
x-ray reflectivity measurements for three high surface-tension
metals: Hg,\cite{Magnussen95} Ga \cite{Regan95} and
In,\cite{Tostmann99} and the low-surface tension
K.\cite{Shpyrko03} The signature of layering in the x-ray
reflectivity curve is the appearance of a \index{quasi-Bragg peak}
quasi-Bragg peak at a wavevector transfer $q_z = 2 \pi / d$, where
$d$ is the atomic spacing between the layers. The peak arises from
constructive interference of waves diffracted by the ordered
subsurface layers. These measurements are complicated by several
technical issues, in particular the need to use UHV conditions to
preserve surface cleanliness for highly reactive liquid metals
surfaces.

Both Ga and In were found to \index{Regan, Michael} Regan et
al.\cite{Regan95} comprising equal-density, periodically-spaced
atomic layers. Liquid Hg\cite{Magnussen95,Dimasi98} shows a more
complicated surface structure. However, unlike Ga and In, the high
vapor pressure of Hg did not allow its study under UHV conditions,
and the possibility that the more complicated structure originates
in chemical interactions with foreign atoms at the surface cannot
be definitively ruled out. For Ga, in particular, the decay length
of the layering into the bulk was found to be slightly larger than
that expected, and found, for other liquid metals. This was
tentatively assigned to the enhanced Ga-Ga pairing
tendency,\cite{Regan95} reflected also in the appearance of a
small shoulder on the first, nearest-neighbor, peak of the bulk
radial distribution function,\cite{Dicicco94,Gong93} and in the
ordering of Ga at the liquid-solid interface.\cite{vanderVeen97} A
similarly strong pairing tendency has been reported also for Sn,
the subject of the present study.\cite{Jovic76, DiCicco96,
Itami03}

The motivation of the present study was to investigate whether
atomic layering exists in liquid Sn, and, if it does, to find out
whether or not the layering follows the classic behavior of Ga and
In, or do new effects appear, e.g. due to the strong pairing. The
relatively low melting temperature, $T_m = 232$~$^{\circ}$C, and
vanishingly low vapor pressure at $T_m$ render such measurements
possible. As shown below, the measured x-ray reflectivity off the
Sn surface indeed exhibits an anomalous feature that is not
present for either Ga or In. Additional experiments were carried
out to rule out the possibility that this anomalous feature is not
intrinsic, but caused by chemical impurities at the surface. The
origin of this structure is identified as a $\sim 10$\%
contraction in the first to second interlayer distance.

\section{Experimental}

The UHV chamber and the preparation procedures of the liquid Sn
sample had been described previously.\cite{Tostmann99, Huber02}
Prior to placing an Sn sample inside the Mo sample pan, the
surface of the pan was sputtered clean. In the sputtering process
the sample's surface is bombarded by Ar$^{+}$ ions, which break up
and sputter away the oxidized material covering the surface.
Ingots of solid Sn with purity of 99.9999 $\%$ were placed in the
Mo sample pan inside a UHV chamber evacuated to $10^{-9}$ Torr and
were then melted by a \index{Boralectric resistive heater}
Boralectric heating element mounted underneath the pan. After the
molten Sn filled the pan it was cooled to the solid phase, a
24-hour \index{bakeout} bakeout process was performed. In the
bakeout process the walls of the chamber and its components are
gradually heated up to $100-200^{\circ}$C and then cooled down to
room temperature. Following this procedure the sample was melted
again and the macroscopic native oxide, as well as any possible
contaminations, at the surface were removed by a mechanical
scraping of the liquid surface with Mo foil wiper. After this, the
residual microscopic surface oxide layer was removed by further
sputtering by the Ar$^{+}$ ion beam for several hours. If there
are impurities in the bulk with surface energies that are lower
than that of Sn the \index{Gibbs!adsorption} Gibbs adsorption rule
implies that they would segregate at the surface. We will
demonstrate below that even though Sn has a relatively high
surface tension impurities are not present at the surface.

\section{X-ray reflectivity measurements}

\begin{figure}[tbp]
\epsfig{file=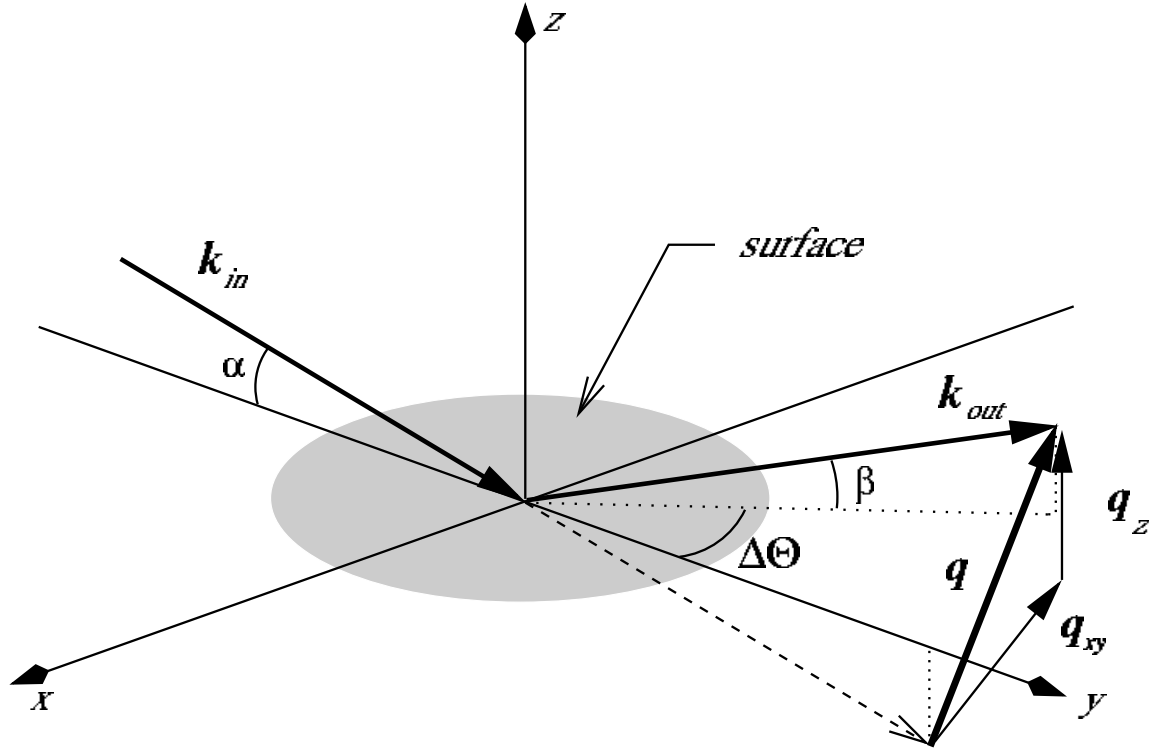, angle=0, width=1.0\columnwidth}
\caption{Kinematics of the x-ray measurement. $k_{in}$\ and
$k_{out}$ are the wavevectors of the incident and detected x-rays,
respectively.} \label{fig:K_kinema}
\end{figure}

The measurements were carried out at the liquid surface
\index{spectrometer} spectrometer facility of the ChemMAT~CARS
beamline at the Advanced Photon Source, \index{Argonne} Argonne
National Laboratory, Argonne, IL. The kinematics of x-ray
measurements described in this work are illustrated in Fig.
\ref{fig:K_kinema}. X-rays with a wavevector
$k_{in}=2\pi/\lambda$, where $\lambda=0.729~\AA$  is the x-ray's
wavelength, are incident on the horizontal liquid surface at an
angle $\alpha$. The detector selects, in general, a ray with an
outgoing wavevector $k_{out}$. The reflectivity $R(q_z)$ is the
intensity ratio of these two rays, when the specular conditions,
$\alpha=\beta$\ and $\Delta\Theta=0$, are fulfilled. In this case
the surface-normal momentum transfer is $q_z=(2\pi/\lambda)(\sin
\alpha+\sin \beta)=(4\pi/\lambda) \sin \alpha$.

\begin{figure}[tbp]
\centering
\includegraphics[width=1.0\columnwidth]{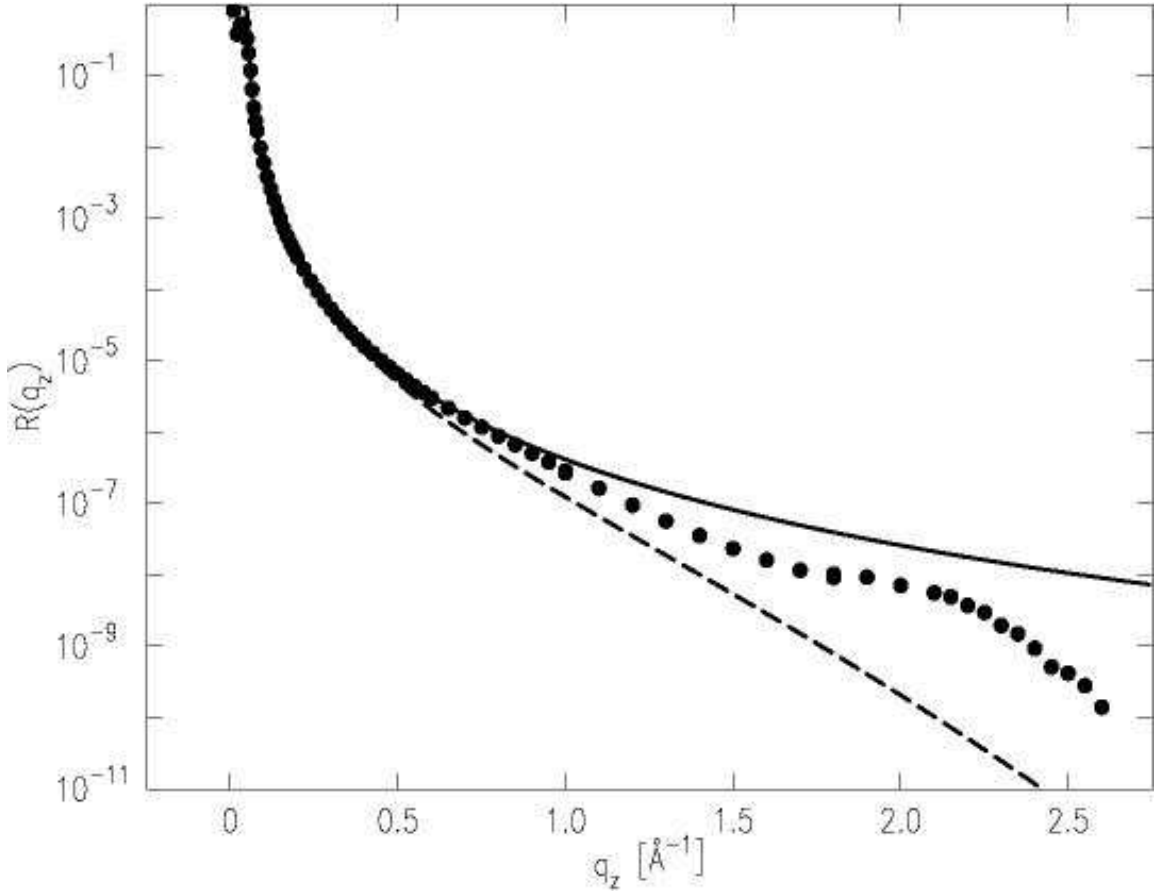}
\caption{\label{fig:Snref} The measured x-ray specular
reflectivity (points) of the surface of liquid Sn. The Fresnel
reflectivity (solid line) of an ideally flat and abrupt surface,
and the reflectivity of an ideal surface roughened by
thermally-excited capillary waves (dashed line) are also shown.}
\end{figure}

The x-ray specular reflectivity shown with circles in
Fig.~\ref{fig:Snref} is the difference between the specular signal
recorded with Oxford scintillation detector at $\alpha=\beta,
\Delta\Theta=0$ and the off-specular background signal recorded at
the same $q_z$, but at $\Delta \Theta=  \pm 0.1^\circ$. The data
is then normalized to the measured incident intensity. This
background subtraction procedure is particularly important because
the bulk scattering function peaks at approximately the same $q_z$
as the \index{quasi-Bragg peak} quasi-Bragg surface layering peak,
since both peaks correspond roughly to the interatomic distance.
As a result the background is particularly strong at the $q_z$
values corresponding to the layering peak. Since the measured
$\Delta \Theta \neq 0$ intensity includes contributions from the
capillary-wave-induced diffuse surface scattering, all of the
theoretical simulations discussed below include a similar
background subtraction procedure. The Fresnel reflectivity curve,
$R_F(q_z)$,  due to an ideally flat and abrupt surface, is shown
in a solid line in Fig.~\ref{fig:Snref}. The dashed line is
$R_F(q_z)$ modified by the theoretically predicted thermal
capillary wave contributions, which depend only on the known
values of the resolution function, surface tension and
temperature. The quasi-Bragg layering peak, as well as the
deviation of the measured $R(q_z)$ from the
capillary-wave-modified $R_F(q_z)$ are unambiguous proof of the
existence of local structure at the surface. The quasi-Bragg peak
at $q_z \approx 2.2$~{\AA}$^{-1}$ corresponds to an atomic
layering of close packed spheres with the $d\approx2.8$~{\AA}
spacing of the atomic diameter of Sn. An additional new feature, a
subtle but significant peak at $q_z \approx 0.9$~{\AA}$^{-1}$ is
not discernible in this figure, and is revealed only upon removal
of the effects of the capillary waves from the measured
Fresnel-normalized reflectivity, as we show below.

\section{The Surface Structure Factor}

\begin{figure}[tbp]
\epsfig{file=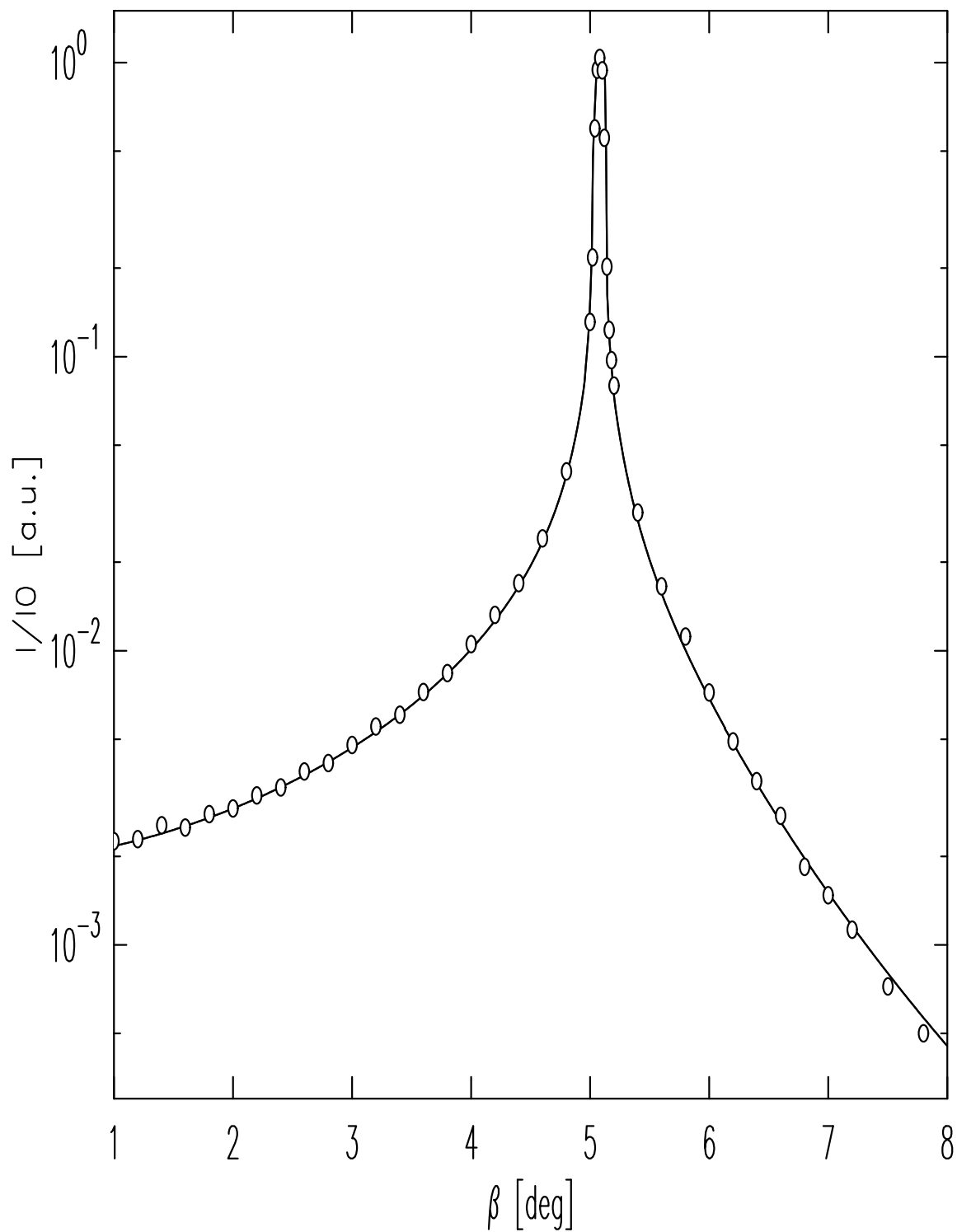, angle=0, width=1.0\columnwidth }
\caption{\label{fig:Sndiff} Diffuse scattering off the liquid Sn
surface measured at a fixed grazing angle of incidence $\alpha=$
5.05$^{\circ}$ (open circles). The peak corresponds to the
specular condition at $q_z=0.48$~{\AA}$^{-1}$. The line is the
capillary wave theory predictions for surface tension
$\gamma=560$~mN/m.}
\end{figure}

To obtain a quantitative measure of the intrinsic surface
structure factor the effects of thermal capillary excitations must
be deconvolved from the reflectivity curve shown in
Fig.~\ref{fig:Snref}. As mentioned above, the method for doing
this has been described in detail in earlier papers on the surface
layering in Ga,\cite{Regan95} In,\cite{Tostmann99} K
\cite{Shpyrko03} and water.\cite{Shpyrko04} The principal result
of this analysis is that the measured reflectivity can be
expressed as
\begin{equation}
R(q_z) = R_F(q_z) \cdot  | \Phi (q_z) | ^2 \cdot CW(q_z)
\label{eq:structure1}
\end{equation}
where $\Phi (q_z)$ is the intrinsic structure factor of the
surface and CW($q_z$) accounts for the effects of the
thermally-excited capillary waves on $R(q_z)$. The surface
scattering cross section yielding the CW($q_z$) term depends on
$T$, $q_z$, the surface tension and the geometric parameters
defining  the reflectometer's resolution function. It is
understood well enough to allow us to fully account analytically
for the effects of capillary waves on the measured $R(q_z)$. This
is demonstrated in Fig.~\ref{fig:Sndiff} where we plot the
background-subtracted and incident-intensity-normalized intensity
scattered by the surface capillary waves measured as a function of
$\beta$ for a fixed incidence angle $\alpha$ in the reflection
plane ($\Delta\Theta=0^\circ$). The agreement between the measured
values (points) and the theoretical cross section (line) is very
good over more than 3 decades in intensity.

For this scan the incident beam is 100 $\mu$m high and the
detector is 0.5~mm high and 1~mm wide. Integration of the
intensity in this figure over the $\beta$-range spanned by the
resolution function of the reflectivity measurements shown in
Fig.~\ref{fig:Snref} yields a value identical with that obtained
in the reflectivity measurements at $q_z=0.48$~{\AA}$^{-1}$. This
further strengthens our claim that the effects of the capillary
waves' scattering on the reflectivity are well understood, and can
be separated out confidently from the measured $R(q_z)$.

The iterative procedure by which the theoretical line in
Fig.~\ref{fig:Sndiff} is calculated is the following. We first
obtain a measure of the structure factor from a best fit of
Eq.~\ref{eq:structure1} to the reflectivity (Fig.~\ref{fig:Snref})
using values for $CW(q_z)$ that are calculated from published
values for the surface tension. This form of the surface structure
factor is then used to fit measured \index{x-ray!diffuse
scattering} diffuse scattering curves similar to that shown in
Fig.~\ref{fig:Sndiff}, with the surface tension as the only
adjustable parameter. When needed, the cycle can be repeated using
now the new surface tension value. In practice, however, a single
cycle was enough, since the best fit value for the surface
tension, $\gamma=560$~mN/m, was in good agreement with the
published values. Thus, we assert that the form of the diffuse
scattering that gives rise to the value of $CW(q_z)$ is fully
determined, and the only unknown quantity in
Eq.~\ref{eq:structure1} is the surface structure factor. This
quantity is the Fourier transform of the surface-normal derivative
of the \index{local electron density} local electron density,
$\rho(z)$, averaged over the surface-parallel directions:
\cite{Braslau88}
\begin{figure}[tbp]
\epsfig{file=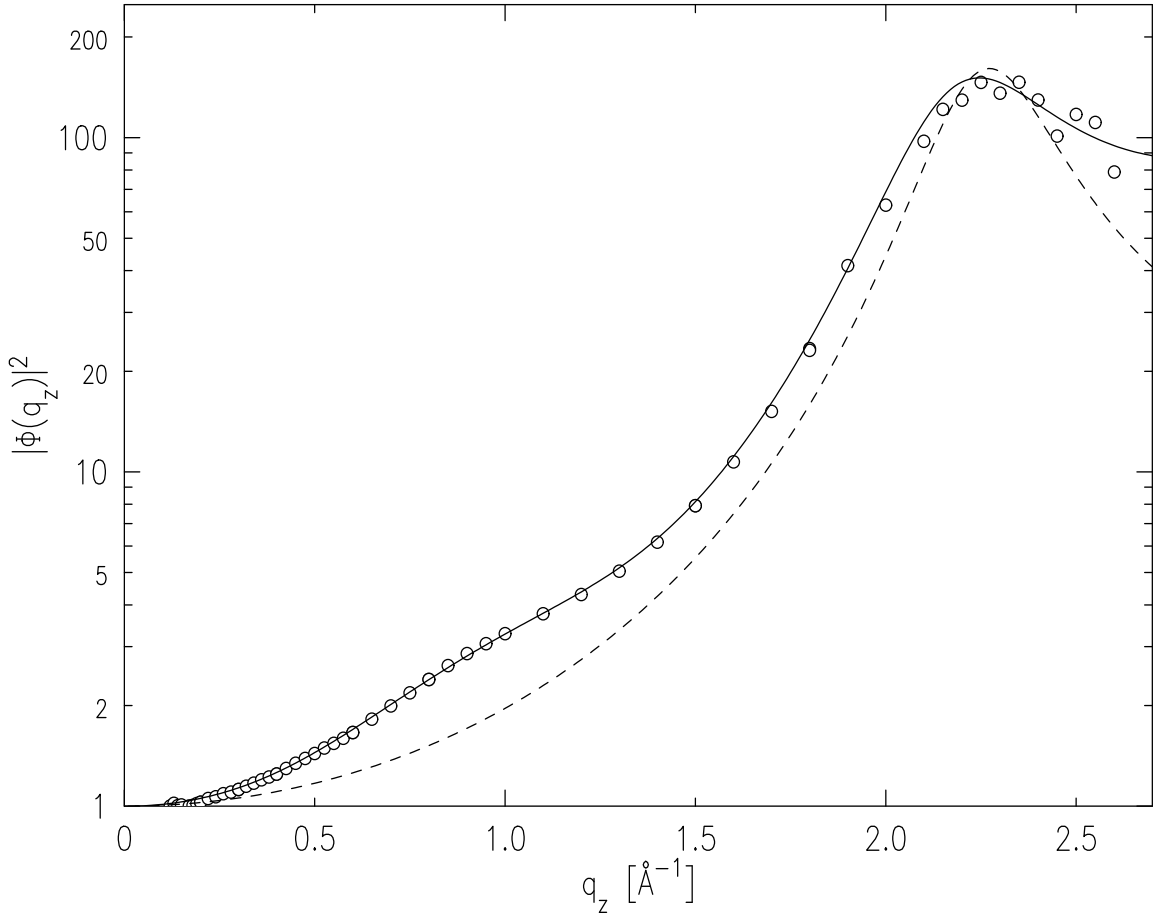, angle=0, width=1.0\columnwidth}
\caption{\label{fig:Sncw} The structure factor (squared) of the Sn
surface as derived from the measured reflectivity $| \Phi (q_z) |
^2 =R(q_z)/[R_F(q_z)CW(q_z)]$ (open circles). The dashed line is
the theoretically expected $| \Phi (q_z) | ^2$ of a simple layered
density profile as found previously for Ga and In. The solid line
is a fit to a model, discussed in the text, where the distance
between the first and second layers is reduced by 10\% relative to
that of the subsequent layers.}
\end{figure}
\begin{equation}
\Phi (q_z) = \frac{1}{\rho_{\infty}} \int dz\frac{\langle d\rho(z)
\rangle }{dz} \exp(\imath q_z z) \label{eq:structure1}
\end{equation}
where $\rho_{\infty}$ is the electron density of the bulk.

The ratio $R(q_z)/[R_F(q_z)CW(q_z)]=| \Phi (q_z) | ^2 $ derived
from the measured reflectivity, is plotted in Fig.~\ref{fig:Sncw}
(open circles). The dashed line is the theoretical reflectivity
calculated using the surface structure factor obtained from the
simple atomic layering model that successfully described the
intrinsic surface structure factor of the pure liquid Ga and In.
As can be seen from the plot, this model does not describe the
measured values well. In particular, it does not exhibit any
feature corresponding to the weak but distinct peak observed in
the measured values at $q_z \approx 0.9$~{\AA}$^{-1}$. As this
peak's position is incommensurate with the large layering peak at
2.2~{\AA}$^{-1}$, it must indicate the existence of a second
length scale, additional to the periodicity of the surface-induced
layering. It is highly unlikely that this second length scale,
$2\pi/0.9 = 6.96$~{\AA}, is associated with a \emph{periodic}
structure, in the same way that the 2.2~{\AA}$^{-1}$ peak is
associated with the surface-induced layering, since such
two-periodicity structure would be very difficult to rationalize
physically. Similarly, the low-$q_z$ peak can not be assigned to
just a single-periodicity layered structure with a top layer which
is denser than the subsequent layers. While such a structure will
yield a subsidiary peak at low $q_z$, this peak would be
commensurate with the high-$q_z$ one, i.e. appear at $q_z=0.5
\times 2.2 \approx 1.1$~{\AA}$^{-1}$\ rather than the observed
$\sim 0.9$~{\AA}$^{-1}$. Thus, we conclude that an appropriate
model needs to include a second, non-commensurate and non-periodic
length scale, with a dense top layer. The solid line shown in
Fig.~\ref{fig:Sncw} is a fit of the measured data by a model
constructed along these lines. As can be seen, a good agreement is
achieved with the measured values. We now proceed to discuss the
details of this model.

\begin{figure}[tbp]
\epsfig{file=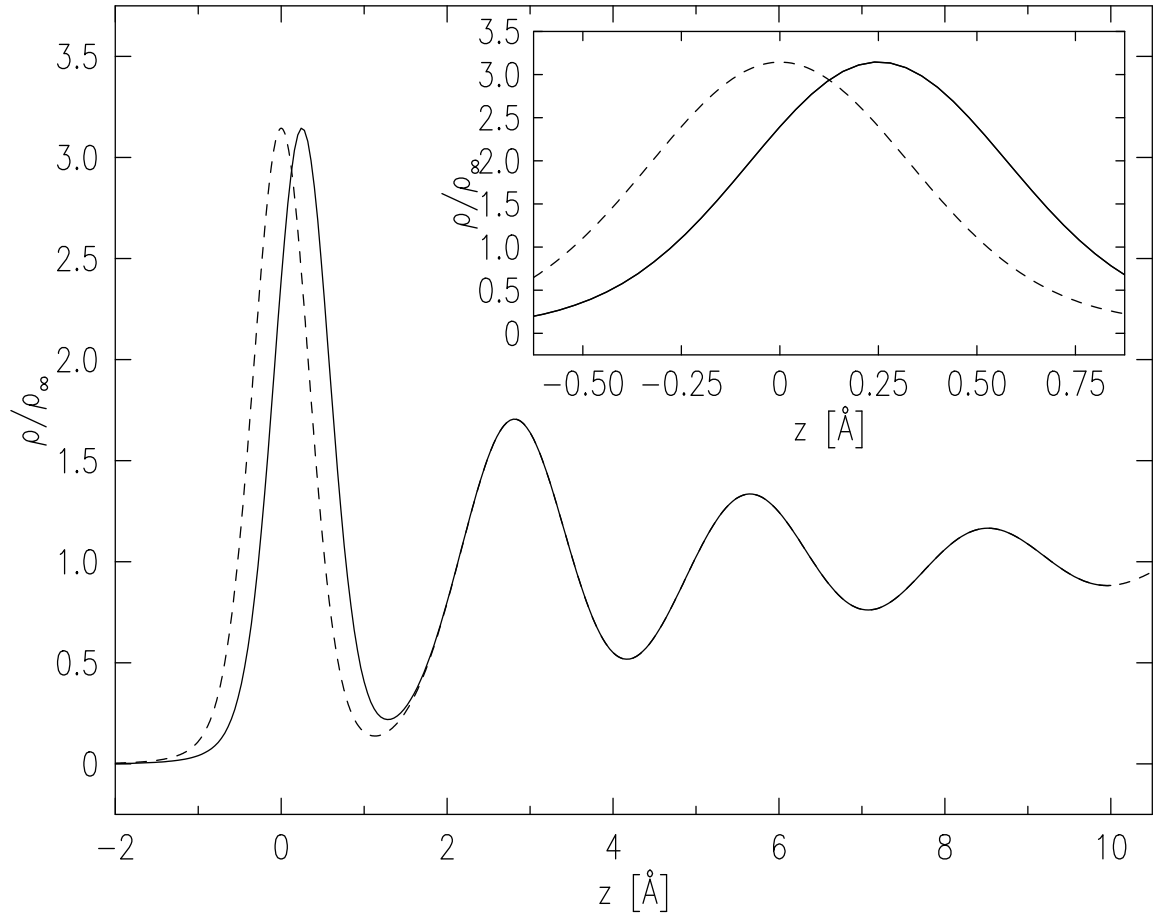, angle=0, width=1.0\columnwidth}
\caption{\label{fig:Snden} Models for the intrinsic surface-normal
electron density profiles of Sn for a simple, equally-spaced
layering model (dashed line), and for a model including a
contraction of the spacing between the first and second layers by
10\% (solid line). The models yield the reflectivities shown by
the same lines in Fig.~\ref{fig:Snref}. The inset is a blow up of
the peak region.}
\end{figure}

The simple density model that has been used to represent the
reflectivity of Ga\cite{Regan95} and In\cite{Tostmann99} consists
of a convolution of an intrinsic density profile and a
\index{Debye-Waller} Debye-Waller-like factor, which accounts for
the smearing of this profile by the thermally-excited capillary
waves. The intrinsic profile is modelled by a semi-infinite series
of \index{gaussian} equally-spaced gaussians, each representing a
single atomic layer. The gaussians have equal integrated areas
(i.e. equal areal electron density) but their widths increase with
depth below the surface. This, in turn, results in a gradual
decrease with depth of the individual peaks and valleys of the
density profile, and an eventual evolution of the density towards
the constant average density of the bulk liquid. The decay length
of the surface layering is typically of the order of just a few
atomic spacings.\cite{Regan95, Tostmann99} The density profile of
this model is shown in a dashed line in Fig.~\ref{fig:Snden}, and
yields the dashed line in Fig.~\ref{fig:Sncw}. The solid line that
runs through the measured values in Fig. \ref{fig:Sncw} is
calculated from a slightly modified model, shown in a solid line
in Fig. \ref{fig:Snden}. The only difference between this model
and the original one is that the distance between the first and
second layers, $2.55$~{\AA}, is smaller by $\sim 10${\%} than the
$2.8$~{\AA} spacing of the subsequent layers. The average density
over the first two layers is thus larger than that of the bulk,
and a second, non-periodic, length scale is introduced. The good
agreement of this minimally-modified model with the measured data,
demonstrated in in Fig.~\ref{fig:Sncw}, strongly supports our
interpretation of the surface structure on liquid Sn.

\section{The Case For Exclusion of Surface Impurities }

Up to this point we have not considered the possibility that the
dense surface layer may be a layer of atoms of a different metal
adsorbed onto the Sn surface. This is precisely what happens, for
example, in the liquid binary alloys GaBi\cite{Tostmann00b,
Huber02} and Ga-Pb,\cite{Yang99} where a (dense) monolayer of the
lower-surface-energy species (Bi,Pb) is found to
\index{Gibbs!adsorption} Gibbs-adsorb at the free surface of the
alloy. We now show experimental evidence that this is not the case
here.

Consideration of the Gibbs rule\cite{Gibbs28} shows that
relatively low concentrations of three metals, Bi ($\gamma$= 378
mN/m, $\rho_e = 2.49$~e/{\AA}$^{3}$), Pb ($\gamma$= 458~mN/m,
$\rho_e = 2.63$~e/{\AA}$^{3}$) and Tl ($\gamma$= 464~mN/m, $\rho_e
= 2.79$~e/{\AA}$^{3}$) could have produced surface electron
densities sufficient to cause the low-$q_z$ peak shown in Fig.
\ref{fig:Sncw}. In order to address this possibility we performed
two independent experiments, \index{x-ray!grazing incidence
diffraction} grazing incidence \index{x-ray!fluorescence} x-ray
fluorescence (GI-XRF) and resonant \index{x-ray!resonant
scattering} x-ray reflectivity, both of which are sensitive to the
presence of sub-monolayer quantities of impurities at the surface.
None of the two measurements found evidence for contamination of
the liquid Sn surface by a foreign species. Therefore, this
supports the conclusion that the anomalous density feature
observed here is in fact an intrinsic property of the liquid Sn
surface.

\subsection{Grazing Incidence Fluorescence Scattering}
\begin{figure}[tbp]
\epsfig{file=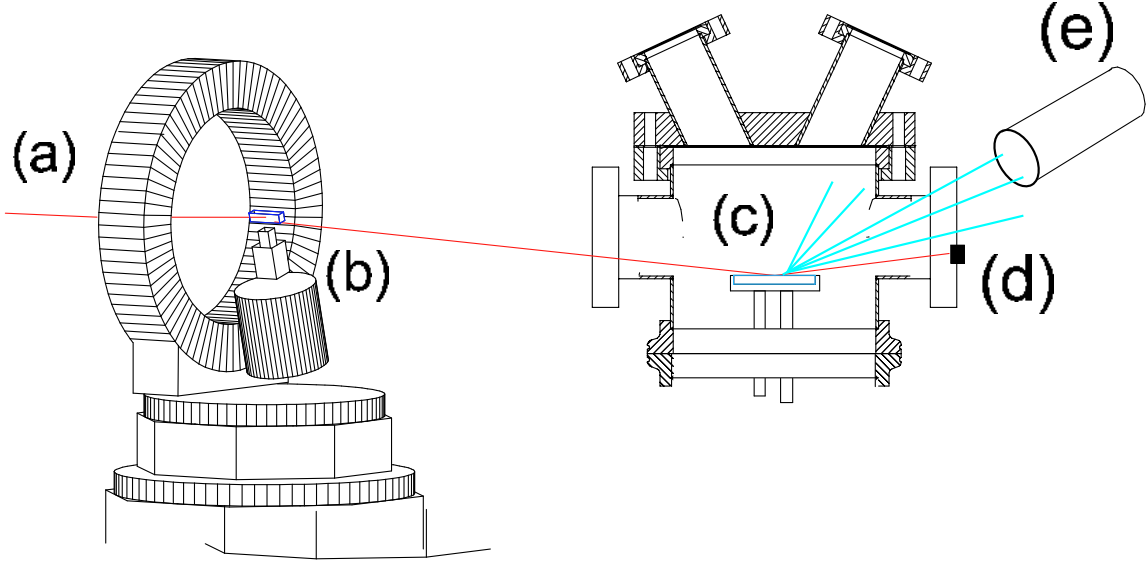, angle=0, width=1.0\columnwidth}
\caption{\label{fig:gid} Experimental setup used in the GI-XRF
measurements: the synchrotron beam (a) is Bragg-reflected from a
monochromator (b) which directs it downwards at the liquid Sn
sample located inside a UHV chamber (c). The specularly-reflected
beam (solid line) is blocked by a beam stop on the exit window (d)
to minimize the background. The fluorescence from the sample's
surface (dotted lines) is detected off-specularly by an
energy-dispersive detector (e) located as close to the chamber as
possible to maximize its solid angle of acceptance.}
\end{figure}

The experimental setup used in these measurements is shown in Fig.
\ref{fig:gid}.
The sample was illuminated by x-rays of wavelength
$\lambda=0.42$~{\AA} (E=29.5~keV), incident at a grazing angle of
$\alpha=0.03^{\circ}$. Since this is approximately one third of
the \index{critical!angle} critical angle of Sn at this energy,
the refracted wave is evanescent. Its penetration depth below the
surface is given by \cite{Tolan99} $\tau =
1/Im[(4\pi/\lambda)\sqrt{\alpha ^2 - \alpha_{crit}
^2-\imath\lambda\mu/2\pi} ]$, where the critical angle for Sn is
$\alpha_{crit}=$0.09$^{\circ}$, and the \index{x-ray!absorption
length} linear absorption coefficient is $\mu=3.26 \times
10^{-6}$~{\AA}$^{-1}$. For these values the incident beam probes
only the uppermost $\tau = 30$~{\AA}\ of the sample. Due to the
high surface tension, the surface of the liquid Sn sample is
curved, which has to be taken into account as well: for an
incident beam height H and a convex sample with a radius of
\index{curvature!radius} curvature $r \approx 10$~m the angle of
incidence $\alpha$ varies over the illuminated area by $\sim
H/(\alpha r)$. In these measurements $H$ was set to 5~$\mu$m, so
that for a nominal angle of incidence $\alpha_0 =0.03^{\circ}$ the
local incidence angle varies by slightly less than $\pm
0.03^{\circ}$. Over this range, $0^{\circ}<\alpha<0.06^{\circ}$,
the penetration length $\tau$ varies from 21~{\AA}\ to 28~{\AA}.
Thus, for our nominal incident angle of $\alpha\approx
0.03^{\circ}$ any detected fluorescence signal comes mostly from
the top 8-10 atomic layers. Depending on the signal-to-noise ratio
and other factors, such as the relative fluorescence yield, with
this geometry it is possible to detect trace amounts of selected
materials to sub-monolayer accuracies, as we show below. Note that
at our incident energy, 29.5~keV, which is well above the Sn K
edge, the K lines of Sn are all excited. However, the K edges of
Bi, Pb and Tl are all $>29.5$~keV, so that only the L lines of
these elements are excited.

Fluorescent x-ray emission from the illuminated portion of the
sample was detected by an energy dispersive intrinsic Ge detector
(area $\sim 9$~mm$^2$) that was mounted about 30~cm away from the
center of the sample, 15 cm above it, and displaced azimuthally by
about 30$^{\circ}$ from the incidence plane. The experimental
setup for this geometry is shown in Fig.~\ref{fig:gid}. The signal
was recorded using a multichannel analyzer. A lead shield was
mounted on the output window as shown in the Fig.~\ref{fig:gid} to
eliminate scattering from the specularly reflected beam into the
detector by the exit window or outside air. Unfortunately, there
was no simple way to shield the detector from scattering or
fluorescence originating in the entrance window of the UHV
chamber. In order to account for this background scattering, a
differential measurement has been performed. The signal detected
with the sample displaced 3 millimeters below the incident beam
was subtracted from the signal that was measured for the beam
striking the sample.
\begin{figure}[tbp] \centering
\includegraphics[width=1.0\columnwidth]{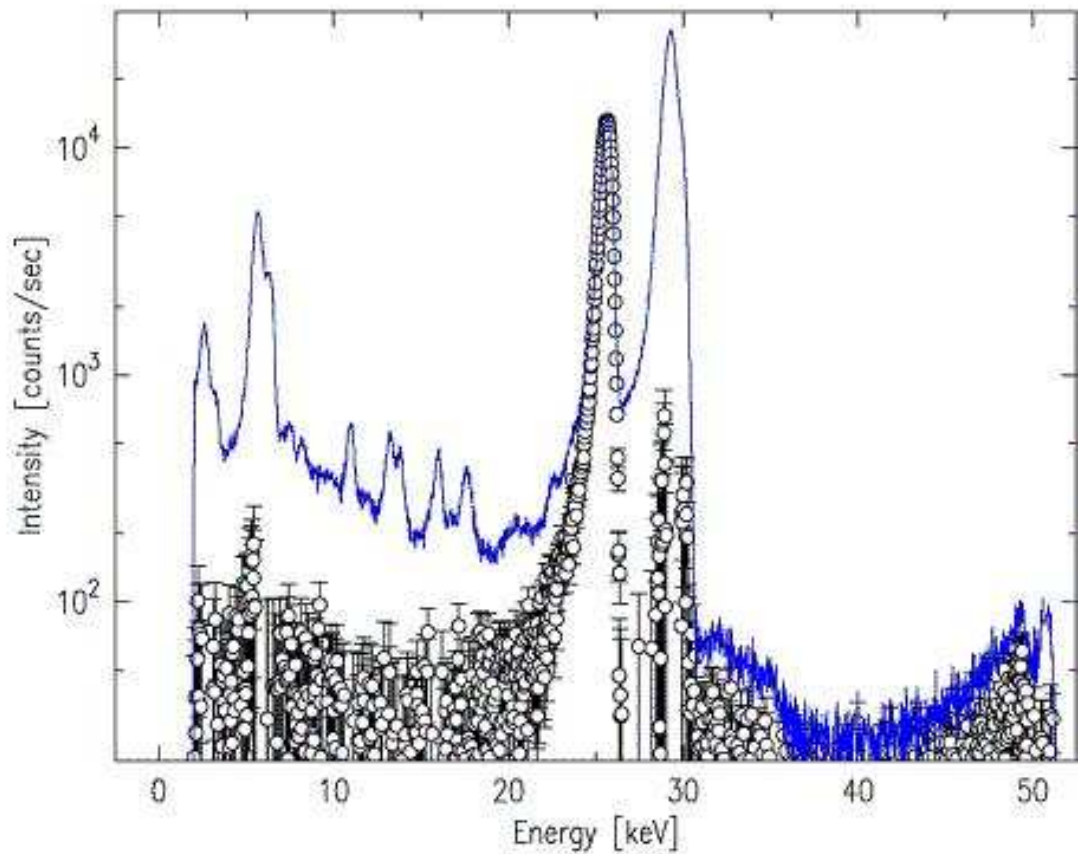}
\caption{\label{fig:Snflu} Raw
(line) and background-subtracted (open circles) measured
fluorescence from the surface of liquid Sn. The sharp lines are
due to the K emissions lines\cite{Bearden67} of Fe (E=6.4~keV), Nb
(E=16.6~keV), Mo (E=17.5~keV) and Sn (E=25.2~keV).}
\end{figure}
The fluorescence data is shown in Fig.~\ref{fig:Snflu}. The solid
line is the as-measured raw fluorescence spectrum. It consists of
fluorescence from both the sample and from the entrance window,
elastic scattering from the incident beam (the strongest peak at
$\sim 29.15$~keV) and the Sn fluorescence (at 25.2~keV). There are
a number of lower-energy peaks originating in the UHV chamber's
components. The open circles show the spectrum obtained after
subtracting from the raw \index{x-ray!spectrum} spectrum the
background spectrum recorded when the sample is moved out of the
beam. The only remaining prominent peak is the Sn fluorescence
observed at 25.2~keV. No other characteristic lines are
discernible above the noise level. The ratio of the Sn
K-fluorescence to the noise level in the 10-15~keV energy region,
where the L-fluorescence lines of the Pb, Bi or Tl are expected,
is $\sim$200. Factoring in the ratio of the K-fluorescence yield
of Sn (0.84) to the L-fluorescence yields of Pb, Bi or Tl
($\sim$0.4) indicates that the presence of as little as 10$\%$ of
a full monolayer of these metals (out of the $\sim$10 atomic
layers of Sn illuminated by the evanescent wave) should be
detectable in this experiment. Since an impurity coverage of the
surface required to generate the $\sim 0.9$~{\AA}$^{-1}$\
reflectivity peak is significantly higher than that, the absence
of an impurity signal in Fig.~\ref{fig:Snflu} rules out with a
high degree of confidence the possibility that a Gibbs adsorbed
layer of impurities is the originator of the $0.9$~{\AA}$^{-1}$
peak.

\subsection{Resonant X-ray Reflectivity}
The effective electron density of an x-ray-scattering atom is
proportional to the scattering \index{form factor} form factor
$f(q)+f'(q,E)$. Here $f(q) \rightarrow Z$ for $q \rightarrow 0$
and $Z$ is the atomic number. When the x-ray energy is tuned
through an \index{x-ray!absorption edge} absorption edge of a
scattering atom, the magnitude of the real part of $f'$ undergoes
a sharp decrease, producing a change in the scattering power of
that specific atom.\cite{Sparks94, Dimasi00} Thus, by measuring
the scattering on- and off-edge it is possible to isolate the
scattering due to the specific atom. We have used this method,
called \index{x-ray!resonant scattering} resonant (or anomalous)
\index{x-ray!anomalous scattering} scattering, in the reflectivity
mode to probe the liquid surface of Sn for the presence of foreign
atoms. This is done by comparing the x-ray reflectivities of the
liquid surface measured with the x-ray energy tuned on and off the
K edge of Sn (29.20 keV). If the low-$q_z$ feature at $q_z \approx
0.9$~{\AA}$^{-1}$ is due a thin surface layer of foreign atoms,
the effective electron density of Sn at the edge will change,
while that of the foreign atoms will not. Thus, the density
contrast will change, and so will the prominence of the low-$q_z$
peak. If, however, the low-$q_z$ peak is due to an intrinsic Sn
structure, the electron density \emph{difference} between
high-density surface layer and the bulk will remain unchanged, and
so will the low-$q_z$ peak.

\begin{figure}[tbp]
\epsfig{file=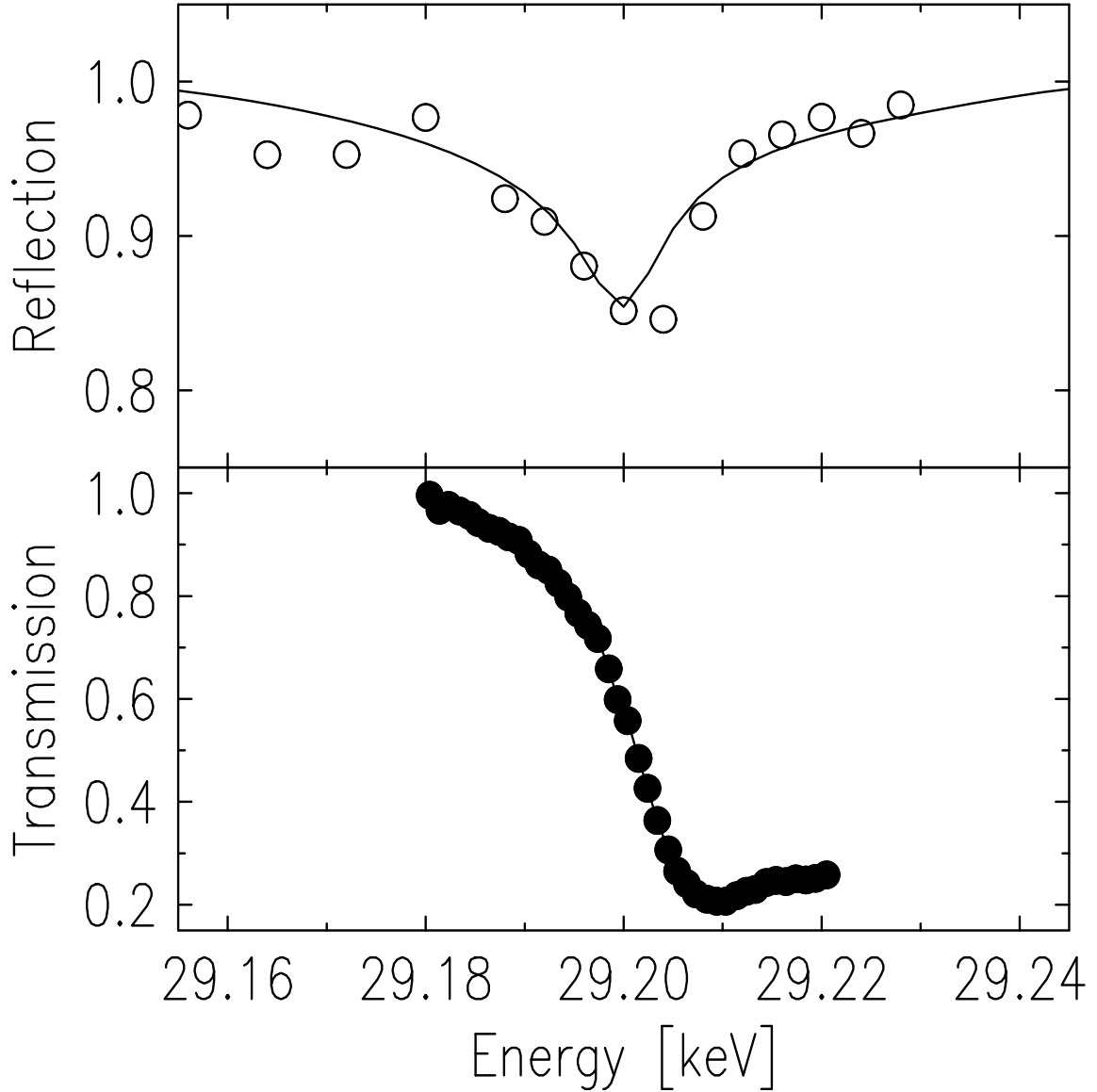, angle=0, width=1.0 \columnwidth}
\caption{\label{fig:Snescan} (Top) Relative energy variation of
the reflectivity off the liquid Sn surface at a fixed
$q_z=0.3$~{\AA}$^{-1}$. The line is the theoretical prediction of
the change in the reflectivity due to the variation of the
\index{dispersion} dispersion correction $f'(E)$ of the scattering
factor near the edge, as calculated by the \index{IFEFFIT}
IFEFFIT\cite{ifeffit} program. The open circles are the measured
values. (Bottom) Measured energy variation of the intensity
transmitted through an Sn foil (points). The K edge of Sn at
E=29.20~keV shows up clearly in both measurements.}
\end{figure}

The 29.20~keV edge of Sn was identified by a transmission
measurement through a Sn foil, shown in Fig. \ref{fig:Snescan}(b),
as well as by the reflectivity from the liquid Sn surface at a
fixed $q_z=0.3$~{\AA}$^{-1}$, shown in Fig. \ref{fig:Snescan}(a).
Since the surface structure factor $\Phi (q_z)$ depends on the
electron density contrast between surface and bulk, the only
energy dependence of $R(q_z)$ away from both structural peaks ( at
$q_z=0.9$~{\AA}$^{-1}$\ and $q_z=2.3$~{\AA}$^{-1}$) is due to the
energy variation of the \index{critical!wavevector} critical
wavevector $q_c$. This leads to a $(Z+f'(E))^2$ energy dependence
of the reflectivity at a fixed $q_z$. The minimum value of
$f'(E)$, estimated from Fig. \ref{fig:Snescan}(a), is consistent
with the theoretically predicted value calculated by the
\index{IFEFFIT} IFEFFIT\cite{ifeffit} program, taking the smearing
due to the lifetime widths of the K and L levels into
consideration. The observed change in reflectivity at the minimum
relative to that 50~eV away is measured to be about 15\%, which
corresponds, in turn, to a 7\% reduction in the effective electron
density of Sn.

\begin{figure}[tbp]
\epsfig{file=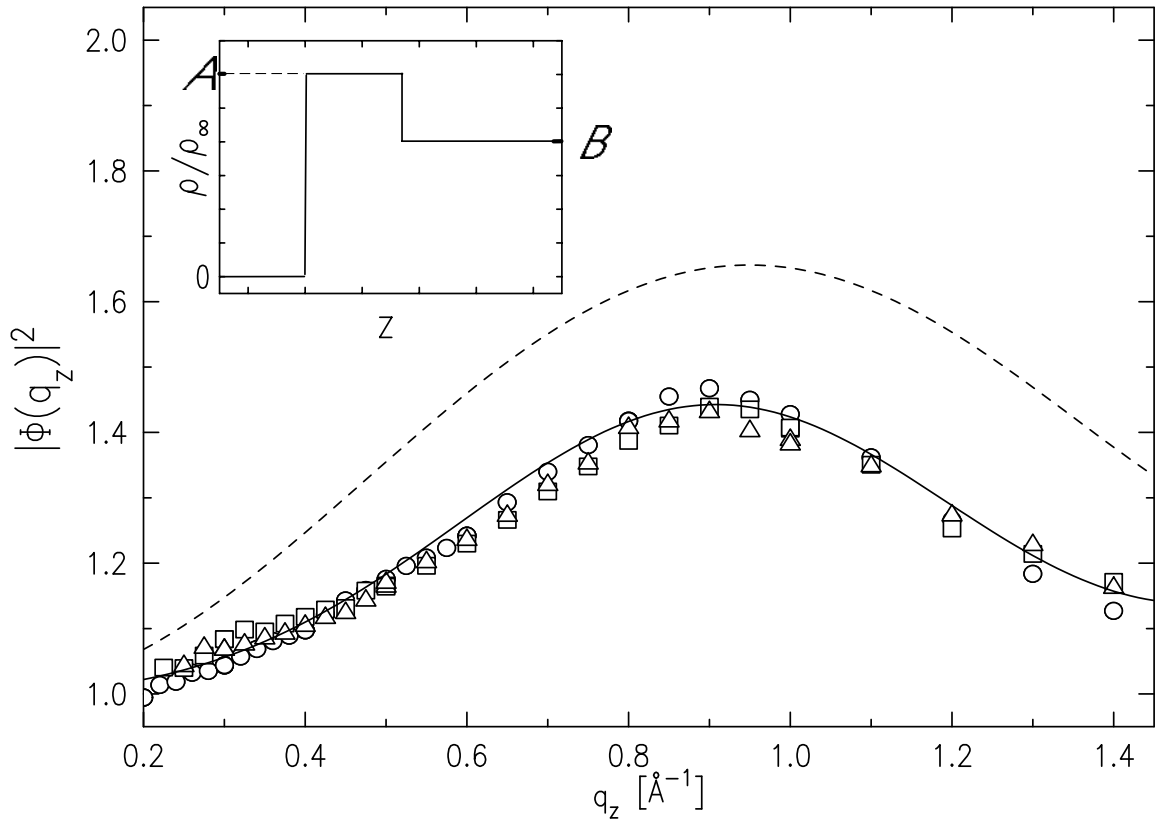, angle=0, width=1.0\columnwidth}
\caption{\label{fig:Sncwnorm} The structure factor (squared) of
the Sn surface as derived from the measured reflectivity, at
(E=29.20~keV, triangles), above ((E=29.22~keV, circles) and below
(E=17~keV, squares) the K-edge. The solid line is a fit to the
density model featuring reduced interatomic spacing for the top
layer by approximately 10 $\%$. The dashed line is calculated for
a model with a high-density surface monolayer of an atomic species
other than Sn. The inset shows the density profile of such a
layer. For a discussion see text.}
\end{figure}

In Fig.~\ref{fig:Sncwnorm} we show the low-$q_z$ normalized
reflectivity data, $R(q_z)/[R_F(q_z)CW(q_z)]=| \Phi (q_z) | ^2 $,
measured at the K edge of Sn (29.20 keV), just above the edge
(29.22 keV) and far below the edge (17 keV). The fact that the
three data sets coincide proves unambiguously  that the
$0.9$~{\AA}$^{-1}$ peak does not arise from contrast between Sn
and a different element. A more quantitative analysis can be done
in two ways. First, we assume a simple box model for the
surface,\cite{Tolan99} shown in the inset to
Fig.~\ref{fig:Sncwnorm}. In this model, the surface layer's
electron density, A, is larger than that of the bulk, B, yielding
$| \Phi (q_z) | ^2 \approx (2A/B-1)^2$\ at the peak. To obtain the
measured peak of $| \Phi (q_z) | ^2 = 1.4$, we have to assume
$A/B=1.1$. We note that only a handful of elements would be
consistent with a model where this 10\% increase in density
originates in a surface layer of impurity atoms. Eliminating the
elements which have a higher surface tension than that of Sn
($\gamma$=560 mN/m), which will not \index{Gibbs!adsorption}
Gibbs-segregate at the surface, the only reasonably-likely
remaining candidates are Bi, Pb and Tl. If we assume that the
surface layer is a monolayer of any one of these elements the
effect of the resonance is that the effective electron density of
the Sn would be reduced by about $7\%$, as mentioned above. The
measured $R(q_z)/[R_F(q_z)CW(q_z)]=| \Phi (q_z) | ^2 $ would
therefore increase at the Sn K-edge to $(2\times 1.1/0.93 -
1)^2=1.86$. The fact that this does not occur proves, again, that
the increased density at the surface is an intrinsic surface
effect, due entirely to a structure comprising Sn atoms only.

A second quantitative argument is the full modelling of the
reflectivity, rather than considering only the maximum of the
$0.9$~{\AA}$^{-1}$ peak. The solid line in Fig.~\ref{fig:Sncwnorm}
is the same fit shown in Fig.~\ref{fig:Snden} to the model where
the density of each atomic layer is kept at the Sn density value,
while the spacing between the first and the second atomic layer is
reduced from the 2.80~{\AA} of subsequent layers to 2.55~{\AA}.
The agreement of all measured reflectivities, in particular that
measured at the edge (triangles), with this model is excellent.
The dashed line is the reflectivity calculated assuming an
additional surface monolayer of a different species having a
density higher than that of Sn. As can be seen, for the on-edge
measurement, this line is expected to lie considerably higher than
off-edge. It clearly does not agree with the on-edge measured data
(triangles).

The results of the two measurements presented in this section rule
out the possibility that the dense surface layer found in the
reflectivity measurements is due to surface segregation of
impurity atoms. Thus, they strongly supports our claim that the
high-density layer present at the surface is the intrinsic
property of the Sn itself.

\section{Conclusion}

We have found that the surface of liquid Sn exhibits atomic
layering, similar to that found in other metallic systems.
However, the surface structure factor exhibits an additional
low-$q_z$ peak, which has not been observed for any of the pure
metals studied to date. The possibility that this structure is due
to a layer of surface-adsorbed contaminants has been addressed by
two experiments, both of which confirm that the observed
reflectivity feature is an intrinsic property of liquid Sn. The
strongest argument comes from the \index{x-ray!resonant
scattering} resonant x-ray reflectivity measurement which finds no
detectable difference between reflectivities measured on- and off-
the Sn K-edge, while model calculations predict a significant
change for a layer of foreign surface atoms. Our analysis shows
that the anomalous feature at low $q_z$ is consistent with a
modified atomic layering density model where the spacing between
the first and second atomic layers is reduced by 10\% relative to
the subsequent ones, thereby increasing the average density at the
surface. The physical reasons for the existence of this denser
layer at the surface of liquid Sn are not clear, since the studies
of other liquid metals and alloys exhibit no evidence for such a
feature. We note, however, that the  density of the liquid Sn
(6970~kg/m$^3$) is in between the densities of the solid alpha
phase Sn (5750~kg/m$^3$), also known as "gray tin" and forming a
cubic atomic structure, and the solid beta phase Sn
(7310~kg/m$^3$), known as "white tin" and forming a tetragonal
atomic structure. It is possible that while the higher-density
beta phase is no longer stable in the bulk beyond the melting
point, it prevails in some form at the surface, where the packing
restrictions are relaxed due to smaller number of nearest
neighbors. Another possibility which may be related to the
previous argument, is that the Sn-Sn pairing, know to be present
in the bulk phase \cite{Jovic76, DiCicco96, Itami03} of liquid Sn,
is stronger near the surface than in the bulk, and induces a
higher density. Clearly, the origin of this increased surface
density in liquid Sn warrants further investigation.


\bibliographystyle{unsrt}

%% file: alloy_all.tex
\chapter{Surface Phases in Binary Liquid Metal Alloys}

 \section{Abstract}

  Surface sensitive x-ray scattering techniques with atomic
scale resolution are employed to investigate the microscopic
structure of the surface of three classes of liquid binary alloys:
(i) Surface segregation in partly miscible binary alloys as
predicted by the \index{Gibbs!adsorption} Gibbs adsorption rule is
investigated for Ga-In. The first layer consists of a supercooled
In monolayer and the bulk composition is reached after about two
atomic diameters. (ii) The Ga-Bi system displays a
\index{wetting!transition} wetting transition at a characteristic
temperature $T_w \approx$ 220$ ^o$C. The transition from a Bi
monolayer on Ga below $T_w$ to a thick Bi-rich \index{nanoscale}
nanoscale \index{wetting!film} wetting film above $T_w$ is
studied. (iii) The effect of attractive interactions between the
two components of a binary alloy on the surface structure is
investigated for two Hg-Au alloys.

\section{Introduction}

Theoretical calculations and computer simulations indicate that
the microscopic structure of the surface of liquid metals (LM) and
alloys is considerably different from the surface structure of
\index{liquid!dielectric} dielectric liquids and mixtures
\cite{alloyref_1,alloyref_2, alloyref_3}. For liquid metals, the
density profile normal to the surface has been predicted
theoretically to show  oscillations with a period of about one
atomic diameter and extending a few atomic diameters into the bulk
\cite{alloyref_2}. This layering normal to the surface has been
confirmed experimentally for liquid Hg  \cite{alloyref_4}, Ga
\cite{alloyref_5} and, most recently, In \cite{alloyref_6}.
Surface induced layering in LM is is due to the drastic changes in
the interactions across the interface from short-ranged
\index{interactions!Coulomb} screened Coulomb interactions in the
liquid phase to long-ranged \index{interactions!Van der Waals} van
der Waals interactions in the gas phase. In
\index{liquid!dielectric}dielectric liquids, on the other hand,
van der Waals interactions prevail both in the liquid and in the
gas phase and theory predicts a monotonic density profile
\cite{alloyref_7}.

Compared to elemental LM, binary LM alloys have an additional
degree of freedom and it is possible  to study the effect of the
second component on surface induced layering. Moreover, binary
mixtures are expected to exhibit a richer array of surface
structures. In  partly miscible mixtures for example, the lower
surface tension component is expected to segregate at the surface
as predicted by the \index{Gibbs!adsorption} Gibbs adsorption rule
of thermodynamics \cite{alloyref_8}.
 For binary liquid metal alloys,
computer simulations predict that the component that segregates at
the liquid-vapor interface forms a nearly pure monolayer
\cite{alloyref_9,alloyref_10,alloyref_11}. The adjacent layer on
the bulk liquid side is slightly depleted of the surface-active
component, and the bulk composition is reached at a distance of
about two atomic diameters into the bulk. For simple dielectric
mixtures on the other hand, theory predicts that the excess
concentration of the segregating component falls smoothly with
distance into the bulk over a range of several atomic diameters
\cite{alloyref_8}. These predictions remain largely untested
experimentally.

The bulk phase behavior of  \index{demixing transition} demixing
leading to a \index{miscibility gap} miscibility gap with a
critical consolute point induces a new class of surface phase
transitions as predicted by \index{Cahn, John} Cahn in 1977
\cite{alloyref_12}. Based on scaling law arguments Cahn postulated
that a phase transition from \index{nonwetting} nonwetting of the
two \index{immiscible phase} immiscible phases to
\index{wetting!complete} complete wetting of one phase by the
other necessarily occurs close to the critical consolute point. In
this transition, a macroscopically thick wetting phase is formed
intruding between the bulk liquid and the vapor phase (or an inert
substrate). This is in contrast to the phenomenon of surface
segregation that takes place on atomic length scales. Experimental
studies of such wetting phase transitions have been restricted
mostly to binary dielectric mixtures \cite{alloyref_13}. Wetting
and \index{prewetting} prewetting phase transitions have been
observed only recently in fluid systems governed by screened
\index{interactions!Coulomb} Coulomb interactions
\cite{alloyref_14}, namely K-KCl \cite{alloyref_15}, Ga-Pb
\cite{alloyref_16} and Ga-Bi \cite{alloyref_17}. These experiments
provide no direct microscopic information on the structure of the
interface. By contrast, the study presented here probes the
structure of the liquid interface with {\AA} resolution using
surface x-ray scattering techniques.

A  characteristic feature of many well-known and technically
important alloys is the  formation of intermetallic phases in the
solid state \cite{alloyref_18}. These phases include stoichometric
Laves phases, broad \index{Hume-Rothery phases} Hume-Rothery
phases and semiconducting \index{Zintl phases} Zintl phases
containing \index{polyanions} polyanions. This phase formation is
due to attractive interactions between the two components. An
interesting question is whether these attractive interactions
affect the surface structure of liquid binary alloys.

In this paper, we will demonstrate how surface x-ray scattering techniques
can be used to study LM alloys exhibiting different types of surface structure.
In Sections 2 and 3 we review the x-ray scattering techniques and
experimental details. In Section 4 we present three different alloy
systems with increasing complexity of bulk phase
behavior and  surface structure: Ga-In, Ga-Bi
and Hg-Au.

\section{Surface X-ray Scattering Techniques}

X-rays are well suited to study the structure of the surface of
liquids with atomic scale resolution. The large dynamic range in
intensity required for such studies can be realized to advantage
by \index{synchrotron!source} synchrotron x-ray sources. In this
section we describe the surface x-ray scattering techniques used
in our studies to determine the structure normal to the surface
and within the surface plane.

\begin{figure}[tbp]
\epsfig{file=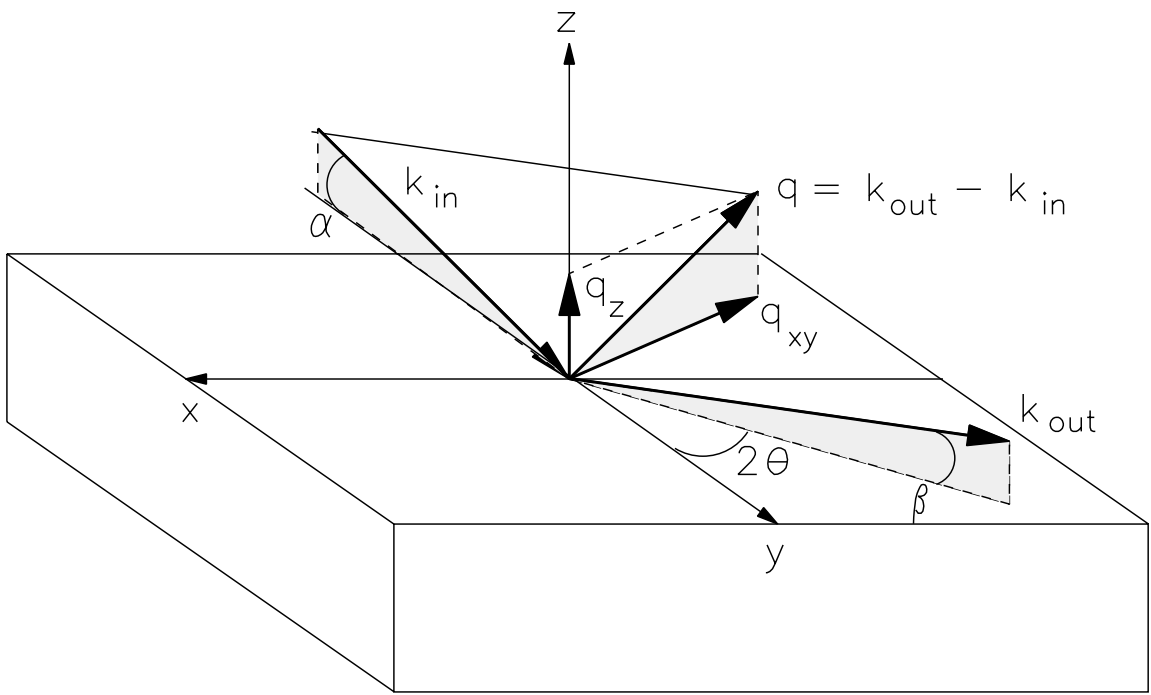, angle=90, width=1.0\columnwidth}
\caption{ Sketch of the geometry of x-ray scattering from the
liquid surface with $\alpha$ and $\beta$ denoting incoming and
outgoing angle, the incoming and outgoing wavevector $k_{in}$ and
$k_{out}$ respectively and the azimuthal angle $2\theta$. The
momentum transfer $q$ has an in-plane component $q_{xy}$ and a
surface-normal component $q_z$.}
 \label{alloy_fig:1}
\end{figure}

{\bf 2.1. Specular X-ray Reflectivity (XR):} X-ray reflectivity
probes the  structure normal to the  interface. Here, the x-ray
intensity is measured at the specular condition where incoming and
outgoing angles $\alpha$ and $\beta$ are equal and the azimuthal
angle $2\theta$ is zero (see Figure~\ref{alloy_fig:1}). In this
case, the scattering  momentum transfer, $\vec{q} =
\vec{k}_{out}-\vec{k}_{in}$, is normal to the surface
\begin{equation}
|\vec{q}| = q_z = \frac{2\pi}{\lambda} (\sin \alpha +  \sin \beta) = \frac{4\pi}{\lambda} \sin
\alpha
\end{equation}
where $\lambda$ is the x-ray wavelength.

If the liquid surface is ideally flat and abruptly terminated
(i.e. has a step function density profile), the reflectivity $R$
is given by the \index{Fresnel reflectivity} Fresnel reflectivity
$R_f$ known from classical optics \cite{alloyref_19}. The
reflectivity of a real surface will deviate from $R_f$. For
example, height fluctuations produced by thermally excited
capillary waves cause phase shifts in x-rays reflected from
different points on the surface. Averaging over these  phase
variations  gives rise to a \index{Debye-Waller} Debye-Waller type
factor, $\exp(-\sigma_{cw}^2 q_z^2)$, characterized by a capillary
wave roughness $\sigma_{cw}$ \cite{alloyref_6}. In addition, the
density profile normal to the surface deviates from a step
function, being oscillatory for elemental LM. For binary alloys
displaying \index{phase separation} phase separation or phase
transitions we expect further modification of the surface.
Therefore, a surface \index{surface!structure factor} structure
factor $\Phi$ has to be taken into account  \cite{alloyref_20}:
\begin{equation}
\label{alloy_master} \Phi (q_z) =  \frac{1}{\rho_{\infty}}
\int_{-\infty}^{+\infty} \frac{d \langle \rho (z) \rangle }{d z}
\exp(\imath q_z z)dz
\end{equation}
with $\langle \rho (z) \rangle$ denoting the electron density average over a microscopic
surface area on the
atomic scale and $\rho_{\infty}$ the bulk electron density.
This factor is analogous to the bulk structure factor $S(\vec{q})$ which is
the Fourier transform of the bulk electron density. The reflectivity from
a real surface is then given by:
\begin{equation}
\label{alloy_final} R(q_z) = R_f \exp(-\sigma_{cw}^2 q_z^2) \left|
\Phi (q_z)\right|^2 .\end{equation} It is not possible to directly
invert the measured reflectivity $R(q_z)$ to the density profile
normal to the surface due to the loss of phase information.
Rather, the common practice is the construction of a physically
plausible model for the density profile which is then inserted in
Eq.~\ref{alloy_master} and fitted to the measured reflectivity
\cite{alloyref_6, alloyref_21}.

{\bf 2.2. Grazing Incidence Diffraction (GID):}

\index{x-ray!grazing incidence diffraction} GID probes the
\index{in-plane ordering} in-plane structure of the surface. The
in-plane momentum transfer $q_{xy}$ is probed by varying the
azimuthal angle $2\theta$ (see Figure~\ref{alloy_fig:1}). This
geometry is surface sensitive if the incident angle $\alpha$ is
kept below the \index{critical!angle} critical angle for total
\index{external reflection} external reflection $\alpha_{crit}$
thereby limiting the x-ray penetration depth to $\simeq 50~{\AA}$
and minimizing the background from bulk scattering
\cite{alloyref_22, alloyref_23}. The GID surface in-plane
structure can be directly compared to the bulk
\index{liquid!structure factor} liquid structure by making similar
measurements for $\alpha > \alpha_c$.

\section{Experimental}

It is essential  to establish a clean oxide free liquid alloy
surface since even microscopic impurities significantly change the
x-ray reflectivity \cite{alloyref_24}. We use slightly different
methods to prepare the three alloys presented in this paper. The
\index{eutectic} eutectic Ga-In alloy is prepared in a High Vacuum
chamber using in situ glow-discharge to clean the sample
\cite{alloyref_21}. The sample is transferred into the x-ray
\index{UHV!chamber} UHV chamber, melted and cleaned by Ar ion
sputtering.

The Ga-Bi alloy is prepared as follows. Bi is  melted under
\index{UHV!conditions} UHV conditions and any residual oxide is
evaporated by heating the Bi to about 600$^{\circ}$\,C. After the
Bi is transferred into an Ar
 glove box, liquid Ga is added so that  the  mole fraction of
Bi is $x_{Bi}\approx 0.3$. This value is chosen
to correspond to the concentration of the bulk solution at the critical point of 265$^{\circ}$\,C .
The sample is then frozen and transferred to the x-ray UHV chamber. The chamber
is baked out and the liquid surface sputter cleaned.

The preparation of  the Hg-Au alloy is quite different since the
high vapor pressure of Hg is not compatible with
\index{UHV!conditions} UHV conditions. Quadruple distilled Hg and
Au powder are mixed in a glove box and transferred to a stainless
steel reservoir attached to a valve with a filling capillary. This
container is attached to a \index{UHV!chamber} UHV chamber with
the valve closed. This chamber is then subjected to a standard UHV
\index{bakeout} bakeout to remove oxygen and water, after which
the valve is opened and the Hg-Au alloy is poured into the sample
pan. Due to the low oxygen partial pressure in the chamber,
residual oxide is instable and disappears from the surface within
a couple of hours.

The sample chamber is placed on an active \index{vibration
isolation} vibration isolation table that effectively quenches all
mechanically induced vibrations \cite{alloyref_6}. The experiments
have been performed at the beamlines X22B and X25 at the
\index{synchrotron!source} National Synchrotron Light Source at
\index{Brookhaven} Brookhaven National Laboratory. The x-ray
energies are 19keV and 10keV at X25 and X22B respectively. A
discussion of the liquid surface \index{spectrometer} spectrometer
can be found elsewhere \cite{alloyref_21}.

\section{Results and Discussion}

\subsection{Surface Segregation in Ga-In}

\begin{figure}[tbp]
\epsfig{file=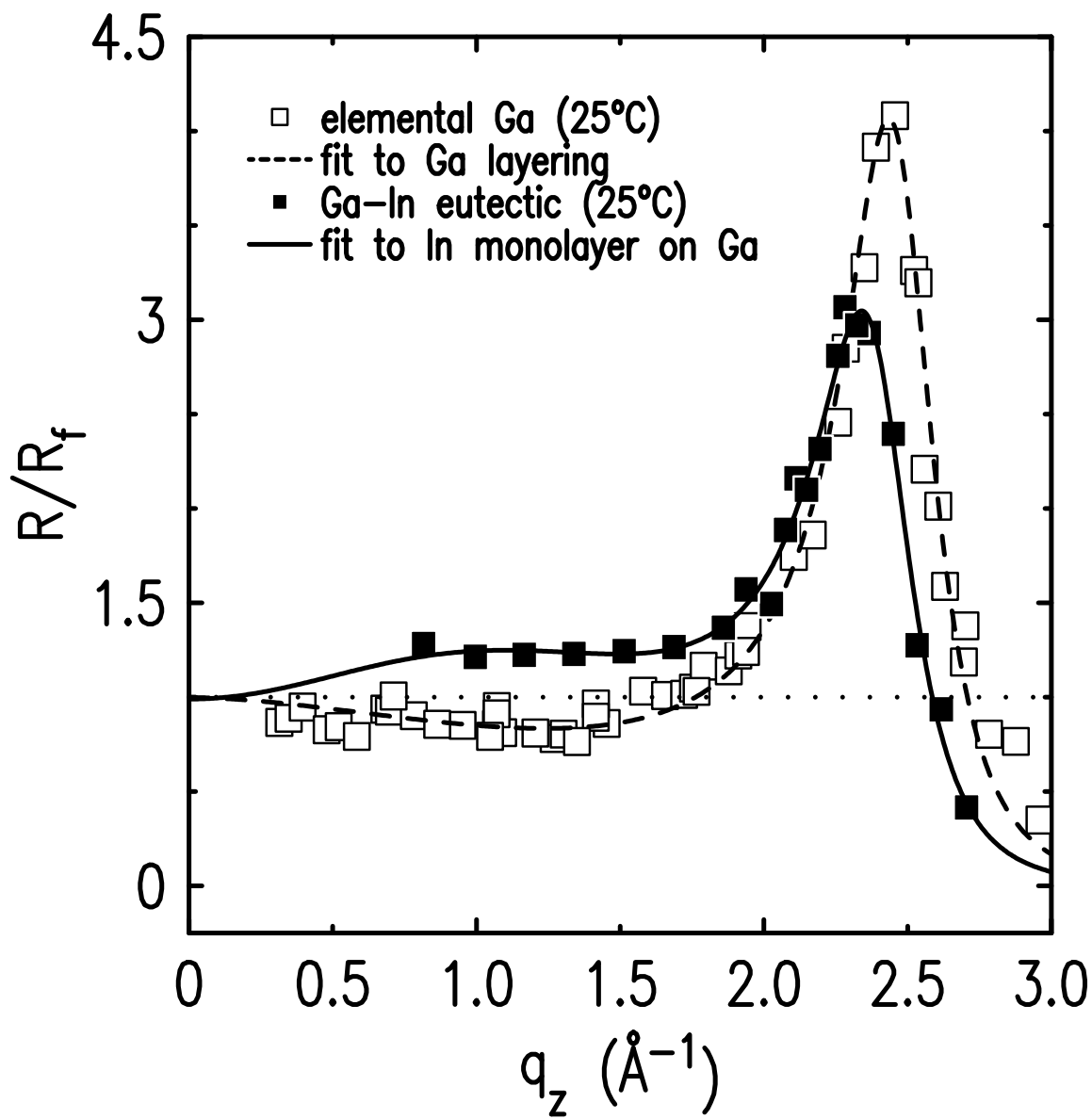, angle=90, width=1.0\columnwidth}
\caption{ X-ray reflectivity from liquid Ga and liquid Ga-In
(16.5\% In) at 25$^{\circ}$\,C. The broken line represents the fit
of the model described in the text to the measured reflectivity
from liquid Ga. The solid line represents the fit of the same
model with one additional high density adlayer to the measured
reflectivity from the liquid  Ga-In \index{eutectic}  eutectic.}
 \label{alloy_fig:2}
\end{figure}

As an example of a simple miscible alloy with a \index{eutectic}
eutectic point we have studied the  Ga-In (16.5\% In) system
\cite{alloyref_21}. The Fresnel-normalized  reflectivity is shown
in Figure~\ref{alloy_fig:2} (filled squares). The data have been
recorded at room temperature and are compared to the normalized
reflectivity from elemental Ga (open squares) at the same
temperature. The most pronounced feature of the Ga reflectivity is
the \index{quasi-Bragg peak}  quasi-Bragg peak at 2.4$~{\AA}^{-1}$
that indicates the presence of surface induced layering
\cite{alloyref_5}. This layering peak is present in the Ga-In
alloy as well, although slightly suppressed. The peak position is
centered at a smaller value of $q_z$ compared to Ga indicating a
larger layer spacing, as expected for a larger In atom. The second
pronounced difference between the Ga-In and the Ga reflectivity is
the observation that the reflectivity from Ga-In is  about 30\%
higher than the Fresnel reflectivity of an ideal Ga-In surface at
intermediate values for $q_z$. This can only be understood if a
higher density adlayer exists at the surface. The measured x-ray
reflectivity is consistent with the segregation of a 94\% In
monolayer on top of the  Ga-In eutectic. The presence of the In
monolayer increases the reflected x-ray intensity at intermediate
$q_z$ due to the higher electron density of In but has a less
pronounced effect at larger $q_z$ values where the reflectivity is
dominated
 by the Ga layering peak.
The density profile that fits the measured XR can be modeled by a
single Gaussian for the In segregation layer plus a semi-infinite
sum of Gaussians for the Ga-In eutectic with half widths that
increase with distance from the surface, representing the decay of
the  layering with increasing distance from the surface
\cite{alloyref_21}. The best fit of this model to the data results
in a density profile normal to the surface that shows the
segregation of an In monolayer with a 2.6~${\AA}$ spacing between
the In monolayer and the underlying eutectic layer. This fit is
represented by the solid line in Figure~\ref{alloy_fig:2}.

These results provide new information about the surface structure
of miscible LM alloys. The constituents of this alloy have similar
size and the same valency. The Ga-In interactions are comparable
to the Ga-Ga and the In-In interactions. As a consequence, Ga-In
is close to being an ideal mixture, for which surface segregation
of the lower surface tension component is predicted. Our
measurements show that this segregation is actually confined to a
single monolayer. An arguably more interesting issue is the
interplay between surface segregation and surface induced
layering. Our measurements show that layering persists, with the
In monolayer comprising the outermost layer. Similar conclusions
were reached by recent Quantum \index{molecular Dynamics
simulations} Molecular Dynamics simulations \cite{alloyref_25}.

\subsection{Wetting phase transition in Ga-Bi}

The Ga-Bi system  is an example of an alloy  with a miscibility
gap. Below the \index{monotectic point} monotectic temperature,
$T_{mono}$= 222$^{\circ}$\,C, a Ga rich liquid coexists with a
solid Bi phase. (see\cite{alloyref_26} for a full description of
the \index{phase diagram} phase diagram). However, due to its
lower surface energy a Bi monolayer is expected to segregate at
the surface of the Ga rich liquid. The wetting phase transition
that is predicted for all binary mixtures with \index{demixing
transition} critical demixing \cite{alloyref_12} occurs at a
characteristic wetting temperature $T_w$ below the critical
temperature $T_{crit}$\, \cite{alloyref_27,alloyref_28}. Optical
studies show that $T_w$ coincides with the \index{monotectic
point} monotectic temperature for Ga-Bi
\cite{alloyref_14,alloyref_17}. Above $T_w$, a macroscopically
thick Bi rich phase is expected to completely wet the lighter Ga
rich phase in defiance of gravity. The Bi concentration in the Ga
rich phase increases with increasing temperature as long as the
liquid coexists with the solid Bi phase.

\begin{figure}[tbp]
\epsfig{file=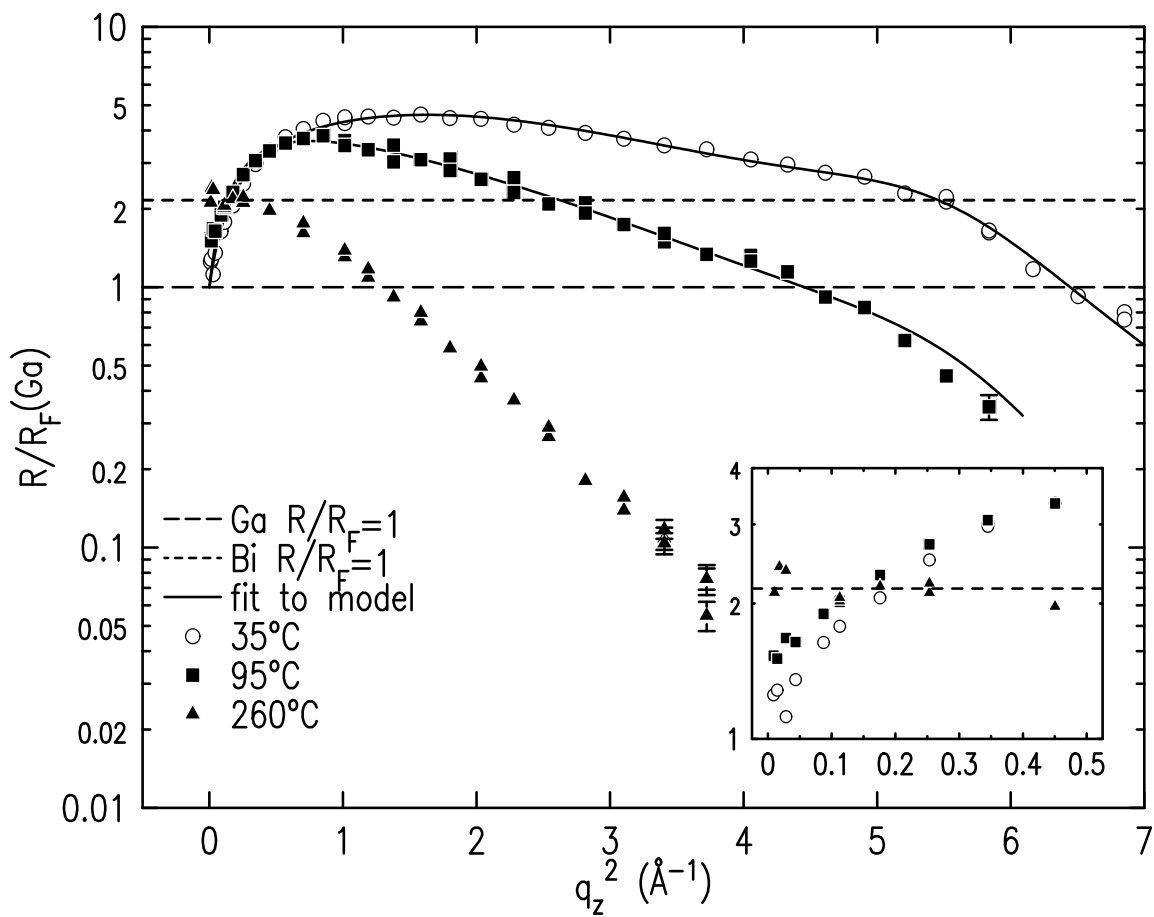, angle=90, width=1.0\columnwidth}
\caption{ X-ray reflectivity from Ga-Bi for three different
temperatures. The solid lines show the fit of the data to a simple
surface segregation model with one Bi monolayer on top of the Ga
rich bulk phase. The difference between the 35$^{\circ}$\,C and
the 95$^{\circ}$\,C data is due to the increased  roughness at
higher temperatures. The 260$^{\circ}$\,C data cannot be fitted by
this model. The inset shows the principal difference in surface
structure between the 35$^{\circ}$\,C and the 260$^{\circ}$\,C
data, indicating a wetting phase transition.}
 \label{alloy_fig:3}
\end{figure}
The normalized x-ray reflectivity spectra, $R/R_{f,Ga}$, for Ga-Bi
at 35$^{\circ}$\,C and 95$^{\circ}$\,C  are shown in
Figure~\ref{alloy_fig:3} versus $q_z^2$. At the lowest values of
$q_z$, the normalized reflectivity approaches the reflectivity of
an ideal Ga surface given by the \index{Fresnel reflectivity}
Fresnel Law of Optics. At values of $q_z$ close to the critical
q-vector $q_{crit}=4\pi/\lambda \sin \alpha_{crit}$, the
reflectivity is not particularly sensitive to the structure or
roughness of the surface (see Eq.(\ref{alloy_final})) and is
simply related to the electron density within $\approx$20~${\AA}$
of the surface. The fact that the reflectivity approaches the
Fresnel reflectivity given by the Ga electron density implies that
the higher density Bi can not be more than a few monolayers thick.
 With increasing $q_z$,
the normalized reflectivity first increases and reaches a maximum
at about $q_z \approx 1~{\AA}^{-1}$. The variation of $R/R_{f,Ga}$
with $q_z$, is similar to that observed by Lei et al.
\cite{alloyref_10}, corresponding to the thickness of a single
atomic layer, again suggesting that only the top layer has a
sizeable bismuth concentration. Compared to the Ga-In alloy
discussed above, this variation is much more pronounced. This
results from the 30\% higher electron density of bismuth compared
with gallium, and only a 5\% higher density for indium.  The
behavior of the same alloy at 260$^{\circ}$C is markedly
different. Here, the reflectivity starts out at $R/R_{f,Ga}
\approx 2$ and falls off monotonically with $q_z$, a behavior
consistent with a thick Bi rich layer terminating the metal/vapor
interface above $T_w$.

\begin{figure}[tbp]
\epsfig{file=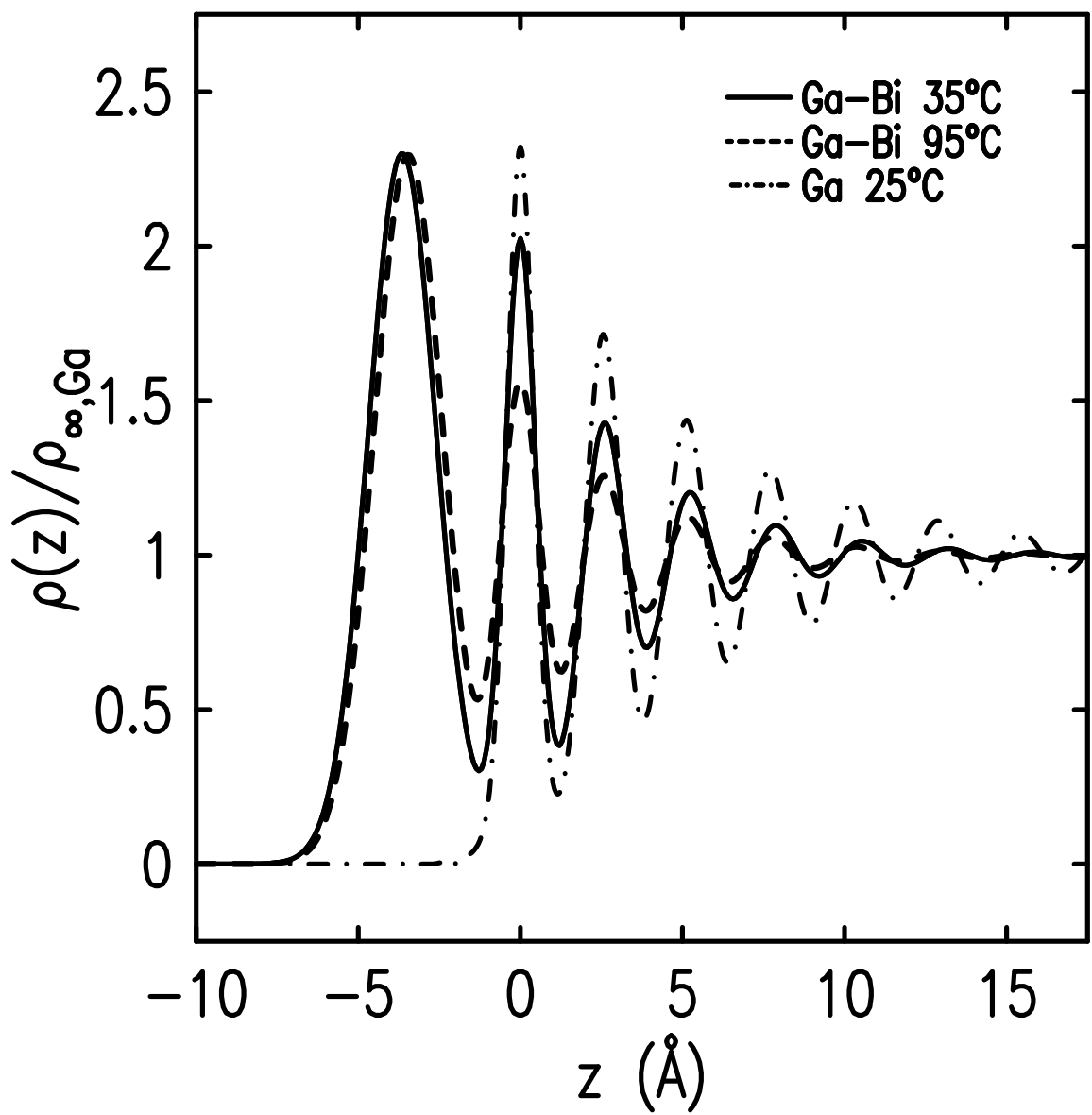, angle=90, width=1.0\columnwidth}
\caption{ Comparison of the intrinsic or local real space density
profiles (thermal broadening removed) for Ga-Bi at two different
temperatures as well as for elemental Ga. The density is
normalized to the bulk density of liquid Ga. The zero position in
$z$ is arbitrarily assigned to the center of the first Ga layering
peak.} \label{alloy_fig:4}
\end{figure}

The reflectivity below $T_w \approx 220^{\circ}$C has been
analyzed in terms of a density profile similar to the one which
describes the Ga-In data \cite{alloyref_21}. The best fits, the
solid lines in Figure~\ref{alloy_fig:3}, provide an excellent
description of the data. The corresponding local density profile,
shown in Figure~\ref{alloy_fig:4} after removing the temperature
dependent capillary wave roughness factor, exhibits a top-layer
density which is about 1.5 times higher than the Ga bulk liquid
density and is consistent with the formation of a complete Bi
monolayer. The 3.6$\pm 0.2~{\AA}$ layer spacing between the
monolayer and the adjacent Ga layer (obtained from the fits) is
much larger than the $2.5\pm 0.1~{\AA}$ layer spacing obtained in
liquid gallium. This also supports the conclusion of a top layer
with a much larger atomic diameter. In addition, the 4.3~{\AA}
exponential decay length of the layering amplitude is
significantly smaller than the 5.5~{\AA} decay length obtained for
pure liquid Ga \cite{alloyref_5}. The suppression of the layering
in the presence of either a bismuth or indium monolayer is much
more apparent in the case of bismuth (compare
Figure~\ref{alloy_fig:3} to Figure~\ref{alloy_fig:2}. A likely
explanation is the larger size difference in Ga-Bi (the Bi
diameter is 30\% larger than that of Ga) compared to Ga-In (In is
15\% larger than Ga). As demonstrated by the solid lines, the
layering model which describes the 35$^{\circ}$C reflectivity
profile also agrees closely with
 the 95$^{\circ}$C profile, without a discernible change in the \index{fitting!parameters} parameters of the density profile,
except for the value of $\sigma_{cw}$, which is scaled, using
capillary wave theory, to correspond to the higher temperature
(see Eq.(\ref{alloy_final})).

\begin{figure}[tbp]
\epsfig{file=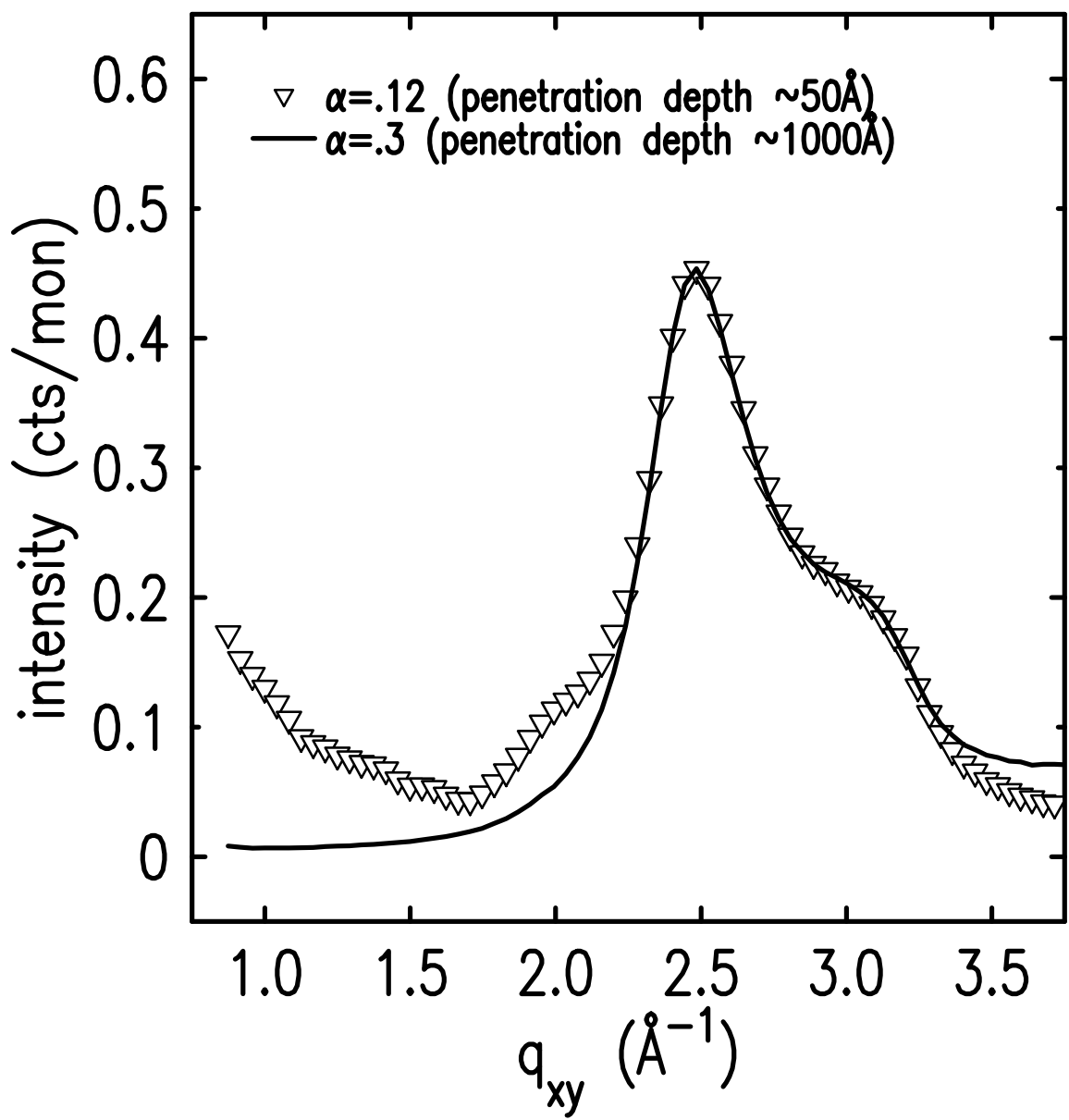, angle=90, width=1.0\columnwidth}
\caption{ Grazing Incidence Diffraction from Ga-Bi at
35$^{\circ}$\,C. The solid line shows the in-plane structure
averaged over the penetration depth of x-rays away from the
\index{critical!angle} critical angle for total \index{external
reflection} external reflection (about 1000~{\AA}). The open
triangles show the diffraction peak obtained for $\alpha <
\alpha_{crit}$ ($\alpha_{crit}$ = 0.14$^{\circ}$ for Ga at
$\lambda = 0.653~{\AA}$) thus probing the surface region
evanescently.}
 \label{alloy_fig:5}
\end{figure}

This picture of a Bi monolayer segregating on top of the Ga rich
bulk phase is supported by GID experiments which probe the
\index{in-plane ordering} in-plane structure of the surface (see
Figure~\ref{alloy_fig:5}). The solid line shows the in-plane
\index{liquid!structure factor} liquid structure factor measured
for $\alpha > \alpha_{crit}$ where the x-rays penetrate the bulk
to a depth $>1000$~{\AA}. The broad peak at $q_{xy} \approx
2.5~{\AA}^{-1}$ and the shoulder on the high-angle side of the
peak agree with bulk liquid Ga \index{surface!structure factor}
structure factor \cite{alloyref_29}. There is no evidence for a
peak or shoulder at the position corresponding to the first peak
of the Bi liquid structure factor at $q_{xy} \approx
2.2~{\AA}^{-1}$. The bulk scattering completely masks the weak
scattering from the bismuth monolayer due to the large penetration
depth of x-rays. The surface sensitivity is drastically enhanced
by keeping the incoming angle $\alpha$ below the
\index{critical!angle} critical angle for total \index{external
reflection} external reflection (0.14$^{\circ}$ for Ga). As the
x-rays now penetrate only to a depth $<50$~{\AA}, the first peak
of the Bi \index{liquid!structure factor} liquid structure factor
is clearly visible at $q_{xy} \approx 2.2~{\AA}^{-1}$ (open
triangles in Figure~\ref{alloy_fig:5}). Still, since the Bi layer
is only a single monolayer thick, the contribution from the
underlying Ga is larger then that of the Bi, as seen from the Bi
being only a shoulder on the Ga structure factor peak in
Figure~\ref{alloy_fig:5}.

While the XR results at 35$^{\circ}$\,C and 95$^{\circ}$\,C are
consistent with a real space model where the Bi is  segregated
only in the top atomic layer, at temperatures above
220$^{\circ}$\,C, corresponding to a Bi mole fraction of about
0.08, this model no longer describes the data. The ratio
$R/R_{f,Ga}$ for the reflectivity taken at 260$^{\circ}$\,C close
to the \index{critical!wavevector} critical wavevector for Ga is
approximately twice as large as the same ratio taken at
35$^{\circ}$\,C and 95$^{\circ}$\,C. The dashed horizontal line,
shown in the inset to Figure~\ref{alloy_fig:3}, represents the
theoretical Fresnel-normalized reflectivity curve for bulk Bi
($R/R_{f,Bi} \approx 2\times R/R_{f,Ga}$).  This curve agrees well
with  the 260$^{\circ}$ data at small $q_z$ where the surface
roughness contribution is minimal, and clearly supports the
conclusion of a thick Bi-rich surface layer. Such a thick wetting
layer has been proposed on the basis of previous optical
experiments \cite{alloyref_14, alloyref_17}.  These findings are
also supported by GID experiments above 220$^{\circ}$\,C which
show a pronounced increase in the intensity of the Bi structure
factor relative to the Ga structure factor  \cite{alloyref_30}.

\subsection{Pair formation in  Hg-Au}

\begin{figure}[tbp]
\centering
\includegraphics[width=0.7\columnwidth]{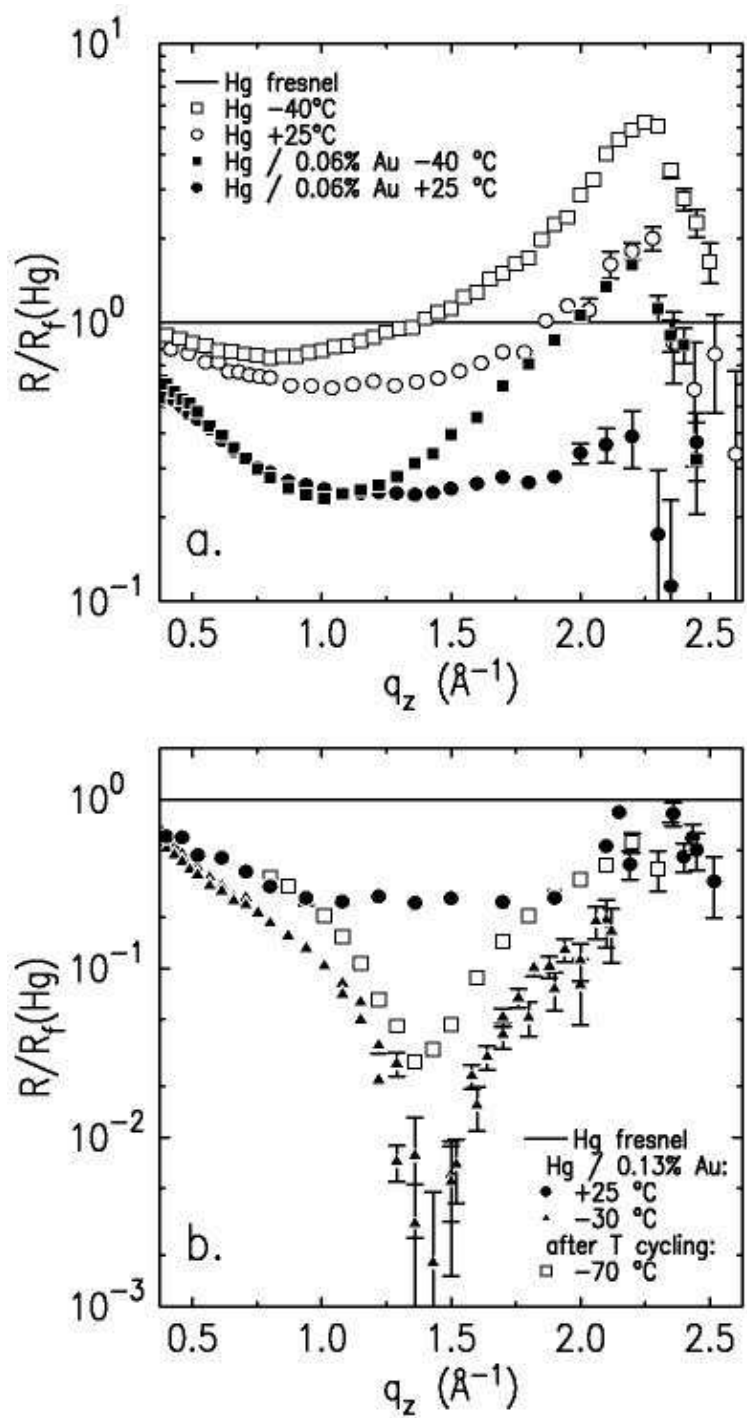}
\caption{ (a) X-ray reflectivity of Hg and Hg/0.06at\%Au
normalized to the Fresnel reflectivity of Hg. (b) Normalized
reflectivity of  a Hg/0.13\% alloy.}
 \label{alloy_fig:6}
\end{figure}

Our x-ray reflectivity measurements of dilute liquid Hg-Au alloys
demonstrate that dramatic differences in surface structure can be
induced by very small changes in concentration.
Figure~\ref{alloy_fig:6}(a) compares the reflectivity of Hg with
that of a solution of 0.06at\% Au in Hg. This composition is
approximately half of the reported \index{solubility limit}
solubility limit for Au in Hg at room temperature
\cite{alloyref_26}. For pure Hg, the \index{surface!layering}
surface layering peak at $q_z=2.2$~${\AA}$ is much more prominent
close to its melting point at $-40^\circ$C than at room
temperature \cite{alloyref_31}. This difference arises primarily
from increased roughening of the surface due to thermally excited
capillary waves with increasing temperatures. The presence of
0.06at\% Au sharply suppresses the surface induced layering of Hg,
as shown by the substantial attenuation of the layering peak. The
temperature dependence in both systems suggests that at this
concentration, the suppression of the surface layering by the Au
has a small dependence on temperature in comparison to the effect
of capillary waves.

Increasing the Au concentration to 0.13at\% yields  a
qualitatively different behavior as shown in
Figure~\ref{alloy_fig:6}(b). At room temperature, layering is
suppressed to a similar extent in both alloys.  However, upon
lowering the temperature of the 0.13at\% solution to $-30^\circ$C,
a deep minimum appears at $q_z=1.4$~\AA$^{-1}$.  This minimum is
produced by destructive interference between x-rays reflecting
from layers of different density near the surface, implying that
one or more phases have formed that are different from the
composition of the bulk. We find significant hysteresis of the
reflectivity under temperature cycling.  In particular, on
returning to room temperature and cooling again, the low
temperature minimum is much less deep than that observed
previously, and saturates at this higher value even when the
sample is supercooled to about $-70^\circ$C. This pronounced
effect has been reproduced and in the absence of detailed modeling
we can only speculate about its origin. As opposed to Ga-In and
Ga-Bi, the interactions between Hg and Au are attractive leading
to intermetallic phase formation in the solid state with
$Au_{2}Hg$ being the best known phase \cite{alloyref_26}. Several
other phases have been reported that are only  stable below
-39$^{\circ}$C  \cite{alloyref_26}. It is known from x-ray studies
on
 FeCo (001)  \cite{alloyref_32} and $ Cu_3Au$ (001)  \cite{alloyref_33} surfaces, that
this pair formation gives rise to mesoscopic surface segregation profiles in the solid state.
As soon as one component segregates at the surface, it is energetically
favorable for the next layer to consist entirely of the other component
due to the pronounced attractive interactions. It is conceivable that a
similar mechanism prevails in liquid Hg-Au. An alternating or more
complicated segregation profile would destroy the Hg layering even
more effectively than the segregation of a monolayer in Ga-In or
Ga-Bi and is consistent with our experimental findings that a small
amount of Au completely suppresses the otherwise very pronounced Hg
layering peak even at low temperatures.

\bibliographystyle{unsrt}

%% file: jop_all.tex
\chapter{Surface Induced Order in Liquid Metals and Binary Alloys}

\section{Abstract}

Surface x-ray scattering measurements from several pure liquid
metals (Hg, Ga and In) and from three alloys (Ga-Bi, Bi-In, and
K-Na) with different \index{interactions!heteroatomic}
heteroatomic chemical interactions in the bulk phase are reviewed.
Surface induced layering is found for each elemental liquid metal.
The surface structure of the K-Na alloy resembles that of an
elemental liquid metal. Surface segregation and a
\index{nanoscale} nanoscale \index{wetting!film} wetting film are
found for Ga-Bi. Bi-In displays pair formation at the surface.

\section{Liquid Metals and Surface Induced Order}

Liquid metals (LM) are comprised of charged ion cores whose
\index{interactions!Coulomb} Coulomb interactions are screened by
a conduction electron sea. At the liquid-vapor interface, this
screened \index{interactions!Coulomb} Coulomb potential gives way
to the weaker \index{interactions!Van der Waals} van der Waals
interactions that prevail in the vapor. Since the potential
changes so substantially across the interface, the potential
gradient is high, producing a force that acts on the ions at the
liquid surface as though they were packed against a \index{hard
wall} hard wall. Analytic calculations and \index{molecular
Dynamics simulations} molecular dynamics simulations predict that
atoms at the LM surface are stratified in layers parallel to the
        interface \cite{rice96}.
By contrast, a monotonic density profile is
predicted for the vapor interface of a nonmetallic liquid.

\begin{figure}[tbp]
\centering
\includegraphics[width=0.7\columnwidth]{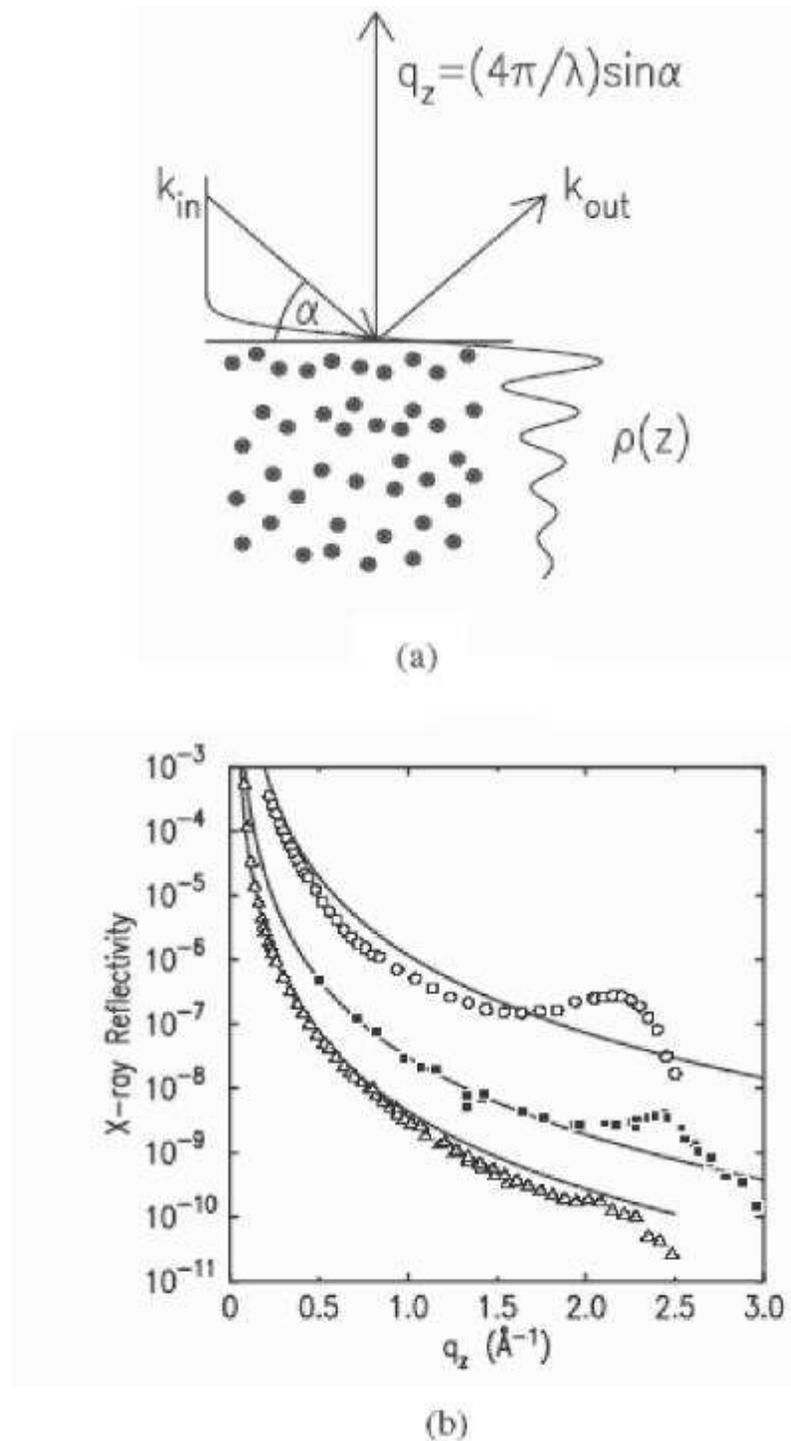}
 \caption{ (a) X-ray reflectivity geometry for
the liquid metal, with layering of ions producing an oscillatory
density profile $\rho(z)$. (b) X-ray reflectivity for liquid Hg
($-35^\circ$C, $\circ$), Ga ($+25^\circ$C) and In ($+170^\circ$C,
$\triangle$). Solid lines: calculated Fresnel reflectivity from a
flat surface. Data for Ga and In are shifted for clarity.
\label{fig:jop_fig1}}
\end{figure}

Observation of surface layering in LM requires an experimental
technique sensitive to the surface-normal density profile that can
resolve length scales of 2--3~\AA. Specular X-ray reflectivity
provides the most direct probe of the surface normal structure.
X-rays incident on the liquid surface at an angle $\alpha$ are
scattered at the same angle within the reflection plane defined by
the incident beam and the surface normal
(Fig.~\ref{fig:jop_fig1}(a)). The reflected intensity is directly
related to the surface normal density profile $\tilde{\rho}(z)$:
\begin{equation}
R ( q_z) \propto \left|  q_z^{-2} ( \partial \tilde{  \rho } (z )
/\partial z) \exp (iq_z z) dz \right|^2 .
\end{equation}
Since $ \partial \tilde{  \rho }(z )/ \partial z$ is nonzero only
near the surface, x-ray reflectivity is sensitive to the
surface-normal structure and not to the structure of the bulk
liquid. For example, surface layering with a spacing $d$  produces
a \index{quasi-Bragg peak} quasi-Bragg peak in the reflectivity,
centered at the surface-normal momentum transfer $q_z = (4 \pi
/\lambda)\sin \alpha \approx 2 \pi /d$
    \cite{bosio,magnu95,regan95}.

\index{x-ray!grazing incidence diffraction} Grazing incidence
diffraction (GID) is sensitive to the \index{in-plane ordering}
in-plane structure of the surface. The in-plane momentum transfer
$q_{\|}$ is probed by varying the azimuthal angle $2\theta$ at
fixed $\alpha$. This geometry is surface sensitive when the
incident angle $\alpha$ is kept below the \index{critical!angle}
critical angle for total external reflection, $\alpha_{c}$,
thereby limiting the x-ray penetration depth
    \cite{eisenberger}.

For these structural studies it is essential to maintain a liquid
metal surface that is flat and clean on an atomic scale. The
sample is contained either in an ultra high vacuum (UHV)
environment, or under a reducing atmosphere such as dry hydrogen
gas,  to prevent oxidation. For low vapor pressure, UHV-compatible
metals such as Ga, Bi, and In, argon ion sputtering is possible,
and this is the most reliable way to produce an atomically clean
surface\,\cite{regan95}.

Surface layering in elemental LM
was first experimentally confirmed by synchrotron x-ray
reflectivity measurements of liquid
        Hg \cite{magnu95}
and
        Ga \cite{regan95}.
Experiments on
        In \cite{tostmann99}
and a number of
        alloys \cite{lei96,lei97,regan97,tostmann98}
followed.
        Fig.~\ref{fig:jop_fig1}(b) shows
experimental reflectivities for  three low melting point elemental
LM. The principle deviation from the Fresnel reflectivity
calculated for a perfectly flat metal surface (solid lines) is in
the broad quasi-Bragg peak centered near $q_z = 2.2$~\AA$^{-1}$.
These reflectivity profiles can be well described by layered
density profiles decaying over several layers, shown schematically
in Fig.~\ref{fig:jop_fig1}(a).

\section{Surface Structure of Binary Liquid Alloys}

In binary alloys properties such as atomic size, surface tension,
and electronic structure can be varied and should affect the
details of the  surface structure, thus  allowing a more
systematic understanding of  surface layering. Also, since binary
alloys form various ordered phases in the bulk, another
interesting question arises: How does the alloy's bulk phase
behavior manifest itself at the surface, where the electronic
structure, atomic coordination and local composition are
different? This question has motivated a number of studies on
alloys, which have found that in general, surface layering
competes with the formation of more complicated surface phases.
For example, in miscible alloys the \index{Gibbs!adsorption} Gibbs
adsorption rule predicts that the species having the lower surface
energy will segregate at the surface. Observations on
    Ga-In \cite{regan97},
    Ga-Sn \cite{lei97}
and Ga-Bi at low Bi concentrations
    \cite{lei96,tostmann98}
have found that surface segregation coexists with surface
layering.  In these alloys the first surface layer is almost
entirely composed of the lower surface tension component (In, Sn
or Bi). By the second or third atomic layer, the bulk composition
has been reached. In the following sections, we describe recent
x-ray results from alloy surfaces which demonstrate a range of
different surface induced structural effects.

\subsection{K-Na}
\index{alkali metals} Alkali metals have a simple electronic
structure which can be described by ideal \index{Fermi!surface}
Fermi surfaces, and are soluble in each other with only a weak
tendency towards phase formation. Since alkali metals have a very
low surface tension, surface fluctuations are enhanced. These
properties are expected to make the alkali metals' surface
structures different from those of the main group metals studied
so far. Ideally alkali metals would be investigated under UHV
conditions due to their high reactivity. However, at the melting
point their high vapor pressures precludes this. By contrast, the
melting point of the \index{eutectic} eutectic $K_{80}Na_{20}$
alloy is sufficiently low to allow UHV conditions. Due to the
almost identical electron densities of the two components, when
probed by x-rays this alloy exhibits the structure of a
homogeneous liquid metal. Here we present preliminary results for
the eutectic $ K_{80}Na_{20}$ alloy.

\begin{figure}[tbp]
\centering
\includegraphics[width=0.7\columnwidth]{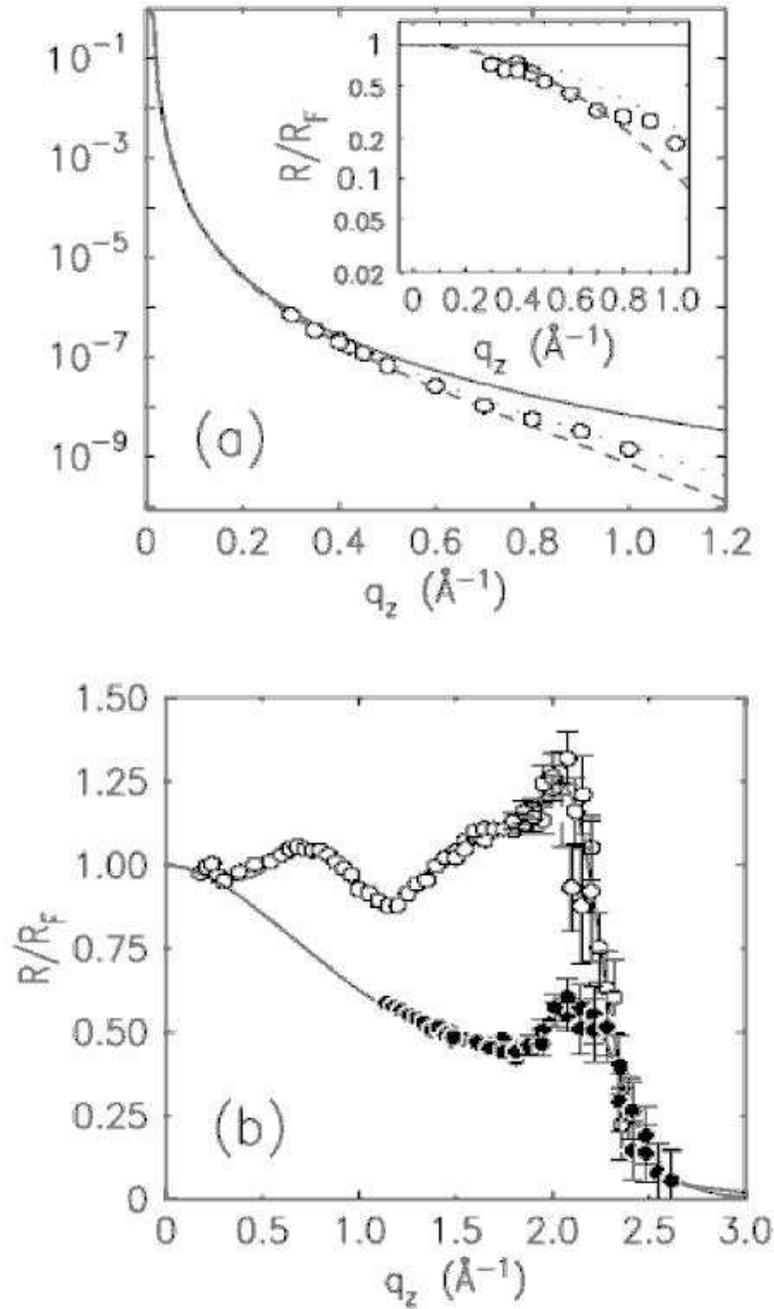}
 \caption{(a) X-ray reflectivity from a $
K_{80}Na_{20}$ alloy measured by integrating over a large range of
$\alpha$ at fixed $\alpha +\beta$. The normalized reflectivity is
shown in the inset.  The dotted lines show a capillary wave
roughness with no layering with $\sigma = 1.2~\AA$ and 1.5~$\AA$,
respectively. (b) Normalized x-ray  reflectivity of liquid In
(+170$^\circ$C, closed circles) and In-22at \%Bi (80 $^\circ$C,
open circles).\label{fig:jop_fig2}}
\end{figure}

Fig.~\ref{fig:jop_fig1}(a) shows the x-ray reflectivity from
$K_{80}Na_{20}$ along with the predicted reflectivity assuming
capillary wave roughness (Gaussian form) of 1.2 and
1.5~\AA\cite{oops}. At all $q_z$ the reflectivity is bounded by
these two curves; at lower $q_z$ it is better described by the
1.5~\AA\ roughness.  On length scales $\stackrel{>}{\scriptsize
\sim} 6$~\AA\ no obvious structural feature is found beyond the
predicted capillary wave roughness. The low surface tension
($\approx 120 dyn/cm$) and the subsequently high roughness,
appears to preclude measurements to $q_z$ large enough to directly
observe a surface layering peak. This is in contrast to
well-defined surface layering peaks observed for Ga, Hg or In
   (see Fig.~\ref{fig:jop_fig1}(b)) \cite{EDHT}.

\subsection{Bi-In}

For systems having significant attractive interactions between
unlike atoms, the surface structure is more complex. This is
especially true of alloys such as Bi-In which form well-ordered
intermetallic phases in the bulk solid. In
Fig.~\ref{fig:jop_fig2}(b) we show the normalized reflectivity for
\index{eutectic} the eutectic composition Bi$_{22}$In$_{78}$,
measured at 80$^\circ$C ($\triangle$) along with the normalized
reflectivity for liquid In at $+170^\circ$C. The alloy exhibits a
well defined layering peak centered at 2.0~\AA$^{-1}$ which
resembles the layering peak found for pure In ({\bf smudges}). In
addition, the reflectivity displays a modulation with a period of
about 0.9~\AA$^{-1}$. This foscillation indicates that ordering
over a short region at the surface occurs with a length scale
nearly twice that of the longer-range layering.  This suggests the
presence of Bi-In pairs at the surface.  A full report on the
phase behavior of three different In-Bi alloys will be given
elsewhere \cite{EDHT}.

\subsection{Ga-Bi}

The Ga-Bi system  is an example of an alloy with repulsive
\index{interactions!heteroatomic} heteroatomic interactions
leading to a bulk \index{miscibility gap} miscibility gap. Below
the \index{monotectic point} monotectic temperature, $T_{mono} =
222^\circ$C, a Ga-rich liquid coexists with a solid Bi phase
    \cite{nattland96}.
However, due to its lower surface energy a Bi monolayer is
expected to segregate at the surface of the Ga-rich liquid. Above
$T_{mono}$, Ga-Bi exhibits a thick \index{wetting!film} wetting
film, as predicted for all binary mixtures with critical
    demixing \cite{cahn77}.
This transition occurs at a characteristic wetting temperature
$T_w$ below the critical temperature
    $T_{crit}$ \cite{nattland96}.
Above $T_w$, a macroscopically thick Bi-rich phase is expected to
completely wet the less dense Ga-rich phase in defiance of
gravity. The Bi concentration in the Ga-rich phase increases with
increasing temperature as long as the Ga-rich liquid coexists with
the solid Bi phase.

\begin{figure}[tbp]
\centering
\includegraphics[width=0.6\columnwidth]{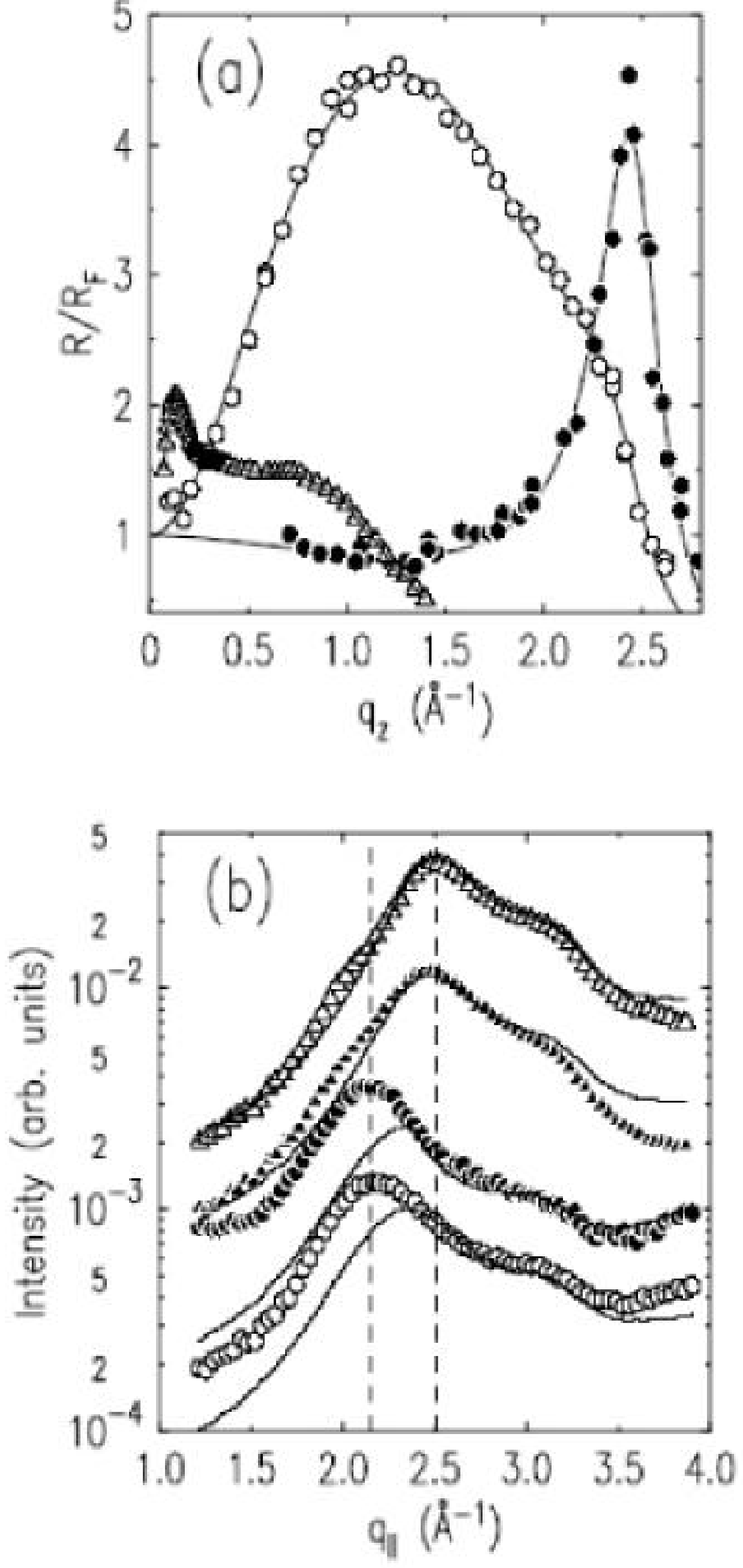}
 \caption{ (a) Normalized x-ray reflectivity
of liquid Ga (+25$^\circ$C, $\circ$), and on the Ga-Bi two-phase
coexistence curve at $35^\circ$C ($\bullet$) and $228^\circ$C.
(b)Grazing incidence diffraction from Ga-Bi at $150^\circ$C
(filled triangles), $205^\circ$C ($\triangle$), $228^\circ$C,  and
$255^\circ$C
 at $\alpha=0.08^\circ$. The solid line shows
corresponding profiles for $\alpha=0.30^\circ$ where the bulk is
predominately sampled. The data was acquired using Soller slits,
0.05~$\AA^{-1}$ FWHM. Still, it was not possible to reliably
subtract the background. \label{fig:jop_fig3}}
\end{figure}

The normalized x-ray reflectivity spectra, $R/R_F$, for Ga-Bi at
$35^\circ$C and $228^\circ$C  are shown in
    Fig.~\ref{fig:jop_fig3}(a)
versus $q_z$, along with the profile for pure Ga at room
temperature. At $35^\circ$C the normalized reflectivity has a
broad maximum at $q_z \approx 1$~\AA$^{-1}$. As suggested by
    Lei et al.\cite{lei96}, this is consistent
with a density profile with a thin, high density monolayer of Bi.

We have fit the reflectivity profiles to simple density profiles
using Eq. (1). The fitted reflectivities are shown in
    Fig.~\ref{fig:jop_fig3}(a) (solid lines).
At $+35^\circ$C the local density profile exhibits a top-layer
density which is about 1.5 times higher than the Ga bulk liquid
density. The $3.4 \pm 0.2 $~\AA\ layer spacing between the surface
and the adjacent Ga layer obtained from the fits is much larger
than the $2.5 \pm 0.1$~\AA\ layer spacing obtained in liquid
gallium. The data show that the surface layer has a higher density
than in the underlying Ga-rich subphase, confirming the surface
segregation of a Bi monolayer.

The behavior of the same alloy at
    $228^\circ$C is markedly
different: a sharp peak in $R(q_z)$ has emerged, centered around
$0.13$~\AA$^{-1}$
    (Fig.~\ref{fig:jop_fig3}(a)).
The peak at small $q_z$ indicates the presence of a thick surface
layer with a density greater than that of the bulk subphase. The
absence of additional oscillations following the sharp peak
suggests that the boundary between the two regions must either be
diffuse or rough. The persistence of the broad maximum at $q_z
\approx 0.75~$\AA $^{-1}$ indicates that Bi monolayer segregation
coexists with the newly formed \index{nanoscale} nanoscale
\index{wetting!film} wetting film. Fits to a simple two-box model
yield a film thickness of 30~\AA\, consistent with
\index{ellipsometry} ellipsometry results
    \cite{nattland96},
and a surface density consistent with the high density liquid
phase of the bulk alloy. The temperature dependent reflectivity
will be reported elsewhere
    \cite{EDHT}.

In Fig.~\ref{fig:jop_fig3}(b) GID data are shown from the same
Ga-Bi alloy in the temperature range from 150 to 255$^\circ$C.
Data at 35$^\circ$C was previously reported
    \cite{tostmann98}.
In the liquid Ga-rich phase the Bi concentration ranges from 3.3
at\% to 17.8at\%. At each temperature data was taken above and
below $\alpha_{crit}= 0.14^\circ $ at $\alpha =0.08^\circ$
(symbols) and at $\alpha=0.30^\circ$(lines). At $\alpha
=0.08^\circ$ the x-ray penetration depth equals $28~\AA$ or about
10 atomic layers.

The solid lines in Fig.~\ref{fig:jop_fig3}(b) at 150 $^\circ$C
show the bulk liquid scattering which is predominately from pure
Ga since the Bi concentration is low. The broad peak at $q_{\|} =
2.5$~\AA$^{-1}$ and the shoulder on the high-angle side of the
peak are in agreement with the bulk liquid Ga structure factor
    \cite{narten72}.
There is no evidence for a peak or shoulder at the position
corresponding to the first peak of the Bi liquid structure factor
at $q_{\|} \approx 2.2$~\AA$^{-1}$.  This is expected since the
surface regime is so much smaller than the bulk volume sampled.
For $\alpha = 0.08^\circ < \alpha_c$, the x-rays penetrate to a
depth of only about 30~\AA . Here a shoulder appears on the
low-$q$ side of the gallium liquid peak, due to enhanced
sensitivity to the Bi surface monolayer. Between $150^\circ$C and
$205^\circ$C ($\bullet$) there is little change in the GID data,
except a slight increase in the shoulder associated with the Bi
monolayer.

Above $T_{mono}$ there is a dramatic change in the GID profiles.
In Fig.~\ref{fig:jop_fig3}(b), GID data is shown at $228^\circ$C
and $255^\circ$C. In both cases, for $\alpha > \alpha_c$ the peak
has shifted to $q_{\|} \approx 2.3$~\AA$^{-1}$ from the
2.5~\AA$^{-1}$ peak position found at lower temperatures.  This
results from the much higher Bi concentration in the bulk at the
higher temperatures and the larger atomic size of Bi.  Even more
dramatic is the shift in the peak position for $\alpha <
\alpha_{crit}$ where the peak is at 2.15~\AA$^{-1}$. Thus, the
surface region contains considerably more Bi than the underlying
bulk alloy. This finding is consistent with the wetting layer
observed in the x-ray reflectivity measurements.

\bibliographystyle{unsrt}

%% file: hgau_all.tex
\chapter{Surface studies of
liquid Hg/Au alloys}

\section{Abstract}

In this chapter we present temperature dependent x-ray
reflectivity measurements of liquid Hg alloyed with 0.06--0.20at\%
Au.  At low Au concentrations, we find temperature dependent
surface induced layering similar to that observed in pure Hg,
except that the presence of Au reduces the layering amplitude.
Upon approaching the \index{solubility limit} solubility limit of
Au in Hg, a new surface phase forms which is 1--2 atomic diameters
thick and has a density of about half that of bulk Hg. We present
a surface \index{phase diagram} phase diagram, summarizing the
evolution of this unexpected  surface structure through
comparatively small changes in temperature and composition.
Possible implications for altering the electronic structure for
control of catalytic and electrochemical reactions at the liquid
metal surface are discussed.

\section{Introduction}

Mercury and gold are among the few metals that can be found in their
native state in nature, and that were known to early
    civilizations.\cite{ancient}
The use of Hg to extract Au and Ag from their ores, known as amalgamation,
is one of the earliest metallurgical processes known to
    humanity.\cite{metallurgy}
It is therefore no surprise that Hg-Au \index{amalgam} amalgams
have captured the interest of alchemists, metallurgists, chemists,
physicists, and
    dentists.\cite{dentist}
Hg and Au are both transition metals having a filled $d$-band, of similar
atomic size, and they form a variety of stable intermetallic phases
in the solid
    state.\cite{rolfe67}
The low cohesive energy of Hg, evident from its low melting point
of $-38.9^\circ$C, enables these \index{amalgam} amalgams to form
readily, simply by bringing Au into contact with liquid Hg under
ambient conditions.

Despite a prevailing interest in Hg-Au amalgams, not many detailed structural
studies of Hg-Au compounds have been performed.
It proves very difficult to establish the Hg-Au phase diagram
due to the high volatility of Hg, which requires the use of sealed tubes at
undetermined pressures for Hg-rich
    alloys.\cite{rolfe67}
Much recent work has focussed on surfaces of Hg-Au \index{amalgam}
amalgams. For example, microscopy and spectroscopy studies have
addressed the morphology and composition of Hg-Au phases formed by
depositing Hg onto Au
    films.\cite{yang94,battis96,nowakowski97}
These studies are complicated by the morphology of the Au substrate, and
the coexistence of several Hg-Au phases in the
    amalgam.\cite{nowakowski97}
Deposition of Hg onto Au electrodes has also been studied through
electrochemical
    techniques.\cite{abruna97}
In-situ surface x-ray diffraction
measurements of underpotential deposition of Hg onto the crystalline
Au(111) electrode have revealed that amalgamation occurs in several steps,
characterized by distinct surface
    phases.\cite{abruna97}
Since some of these phases are modified by coadsorbed anions from
the electrolyte, and since kinetic effects produce further
complications, it is difficult to obtain basic structural
information about the solid Hg-Au \index{amalgam} amalgam through
such studies.

In light of the knowledge about Hg-Au phase formation, it is surprising that
Au was once regarded as an ``inert'' material that could be used as part
of a liquid Hg electrode in electrochemical
    studies.\cite{kemula59}
However, it was soon recognized that the potential of the amalgam electrode
differed from that of pure Hg.  A more interesting observation was that
small amounts of Au dissolved into the Hg droplet can strongly affect
the electrochemical behavior by forming intermetallic compounds
with other dilute metals such as Cd and Zn that are present in the
    liquid Hg.\cite{kemula59}
Dilute metal impurities in liquid Hg have also been observed to affect
the activation energies of reactions catalyzed by the liquid metal
    surface.\cite{catalysis}
It is not known whether such effects are due to changes in
the electronic properties of liquid Hg, modification of the surface structure,
or the formation of intermetallic phases at the surface.

\begin{figure}[tbp]
\centering
\includegraphics[width=0.5\columnwidth]{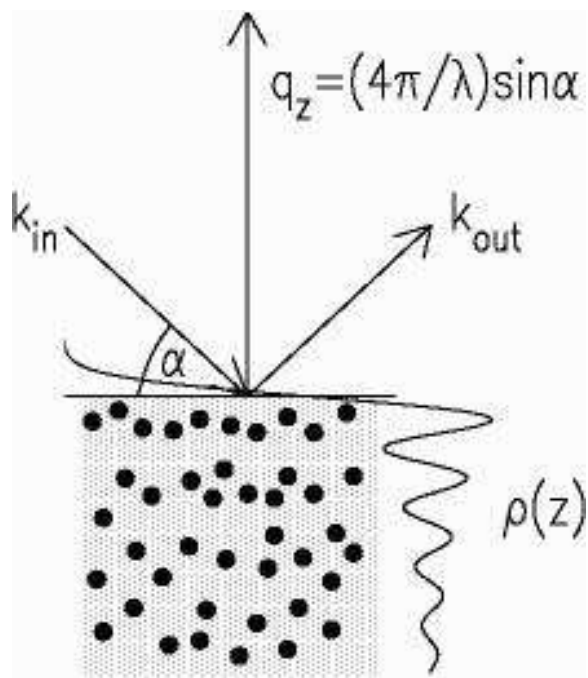}
 \caption{ Geometry for x-ray
reflectivity from a layered liquid metal surface. Maxima in the
oscillatory surface-normal density profile $\rho(z)$ correspond to
layers of atoms parallel to the liquid--vapor interface. }
\label{hgau_fig:geo}
\end{figure}

Until very recently,
atomic-scale structural measurements on liquid metal surfaces have
not been available, leaving such questions unresolved.  It is now
known that a wide variety of structures form at the surfaces of liquid
metals and alloys.  Elemental Hg, Ga and In exhibit surface induced layering,
in which atoms are stratified parallel to the liquid--vapor
    interface,\cite{magnu95,dimasi98,regan95,regan96,tostmann99}
a result long predicted by
    theory.\cite{rice96}
This stratification of atoms, with the corresponding oscillatory
surface-normal density profile, is shown schematically in
    Figure~\ref{hgau_fig:geo}.
In liquid metal alloys, surface layering may compete with the formation
of surface phases, making the structure more complex.
Prominent examples are the monolayer surface segregation observed in
    Ga-In\cite{regan97} and Ga-Sn\cite{lei97},
and a concentration dependent wetting transition in
    Ga-Bi.\cite{tostmann98}
In all these cases, surface layering persists in the alloys, but
is modified to varying extents by the surface phase. Competition
between layering and surface phase formation is expected to be
particularly important for systems such as Hg-Au, where attractive
\index{interactions!heteroatomic} heteroatomic interactions
dictate the formation of intermetallic phases in the bulk solid.
Such well ordered phases in the solid typically give way upon
melting to homogeneous mixtures in the bulk liquid. However, one
may expect that the region close to the liquid--vapor interface,
where order is induced in elemental liquid metals, may support the
formation of intermetallic phases in the liquid alloy. Structural
information for such systems is of fundamental interest, and may
well shed light on reactions occuring at liquid alloy surfaces,
relevant to catalysis or electrochemistry.

In this paper we report x-ray reflectivity measurements of liquid Hg-Au
alloys near the room temperature saturation
    limit\cite{rolfe67,gumin91}
of 0.14~atomic(at)\%~Au in Hg,
at temperatures between $+25^\circ$C and $-39^\circ$C.
We construct a temperature--concentration surface phase diagram and
identify two distinct regions of surface phase behavior.
At high $T$ and low Au concentrations, surface
layering similar to that of pure Hg is observed.
By contrast, at low $T$ and comparatively higher Au concentrations,
we find evidence for the formation of a more complicated
surface phase, where a new length scale for surface layering emerges
along with a low-density layer at the interface.

\section{Experimental Details}

X-ray reflectivity measurements were carried out using the
Harvard--BNL liquid surface \index{spectrometer} spectrometer at
beamline X22B at the National \index{synchrotron!source}
Synchrotron Light Source, with an x-ray wavelength of 1.24~\AA\
and a detector resolution of
    $\Delta q_z = 0.035$~\AA$^{-1}$
within the reflection plane.
The background intensity, due mainly to scattering from the bulk
liquid, was subtracted from the specular signal by displacing
the detector out of the reflection plane.
The design of the liquid
    spectrometer\cite{brasl88}
and a review of the measurement
    technique\cite{encyc}
are given elsewhere.

Samples with nominal concentrations of 0.06, 0.10, 0.13 and 0.20at\% Au
were produced by adding Au powder (Johnson Matthey, 20 mesh, 99.9995\%) to
liquid Hg (Bethlehem Apparatus Company, quadruple distilled, 99.9999\%),
several days or weeks before the x-ray experiments.
The main experimental problem in surface measurements of metals is to
ensure that a contaminant-free surface is obtained.
The most reliable way to produce atomically clean metal surfaces is to
keep them under ultra high vacuum (UHV) conditions,
and to remove any residual oxide through argon
ion sputtering.  These UHV techniques, which we found to be
very successful for x-ray scattering measurements of Ga, In, and
low vapor pressure
    alloys,\cite{regan95,tostmann99,tostmann98}
must be modified in the case of Hg and its alloys due to the high
vapor pressure of Hg ($\sim 10^{-2}$~Torr at room temperature).
For the measurements presented here, each sample was poured into a
stainless steel reservoir within an argon-filled glove box
($<5$~ppm oxygen, $<2$~ppm water). The reservoir is connected to a
stainless steel valve, filling capillary, and UHV flange and is
\index{UHV!chamber} affixed to a UHV compatible chamber. The empty
chamber is evacuated to $10^{-7}$~Torr and baked out, after which
the liquid Hg alloy is dropped from the reservoir through the
valve and capillary into a ceramic sample pan within the chamber,
as was done previously for measurements of pure
    Hg.\cite{dimasi98}
Immediately after opening the valve to the reservoir,
the Hg vapor pressure determines the total pressure in the chamber, while the
partial pressures of possible contaminants such as water and oxygen remain
very low.
The resulting liquid alloy surfaces were found to be stable for over a week,
due to two mechanisms.
First, Hg oxide introduced from pouring the sample is unstable and decomposes
under these low oxygen partial
    pressures.\cite{pourbaix,decompose}
In addition, slight but continual evaporation constantly refreshes the
    surface.\cite{renew}
Preliminary reflectivity
measurements utilizing a glass sample chamber evacuated to
$10^{-4}$~Torr and backfilled with dry hydrogen gas were qualitatively similar,
but not sufficiently reproducible.

The samples were cooled with a liquid nitrogen cold finger
beneath the sample pan.
Temperatures were monitored at the sample pan and calibrated to that of the
liquid surface in separate experiments using thermocouples immersed in the
samples.  The calibration has an uncertainty of $\pm 1^\circ$C.

\section{X-ray Reflectivity and Modelling}

X-ray reflectivity is a powerful technique for investigating structure
normal to surfaces and interfaces on atomic length
    scales.\cite{encyc,xr1,xr2,xr3}
In these measurements, the scattered intensity is measured as
a function of momentum transfer
    $q_z = k_{\mbox{\scriptsize out}} - k_{\mbox{\scriptsize in}}$
perpendicular to the surface
    (Figure~\ref{hgau_fig:geo})
and normalized to the incident photon intensity.

\begin{figure}[tbp]\epsfig{file=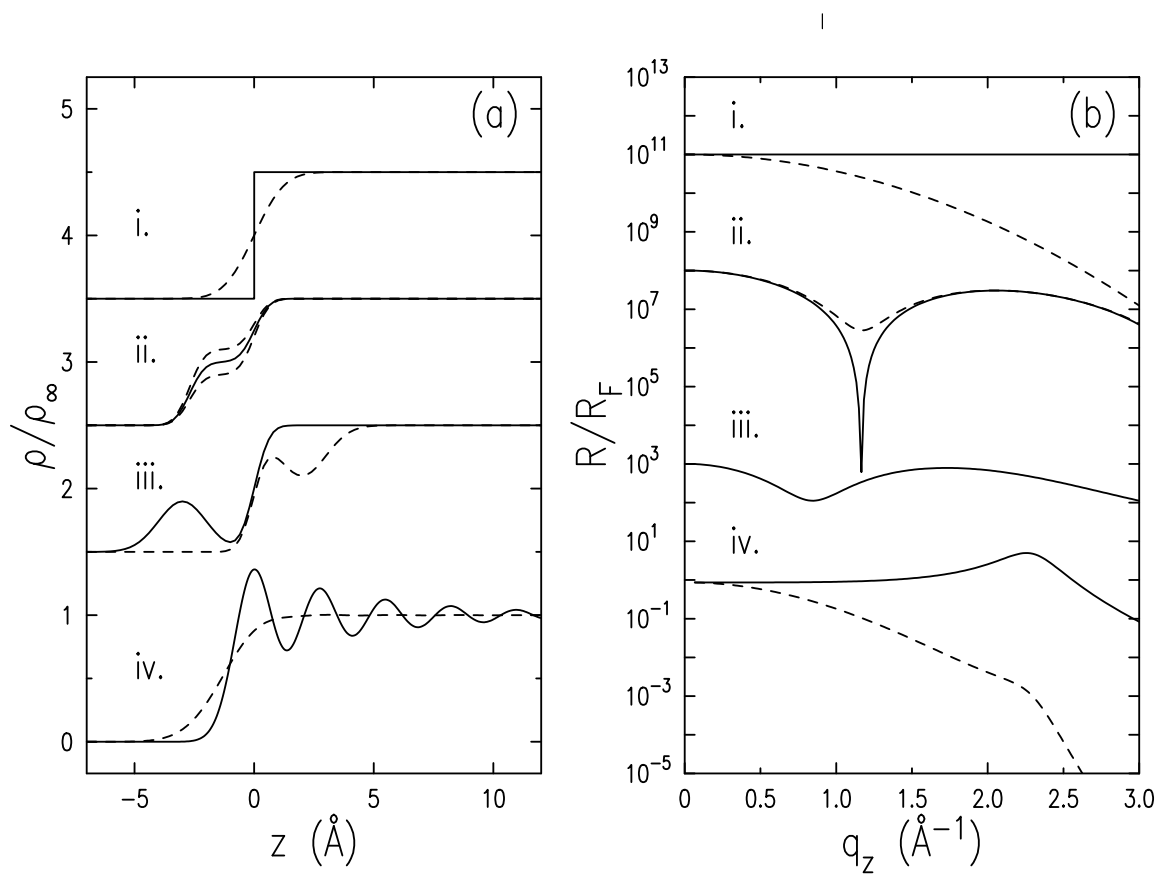, angle=90,
width=1.0\columnwidth}
 \caption{ (a) Model density profiles and
(b) corresponding Fresnel-normalized reflectivities as described
in text.  Curves i--iii in both panels are shifted for clarity. }
\label{hgau_fig:model}
\end{figure}

The simplest interface is a step
function describing a sharp truncation of a homogeneous bulk density
$\rho _\infty$
    (Fig~\ref{hgau_fig:model}(a)i, solid line).
Scattering from this sharp interface takes the form of the Fresnel
reflectivity
\begin{equation}
    R(q_z) = \left[ \frac{q_z}{q_c}  + \sqrt{
    \left( \frac{q_z}{q_c} \right) ^2 - 1 } \; \right] ^{-4}
    \equiv R_F \: .
\end{equation}
In this expression $R_F$ depends on the bulk density through the
\index{critical!wavevector}  critical wavevector $q_c = \sqrt{16
\pi \rho_\infty r_0}$, where $r_0 = e^2/mc^2 = 2.82\times
10^{-5}$~\AA\ is the classical electron radius, and
\index{x-ray!absorption} absorption has been neglected. Additional
structure in the surface-normal density profile $\rho(z)$ modifies
the reflectivity from the Fresnel form. When the scattered
intensity is much less than the incident intensity (generally the
case as long as $q_z \stackrel{\scriptstyle >}{\scriptstyle \sim}
5 q_c$), the kinematic approximation is valid and the reflectivity
may be written
\begin{equation}
\label{hgau_phase}
    R(q_z) = R_F \left| \frac{1}{\rho _\infty}
    \int _{-\infty} ^{\infty} (\partial \rho / \partial z)
    \exp(iq_z z) dz \right| ^2 \: .
\end{equation}
Since the phase information is not accessible to the experimental
intensity measurement,
    Eq.~\ref{hgau_phase}
can not be inverted to determine the density profile directly.
For this reason the usual practice is to construct a model density
profile and compare its calculated reflectivity to the experiment.
Because of this phase problem, distinct structural models can produce
essentially identical reflectivity curves,
lending some ambiguity to the analysis.

The simplest physically reasonable modification of the step function
profile takes surface roughness into account.
In a liquid, thermally excited capillary waves produce height
variations across the surface, which are averaged in the scattering
measurement over length scales determined by the resolution of the
spectrometer.
These occur in addition to any roughness intrinsic
to the local surface-normal profile.  Detailed discussions of the effect
of capillary waves on the scattering cross section are given
    elsewhere.\cite{tostmann99,brasl88}
The capillary wave roughness
    $\sigma _{cw}$
is a function of the temperature $T$ and
the surface tension $\gamma$, and is given by
\begin{equation}
\label{hgau_eq:cw}
    \sigma _{cw} ^2 = \frac{k_B T}{2 \pi \gamma}
    \ln \left(
    \frac{q_{\mbox{\scriptsize max}}}{q_{\mbox{\scriptsize res}}}
    \right) \: .
\end{equation}
This expression has been arrived at by integrating over
those capillary modes having wavevectors less than
$q_{\mbox{\scriptsize max}} = \pi/a$, where $a$ is the atomic diameter,
and greater than
$q_{\mbox{\scriptsize res}}$, determined by the instrumental
    resolution.\cite{error1}

We will consider a more general broadening of the density profile, by
$\sigma _T ^2 = \sigma _{cw} ^2 + \sigma _i ^2$,
to incorporate the capillary wave roughness
$\sigma_{cw}$
along with any other roughness
$\sigma _i$  intrinsic to the surface.
This profile, having
$$
    \partial \rho /\partial z = \exp (-\sigma_T ^2 z^2 /2)  \: ,
$$
is shown by dashed lines in
    Figure~\ref{hgau_fig:model}(a)i
for $\sigma _T=1.0$~\AA .
The density profile can be written in terms of the error function:
\begin{equation}
    \rho(z) = (\rho _\infty / 2)
    \left[ \mbox{erf} (z / \sigma _T \sqrt{2} ) +1 \right]
    \equiv \rho _{\mbox{\scriptsize erf}}(z) \: ,
\end{equation}
yielding for the reflectivity
\begin{equation}
\label{hgau_eq:dw}
    R = R_F \exp (-q_z ^2 \sigma _T ^2 )  \: .
\end{equation}
The resulting Fresnel-normalized reflectivity is shown in
    Figure~\ref{hgau_fig:model}(b)i, dashed line.
The exponential factor in
    Eq.~\ref{hgau_eq:dw})
is analogous to the \index{Debye-Waller} Debye-Waller term used to
describe the reduction of the diffracted intensity from crystals
due to thermal displacements of the atoms from the lattice sites.

When more complicated surface-normal density profiles must be constructed,
two straightforward methods are commonly used.
Regions of differing electron density can be represented by a combination
of error functions, parameterizing the thickness, density, and roughness
between each region.  A general profile of this type may be written
\begin{equation}
    \rho(z) = \sum _{n=0} ^N
    \left( \frac{\rho _n - \rho_{n+1}}{2} \right)
    \left[ \mbox{erf}
    \left( \frac{z - z_n}{\sigma _n \sqrt{2} } \right) \right] \: ,
\end{equation}
with $\rho _0$ equal to the bulk density $\rho _\infty$,
$\rho _{N+1} = 0$ representing the vapor, and
$\sigma _n$ describing the broadening between regions $n$ and $n+1$
at position $z_n$.
The reflectivity for this model is
\begin{equation}
    R / R_F = \left|
    \sum _{n=0} ^N \left( \frac{\rho _n - \rho _{n+1} }{\rho _\infty}
    \right) \left[ \exp ( -q_z ^2 \sigma _n ^2 /2 + i q_z z_n )
    \right] \right| ^2 \: .
\end{equation}
    Figure~\ref{hgau_fig:model}(a)ii
shows three such profiles, having one additional slab
of density between the vapor and the bulk liquid.
Here $\sigma _0 = \sigma _1 = 0.5$~\AA\
and $| z_1 | = 2.7$~\AA .
With the same roughness for both interfaces, the reflectivity at
$q_z = \pi / | z_1 |$ has an amplitude proportional to
$(1 - 2 \rho _1 / \rho _0) ^2$.
When $\rho _1 / \rho _0 = 1/2$
    (Figure~\ref{hgau_fig:model} (a)ii, solid line),
the result is total destructive interference at
$q_z = \pi / | z_1 |$,
producing a deep notch in the reflectivity
    (Figure~\ref{hgau_fig:model}(b)ii, solid line).
For values of $\rho _1$ symmetrical about 1/2, identical reflectivity
curves are obtained.
    Dashed lines in
    Figure~\ref{hgau_fig:model}(a)ii
show profiles having $\rho _1 / \rho _0 = 0.4$ and 0.6,
which both produce the reflectivity curve shown as a dashed line in
    Figure~\ref{hgau_fig:model}(b)ii.
This is an example of the unavoidable ambiguity in
interpreting reflectivity measurements even for  simple models.

An additional method of constructing a structured surface profile is to
define density through Gaussian functions, as shown in
    Figure~\ref{hgau_fig:model}(a)iii.
For the example shown,
\begin{equation}
\label{hgau_eq:lump}
    \rho(z) = \rho _{\mbox{\scriptsize erf}} +
    \frac{h _g / \sigma _g}{\sqrt{2 \pi}}
    \exp \left[ -(z - z_g)^2 / 2 \sigma _g ^2 \right]
\end{equation}
and
\begin{equation}
\label{hgau_eq:gaus}
    R / R_F = \left|
    \left[ \exp (-q_z ^2 \sigma _T ^2 / 2) \right] ^2
    + \frac{h _g}{\rho _\infty} q
    \exp (-q_z ^2 \sigma _g ^2 /2)
    \exp \left[ i(q_z z_g - \pi /2) \right] \right| ^2 \: .
\end{equation}
The Gaussian position $z_g$ and the sign of its amplitude $h_g$ enter
only in the cross term of
    Eq.~\ref{hgau_eq:gaus},
which is proportional to $h_g \sin(q_z z_g)$.
For this reason, the reflectivity measurement does not distinguish
between a positive Gaussian density added to the vapor side at position
$z=z_g$
    (Figure~\ref{hgau_fig:model}(a)iii, solid line)
and density subtracted from the liquid side at $z=-z_g$
    (Figure~\ref{hgau_fig:model}(a)iii, dashed line).
These two rather different looking density profiles both produce the
reflectivity curve shown in
    Figure~\ref{hgau_fig:model}(b)iii.
The effect of this Gaussian term can be similar to that of a shelf of
density: both produce a minimum in the reflectivity at low $q_z$, as
shown by comparison of
    Figure~\ref{hgau_fig:model}(b)iii
and
    Figure~\ref{hgau_fig:model}(b)ii, dashed line.

The models discussed so far describe a liquid--vapor interface having
structure near the surface.  To describe the surface induced layering
of liquid metals, which extends further into the bulk,
a damped oscillatory surface-normal density profile must be constructed.
A convenient way to do this is to model layers of
atoms parallel to the surface by a series of Gaussian terms,
representing mean-squared displacements of atoms assigned to each
layer at spacing $d$:
\begin{equation}
\label{hgau_eq:ripl}
    \rho(z) =  \rho_\infty
    \sum _{n=0} ^\infty \frac{d / \sigma _n}{\sqrt{2 \pi}}
    \exp \left[ -(z-nd)^2 / 2 \sigma _n ^2 \right] \: .
\end{equation}
With the choice of
$\sigma _n ^2 = \sigma _T ^2 + n \overline{\sigma} ^2$,
we parameterize two effects in a simple way.
First, the surface layers become less well defined with increasing
depth into the bulk liquid, at a rate controlled by $\overline{\sigma}$.
Second, the effect of increasing $\sigma _T$ is to reduce the overall
layering amplitude, an effect expected both from capillary waves and from
any roughness intrinsic to the local surface-normal profile.

The corresponding reflectivity is given by
\begin{equation}
    R / R_F = \left[ F_Z(q_z) \right] ^2
    (q_z d)^2 \exp(- \sigma _T ^2 q_z ^2)
    \left[1 - 2\exp(-q_z ^2 \overline{\sigma} ^2 /2)\cos(q_z d)
    + \exp(-q_z ^2 \overline{\sigma} ^2) \right] ^{-1} \: .
\end{equation}
Since each Gaussian now explicitly represents a collection of
scattering atoms, here it is meaningful to incorporate the reduced
atomic \index{form factor} form factor
$F_Z(q_z)=[f_Z(q_z)+f'_Z]/Z$ for atoms having atomic number
    $Z$.\cite{error2}
    Figure~\ref{hgau_fig:model}(a)iv (solid line)
illustrates a layered profile with
$d=2.7$~\AA , $\sigma _T = 0.8$~\AA , and
$\overline{\sigma} = 0.45$~\AA .
The reflectivity
    (Figure~\ref{hgau_fig:model}(b)iv, solid line)
shows the broad \index{quasi-Bragg peak} quasi--Bragg peak
produced by the oscillatory density profile.  In the limit where
$\sigma _T$ is large, the profile approaches the error function
form, though with its argument shifted from $z=0$ to $z=-d/2$ due
to the choice of origin in
    Eq.~\ref{hgau_eq:ripl}.
A layered profile having $\sigma _T = 1.5$~\AA\ is shown by
    dashed lines in Figure~\ref{hgau_fig:model}(a)iv.
Although the intensity falls off quickly for large $q_z$
    (Figure~\ref{hgau_fig:model}(b)iv, dashed line),
constructive interference in the region of the layering peak is
still evident.

Finally, this profile can be modified by additional Gaussian terms,
and by changing the amplitudes, positions and widths of the terms in
    Eq.~\ref{hgau_eq:ripl}.
The most general model that we will present in this study takes the form
\begin{equation}
\label{hgau_eq:model}
    \rho(z) / \rho_\infty =
    \left( \frac{w_A d}{\sigma_A \sqrt{2 \pi}} \right)
    \exp \left[ -(z-z_A) ^2 / \sigma _A ^2 \right] +
    \sum _{n=0} ^\infty
    \left( \frac{w_n d}{\sigma_n \sqrt{2 \pi}} \right)
    \exp \left[ -(z-nd) ^2 / \sigma _n ^2 \right] \: ,
\end{equation}
where the Gaussian scaled by $w_A$ is typically much broader
than those controlled by $\sigma _n$ in the sum, and
$w_n$ may deviate from unity.
Since the effect of the extra Gaussian is similar to that of the
Gaussian plus error function model described by
    Eq.~\ref{hgau_eq:lump},
placing this term into the
vapor side of the interface has an effect which can be difficult to
distinguish from that of decreasing the weights $w_n$ of the first few
terms in the sum.

\section{Experimental Results}

\begin{figure}[tbp]\unitlength1cm \epsfig{file=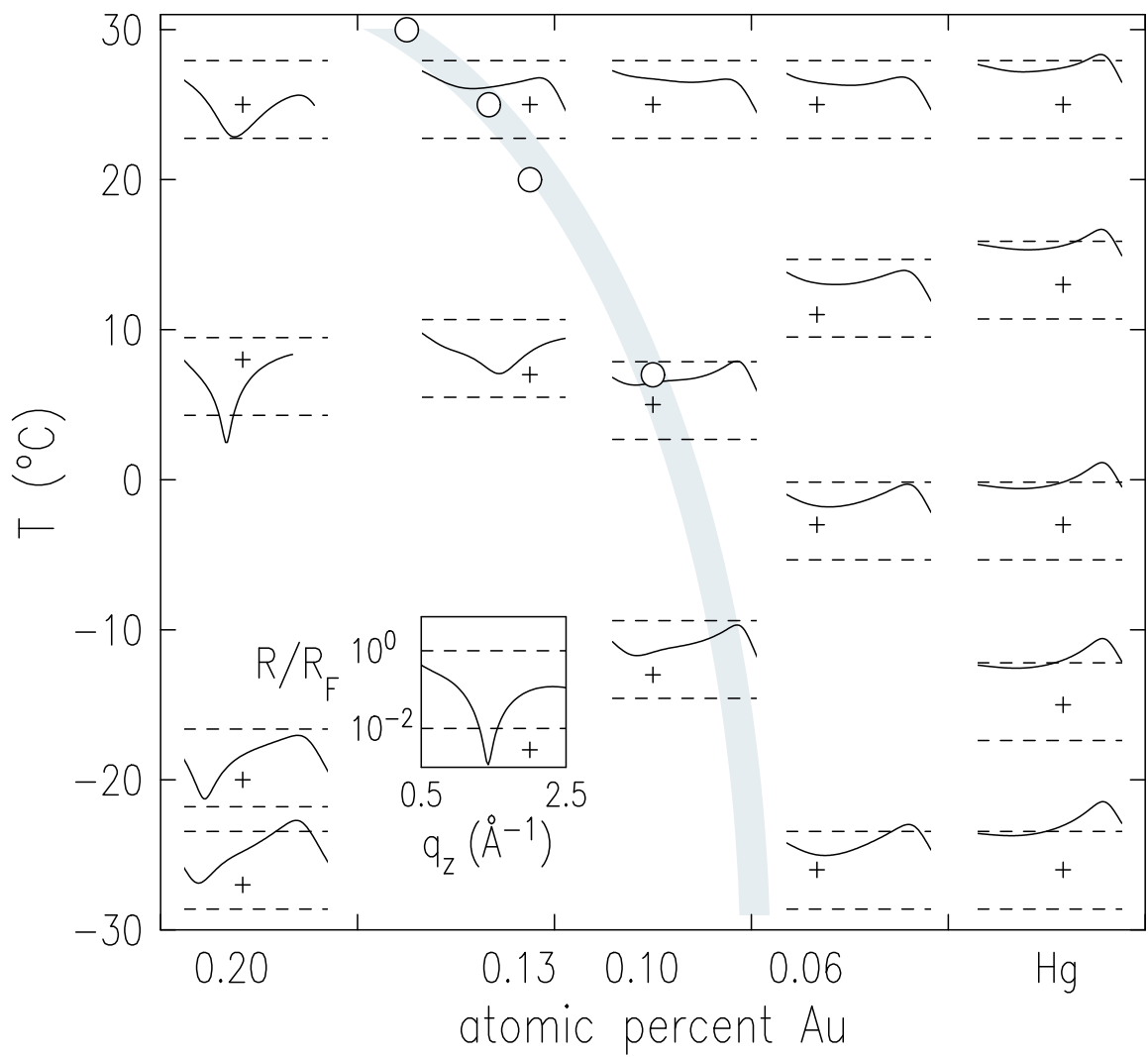, angle=90,
width=1.0\columnwidth}
 \caption{ Surface phase diagram for liquid
Hg-Au alloys, with nominal Au concentrations from 0--0.20\% and at
temperatures between  $-28$ and $+25^\circ$C indicated by crosses
(+). $R/R_F$ curves are shown on semilog scales, all with
identical axes as marked for the 0.13\%, $-18^\circ$C alloy.
($\circ$): Reported \index{solubility limit} solubility limit of
Au in Hg. The shaded band is a guide for the eye. }
\label{hgau_fig:grid}
\end{figure}

Distinctly different reflectivities are obtained depending on whether
the Au concentration exceeds the solubility limit.
The solubility of Au in Hg is plotted as a function of temperature in
    Figure~\ref{hgau_fig:grid}
(open circles; the shaded band is a guide for the
    eye).\cite{gumin91,solubility}
Normalized reflectivity curves are shown for a selection of
temperature and concentration points, identified by small crosses
on the graph. All $R/R_F$ curves are shown on a log scale with
ranges identical to those shown for 0.13at\% Au, $-18^\circ$C.
This \index{phase diagram} phase diagram provides an overview of
our results, which will be discussed in more detail further below.

\begin{figure}[tbp]
\centering
\includegraphics[width=1.0\columnwidth]{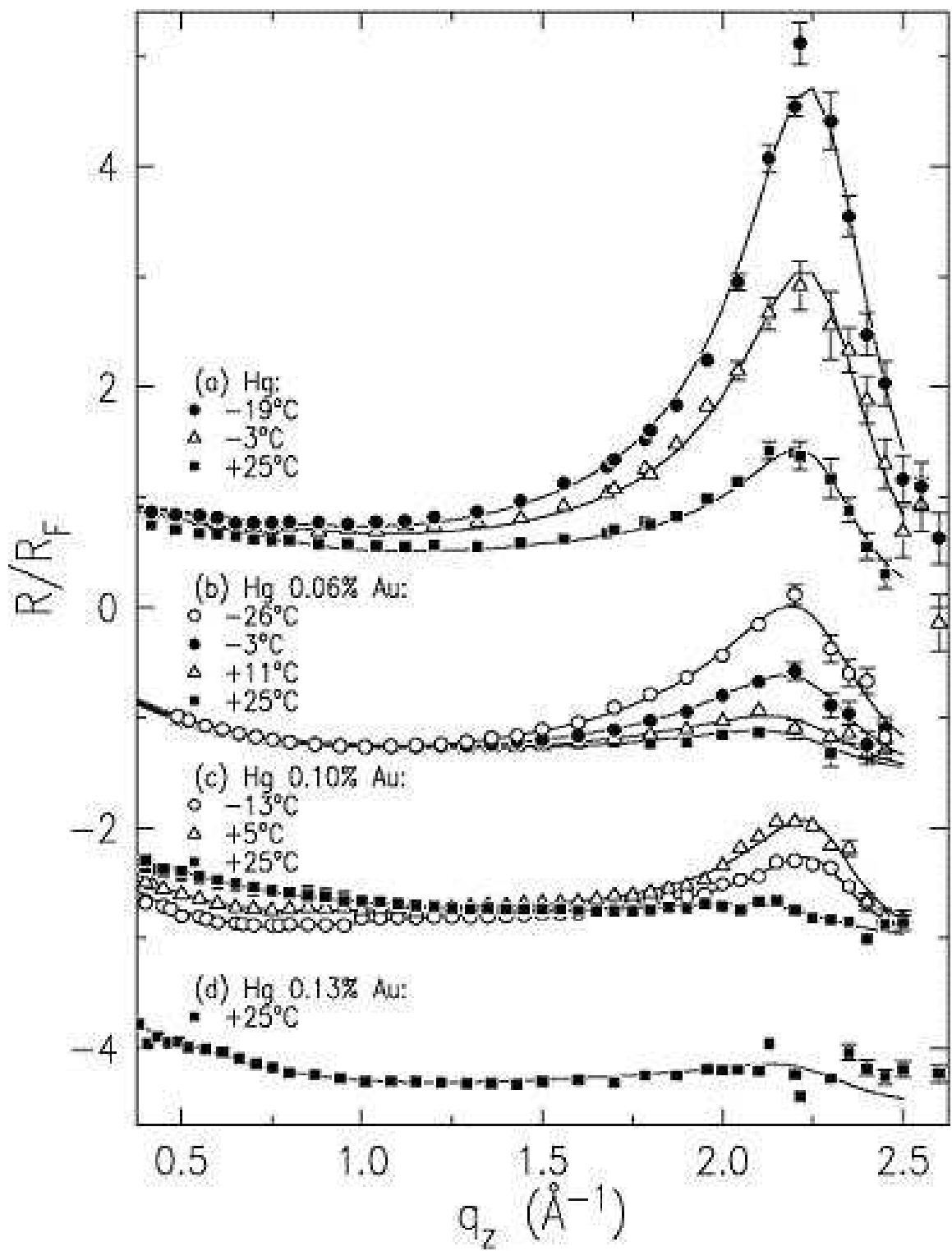}
\caption{ Fresnel-normalized reflectivity
of (a) Hg 0.06at\% Au and  (b) Hg 0.10at\% Au at all measured
temperatures, and  (c)  Hg 0.13at\% Au at room temperature. The
curves for 0.10at\% and 0.13at\% alloys are shifted for clarity.
Solid lines are fits as discussed in text. } \label{hgau_fig:poor}
\end{figure}

At low Au concentrations and high temperatures (upper right region of
    Figure~\ref{hgau_fig:grid}),
the reflectivity is characterized mainly by the layering peak
    at $q_z \approx 2.2~$\AA$^{-1}$.
These reflectivity data are shown in
    Figure~\ref{hgau_fig:poor}.
Solid lines are fit curves that will be described in the Discussion section.
At the lowest Au concentration, 0.06at\% Au, the reflectivity is very
similar to that of pure Hg.  As for Hg, the effect of increased temperature
is to monotonically reduce the amplitude of the interference peak,
from $+25^\circ$C down to the lowest temperature measured, $-26^\circ$C.
The data are consistent with scattering from a layered liquid metal surface,
where thermal excitations partially disrupt the layering.
For 0.10at\% Au in Hg, comparable behavior is observed at
$+25^\circ$C and $+5^\circ$C.
On cooling further to $-13^\circ$C, however, the amplitude
of the layering peak decreases, showing that surface layering is being
suppressed at lower temperatures.
The reflectivity in the region $q_z \sim 0.3$--1.0~\AA$^{-1}$ also
changes with $T$, unlike the case of the more dilute alloy.
This means that the structure within 2~\AA\ or so of the interface,
which for pure Hg and for Hg  0.06at\% Au is temperature independent,
is being modified by $T$ at the higher concentration of 0.10at\% Au.
The behavior of the 0.10at\% alloy in this temperature range
marks the transition between
viability of the surface layering, which is enhanced at lower temperatures,
and the formation of a competing surface structure.
Similarly, the reflectivity from Hg 0.13at\% Au can only be described
by a simple surface layering model at room temperature, and at lower
temperatures is substantially different.

\begin{figure}[tbp]
\centering
\includegraphics[width=1.0\columnwidth]{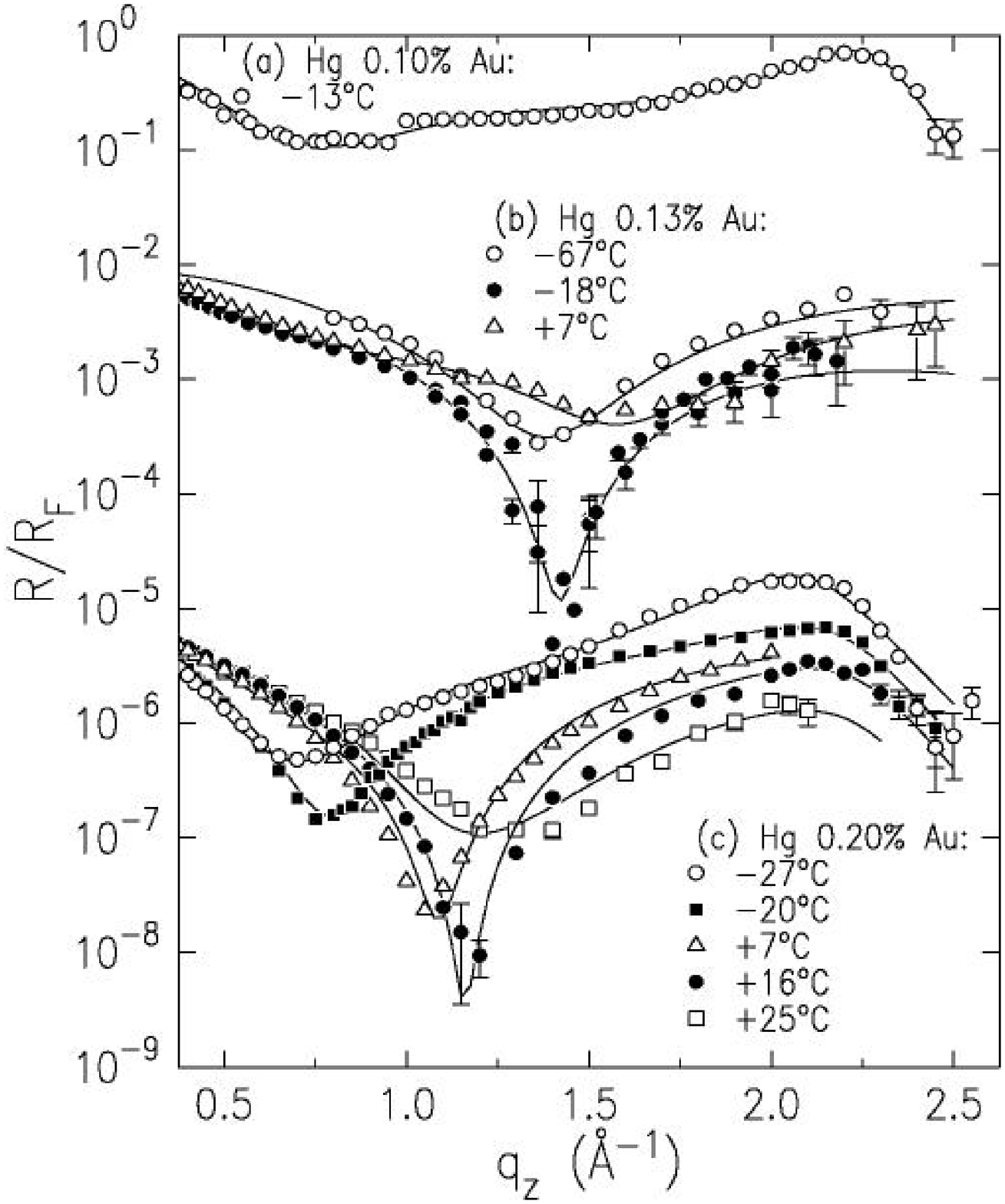}
     \caption{ Fresnel-normalized reflectivity
of (a) Hg 0.10at\% Au at $-13^\circ$C, (b)  Hg 0.13at\% Au below
room temperature, and (c) Hg 0.20at\% Au for the first cooldown
from room temperature. The curves for 0.13at\% and 0.20at\% alloys
are shifted by factors of $10^{-2}$ and $10^{-5}$ respectively.
Solid lines are fits as discussed in text. } \label{hgau_fig:rich}
\end{figure}

Data from the low temperature, Au-rich phase (bottom left region in
    Figure~\ref{hgau_fig:grid})
are qualitatively different.
For 0.13at\% Au in Hg, below room temperature, the reflectivity
exhibits a deep minimum, which shifts with decreasing temperature, from
$q_z \approx 1.6$ at $+7^\circ$C to
$q_z \approx 1.4$~\AA$^{-1}$ at $-67^\circ$C
    (Figure~\ref{hgau_fig:rich}).
The position and depth of this minimum suggest the formation of a
surface region having about half the density of the bulk, which is
about one atomic diameter thick and which grows slightly thicker
at lower temperatures. This trend with temperature is confirmed
more dramatically in reflectivity measurements of 0.20at\% Au in
Hg, where the Au concentration exceeds the \index{solubility
limit}  solubility limit at all temperatures. The notch in the
reflectivity is present at room temperature,  and shifts in $q_z$
from 1.3~\AA$^{-1}$ at room temperature to 0.6~\AA$^{-1}$ at
$-27^\circ$C
    (Figure~\ref{hgau_fig:rich}(c)).

\begin{figure}[tbp]
\centering
\includegraphics[width=1.0\columnwidth]{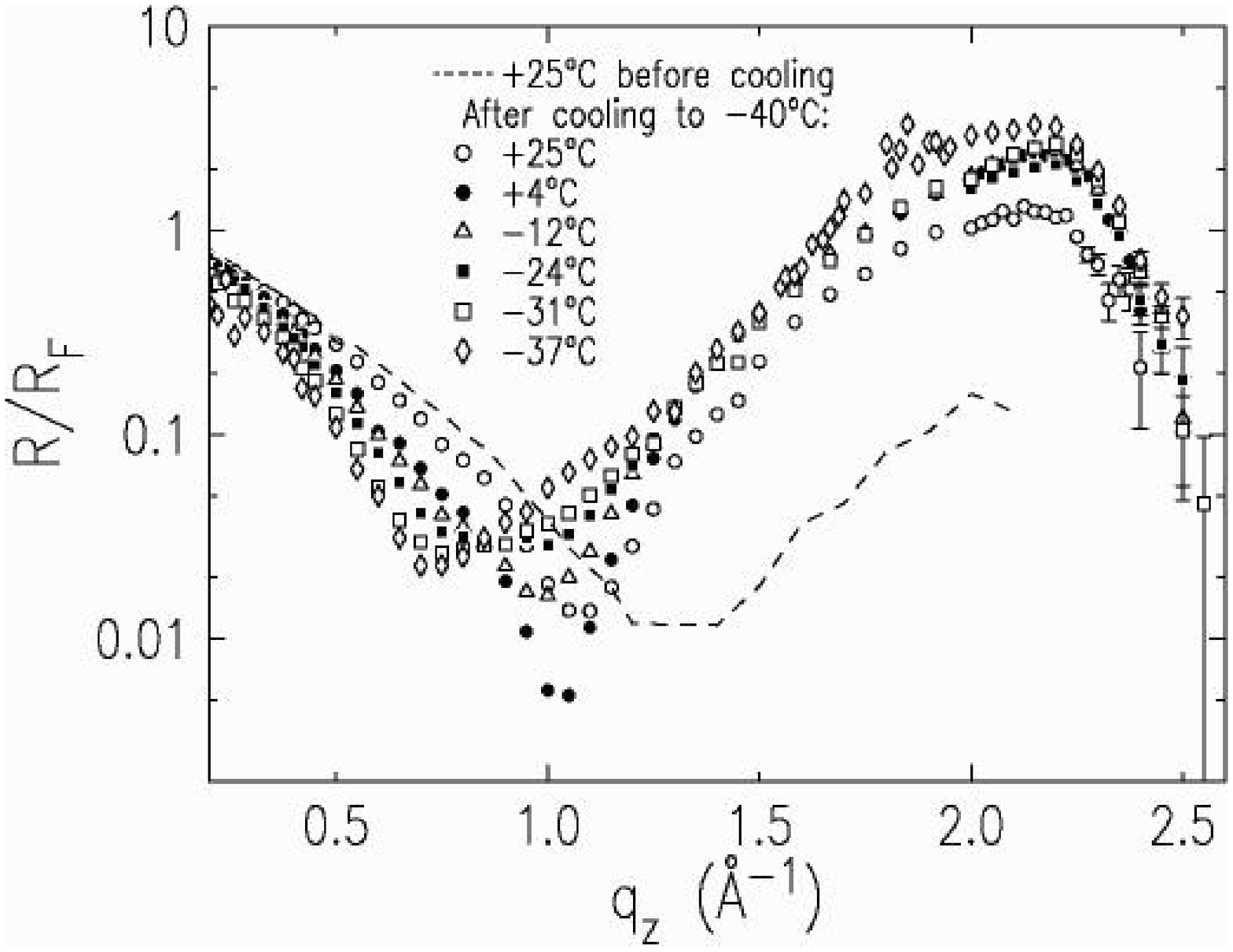}
\caption{ Fresnel-normalized reflectivity for Hg 0.20 \%Au  for
the second cooldown from room temperature. $+4^\circ$C
($\bullet$), $-12^\circ$C ($\triangle$), $-24^\circ$C (\protect
\rule{1.4ex}{1.4ex}), $-31^\circ$C ($\Box$), $-37^\circ$C
($\diamond$), (- - -): $+25^\circ$C experimental data from first
cooldown. } \label{hgau_fig:20}
\end{figure}

Even more interesting, at the lowest temperatures a new layering peak has
appeared, at measurably smaller $q_z$ than that of Hg.
At these two highest Au concentrations, temperature cycling results in
hysteresis  and long equilibration times.
After cooling the 0.13at\% alloy to nearly the melting point, approximately
ten hours at room temperature were required to reproduce the reflectivity.
With 0.20at\% Au, the original room temperature reflectivity was never
recovered after cooldown, even after fifty hours at $+25^\circ$C.
Reflectivity data for the second cooldown of this alloy are shown in
    Figure~\ref{hgau_fig:20}.
The reflectivity curves for the second cooldown all exhibit the notch
at lower $q_z$ for each temperature than the first measurements, until the
melting point is approached.
Au expelled from the bulk,
presumably incorporated into the surface structure, apparently
can not readily resolubilize at these concentrations.

\section{Analysis and Discussion}

To our knowledge, this work is a unique application of surface x-ray
scattering to probe the bulk solubility behavior of a liquid alloy.
This is of particular importance, since the most reliable methods for
determining solubilities in dilute alloys are electroanalytical techniques.
These techniques are not applicable to alloys such as Au in Hg, where the
solute is more noble than the
    solvent,\cite{gumin91}
unless they are modified by probing the formation of intermetallic
compounds with a second added solute such as Zn. For the case of
Hg-Au, where formation of a surface phase signals that the
\index{solubility limit} solubility limit has been exceeded,
atomic-scale surface structural measurements may be of special
value.

By fitting reflectivity curves calculated from model
surface-normal density profiles to the data, we can obtain more
specific information about the surface structure of these alloys.
Nearly all of the experimental reflectivity curves exhibit an
interference peak along with some additional structure at lower
$q_z$. We have fit these measurements to an oscillatory model
profile with an additional \index{gaussian} gaussian term
positioned near the liquid--vapor interface, as described by
    Eq.~\ref{hgau_eq:model}.
Since this is the parameterization used previously in studies of pure
liquid Hg, we can compare the pure and alloyed systems quantitatively.
The layer spacing $d$, the thermal roughness $\sigma _T$ and the decay
parameter $\overline{\sigma}$ are parameters which characterize the
layered subsurface structure and are determined well by the fits.
Other parameters affect the structure within about 2~\AA\ of the interface,
and are somewhat less reliably determined.
Some reflectivity data, lacking a prominent layering peak,
have been fit with a simpler form consisting of an error
function profile along with an extra gaussian term, as in
    Eq.~\ref{hgau_eq:lump}.
Because of the low Au concentrations, the nearly equal scattering
factors of Hg and Au, and the limited $q_z$ range, we will treat
our model profiles in this study as though they were
entirely composed of Hg.  All models incorporate the
bulk density and atomic form factor of Hg, and therefore represent
an average surface-normal electron density without providing any
direct information about the Au composition as a function of depth.

\subsection{Dilute Au limit and Capillary Wave Roughness}

\begin{figure}[tbp]
\centering
\includegraphics[width=1.0\columnwidth]{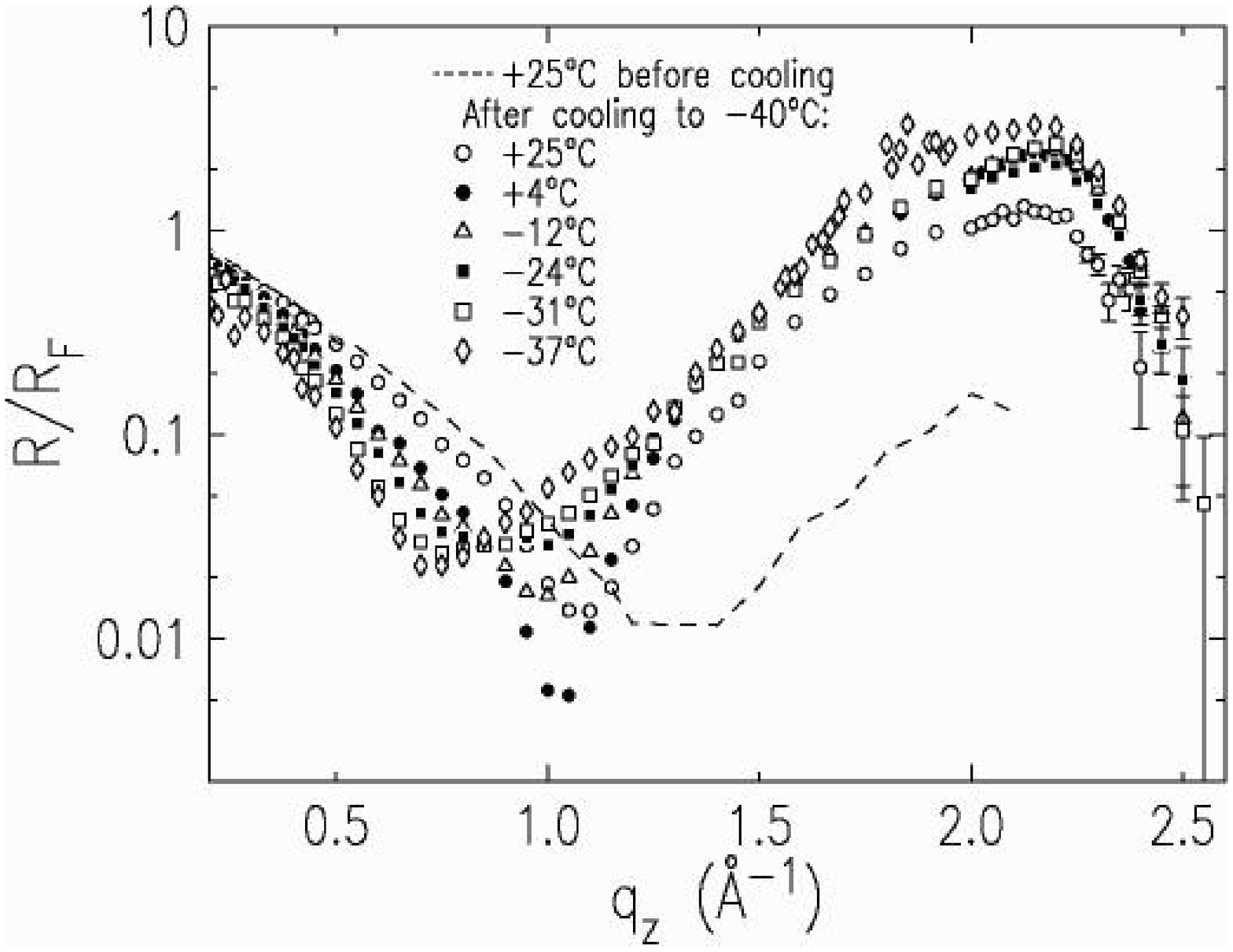}
 \caption{ Model density profiles calculated
from parameters listed in Table~1. Solid lines are from room
temperature fits, and shorter dash length corresponds to lower
temperatures. (a) Hg (b) Hg 0.06\% Au (c) Hg 0.10\% Au Inset:
temperature dependence of the excess mean-squared surface
roughness $(\sigma _T ^2 - \sigma_{cw} ^2)$ for Hg (\protect
\rule{1.2ex}{1.2ex}), Hg 0.06at\% Au ($\circ$), and Hg 0.10at\% Au
(closed triangles). } \label{hgau_fig:rho}
\end{figure}

At low Au concentrations, the main effect of the Au is to partially disrupt
the layered profile near the surface, compared to the structure of Hg.
Model Hg profiles, from studies reported
    previously,\cite{dimasi98}
are shown in
    Figure~\ref{hgau_fig:rho}(a).
The oscillatory profile has a layer spacing $d=2.72$~\AA ,
a decay parameter $\overline{\sigma} = 0.46$~\AA ,
and a temperature dependent roughness $\sigma _T$, given in
    Table~\ref{hgau_tab:params},
that determines the amplitude of the layers in the density profile.
The model also incorporates a slight tail of density extending towards
the vapor, described by the term proportional to $w_A$ in
    Eq.~\ref{hgau_eq:model}.

\begin{table}[tcp]

\begin{tabular}{ccccccccc}
ppm  & $T$ ($^\circ$C) & $\sigma_T$ (\AA)    & $\overline{\sigma}$
(\AA)
    & $w_1$     & $w_0$
    & $w_A$     & $\sigma_A$ (\AA)  & $z_A$ (\AA)   \\ \hline
0   & $+25$     & 1.00$\pm$0.01 & 0.46$\pm$0.05
    & 1.0       & 1.0
    & 0.2$\pm$0.1   & 1.5$\pm$0.5   & $-2$$\pm$1        \\
0   & $-3$      & 0.87$\pm$0.01 & 0.46$\pm$0.05
    & 1.0       & 1.0
    & 0.2$\pm$0.1   & 1.5$\pm$0.5   & $-2$$\pm$1        \\
0   & $-19$     & 0.82$\pm$0.01 & 0.46$\pm$0.05
    & 1.0       & 1.0
    & 0.2$\pm$0.1   & 1.5$\pm$0.5   & $-2$$\pm$1        \\
0   & $-36$     & 0.80$\pm$0.01 & 0.46$\pm$0.05
    & 1.0       & 1.0
    & 0.2$\pm$0.1   & 1.5$\pm$0.5   & $-2$$\pm$1        \\ \hline
6    & $+25$     & 1.04$\pm$0.01 & 0.54$\pm$0.04
    & 0.95  & 0.80
    & 0.24$\pm$0.10 & 1.6$\pm$0.5   & $-2$$\pm$1        \\
6    & $+11$     & 1.02$\pm$0.01 & 0.53$\pm$0.02
    & 0.96  & 0.78
    & 0.24$\pm$0.10 & 1.6$\pm$0.5   & $-2$$\pm$1        \\
6    & $-3$      & 0.97$\pm$0.01 & 0.50$\pm$0.02
    & 0.97  & 0.78
    & 0.24$\pm$0.10 & 1.6$\pm$0.5   & $-2$$\pm$1        \\
6    & $-26$     & 0.91$\pm$0.01 & 0.50$\pm$0.02
    & 0.98  & 0.78
    & 0.24$\pm$0.10 & 1.6$\pm$0.5   & $-2$$\pm$1        \\ \hline
10    & $+25$     & 1.07$\pm$0.02 & 0.52$\pm$0.02
    & 0.96  & 0.84
    & 0.1$\pm$0.1   & 2$\pm$1   & $-2.6$$\pm$0.5    \\
10    & $+5$      & 0.99$\pm$0.01 & 0.44$\pm$0.02
    & 0.96 & 0.84
    & 0.4$\pm$0.2   & 2$\pm$1   & $-2.6$$\pm$0.5    \\
10    & $-13$     & 1.02$\pm$0.01 & 0.43$\pm$0.02
    & 0.96  & 0.78
    & 0.5$\pm$0.2   & 2$\pm$1   & $-2.6$$\pm$0.5
\end{tabular}
\caption{Parameters for layered density profiles having the form
given in  Eq.~\ref{hgau_eq:model}, where $\sigma _n ^2 = n
\overline{\sigma} ^2 + \sigma _T ^2$ and $w_n=1$ except as
indicated in the table. Data for pure Hg, measured in the same UHV
chamber, are from Ref.~12. For all fits, $d=2.72\pm0.02$~\AA . }
\label{hgau_tab:params}
\end{table}

The surface-normal profile of Hg 0.06at\% Au has the same layer spacing
of 2.72~\AA , but the layering is less well defined
    (Figure~\ref{hgau_fig:rho}(b)).
For comparable
temperatures, the overall roughness $\sigma _T$ is about
0.1~\AA\ greater  than that of Hg, and layering decays over a slightly
shorter length scale for the alloy.
A notable difference between the structures of Hg and Hg 0.06at\% Au is that
for the alloy, the amplitude of the topmost surface layer is decreased
by a factor of $w_0=0.8$ compared to pure Hg, where $w_0=1$.
The reduction of the second layer amplitude by $w_1 \approx 0.95$--0.98
has a much smaller effect on the calculated reflectivity.
As for Hg,
the 0.06at\% alloy data are consistent with additional density in the
near surface region,
described by the Gaussian term proportional to $w_A$.
This is in contrast to the sharper surface profiles, such as that
illustrated by the solid line in
    Figure~\ref{hgau_fig:model}(a)iv
(having $w_A=0$) that were found to describe liquid Ga and In.

The effect of temperature on the surface structure of Hg 0.06\% Au is
very similar to that of Hg.
While slight variations in several parameters were required to optimize
the fits  at different $T$ for the 0.06\% alloy,
it is apparent that the main effect of
increased temperature is to reduce the amplitude of the density oscillations.
Capillary wave theory, which describes the broadening of the surface profile
due to thermally excited surface waves, gives a specific prediction for such
a mechanism through the temperature dependent roughness $\sigma_{cw}$
defined in
    Eq.~\ref{hgau_eq:cw}.
Previous studies of the temperature dependent surface roughness of liquid
Hg found that thermally excited surface waves were not able to account for
all of the temperature dependence
    observed.\cite{dimasi98}
This is in contrast with the surface roughnesses of liquid Ga and In,
for which capillary wave theory accounted for all measured temperature
    dependence.\cite{regan96,tostmann99}
The temperature dependent reflectivity of Hg 0.06\% Au exhibits
essentially the same behavior as that of Hg.
In the inset of
    Figure~\ref{hgau_fig:rho}
we compare the excess roughness
$\sigma _T ^2 - \sigma _{\mbox{\scriptsize cw}} ^2$ of Hg with that
of Hg 0.06at\% Au.
The 0.06at\% alloy has a larger roughness, but the slope of the
excess roughness versus $T$ is the same.
By alloying Hg with this small amount of Au, it appears that
the total surface roughness has increased,  without changing its
temperature dependence.

The fact that the deviation from capillary wave behavior is the same
for elemental Hg as for the 0.06at\% alloy, which have distinctly
different intrinsic surface-normal structure, suggests that the deviation
from capillary wave theory may not arise from a temperature dependence
of the local profile.
Instead, the apparently anomalous temperature dependence most likely arises
from height fluctuations across the surface which are not accounted for
properly by the capillary wave model.
In our analysis, thermal capillary waves are dependent on a surface tension
which is defined as a macroscopic quantity, despite that fact that we integrate
over capillary wave modes extending down to atomic length scales.
It is very likely that the concept of a constant surface tension is not
applicable to these microscopic length scales.
The success of capillary wave theory in the description of other liquids
may stem from the fact that the modes at atomic length scales make up only
a small part of the entire capillary wave spectrum, and hence the
contribution of these modes in affecting the overall roughness
may differ from system to system.
Comparison of diffuse surface scattering from Hg and Hg-Au alloys may
provide more information on this question.

The Au concentration of 0.10at\% marks the transition between the dilute-Au
regime, where a simple layered structure previals, and the Au-rich phase.
Here the more pronounced structure at low $q_z$ and the attenuated layering
peak complicate the modelling, and the analysis is more ambiguous.
For the layered model given, acceptable
parameters lie in the ranges indicated in
    Table~\ref{hgau_tab:params}.
The corresponding fit curves are shown in
    Figure~\ref{hgau_fig:poor}(b) (solid lines).
Relative to Hg, the 0.10at\% alloy model density profiles
exhibit a reduction in oscillation amplitude and surface layer density
very similar to that found in the 0.06at\% alloy, as shown in
    Figure~\ref{hgau_fig:rho}(c).
Unlike the more dilute alloy, however, a more fundamental change
in surface structure occurs as temperature is reduced beyond the
\index{solubility limit} solubility limit. Upon cooling from room
temperature, the surface roughness first decreases, then increases
again, so that the layering peak is less pronounced at
$-13^\circ$C than it was at $+5^\circ$C. The resulting
non-monotonic dependence of $\sigma _T$ on $T$ is shown in the
inset of
    Figure~\ref{hgau_fig:rho} (closed triangles).

In addition, as $T$ decreases there is a continuous drop in the reflectivity
at low $q_z$.
This indicates a further decrease in the surface layer density, an enhanced
density bump further towards the vapor region,
or a combination of such effects.
The density profiles shown in
    Figure~\ref{hgau_fig:rho}(c)
model this effect as a lump of density
extending into the vapor region.
The density tail is independent of temperature for Hg and Hg 0.06\% Au,
but for Hg 0.10\% Au climbs steeply as temperature decreases.
In this concentration and temperature range, the effect of
decreased temperature is not
a reduction in surface roughness but the establishment of a low-density
surface region that destabilizes surface layering.

\subsection{Au-rich limit and Surface Phase Formation}

\begin{figure}[tbp]\epsfig{file=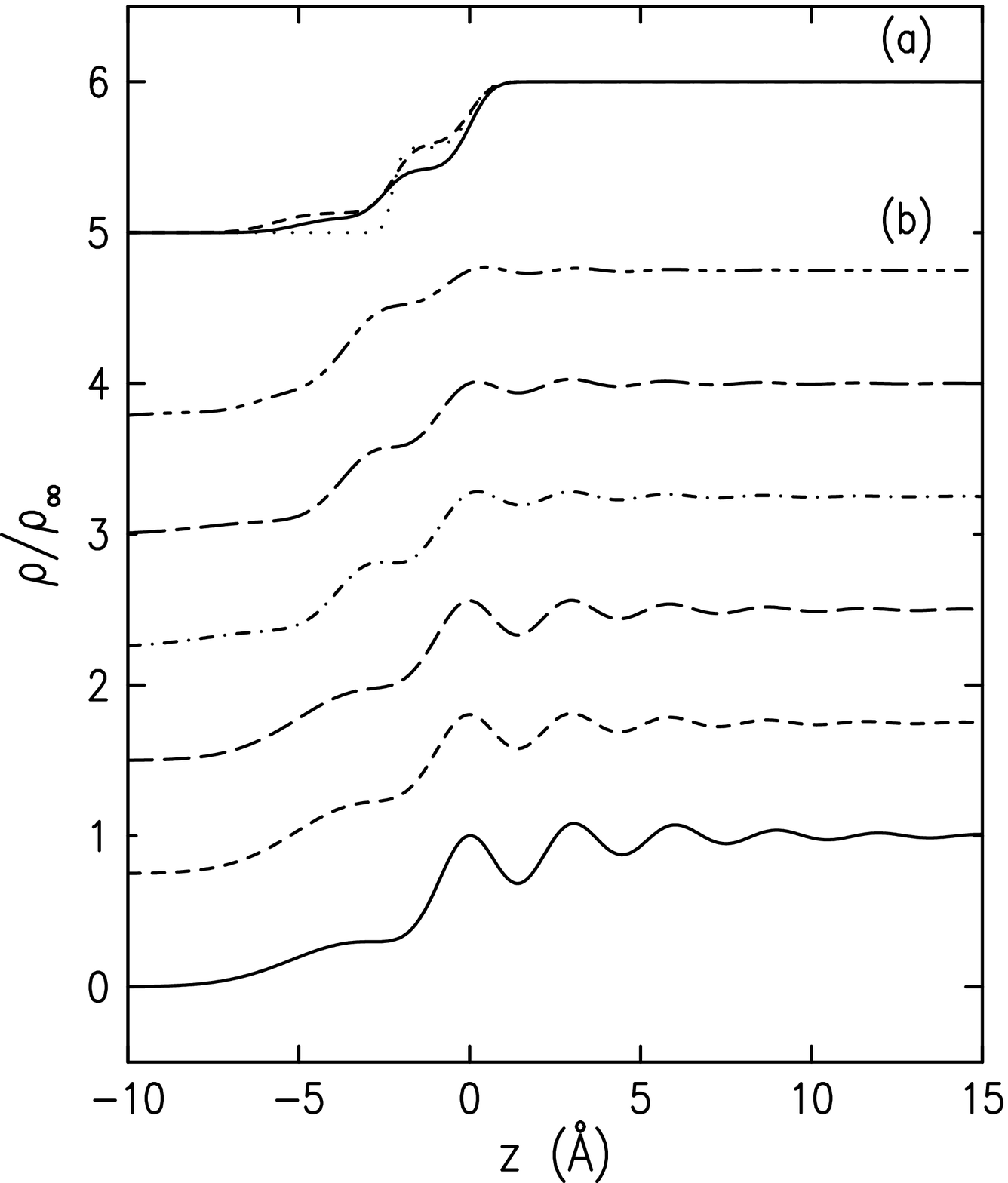, angle=0,
width=1.0\columnwidth}
 \caption{ Model density profiles. (a) Hg
0.13\% Au: $+7^\circ$C (---), $-18^\circ$C (- - -), $-67^\circ$C
($\cdots$). (b) Hg 0.20\% Au: $+25^\circ$C (--- - - ---),
$+16^\circ$C (--- - ---), $+8^\circ$C (- $\cdot$ - $\cdot$ -),
$-20^\circ$C (--- ---), $-24^\circ$C (- - -), $-27^\circ$C (---).
} \label{hgau_fig:rho20}
\end{figure}

The 0.13at\% Au alloy, near the room temperature saturation limit,
could be modelled with a layered density profile only
for our measurement at $+25^\circ$C.
Reflectivity data at lower temperatures exhibit a minimum at
$q_z \sim 1.4$--1.6~\AA$^{-1}$,
along with a slight dip at
$q_z \sim 0.5$~\AA$^{-1}$.
These features can be fit with very simple shelf-type models, as shown in
    Figure~\ref{hgau_fig:rho20}(a).
The relatively high reflectivity at the highest $q_z$ requires that
these profiles have step edges that, in comparison to roughnesses
encountered in other liquid metals and alloys, seem unphysically sharp.
While this low surface roughness is a possibility,
a much more likely explanation of the reflectivity  is that some
surface layering prevails in the 0.13at\% alloy, too weak to produce a
layering peak but substantial enough to produce constructive interference
at high $q_z$ and augment the reflectivity which would otherwise be
reduced in the washed-out step profile.
For this reason, we think that the very simple shelf profiles are likely
not the best description of the reflectivity, and models such as those
presented below for the 0.20at\% alloy may be more realistic.
Despite this, attempts to fit the 0.13at\% alloy with simple layered models
were unsuccessful, resulting in unphysically small layer spacings of
$d \approx 2.1$~\AA\ to position the reflectivity minimum properly.

The surface roughness in the step-type profile was especially low
for the Hg 0.13at\% Au
sample supercooled to $-67^\circ$C, the lowest temperature obtainable in
our chamber.
The 13\% Au  alloy is the only one for which we observed deep
supercooling: all other alloys investigated froze at temperatures
close to the melting point of pure Hg, $-38.9^\circ$C.
We cannot tell whether the
supercooling is particular to the 0.13\% Au composition, or is due instead
to an especially clean sample environment.
It has been suggested before that supercooling at an ordered liquid metal
surface may  be enhanced due to the reduction of the difference in
entropy between the liquid and solid phases relative to a disordered
    surface.\cite{turnbull52}
However, without more systematic studies of supercooled samples it is not
possible to attach much significance to our observation.

The concentration of 0.20at\% Au in Hg exceeds the
\index{solubility limit} solubility limit for all temperatures
investigated.  Reflectivity data for this alloy exhibit minima at
all $T$, along with a layering peak at the lowest $T$. We have fit
these data to the simple layered profiles shown in
    Figure~\ref{hgau_fig:rho20}(b).
To keep the number of parameters relatively small, we have used the form of
    Eq.~\ref{hgau_eq:model}
and fit the same number of paramters as were used for the dilute alloys.
At the lowest temperatures, the surface layer spacing of 2.9~\AA\
for Hg 0.20at\% Au
is unambiguously distinguished from the 2.72~\AA\ layer spacing obtained
for pure Hg and the dilute Hg-Au alloys.
Since the atomic radius of Au (2.8~\AA ) is smaller than that of Hg
    (3.0~\AA ),\cite{iida}
this increase in the layer spacing for the Au-rich alloy indicates
that some significant change in packing has occurred in the
surface layers. At higher temperatures, evidence for layering lies
mainly in the relatively large reflectivity obtained for $q_z \geq
2$~\AA$^{-1}$. Here the fits are less satisfactory, and the
details of the models show no clear systematic dependence on
temperature.  It is difficult in this region of the \index{phase
diagram} phase diagram to assess the relationship between the
details of the surface profile and the underlying layering. While
the oscillations in the density profile clearly grow as
temperature is reduced, it is not possible to say to what extent
this is due to reduction of surface fluctuations rather than to
changes in Au content at the surface. Measurement techniques with
elemental specificity, such as \index{x-ray!resonant scattering}
resonant x-ray scattering or surface \index{x-ray!absorption fine
structure} x-ray absorption fine structure, will be required to
obtain this chemical information. Because of the high vapor
pressure of liquid Hg, \index{Auger electron spectroscopy} Auger
electron spectroscopy will probably not be possible.

All model profiles for the Au-rich phase incorporate a shelf feature or
surface layer that has a density of about half that of the bulk; we were
unable to model the notch in the reflectivity without such a feature.
Presumably, some fraction of the Au which is insoluble in the bulk is
incorporated into this surface phase, rather than precipitating out of
the solution in solid form as might have been expected.
It is difficult to explain how a monolayer composed entirely of Hg and/or Au
might achieve this low density.
All known Hg-Au phases are close-packed,
with densities intermediate between those of elemental Au and
    Hg.\cite{rolfe67}
A uniform submonolayer of Au with a greatly expanded lateral
spacing is one model which  could produce the observed density
profile.  Typically, expanded liquids are created at pressures and
temperatures high enough to approach the critical point. In this
case, the putative liquid Au monolayer would have to be stabilized
at temperatures far below the melting point of bulk Au.  Such an
expanded Au layer would have a very high surface corrugation, and
the tendency of the surface tension to reduce the surface area
would be expected to destabilize such a structure. While it may
not be clear how to extrapolate quantities such as the surface
tension of Au down to these temperatures, the macroscopic contact
angles and \index{curvature} curvature of our samples certainly
suggest that the surface tension is comparable to that of pure Hg.
One possibile reason for an expanded spacing is that large changes
in the concentration of Au, which has a lower valence than Hg,
could reduce the bandfilling in the surface layers. If this serves
to greatly reduce the favored atomic coordination, effectively
increasing the covalency, the surface atoms might be forced to
adopt a more open structure. This picture is also consistent with
the expansion of the surface-normal layer spacing in the Au-rich
phase. Previous studies of liquid  Ga and In surfaces  showed that
compared to the more covalent Ga, the free-electron-like In had a
more compressed surface layer spacing, indicating a structure
closer to hard sphere
    packing.\cite{tostmann99}


Another possible explanation is that patches of more closely packed
Au-rich monolayers
float on the surface, comprising a coverage of about 50\%.  We believe that
several points argue against this model.
Liquid Hg has been observed to amalgamate continuously with Au
    substrates,\cite{abruna97}
suggesting that it would be difficult to maintain a laterally inhomogeneous
distribution of Au at the surface.  Another objection is that if Au were
precipitating out of solution and into a separate surface monolayer,
we would expect a more monotonic change in coverage with temperature or
concentration.  Instead, the density seems to drop quickly to 50\% of the
bulk value and remain at that value.
Finally, we find no evidence that the surface layer is solid.
The surface profile is not particularly sharp, nor is there any sign of
the long-range lateral order that would be expected for a crystalline
    monolayer.\cite{gid}

Studies of underpotential deposition of Hg on Au found that surface
amalgams with open packing always incorporated coadsorbed anions from the
    electrolyte.\cite{abruna98}
In the liquid Hg-Au alloys studied here, it is also possible that
anions, perhaps formed from impurities dissolved within the Hg,
may be incorporated into the outermost atomic layer of the liquid
metal, producing the low-density surface region we observe.
However, it is important to emphasize that such impurities would
also have been present in the pure Hg samples studied previously.
While the pure Hg reflectivity provides evidence for a
sample-dependent tail of density near the surface, as discussed
above, nothing like the shelf feature, or layer having half the
bulk density, was ever observed. In the case of the Au-rich liquid
alloys presented here, the shelf feature depends entirely on the
presence of Au at the surface: only at temperatures and
concentrations beyond the \index{solubility limit} solubility
limit does this structure appear. In contrast to pure Hg, the
Au-rich amalgam surface may be capable of forming a passivated
oxide, even though elemental Au is quite nonreactive, and bulk Hg
oxide is unstable under the low oxygen partial pressure. This
result may be relevant to reactions catalyzed by liquid metal
surfaces. Studies of the rate of formic acid decomposition over
liquid Hg having different concentrations of dissolved metals have
found that the addition of higher-valence metals decreases the
activation energy of the
    reaction.\cite{catalysis}
This was interpreted as indicating that electrons are localized in covalent
bonds at the alloy surface, essentially a statement that the primary
effect of alloying was electronic.
Our observations on the Hg-Au alloys indicate that the formation of
intermetallic phases at the surface may be much more important than
previously thought, especially since the surface composition can differ so
substantially from that of the bulk.
At the same time, the surface induced layering intrinsic to all liquid metals
may be less important for surface reactions, except for the extent to which
the layering competes with the formation of the reactive surface phase.

%% file: biin_all.tex
\chapter{Resonant x-ray study of liquid Bi-In
surface}

 \section{Abstract}
\index{x-ray!resonant scattering} Resonant x-ray reflectivity
measurements from the surface of liquid Bi$_{22}$In$_{78}$ find
only a modest surface Bi enhancement, with 35~at\% Bi in the first
atomic layer. This is in contrast to the \index{Gibbs!adsorption}
Gibbs adsorption in all liquid alloys studied to date, which show
surface segregation of a complete monolayer of the low surface
tension component. This suggests that surface adsorption in Bi-In
is dominated by attractive interactions that increase the number
of Bi-In neighbors at the surface. These are the first
measurements in which resonant x-ray scattering has been used to
quantify compositional changes induced at a liquid alloy surface.

 \section{Introduction}

Current treatments of the thermodynamics of surface phenomena in
solutions rely heavily on the original works by Gibbs in 1878, and
one of the most familiar corollaries is the
\index{Gibbs!adsorption} Gibbs adsorption rule. In its simplest
invocation, the Gibbs rule states that in a binary liquid, the
species having the lower surface tension will segregate
preferentially at the surface. This apparent simplicity is
deceptive: a survey of the literature reveals a hundred years'
debate over the application of the Gibbs adsorption
    rule,\cite{cahn58} 
not to mention its extension to multicomponent
    systems\cite{widom79}
and crystalline
    surfaces,\cite{rusanov96} 
and its connection to atomistic
    models.\cite{speiser87}

Experimental investigations of the validity of the Gibbs rule encompass
measurements of adsorption
    isotherms,\cite{adamson67}
    surface tension,\cite{bhatia78} 
and surface
    composition\cite{hardy82} 
in a variety of systems. Unfortunately, many of the liquids
studied are too complicated for the simplest formulations of the
Gibbs rule. Liquid metal alloys are in many ways ideal for such
studies. Miscible alloys exist which behave as ideal liquids,
while in other systems strongly attractive or repulsive
\index{interactions!heteroatomic} heteroatomic interactions can be
studied. Perhaps an even more important advantage of liquid metals
is that the compositionally inhomogeneous region at the surface is
known in some cases to be confined to an atomic layer. This is
commonly assumed in calculations of Gibbs adsorption that take a
model of a physical surface as their starting point.

For example, x-ray reflectivity, ion scattering, and \index{Auger
electron spectroscopy} Auger electron spectroscopy measurements of
liquid Ga$_{84}$In$_{16}$ found a 94\% In surface monolayer, as
expected given this alloy's positive heat of
    mixing.\cite{regan97,dumke83}
Subsequent layers have the bulk composition.
Similar studies of dilute liquid Bi-Ga ($<0.2$~at\% Bi)
likewise found surface segregation of a pure Bi
    monolayer.\cite{lei96}
Even when the repulsive interactions between Ga and Bi cause more
Bi-rich alloys to undergo additional \index{phase separation}
phase separation above 220$^{\circ}$C, where a 65~\AA\ thick
inhomogeneous Bi-rich region forms, the pure Bi surface monolayer
    persists.\cite{srn99,gabi00}
In Ga-Bi, then, repulsive \index{interactions!heteroatomic}
heteroatomic interactions substantially change the surface
composition profile, but do not defeat the Gibbs adsorption.

The effect of attractive heteroatomic interactions remains an open question.
In alloys such as Bi-In, attractive forces
between the two species produce a number of compositionally ordered phases
in the bulk solid.
It is therefore conceivable that Bi-In pairing may exist at the liquid
surface, and compete with surface segregation.
Our recent temperature dependent x-ray reflectivity measurements
of liquid Bi-In alloys having 22, 33, and 50~at\% Bi
revealed structural features not found in
elemental metals or in the Ga alloys discussed
    above.\cite{ben00}
As we will show, those data were suggestive of Bi-In pair formation along
the surface-normal direction.  However, since the technique did not measure
the Bi surface concentration directly, other interpretations of the data
were also possible.

 \section{Experimental Setup}
A complete characterization of surface composition requires both
elemental specificity and {\AA}-scale structural resolution along
the surface-normal direction, which is difficult to achieve
experimentally. \index{Auger electron spectroscopy}Auger electron
spectroscopy, which satisfies the first of these requirements, is
hampered by contributions from the bulk liquid.\cite{hardy82}
X-ray reflectivity by contrast is a surface-sensitive probe. In
the kinematic
    limit\cite{born}
the reflected intensity, measured as a function of momentum transfer
$q_z$ normal to the surface, is
proportional to the Fresnel reflectivity $R_F$ of a homogeneous
    surface:\cite{refl}
\begin{equation}
\label{biin_eq:refl}
    R(q_z) = R_F \left| (1/\rho_\infty)
        \int_{-\infty}^\infty
        (\partial \rho_{\mbox{\scriptsize eff}} / \partial z)
        \exp(i q_z z) \, dz \right|^2 \: .
\end{equation}
Here $\rho_{\mbox{\scriptsize eff}}$ represents an effective
electron scattering amplitude that combines the electron density
profile with the scattering \index{form factor} form factor, and
$\rho_\infty$ is the density of the bulk. The electron density
variations that produce modulations in the reflectivity may result
from changes in either the composition or the mass density. Thus,
inference of surface composition from the measured reflectivity is
sometimes ambiguous.

 \section{Resonant X-ray Scattering}
This disadvantage can be overcome with the application of resonant
x-ray scattering. The effective electron density of a scattering
atom depends on the scattering form factor $f(q)+f^{\prime}(q,E)
\approx Z+f^{\prime}(E)$. When the x-ray energy is tuned to an
\index{x-ray!absorption edge} absorption edge of a scattering
atom, the magnitude of $f^{\prime}$ becomes appreciable, producing
changes in contrast between unlike
    atoms.\cite{resonant}
With one exception,\cite{hgau00} \index{x-ray!resonant scattering}
resonant x-ray scattering measurements reported in the past have
been confined to studies of solids and bulk liquids, due to the
difficulty of the experiment. The present report is the first to
find compositional changes induced at a liquid surface.

\begin{figure}[tbp]
\centering
\includegraphics[width=1.0\columnwidth]{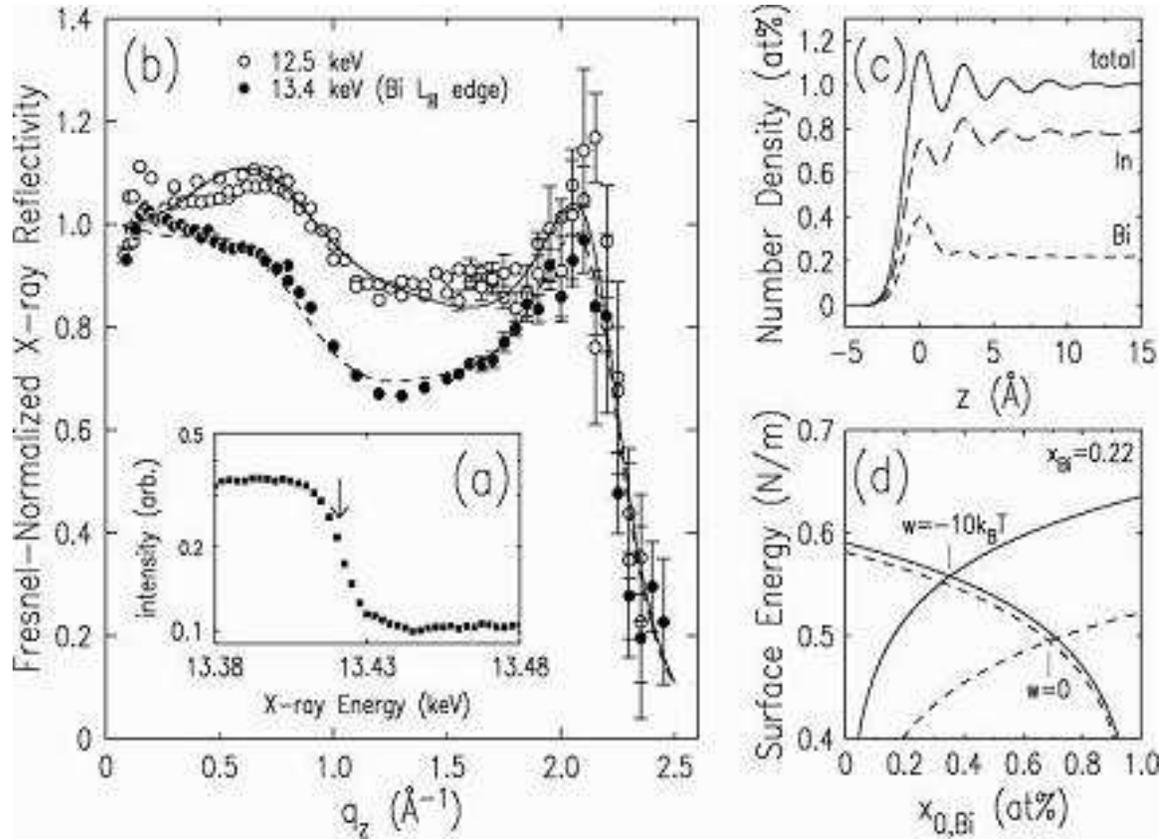}
\caption{ (a) Energy scan in transmission through Bi foil. (b)
Fresnel-normalized x-ray reflectivity of Bi$_{22}$In$_{78}$.
$\circ$: 12.5 keV (two independent measurements); $\bullet$: 13.4
keV; (---): 12.5 keV fit; (- - -) 13.4 keV fit. (c) Best fit real
space number density profile relative to bulk atomic percent.
(---): Total number density; (-- --): In density; (- - -): Bi
density. (d) Surface energy as a function of surface Bi
concentration $x_{0,\mbox{\scriptsize Bi}}$ for the bulk
concentration $x_{\mbox{\scriptsize Bi}} = 0.22$, according to
Eq.~\ref{biin_eq:refl}. (- - -): $w=0$ .  (---): $ w = - 10 k_B
T$. } \label{biin_fig:1}
\end{figure}

 \section{Sample Preparation}
The molten Bi$_{22}$In$_{78}$ sample was maintained at $T=
101^\circ$C and $P = 5 \times 10^{-10}$~Torr within an ultra high
vacuum chamber, and periodically sputter cleaned with Ar$^+$ ions.
Reflectivity measurements were performed at beamline X25 at the
National \index{synchrotron!source} Synchrotron Light Source. The
\index{spectrometer} spectrometer has been described
    previously,\cite{indium}
except that here a double Si(111) crystal monochromator
was used to provide an energy resolution of 9~eV.  For the present
work, we compare reflectivity measured at 12.5 keV to measurements made at
the Bi $L_{III}$ edge at 13.421 keV. The energy was calibrated by
transmission through a Bi foil, shown in
    Figure~\ref{biin_fig:1}(a).
At the inflection point indicated by the arrow,
$f^{\prime}_{\mbox{\scriptsize Bi}}$ has its largest magnitude of $-24.7$
electrons.
Uncertainties in $f^\prime_{\mbox{\scriptsize Bi}}$
may arise from inaccuracies in the calculation,
incorrect establishment of the incident energy, and the energy resolution.
To account for these possibilities, the analysis was performed for deviations
in $f^\prime_{\mbox{\scriptsize Bi}}$ of about
20\% (i.e., $f^\prime_{\mbox{\scriptsize Bi}}=-19.7$ and $-29.7$ electrons).
The results were incorporated into the error ranges tabulated below.

 \section{X-Ray Reflectivity Data}
Reflectivity data at both energies are shown in
    Figure~\ref{biin_fig:1}(b)
(symbols).  These results were found to be both reproducible in energy
and stable over time by measuring two 12~keV data sets, prior to and
following the 13.4~keV measurements.  The two 12~keV data sets are shown
together as open circles in
    Figure~\ref{biin_fig:1}(b).
The interference peak at $q_z = 2.1$~\AA$^{-1}$
is due to stratification of the atoms in planes parallel to the surface,
a well established feature of liquid
    metals.\cite{rice}
For $q_z < 1.8$~\AA$^{-1}$, the data exhibit a modulation
indicative of a structural periodicity roughly twice that of the
\index{surface!layering}   surface layering.  This feature is
consistent with Bi-In dimers oriented along the surface normal,
which could also give rise to alternating Bi and In layers
(bilayers) at the surface. We also find that the low-$q_z$
reflectivity is strongly decreased when measured at the Bi
$L_{III}$ edge. The reflectivity decrease itself  varies smoothly
with $q_z$ and does not exhibit a bilayer-type modulation, instead
suggesting that the surface Bi concentration is larger than that
of the bulk.

 \section{Electron Density Model}
To investigate these possibilities,
we calculate the reflectivity of a model density profile according to
    Eq.~\ref{biin_eq:refl},
which is refined simultaneously against the data taken at both
energies. \index{x-ray!resonant scattering} Resonant effects are
included by combining the model structure and the scattering
amplitude into an effective electron density profile
$\rho_{\mbox{\scriptsize eff}}(z)$. The energy dependence enters
the analysis through the Bi concentration defined in
$\rho_{\mbox{\scriptsize eff}}(z)$, and also through the Fresnel
reflectivity $R_F$, which is a function of the energy dependent
mass absorption coefficient $\mu^{-1}$ and the effective bulk
electron density $\rho_\infty \propto (Z-f^{\prime})$ that defines
the \index{critical!angle} critical angle $q_c$.
    Table~\ref{biin_table:pars}
shows the values of these quantities used in our models.
Reflectivity data acquired at each
energy were normalized to the appropriate energy dependent Fresnel function.

\begin{table}[tcp]

\begin{tabular}{ccccccccc}
    & E  & $\mu / \rho_m $$^a$ & $\rho_m $$^b$
    & $\mu ^{-1} $ & $f^{\prime \; c}$
    & $Z-f^{\prime}$ & $\rho _\infty$  & $q_c$  \\

    & (keV) & (cm$^2$/g) &  (g/cm$^3$) & ($\mu$m) & &
    & ($e^- /$\AA$^3$) & (\AA$^{-1}$) \\ \hline
In: & 12.5  & 74.1 & 7.0 & 19.3 & -0.2 & 48.8 & 1.788 & --- \\
& 13.4  & 60.5 &  & 23.6 &  &  &  & --- \\ \hline
Bi: & 12.5  & 75.2 & 10.0 & 13.3 & -7.3 & 75.7 & 2.180 & --- \\
& 13.4  & 155.0 &  & 6.45 & -24.7 & 58.3 & 1.679 & --- \\
\hline Bi$_{22}$In$_{78}$: & 12.5  & 74.5 & 7.66 & 17.5 & --- &
--- & 1.874 &
0.05154 \\
& 13.4  & 92.6 &  & 14.1 & --- & --- & 1.764 & 0.05000
\end{tabular}
\\

$^a$C.~H.~MacGillavry and G.~D.~Rieck, eds., {\em International
Tables for X-ray Crystallography} Vol.\ III, Kynoch Press,
Birmingham, England (1962). $^b$R.~C.~Weast, ed., {\em CRC
Handbook of Chemistry and Physics\/} (see Ref.\ 20).
$^c$B.~L.~Henke, E.~M.~Gullikson, and J.~C.~Davis, Atomic Data and
Nuclear Tables {\bf 54} (1993) 181. \caption{ Parameters used to
calculate energy dependent x-ray reflectivity. }
\label{biin_table:pars}
\end{table}

Following past
    practice,\cite{indium}
our model incorporates layers of atoms having a Gaussian distribution of
displacements
from idealized positions $nd$ along the surface normal direction:
\begin{equation}
    \rho_{\mbox{\scriptsize eff}} (z)
        =\rho _{\infty } \sum_{n=0}^{\infty }F_{n}
    \frac{d}{\sigma _{n}\sqrt{ 2\pi }}\exp
    \left[ -(z-nd)^{2}/\sigma _{n}^{2}\right] .
\label{biin_eq:layers}
\end{equation}
The roughness $\sigma _n$ arises from both static and dynamic contributions:
\begin{equation}
    \sigma _{n}^{2}=n\overline{\sigma }^{2}+\sigma _{0}^{2}+
    \frac{k_{B}T}{2 \pi \gamma }\ln
    \left( \frac{q_{\mbox{\scriptsize max}}}{q_{\mbox{\scriptsize res}}}
    \right) .
\label{biin_eq:sigma}
\end{equation}
Here $\overline{\sigma }$ and $\sigma _{0}$ are related to
the surface layering coherence length and the amplitude of density
oscillations at the surface. The last term accounts for height
fluctuations produced by capillary waves, and depends on the
temperature $T=101^{\circ }$C,
    the surface tension $\gamma =0.50$~N/m,\cite{gamma}
and wavevector cutoffs $q_{\mbox{\scriptsize max}}=0.99$~\AA $^{-1}$ and
$q_{\mbox{\scriptsize res}} \sim 0.024$~\AA $^{-1}$
(a slowly varying function of $q_z$),
as detailed
    elsewhere.\cite{indium}

The scattering amplitude of each layer depends on the \index{form
factor} form factor and the effective electron density in each
layer relative to the bulk, dependent on energy and Bi
concentration:
\begin{equation}
    F_{n}=w_{n}\times \frac{x_{n,\mbox{\scriptsize  Bi}}
    \left[ f_{ \mbox{\scriptsize  Bi}}(q_z)+
    f_{\mbox{\scriptsize  Bi}}^{\prime }(E)\right]
    +(1-x_{n,\mbox{\scriptsize Bi}})
    \left[ f_{\mbox{\scriptsize  In}}(q_z)+
    f_{\mbox{\scriptsize  In}}^{\prime }\right] }
    {x_{\mbox{\scriptsize  Bi}}\left[ Z_{\mbox{\scriptsize  Bi}}+
    f_{\mbox{\scriptsize  Bi}}^{\prime }(E)\right]
    +(1-x_{\mbox{\scriptsize Bi}})\left[ Z_{\mbox{\scriptsize  In}}+
    f_{\mbox{\scriptsize  In}}^{\prime }\right] }\times \frac{x_{n,
    \mbox{\scriptsize  Bi}}\rho _{\infty ,\mbox{\scriptsize  Bi}}+
    (1-x_{n,\mbox{\scriptsize Bi}})\rho _{\infty ,\mbox{\scriptsize  In}}}
    {\rho _{\infty,\mbox{\scriptsize  bulk}}} \: .
\label{biin_eq:prefix}
\end{equation}
For the first few layers ($n$=0,1,2), the weight $w_{n}$ may differ from
unity and the Bi fraction $x_{n,\mbox{\scriptsize  Bi}}$ can vary from the
bulk value $x_{\mbox{\scriptsize  Bi}}=0.22$.

\begin{figure}[tbp]
\centering
\includegraphics[width=1.0\columnwidth]{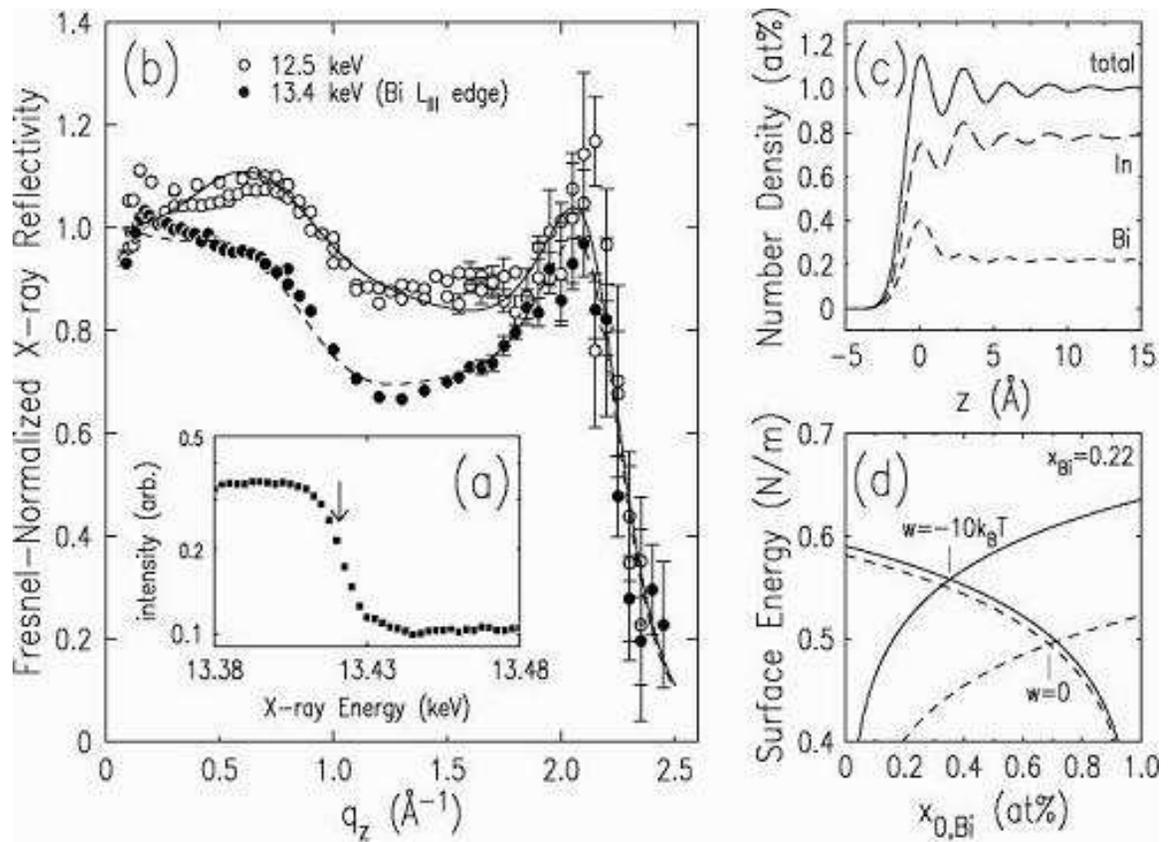}
\caption{ (a) Model surface-normal number density profiles,
relative to bulk atomic percent. (---): total number density; (--
--): In density; (- - -): Bi density. (b) Calculated
Fresnel-normalized reflectivity curves. (---): 12.5 keV; (- - -):
13.4 keV. i: Surface layering with uniform composition
($x_{\mbox{\scriptsize Bi}}=0.22$). ii: Surface layering, with Bi
enhancement in the first atomic layer ($x_{0,\mbox{\scriptsize
Bi}}=0.35$). } \label{biin_fig:2}
\end{figure}

 \section{Data Fitting Analysis}
\index{fitting!function}We now describe the ingredients that are
required to fit the data with this model.  The interference peak
at $q_z = 2.1$~\AA$^{-1}$ can be reproduced by a simple layered
profile in which $x_{n,\mbox{\scriptsize Bi}} =
x_{\mbox{\scriptsize Bi}} $ for all $n$. The Bi and In number
densities for such a model are shown in
    Figure~\ref{biin_fig:2}(a)i,
along with their sum, the total atomic fraction relative to the bulk.
The corresponding reflectivity
curves calculated for both x-ray energies are compared to the
experimental data in
    Figure~\ref{biin_fig:2}(b)i.
Since the Bi concentration is uniform, the
energy dependence is so slight that the curves overlap  almost completely
on the scale of the figure.
Turning again to the data, the reduced intensity in the region
$q_z < 1.8$~\AA$^{-1}$ when measured at the Bi L$_{III}$
edge indicates a reduction in the scattering amplitude at the surface.
This implies that the Bi concentration is enhanced there.
Increasing the Bi fraction from 22~at\% to
$\approx 35$~at\% in the first layer ($n=0$) produces an appropriate energy
dependence
    (Figs.~\ref{biin_fig:2}(a)ii and \ref{biin_fig:2}(b)ii).

Although at this point the frequency of the low-$q_z$ modulation is not
well described, the fit is considerably improved
by allowing the Bi fraction and total number densities to vary for the
first three surface layers ($n$=0,1,2).
We find that the Bi fraction for $n$=1,2 is essentially equal to the bulk
value of 22~at\%, while the total number densities for $n$=0,1,2 have
values of 0.98, 1.01, and 0.98, respectively
    (Figs.~\ref{biin_fig:1}(b),(c)).
Thus, the detailed shape of the low-$q_z$ modulation is modelled
by a very slight density wave of about 2\% affecting the
amplitudes of the first few surface layers. Increasing the number
of model \index{fitting!parameters} parameters to allow for shifts
in positions and widths of the surface layers resulted in
marginally better fits, but at the expense of high frequency
Fourier components appearing as small wiggles in the calculated
reflectivity.  Variations in these extra fit parameters are
extremely slight, and we doubt whether they have any physical
basis. \index{fitting!parameters} Parameters for all models are
shown in
    Table~\ref{biin_table:2}.

\begin{table}[tbp]
\begin{tabular}{cccccccccc}
Figure & $d$ & $\overline{\sigma}$ & $\sigma_0$ & $w_0$ &
$x_{0,\mbox{\scriptsize Bi}}$ & $w_1$ & $x_{1,\mbox{\scriptsize
Bi}}$ & $w_2$ &
$x_{2,\mbox{\scriptsize Bi}}$ \\
\hline
1 & 2.81 & 0.54 & 0.64 & 0.98 & 0.35 & 1.01 & 0.22 & 0.98 & 0.23 \\
2i & 2.81 & 0.48 & 0.64 & 1.0 & 0.22 & 1.0 & 0.22 & 1.0 & 0.22 \\
2ii & 2.81 & 0.54 & 0.64 & 1.0 & 0.35 & 1.0 & 0.22 & 1.0 & 0.22
\end{tabular}
\caption{Fit parameters for model profiles, identified by the
figure in which they appear. The length scales $d$,
$\overline{\protect\sigma}$, and $\protect\sigma_0$ are in units
of \protect\AA .} \label{biin_table:2}
\end{table}

This analysis demonstrates that
to model the essential features of our data, there is no need to invoke
long-range compositional ordering on a second length scale in the
surface-normal
direction, which we had suggested based on the previous non-resonant
reflectivity
    measurements.\cite{srn99,ben00}
Still, we thought it important to
investigate additional, specific models based on Bi-In pairs oriented
along the surface normal.
To test for alternating Bi-rich and In-rich layers, we attempted fits
in which we forced the Bi and In compositions to
be substantially different from the results shown above. We also described
Bi-In pairing by allowing the positions, but not the densities, of the
surface layers to vary.
None of these profiles successfully described the data.

 \section{Summary}
Our principal finding is the Bi enrichment of 35~at\% in the surface
layer, compared to the bulk value of 22~at\%.  This is considerably less
Bi than would be expected in the absence of attractive Bi-In interactions,
which can be estimated from the surface free energy
    requirement:\cite{gugg}
\begin{equation}
\label{biin_eq:gugg}
    \gamma_{\mbox{\scriptsize In}} + \frac{k_BT}{a}
    \ln \left( \frac {1-x_{0,\mbox{\scriptsize Bi}}}
        {1-x_{\mbox{\scriptsize Bi}}} \right)
    - \frac{1}{4} (x_{\mbox{\scriptsize Bi}})^2 \frac{w}{a}
    = \gamma_{\mbox{\scriptsize Bi}} + \frac{k_BT}{a}
    \ln \left( \frac {x_{0,\mbox{\scriptsize Bi}}}
        {x_{\mbox{\scriptsize Bi}}} \right)
    - \frac{1}{4} (1-x_{\mbox{\scriptsize Bi}})^2 \frac{w}{a} \: .
\end{equation}
The quantity $w$ is the excess interaction energy of Bi-In pairs over
the average of the Bi-Bi and In-In interaction energies; for an ideal
mixture, $w=0$.
This analysis assumes that the inhomogeneous region is confined to a
single atomic layer, the atoms are close-packed and take up an area $a$,
and $w$ is small.
Extrapolating the measured surface tensions to
    100$^\circ$C,\cite{gamma}
$\gamma_{\mbox{\scriptsize In}} = 0.56$~N/m and
$\gamma_{\mbox{\scriptsize Bi}} = 0.41$~N/m.
Using the Bi atomic size, $a =  \pi(3.34/2)^2$~\AA$^2$
    (for In, the atomic diameter is 3.14~\AA ),\cite{iida}
and the bulk composition $x_{\mbox{\scriptsize Bi}} = 0.22$.
The equilibrium surface composition $x_{0,\mbox{\scriptsize Bi}}$ is
found from the intersection of plots of both sides of
    Eq.~\ref{biin_eq:gugg}.
For $w=0$, this analysis predicts a surface segregation of 69~at\% Bi
    (Figure~\ref{biin_fig:2}(d), dashed lines).
To reproduce our experimental finding that
$x_{0,\mbox{\scriptsize Bi}} = 35$~at\%,
$w$ must be negative, with a magnitude of $\sim 10 k_B T$
    (Figure~\ref{biin_fig:2}(d), solid lines).
Although this large value of $w$ is most likely outside the range of
validity of
    Eq.~\ref{biin_eq:gugg},
the analysis certainly illustrates the qualitative effect of
attractive \index{interactions!heteroatomic} heteroatomic
interactions on the surface composition. In this Bi-In alloy,
pairing does in fact defeat \index{Gibbs!adsorption} Gibbs
adsorption in the sense that the surface energy is optimized not
by segregating a large fraction of Bi, but by forming larger
numbers of Bi-In neighbors in the surface layer. Exactly how this
balance plays out in Bi-In alloys with the stoichiometric bulk
compositions BiIn and BiIn$_2$ remains to be seen.


%% file: GaBi_prl_all.tex
\chapter{Microscopic Structure of the Short-Range Wetting Film at the Surface of Liquid Ga-Bi Alloys}

\section{Abstract}
 X-ray reflectivity measurements of the  binary liquid Ga-Bi  alloy
 reveal a dramatically different surface structure
 above and below the \index{monotectic point} monotectic temperature $T_{mono}=222^{\circ}$\,C.
 A \index{Gibbs!adsorption} Gibbs-adsorbed Bi monolayer resides at the surface
 at both regimes. However,
 a 30~{\AA} thick, Bi-rich \index{nanoscale} nanometer scale wetting film intrudes
 between the Bi monolayer and the Ga-rich bulk for $T > T_{mono}$.
 The  internal structure of the wetting film
 is determined with {\AA} resolution,
 showing a theoretically unexpected concentration gradient
 and a highly diffuse interface with the bulk phase.

\section{Introduction}
A thick \index{wetting!film} wetting film may be stable at the
free surface of a binary immiscible liquid mixture and the
temperature dependent formation of this surface film is strongly
influenced by the bulk critical \index{demixing transition}
demixing of the underlying bulk phase.\cite{cahn} In contrast to
surface segregation, where a monolayer of the low surface tension
component forms at the interface of the binary mixture, wetting is
a genuine phase transition occurring at the interface resulting in
the formation of a  mesoscopically or macroscopically thick
film.\cite{dietrich} To date, nearly all studies of wetting
phenomena have been carried out with \index{liquid!dielectric}
dielectric liquids dominated by long-range van
\index{interactions!Van der Waals} der Waals
interactions\cite{domb} and the experimental techniques used
achieve mesoscopic resolution only. For binary alloys dominated by
screened \index{interactions!Coulomb} Coulomb interactions,
evidence for a \index{wetting!transition} wetting transition has
so far been obtained  only for Ga-Bi\cite{nattland} and
Ga-Pb.\cite{chatain} In the case of Ga-Bi,  the formation of a
Bi-rich \index{wetting!film} wetting film has been detected by
\index{ellipsometry} ellipsometry in the temperature range from
220$^{\circ}$\,C to 228$^{\circ}$\,C.\cite{nattland} However,
since the film thickness is much smaller than the wavelength of
visible light, \index{ellipsometry} ellipsometry cannot provide
{\AA}-resolution structural information. By contrast, x-ray
surface scattering techniques allow  determination of the
structure of the wetting film with atomic resolution and address
issues such as the internal structure of the film and its
evolution from molecular to mesoscopic thickness, none of which
was hitherto addressed by any of the previous measurements.

\begin{figure}[tbp]
\epsfig{file=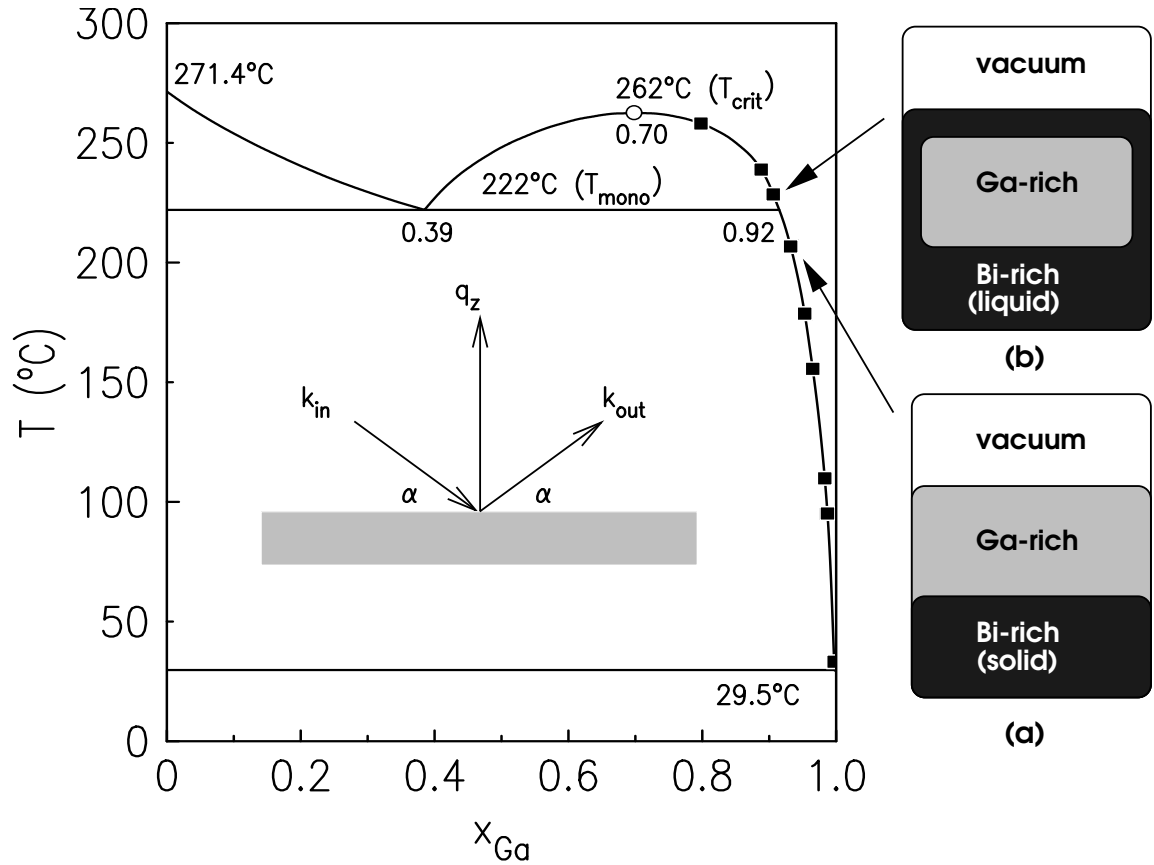, angle=0, width=1.0\columnwidth}
\caption{ \label{fig:fig1} \index{phase diagram} Phase diagram of
Ga-Bi (Ref.\,8). Between 29.5$^{\circ}$\,C and 222$^{\circ}$\,C,
solid Bi is in equilibrium with a Ga-rich liquid phase. Between
222$^{\circ}$\,C and 262$^{\circ}$\,C, a liquid Bi-rich phase is
in coexistence with a liquid Ga-rich phase. The surface structure
of this alloy has been measured between 35$^{\circ}$\,C and
258$^{\circ}$\,C  at selected points along the coexistence line
(\protect\rule{1.5mm}{1.5mm}). For $T<$222$^{\circ}$\,C   partial
wetting occurs (a), whereas \index{wetting!complete} complete
wetting is found  for $T>$222$^{\circ}$\,C (b). The inset depicts
the geometry for specular x-ray reflectivity. }
\end{figure}

The wetting phase transition is depicted schematically in Fig.\,1.
Below the bulk critical point of demixing, $T_{crit}$, the bulk
\index{phase separation} phase separates into two
\index{immiscible phase} immiscible phases, the high density
Bi-rich phase and the low density Ga-rich phase. Below the
characteristic wetting temperature $T_w < T_{crit}$, the high
density phase is confined to the bottom of the container as
expected (see Fig.~\ref{fig:fig1}(a)). As discussed in more detail
below, for Ga-Bi, the high density phase is solid in this case and
$T_w$ coincides with the \index{monotectic point} monotectic
temperature $T_{mono}$. In contrast, above $T_w$ the high density
phase completely wets the free surface by intruding between the
low density phase and the gas phase in defiance of gravity (see
Fig.~\ref{fig:fig1}(b)).\cite{dietrich,rowlinson}

\section{Sample Preparation}
The Ga-Bi alloy was prepared
in a inert-gas box  using at least 99.9999\% pure metals.
A solid Bi ingot was placed in a Mo pan and supercooled liquid Ga
was added to cover the Bi ingot.
At room temperature,  the solubility of Bi in Ga is less than 0.2at\%.
Increasing the temperature results in  continuously dissolving  more Bi in the Ga-rich phase with
 solid Bi remaining at the bottom of the pan  up to the \index{monotectic point} monotectic
temperature, $T_{mono}$= 222$^{\circ}$\,C. The initial Bi content
was chosen to be high enough  that the two phase equilibrium along
the \index{miscibility gap} miscibility gap could be followed up
to few degrees below $T_{crit}$ without crossing into the
homogeneous phase region (see Fig.~\ref{fig:fig1}). The
temperature on the sample surface was measured with a Mo coated
thermocouple. The alloy was contained in an  ultra high vacuum
chamber and the residual oxide on the sample was removed by
sputtering with $Ar^+$ ions.\cite{indium} Surface sensitive x-ray
reflectivity (XR) experiments were carried out using the liquid
surface \index{spectrometer} spectrometer at beamline X22B  at the
National \index{synchrotron!source} Synchrotron Light Source with
an x-ray wavelength $\lambda=$1.24~{\AA} and a detector resolution
of $\Delta q_z=$0.03~${\AA}^{-1}$. The background intensity, due
mainly to scattering from the bulk liquid, was subtracted from the
specular signal by displacing the detector out of the reflection
plane.\cite{indium}

\begin{figure}[tbp]
\centering
\includegraphics[width=0.7\columnwidth]{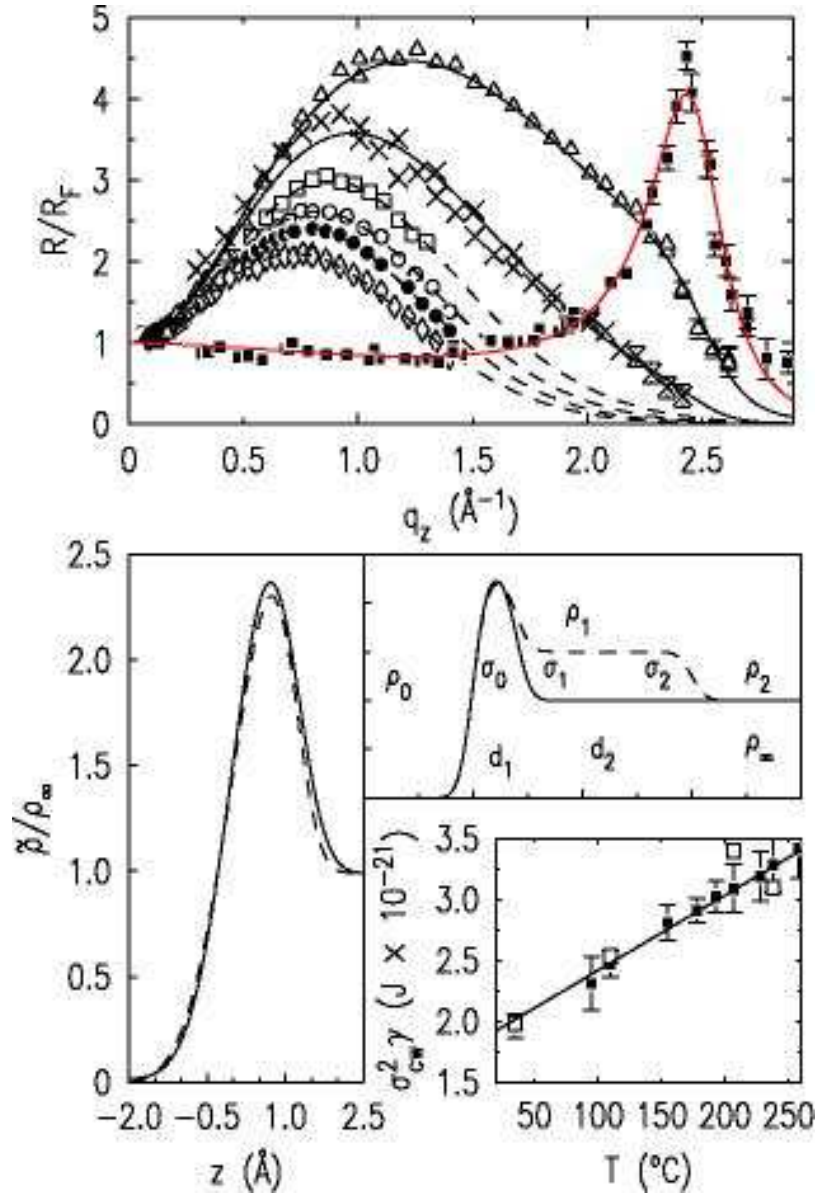}
 \caption{ \label{fig:fig2}(a) Fresnel normalized
reflectivity \protect$R/R_F$ from  Ga-Bi alloys for \protect$T <
T_{mono}$: (\protect\rule{1.5mm}{1.5mm}): 100\,at\% Ga;
 (\protect$\triangle$): 99.7\,at\% Ga
(35\protect$^{\circ}$\,C); (\protect$\times$): 98.7\% Ga (95$^{\circ}$\,C);
(\protect$\Box$): 98.4\% Ga (110$^{\circ}$\,C );
(\protect$\circ$): 96.5\% Ga (155$^{\circ}$\,C );
(\protect$\bullet$): 95.3\% Ga (178$^{\circ}$\,C ) and
(\protect$\diamond$): 93.2\% Ga (207$^{\circ}$\,C ).
Shaded line: fit to layered density profile of pure Ga (see Ref.\,9);
solid lines: fit to layered density profile of Ga plus Bi monolayer (see Ref.\,10);
broken lines: fit to Bi monolayer (this work).
(b) Intrinsic   density profiles
normalized to \protect$\rho_{\infty}$
for Ga-Bi for 110$^{\circ}$\,C $\leq T \leq$ 207$^{\circ}$\,C
using the one-box model described in the text. The profiles essentially
fall on top of each other and two representative fits are shown:
110$^{\circ}$\,C (solid line)
and 207$^{\circ}$\,C (broken line).
(c): Schematic representation  of the (two-)box model.
(d): CW roughness $\sigma_{cw}$ for Ga-Bi alloys as
a function of temperature using  $\gamma$ from
our fit (\protect\rule{1.5mm}{1.5mm}) and from macroscopic
surface tension measurements (\protect$\Box$, see Ref.\,14).
}
\end{figure}

\section{Experimental Techniques}
The  intensity reflected from the surface, $R(q_z)$, is measured
as a function of the normal component $q_z =
(4\pi/\lambda)\sin\alpha$ of the momentum transfer. The XR
geometry is depicted schematically in the inset of
Fig.~\ref{fig:fig1}. $R (q_z)$ therefore yields information about
the surface-normal structure as given by
\begin{equation}
    R(q_z) =  R_F(q_z)
        \left| \Phi (q_z) \right| ^2
    \exp [ - \sigma _{\mbox{\scriptsize cw}}^2q_z^2 ],
\label{eq:circle}
\end{equation}
where $R_F (q_z)$ is the \index{Fresnel reflectivity} Fresnel
reflectivity of a flat, infinitely sharp  surface and $\Phi (q_z)$
is the Fourier transform of the local surface-normal  density
profile $\tilde{\rho} (z)$:\cite{indium}
\begin{equation}
\Phi (q_z) = \frac{1}{\rho_{\infty}} \int dz \frac{d\tilde{\rho} (z) }{dz}
\exp(\imath q_z z)
\label{eq:structure}
\end{equation}
with the bulk electron density, $\rho_{\infty}$.
The exponential factor in Eq.\,\ref{eq:circle} accounts for roughening
of the intrinsic density profile $\tilde{\rho} (z)$ by  capillary waves (CW):
\begin{equation}
\sigma _{\mbox{\scriptsize cw}}^2 = \frac{k_B T}{2\pi \gamma }
    \ln \left(
    \frac{q_{\mbox{\scriptsize max}}}{q_{\mbox{\scriptsize res}}
    }\right)  .
\label{eq:cw}
\end{equation}
The CW spectrum  is cut off at small $q_z$ by the detector
resolution  $q_{res}$ and at large $q_z$ by the atomic size $a$
with $q_{max} \approx \pi/a$.\cite{indium} However, $\tilde{\rho}
(z)$, can not be obtained directly from Eq.\,\ref{eq:structure}
and we resort to the widely accepted procedure of adopting a
physically motivated model for $\tilde{\rho}(z)$ and
\index{fitting!function} fitting its Fourier transform  to the
experimentally determined $R (q_z)$ as will be shown further
below.\cite{indium}

The reflectivity of pure liquid Ga exhibits a pronounced
interference peak indicating surface-induced layering of ions near
the surface (see Fig.~\ref{fig:fig2}(a)).\cite{gallium} Previous
experiments on Ga-Bi at low temperatures\cite{bunsen,rice} show
that the layering peak is suppressed in the liquid alloy and $R
(q_z)$ is dominated by a broad maximum consistent with a single Bi
monolayer segregated at the surface as expected since $\gamma$ is
considerably lower for Bi than for Ga.\cite{rice}

\section{Results}
Here,  we report x-ray measurements from liquid Ga-Bi  for $T$ up
to 258$^{\circ}$\,C encompassing the formation of the
\index{nanoscale} nanoscale \index{wetting!film} wetting film at
$T >T_{mono}$, the lowest temperature at which the liquid Bi-rich
phase is stable. The actual wetting transition likely lies below
$T_{mono}$ but does not manifest itself since the wetting film is
solid for $T < T_{mono}$. Experimental evidence for this
assumption has been found for the similar systems
Ga-Pb\cite{wynblatt} and K-KCl.\cite{kkcl} The discussion of the
data is divided into two parts: (i) $T < T_w \simeq T_{mono}$
(110$^{\circ}$\,C-207$^{\circ}$\,C) and (ii) $T > T_w$
(228$^{\circ}$\,C-258$^{\circ}$\,C). We present the results for $T
< T_{mono}$ (scenario (a) in Fig.~\ref{fig:fig1}) first since this
corresponds to the simpler surface structure with monolayer
segregation but no wetting film.

{\bf  Below $T_{mono}$:} Fig.~\ref{fig:fig2}(a) shows  $R (q_z)$
from liquid Ga-Bi at selected points along the coexistence line
between 110$^{\circ}$\,C  and 207$^{\circ}$\,C. Since subsurface
layering is impossible to resolve at these temperatures, we simply
model the near-surface density, which is dominated by the Bi
monolayer segregation, by one box of density $\rho_1$ and width
$d_1$. The density profiles are shown in Fig.~\ref{fig:fig2}(b)
and the box model is shown schematically in Fig.~\ref{fig:fig2}(c)
(solid line). The best fit to $R(q_z)$ that corresponds to this
density profile is represented by the broken lines in
Fig.~\ref{fig:fig2}(a). The mathematical description of the
general box model  with a maximum number  of two boxes is: {\small
\begin{equation}
\tilde{\rho}(z) = \frac{\rho_0}{2} \left\{ 1 + {erf} \left(
\frac{z}{\sigma_{0}} \right) \right\}
 - \sum_{i=1}^2
\frac{\rho_i}{2} \left\{ 1 + {erf}
\left(\frac{z-d_i}{\sigma_{i}}\right)\right\} \label{eq:box}
\end{equation}
} where $\sigma_0$ and $\sigma_{i}$ are the intrinsic
 roughnesses of each of the three interfaces.
The total roughness, $\sigma$, of the  interface between the vapor
and the outermost surface layer is  given by $\sigma^2 =
\sigma^2_{0} + \sigma^2_{cw}$ with $\sigma_0$ = 0.78$\pm
0.15~{\AA}$ for $T < T_{mono}$. As predicted by capillary wave
theory, the product $\sigma_{cw}^2 \gamma$ (with $\gamma$ from our
fits) depends linearly on $T$ (see Eq.\,\ref{eq:cw} and
(\protect\rule{1.5mm}{1.5mm}) in Fig.~\ref{fig:fig2}(d)) over the
entire temperature range.
 This variation is essentially the same if
$\gamma$ is not taken from our fits but from macroscopic surface
tension measurements (\protect$\Box$).\cite{tschirner} Apart from
the increasing surface roughness, the structure of the surface
does not change over the entire temperature range from
110$^{\circ}$\,C to 207$^{\circ}$\,C, as witnessed by the
intrinsic density profiles that fall right on top of each other
once the  theoretically predicted temperature dependence of the CW
roughness has been corrected for (Fig.~\ref{fig:fig2}(b)). Between
212$^{\circ}$\,C and 224$^{\circ}$\,C, we observed rapid and
random changes in the scattered intensity  over the entire $q_z$
range. This unstable behavior of the surface is most likely due to
the coexistence of patches of different film thickness induced by
the temperature gradient of about 6K normal to the  the sample
pan.

\begin{figure}[tbp]
\centering
\includegraphics[width=1.0\columnwidth]{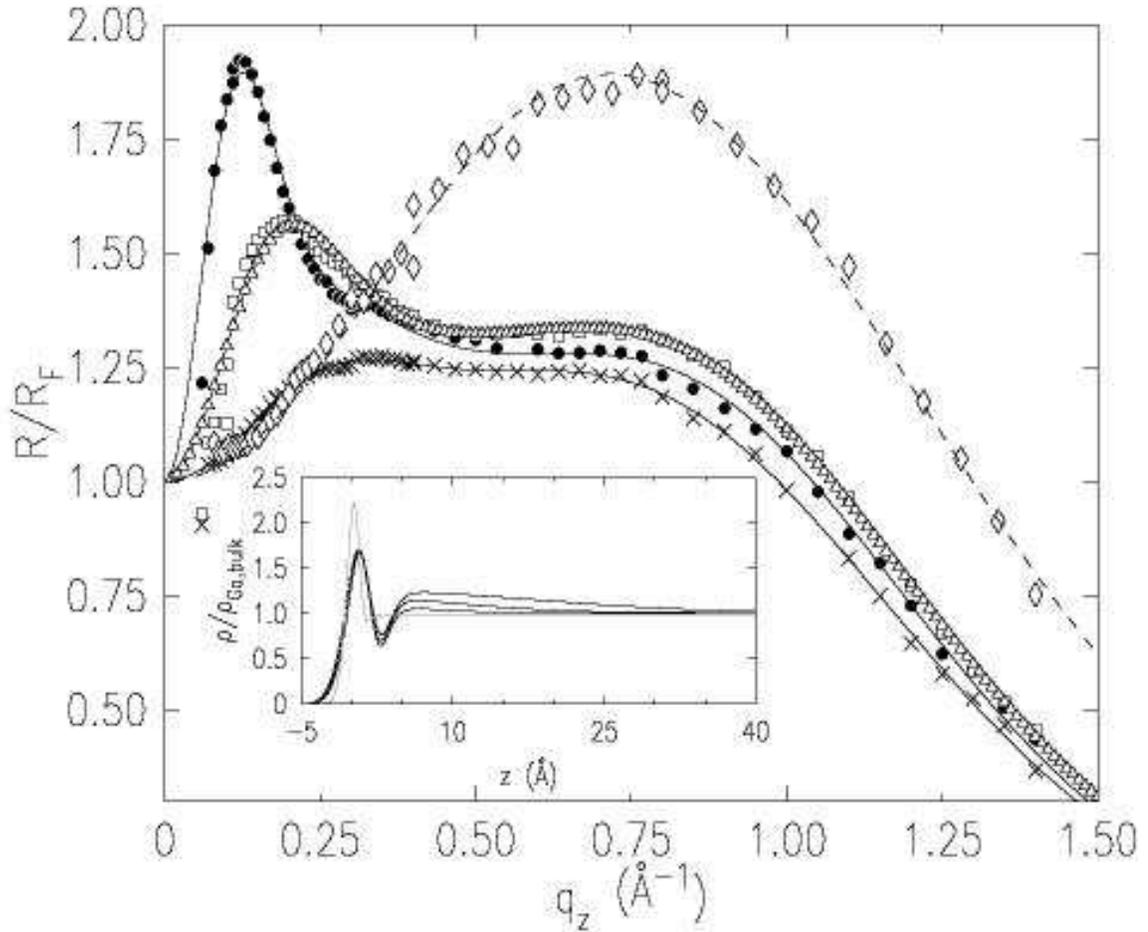}

\caption{ \label{fig:fig3} $R/R_F$ from Ga-Bi alloys for $T >
T_w$: ($\bullet$): 90.6\,at\% Ga (228$^{\circ}$\,C); ($\Box$):
88.8\% Ga (238$^{\circ}$\,C); ($\triangle$): ditto, but after
heating to 258 $^{\circ}$\,C and cooling down; ($\times$): 79.1\%
Ga (258$^{\circ}$\,C); (for comparison, see:($\diamond$),
207$^{\circ}$\,C). Solid lines fit to two-box model; broken line:
fit to one-box model. Inset: intrinsic density profiles,
$\tilde{\rho}(z)$, for Ga-Bi alloys normalized to the bulk
density, $\rho_{Ga,bulk} = \rho_{\infty}$, at $T > T_w$: solid
lines ordered  in decreasing wetting film density:
228$^{\circ}$\,C,  238$^{\circ}$\,C and 258$^{\circ}$\,C. Compare
to the  one-box density profile of Ga-Bi at 207$^{\circ}$\,C
(shaded line). }
\end{figure}

{\bf Above $T_{mono}$:} The surface stabilized  above  $T_{mono}$
at about 228$^{\circ}$\,C and and a sharper peak in $R(q_z)$
appeared centered around 0.13~${\AA}^{-1}$ (see
Fig.~\ref{fig:fig3}), indicating a thick film of high density
forming near the surface. The persistence of the  broad maximum
centered around $q_z \approx 0.75$~{\AA}$^{-1}$ indicates that Bi
monolayer segregation is still present along with the newly formed
thick \index{wetting!film} wetting film. Several models were used
to fit $R\,(q_z)$ but they all result in essentially the same
density profiles. Here, we use the simple two-box model (see
broken line in Fig.~\ref{fig:fig2}(c) and Eq.\,\ref{eq:box}). As
can be seen in Fig.~\ref{fig:fig3}, this simple model gives an
excellent description of the experimentally obtained reflectivity.
The pertinent density profiles describing the surface-normal
structure of Ga-Bi alloys for $T > T_{mono}$  are shown in the
inset of Fig.~\ref{fig:fig3} and  have the following  features
that can be compared to present predictions of wetting theory: (i)
density, (ii) thickness and (iii) roughness of the wetting film.

(i) The concentration profile is highly inhomogeneous with a high
density region at the outermost surface layer.\cite{dip} The
integrated density of this adlayer is consistent with a monolayer
of the same density as the monolayer found for $T < T_{mono}$.
However, the Bi monolayer segregated on top of the
\index{wetting!film} wetting film is rougher and broader than the
Bi monolayer segregated on top of the Ga-rich bulk phase for $T <
T_{mono}$. This change in surface structure corresponds to a
change in the intrinsic roughness of the interface liquid/vacuum,
$\sigma_0$, from 0.78$\pm$0.15~{\AA} below $T_{mono}$  to
1.7$\pm$0.3~{\AA} above $T_{mono}$. However, the CW roughness
still follows the predicted T-dependence of Eq.\,3 (see
Fig.~\ref{fig:fig2}(d)).

This change of $\sigma_0$ upon crossing $T_{mono}$ is possibly
related to the fact that for $T > T_{mono}$, the pure Bi monolayer
segregates against a Bi-rich wetting film made out of atoms of
different size and with repulsive
\index{interactions!heteroatomic} heteroatomic interactions. By
contrast, for $T < T_{mono}$, this segregation takes place against
almost pure Ga. The fact that the Bi monolayer segregation
persists from partial wetting to \index{wetting!complete} complete
wetting means that the \index{nanoscale} nano wetting film does
not intrude between the Ga-rich bulk phase and vacuum, as assumed
by wetting theory, but between the Bi monolayer already segregated
at the surface and the Ga-rich bulk phase. This should have a
pronounced influence on the energy balance at the surface which
ultimately governs the wetting phase transition. Even though the
possibility of an unspecified concentration gradient is included
in \index{Cahn, John} Cahn's general theory\cite{cahn}, to our
knowledge, an inhomogeneous density profile is neither treated
explicitly in theoretical calculations (see the homogeneous
profiles displayed in\,\cite{profile,beysens}) nor found
experimentally.
 At 228$^{\circ}$\,C the density of the thick wetting film is $\rho_2 =(1.25\pm 0.03)\rho_{\infty}$
where $\rho_{\infty}$ is the  density of the Ga-rich bulk phase.
This agrees very well with the ratio of the densities of the
coexisting  Bi-rich and  Ga-rich phases in the bulk, calculated
from the \index{phase diagram} phase diagram to be 1.23. The
density of the wetting film reaches $\rho_{\infty}$
 with increasing T ($\rho_2/\rho_{\infty}=$ 1.18$\pm 0.03$ at
238$^{\circ}$\,C and 1.08$\pm 0.02$ at 258$^{\circ}$\,C)
as the  densities of the two  bulk phases converge upon
 approaching $T_{crit}$  (their density  ratio is 1.20 at 238$^{\circ}$\,C and
1.08 at 258$^{\circ}$\,C). This  strongly supports the conclusion
that the thick \index{wetting!film} wetting film is the Bi-rich
bulk phase as predicted by wetting theory.

(ii)  The thickness of the wetting film, $d_2$, at
228$^{\circ}$\,C  is determined from the density profile to be
$\sim$30~{\AA}, consistent with a model-dependent estimate from
\index{ellipsometry} ellipsometry results.\cite{nattland} In
addition, the thickness of the  wetting film is corroborated by
independent \index{x-ray!grazing incidence diffraction} grazing
incidence diffraction experiments resolving the \index{in-plane
ordering} in-plane structure which will be reported
elsewhere.\cite{iop} The \index{wetting!film} wetting film in this
Coulomb liquid with short-range interactions is considerably
thinner than wetting films that have been observed in
\index{liquid!dielectric} dielectric liquids with long-range
interactions.\cite{domb,kayser} An important question concerning
the thickness of a wetting film is whether the
\index{wetting!film} wetting film has been investigated in
equilibrium. The fact that identical reflectivities were measured
at 238$^{\circ}$\,C taken 24 hours apart (Fig.~\ref{fig:fig3}), is
strong evidence that the film thickness is in equilibrium in our
study.\cite{kayser}

(iii) The roughness between the Bi-rich wetting film and the
Ga-rich bulk film is much higher that the roughness of the free
interface LM/vacuum. This is be expected since the interfacial
tension between two similar liquids is generally much lower than
the liquid/gas surface tension.\cite{israel} On the other hand, it
is not clear how sharp a concentration gradient should be expected
between the Ga-rich and the Bi-rich phase and the interface may be
essentially diffuse, independent of CW roughness. With increasing
temperature, the wetting film becomes less well defined and it is
not obvious whether this is due to the fact that the film is
becoming slightly thinner, or if only the interface between two
converging phases gets more diffuse. At any rate, we are far
enough away from $T_{crit}$ that the predicted thickening of the
\index{wetting!film} wetting film  due to the increasing
\index{correlation length} correlation length of the concentration
fluctuations should not play a role. Once the thick wetting film
has formed, it is not possible to cool the sample below $T_{mono}$
since the Bi-rich wetting film remains at the surface and freezes.

\section{Summary}
In summary, we investigated the structural changes occurring on
atomic length scales at the  surface of  liquid Ga-Bi during the
wetting transition. In the case of partial wetting ($T <
T_{mono}$), a Bi monolayer segregates at the surface to lower the
surface energy. Above the \index{monotectic point} monotectic
temperature \index{wetting!complete} complete wetting is found and
a  30~{\AA} thick wetting film intrudes between this monolayer and
the Ga-rich bulk phase. This is the first time that the
microscopic structure of a wetting film has been studied and that
the concentration profile of the wetting film has been shown to
vary on a level of several atomic diameters.

\bibliographystyle{unsrt}

%% file: GaBiMono_all.tex
\chapter{Wetting film at GaBi surface}

\section{Abstract}
Here we present x-ray reflectivity measurements from the free
surface of a liquid gallium-bismuth alloy (Ga-Bi) in the
temperature range close to the bulk \index{monotectic point}
monotectic temperature $T_{mono}=222%
{{}^\circ}%
C$. Our measurements indicate a continuous formation of a thick
wetting film at the free surface of the binary system driven by
the first order transition in the bulk at the monotectic point. We
show that the behavior observed is that of a
\index{wetting!complete} complete wetting at a
\index{wetting!tetra-point} tetra point of
solid-liquid-liquid-vapor coexistence.

\section{Introduction}

Interfacial phenomena at the surface of critical systems,
particularly at the surfaces of critical binary liquid mixtures,
have attracted interest ever since the seminal paper by J.W. Cahn
\cite{cahn1977} demonstrating the existence of a wetting line,
starting at the wetting temperature, $T_{W}$, below the critical
point of demixing, $T_{crit}$. Initially, experimental efforts
were mainly focused on systems dominated by long-range
\index{interactions!Van der Waals} van-der-Waals interactions like
methanol-cyclohexane and other organic materials.\cite
{Dietrich1988}\cite{Law2001} Only recently have experiments probed
the wetting behavior in binary metallic systems, most prominently
in gallium-lead (Ga-Pb) \cite{Chatain1996}, gallium-thallium
(Ga-Tl) \cite {Shim2001}, and gallium-bismuth (Ga-Bi)
\cite{Nattland1996}. Those systems allow the study of the
influence of interactions characteristic of metallic systems on
the wetting behavior. By the same token, they allow one to obtain
information on the dominant interactions in metallic systems by
measurements of the thermodynamics and structure of the wetting
films.

This paper reports measurements of the wetting behavior of Ga-Bi close to
its \index{monotectic point} monotectic temperature $T_{mono}=222%
{{}^\circ}%
C$. We first present the bulk \index{phase diagram} phase diagram
of Ga-Bi and relate it to the structures at the free surface
(liquid-vapor interface) known from optical \index{ellipsometry}
ellipsometry \cite{Nattland1996} and recent x-ray reflectivity
measurements. \cite{Lei1996}\cite{Tostmann2000} The experimental
setup is then described, along with some basic principles of the
x-ray reflectivity experiment and the data analysis. Finally, our
results and the conclusions emerging therefrom are discussed.

\section{Bulk and Surface Structure}

\begin{figure*}[tbp]
\centering
\includegraphics[width=1.0\columnwidth]{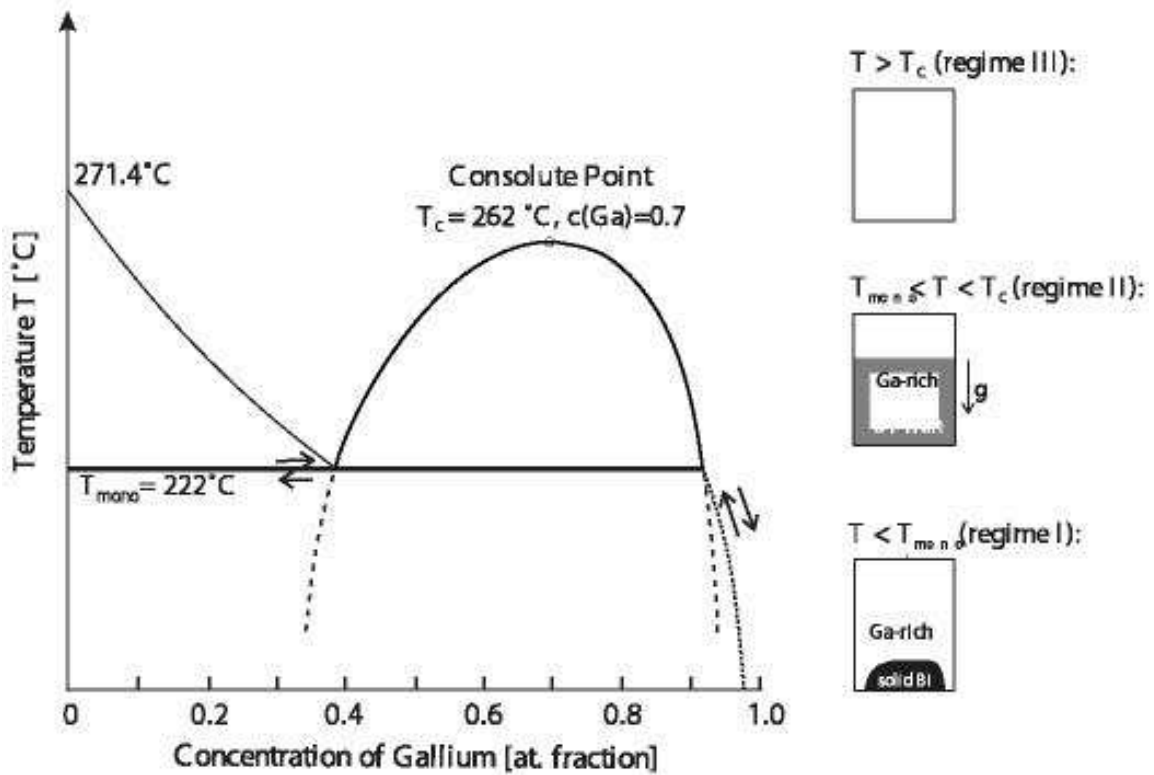}
\caption{\label{fig:Fig1}Bulk \index{phase diagram} Phase diagram
of Ga-Bi (Ref. \cite{Predel1960}). Bold solid line: liquid-liquid
coexistence, dashed line: liquid-liquid coexistence metastable,
dotted line: liquid-solid coexistence. On the right: Schematic
bulk behavior in the different temperature regimes.}
\end{figure*}

The bulk phase diagram of Ga-Bi has been measured by Predel using
calorimetric methods \cite{Predel1960}, and is shown in
Fig.~\ref{fig:Fig1}. Assuming an
overall concentration of 70\% Ga, the following behavior is found: Below $%
T_{mono}$ (regime I) and at temperatures higher than the melting point of Ga
$T_{m}(Ga)=29.5%
{{}^\circ}%
C$, solid Bi coexists with a Ga-rich liquid phase. In this regime,
previous x-ray measurements have shown that a
\index{Gibbs!adsorption}Gibbs-adsorbed Bi monolayer resides at the
free surface.\cite{Lei1996}\cite{Tostmann2000} For
$T_{mono}<T<T_{crit}$ (regime II), the bulk \index{phase
separation} phase separates into two \index{immiscible phase}
immiscible phases, a high density Bi-rich phase and a low density
Ga-rich phase. The heavier Bi-rich phase is macroscopically
separated from the lighter Ga-rich phase and sinks to the bottom
of the sample pan. In temperature regime II, the high density,
Bi-rich phase wets the free surface by intruding between the low
density phase and the Bi monolayer, in defiance of
gravity.\cite{Tostmann2000} Considering that pure Bi has a
significantly lower surface tension than pure Ga, the segregation
of the Bi-rich phase at the surface is not too surprising. In
fact, the thickness of the wetting layer in such a geometry is
believed to be limited only by the extra gravitational potential
energy paid for having the heavier phase at the top
\cite{DeGennes1985}. In regime III, above the consolute point, the
bulk is in the homogeneous phase and only a
\index{Gibbs!adsorption} Gibbs-adsorbed Bi monolayer has been
found at the free surface.\cite {Huber2001}

Despite extensive experimental efforts to relate the bulk phases
to the corresponding surface phases, two important question remain
unanswered: What kind of transitions take place at the free
surface between regime I and regime II and between regime II and
regime III. Here, we focus on the transition between regime I and
regime II. Note that these temperature regimes are separated by
$T_{mono}$, a temperature characterized by the coexistence of the
solid Bi, the Bi-rich liquid, the Ga-rich liquid and the vapor.
$T_{mono}$ is therefore a solid-liquid-liquid-vapor
\index{wetting!tetra-point} tetra point, at which the
solidification of pure bulk Bi takes place. The question is, then,
how is this first order bulk transition related to the changes at
the surface?

\section{Experimental Techniques}

The Ga-Bi alloy was prepared in an inert-gas box using metals with purities
greater than 99.9999\%. A solid Bi ingot was placed in a Mo pan and oxide
formed at the surface was scraped away. Liquid Ga was then added to
completely cover the Bi ingot. The sample with an overall Ga amount of 70
at\% was transferred through air into an ultrahigh vacuum chamber. After a
one day of bake-out, a pressure of 10$^{-10}$torr was achieved and the
residual oxide on the liquid sample was removed by sputtering with Ar$^{+}$
ions, at a sputter current of a 25 microamps and a sputter voltage of 2 kV.

To avoid temperature gradients between the bulk and the surface
induced by thermal radiation, a temperature-controlled
\index{thermal radiation shield} radiation shield was installed
above the sample. The temperature was measured with thermocouples
mounted in the bottom of the sample pan and on the radiation
shield. The sample pan and the radiation shield temperature were
controlled by Eurotherm temperature
controllers yielding a temperature stability of $\pm 0.05%
{{}^\circ}%
C$.

Surface-specific x-ray reflectivity measurements were carried out
using the liquid surface reflectometer at beamline X22B at
\index{Brookhaven} National \index{synchrotron!source} Synchrotron
Light Source with an x-ray wavelength $\lambda =1.54\AA .$
Background and bulk scattering were subtracted from the specular
signal by displacing the detector out of the reflection plane. The
intensity $R(q_{z})$, reflected from the surface, is measured as a
function of the normal component $q_{z}$ of the momentum transfer
and yields information about the surface-normal structure of the
electron density as given by

\begin{equation}
 R(q_{z})=R_{F}(q_{z})\mid \Phi (q_{z})\mid ^{2}\exp [-\sigma
_{cw}^{2}q_{z}^{2}
\label{eq:GaBiMono1}
\end{equation}

\noindent where $R_{F}(q_{z})$ is the Fresnel reflectivity from a flat,
infinitely sharp surface, and $\Phi (q_{z})$ is the Fourier transform of the
local surface-normal density profile $<\widetilde{\rho }(z)>$\cite
{Pershan1984}:

$\bigskip $%
\begin{equation}
\Phi (q_{z})=\frac{1}{\rho _{\infty }}\int dz\frac{d<\widetilde{\rho }(z)>}{%
dz}\exp (iq_{z}z)\text{ \ \ (2),}\ \label{eq:GaBiMono2}
\end{equation}

\noindent with the bulk electron density $\rho _{\infty }$ and the
\index{critical!wavevector}  critical wave vector $q_{crit}$. The
exponential factor in Eq.~\ref{eq:GaBiMono2} accounts for
roughening of the intrinsic density profile $<\widetilde{\rho
}(z)>$ by capillary waves:

$\smallskip $%
\[
\sigma _{cw}^{2}=\frac{k_{B}T}{2\pi \gamma }\ln (\frac{q_{\max }}{q_{res}})%
\text{ \ \ (3),}
\]
\

\noindent where $\gamma $ is the macroscopic surface tension of the free
surface, and $\sigma _{cw}$ is the roughness due to thermally excited
capillary waves (CW) . The CW spectrum is cut off at small $q_{z}$ by the
detector resolution $q_{res}=0.03$\AA $^{-1}$ and at large $q_{z}$ by the
inverse atomic size $a$ , $q_{max\bigskip }\approx \pi /a$.\cite{Ocko1994}.

In Expression (1) the validity of the \index{Born approximation}
Born approximation is tacitly assumed. Since the features in
$R/R_{F}$ characteristic of the thick wetting film appear close to
the \index{critical!wavevector} critical wavevector
$q_{c}=0.049$\AA $^{-1}$ of the Ga-rich subphase, where the Born
approximation is no longer valid, we had to resort to
\index{Parratt formalism} Parratt's dynamical formalism
\cite{Parratt1954} for $q_{z}$ less than $0.25\text{\AA }^{-1}$.
Details of this analysis will be reported
elsewhere.\cite{Huber2001}

\section{Results}

\begin{figure}[tbp]
\centering
\includegraphics[width=0.6\columnwidth]{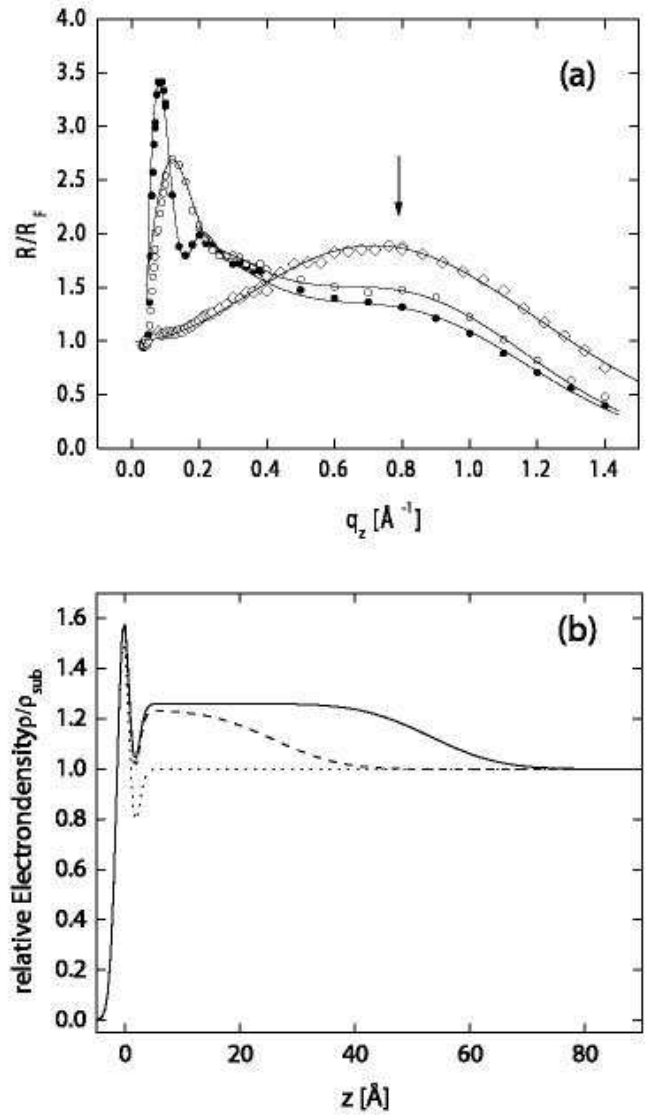}
\caption{\label{fig:Fig2} (a) Fresnel normalized x-ray
reflectivity $R/R_{F}$ from the surface
of Ga-Bi: (diamonds) $T=205%
{{}^\circ}%
C$, (open circles) $T=220%
{{}^\circ}%
C$, (solid circles) $T=222.5%
{{}^\circ}%
C$. solid lines: fit to $R/R_{F}$ with density profiles depicted
in Fig.~\ref{fig:Fig2}(b). The arrow indicates the
$q_{z}$-position for the temperature-dependent reflectivity
measurements at fixed $q_{z}=0.8\AA ^{-1}$ (T-scan) shown in
Fig.~\ref{fig:Fig3}; (b): Electron density profiles for $T=205
{^\circ}C$ (dotted line), $T=220{^\circ}C$ (dashed line) and
$T=222.5 {^\circ} C$ (solid line), normalized to the bulk electron
density of the Ga-rich subphase.}
\end{figure}

Fig.~\ref{fig:Fig2}(a) shows the reflectivity at three
temperatures: well below $T_{mono}$
at $T=205%
{{}^\circ}%
C$ (regime I), well above $T_{mono}$ at $T=222.5%
{{}^\circ}%
C$ (regime II) and at an intermediate temperature $T=220%
{{}^\circ}%
C.$ Due to the loss of phase information (Eq.~\ref{eq:GaBiMono1})
the interesting electron density profiles cannot be obtained
directly from the measured reflectivity. One has to resort to the
widely accepted procedure of adapting a physically motivated model
for the electron density profile and fitting its Fourier transform
to the experimentally determined $R/R_{F}$. The resulting density
profiles for the different temperatures are depicted in
Fig.~\ref{fig:Fig2}(b) and the corresponding fits to $R/R_{F}$ can
be found as solid lines in Fig.~\ref{fig:Fig2}(a).

At $T=205%
{{}^\circ}%
C$ $R/R_{F}$ (Fig.~\ref{fig:Fig2}(a), diamonds) shows a pronounced maximum centered around $%
q_{z}=0.8\AA ^{-1}$. This maximum is indicative of a high electron density
at the surface of the alloy and a previous analysis shows that it is
compatible with the segregation of a monolayer of pure Bi at the
liquid-vapor interface of the bulk alloy.\cite{Tostmann2000}

The normalized reflectivity at $T=222.5%
{{}^\circ}%
C$ (Fig.~\ref{fig:Fig2}(a), solid circles) shows two peaks
(Kiessig fringes) at low $q_{z}$ characteristic of a Bi-rich
wetting film $\approx 50\AA $\ thick. This thickness is in good
agreement with ellipsometric measurements on Ga-Bi alloy surfaces
in the same temperature regime.\cite{Nattland1996} The interfacial
roughness between the Bi-rich and the Bi-poor phase, indicated
by the decay of the Kiessig fringes, is $\approx 12\AA $. Additionally, $%
R/R_{F}$ continues to exhibit an increased intensity around $q_{z}=0.8\AA
^{-1}$: The monolayer of pure Bi is still present at the surface in regime
II, while the Bi-rich wetting film has intruded between it and the Ga-rich
subphase.

The reflectivity at the intermediate temperature $T=220.0%
{{}^\circ}%
C$ (Fig. ~\ref{fig:Fig2}(a), open circles) still exhibits a peak
at low $q_{z}$. However,
compared to the peak at $T=222.5%
{{}^\circ}%
C,$ it is now shifted to higher $q_{z}$. The corresponding
electron density profile, depicted in Fig.~\ref{fig:Fig2}(b) as
the dashed curve, indicates a highly diffuse, thin film of a Bi
enriched phase between the monolayer and the Ga-rich subphase.

To obtain further information about the temperature dependent
transition between the different surface regimes, we performed
temperature-dependent x-ray reflectivity measurements at a fixed
$q_{z}$ in a temperature range close to T$_{mono}$. These
measurements are referred to in the following as T-scans. As can
be seen in Fig.~\ref{fig:Fig2}(a) the reflectivities in the two
regimes differ most distinctly in the low $q_{z}$ part, and
somewhat less, but still significantly in the higher $q_{z}$-part.
Due to the high sensitivity of the reflectivity to temperature
dependent sample height changes at low $q_{z}$, we carried out
measurements at a relatively high $q_{z}=0.8\AA ^{-1}$.

\begin{figure}[tbp]
\centering
\includegraphics[width=1.0\columnwidth]{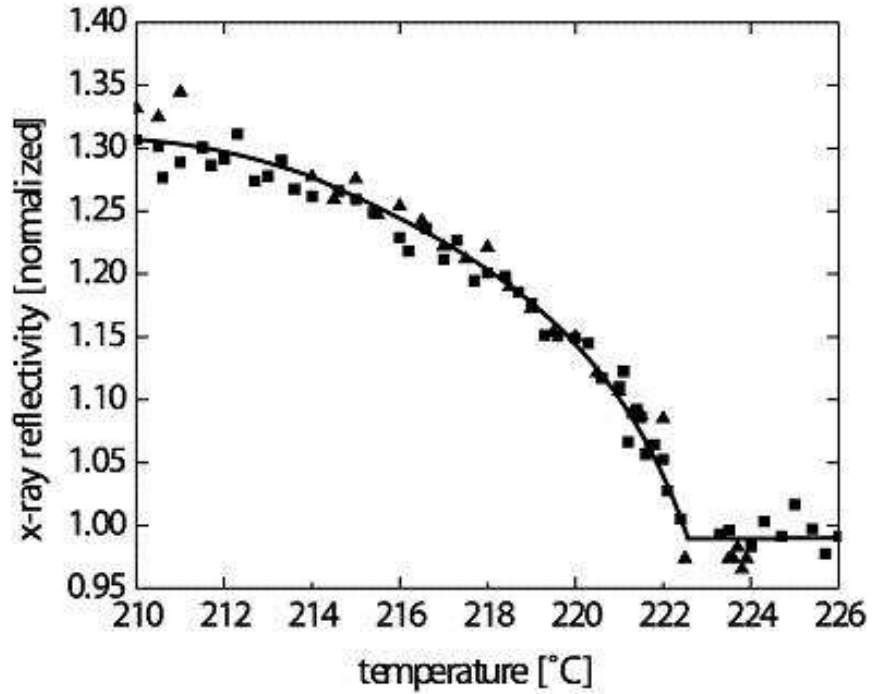}
\caption{\label{fig:Fig3}temperature-dependent x-ray reflectivity
at $q_{z}=0.8\AA ^{-1}$
normalized to $R/R_{F}$ at $T=222.5%
{{}^\circ}%
C$. (squares) decreasing temperature, (triangles) increasing
temperature. line: guide for the eye.}
\end{figure}

A T-scan while cooling (heating) with a cooling (heating) rate of 1$%
{{}^\circ}%
C$/hour between $226%
{{}^\circ}%
C$ and $210%
{{}^\circ}%
C$ is shown in Fig.~\ref{fig:Fig3}. For temperatures higher than
$T_{mono}$ the reflectivity stays constant. Upon $T$ reaching
$T_{mono}$ the intensity
starts rising continuously. When reaching $210%
{{}^\circ}%
C$ the x-ray reflectivity has increased by about 30\% as compared
to its value above $T_{mono}$ - as one would expect from
Fig.~\ref{fig:Fig2}(a) for the change from regime I to regime II
at that $q_{z}$-position. While heating, we observed the inverse
behavior, the intensity decreases until it reaches its lowest
value at $T_{mono}$. No hysteresis was observed between cooling
and
heating for cooling (heating)\ rates smaller than $2%
{{}^\circ}%
C$ per hour. We also carried out one temperature scan with a cooling rate of
10%
${{}^\circ}$%
C per hour. This showed a hysteresis of $4%
{{}^\circ}%
C$ and also the formation of a solid film at the surface. We attribute this
to the expected diffusion-limited growth and the corresponding long
equilibration times in such wetting geometries\cite{Lipowsky1986} and to a
temperature gradient between bulk and surface rather than to an intrinsic
feature of the transition.

\section{Conclusions}

The behavior of the reflectivity in the T-scan shows that the
Bi-rich wetting film forms while approaching $T_{mono}$ from below
and vanishes in the same way on cooling, i.e. without hysteresis.
Hence, the surface sensitive x-ray reflectivity measurements
suggest that the first order bulk transition at the
\index{monotectic point} monotectic point drives a continuous
structural transition at the surface. A similar behavior has been
reported for Ga-Pb, a system with an analogous \index{phase
diagram} phase diagram (monotectic point + consolute point ), by
\index{Wynblatt, Paul} Wynblatt and \index{Chatain, Dominique}
Chatain \cite{Chatain1996}. The topology of the phase diagram
of Ga-Bi forces the approach of the liquid-liquid coexistence line at $%
T_{mono}$ from off coexistence while heating. Conversely, it
enforces a path leading off coexistence while cooling below
$T_{mono}$- see arrows in Fig.~\ref{fig:Fig1}. Therefore, as
\index{Dietrich, Siegfried} Dietrich and \index{Schick, Michael}
Schick \cite{Dietrich1997} have pointed out, one measures along a
path that probes a \index{wetting!complete} complete wetting
scenario (according to the wetting nomenclature). Since the path
ends in the \index{monotectic point} monotectic point with its
four phase coexistence, the phenomenon is readily described as a
complete wetting at a \index{wetting!tetra-point} tetra point of
four-phase coexistence. It should, however, be noted that
$T_{mono}$ is not the wetting temperature, $T_{w}$ of the
liquid-liquid system. $T_{w}$ is defined as the temperature where
the formation of the wetting film can be observed on coexistence
between the two liquids. This temperature has to be below
$T_{mono}$, but it would be experimentally accessible only if the
pure Bi can be supercooled sufficiently to reach the liquid-liquid
coexistence, albeit in a metastable condition.\newline

The observed transition at the surface is closely related to
\index{wetting!triple point} triple point wetting phenomena
\cite{Pandit1983} which have been extensively studied for one
component systems like Krypton on graphite \cite{Zimmerli1992}.
There, one observes that the liquid wets the interface between
substrate and vapor, while the solid does not. Hence, there is a
wetting transition pinned at the bulk \index{wetting!triple point}
triple point $T_{3}$. In the triple point wetting scenario one
walks off the liquid-vapor coexistence line following the
solid-vapor sublimation line.

Another aspect of the wetting behavior close to the
\index{monotectic point} monotectic point is the
possible nucleation of solid Bi at the surface while cooling below $T_{mono}$%
. Systematic studies on that aspect were reported only recently
for different Ga-Bi alloys, concluding that Ga-Bi shows a
phenomenon that can be described as \index{surface!freezing}
surface freezing. The free surface acts as substrate for wetting
of the liquid by the forming solid Bi \cite{Wang2000}\cite
{turchanin2001}. We did also observe the formation of solid Bi at
the
surface while cooling below $200%
{{}^\circ}%
C$, but we could not study this effect in detail, since the forming solid
film destroyed the flat surface of the liquid sample hampering further
reflectivity measurements.

A more detailed experimental study of the wetting transition at $T_{mono}$
has already been performed \cite{Huber2001}. It has revealed the evolution
of the thick film via intermediate film structures dominated by strong
concentration gradients - as reported here for $T=220%
{{}^\circ}%
C$. This behavior is in agreement with density functional
calculations for wetting transitions of binary systems at
\index{hard wall} hard walls \cite{Davis1996}, which also reveal
concentration gradients. Conversely, the study shows that the
liquid vapor interface of a metal system acts not only as a
\index{hard wall} hard wall for one component systems
\cite{Harris1987} forcing the ions into ordered layers parallel to
that surface \cite{Magnussen1995},\cite{Regan1995}, but also
affects other structure and thermodynamic phenomena at the
surface, e.g. the wetting behavior in a binary liquid metal system
discussed here.

\bibliographystyle{unsrt}

%% file: GaBi_all.tex
\chapter{Short-Range Wetting at Alloy Surfaces: Square Gradient Theory}

\section{Abstract}\index{wetting!short-range}
In this chapter of the thesis we present an x-ray reflectivity
study of wetting at the free surface of the binary liquid metal
gallium-bismuth (Ga-Bi) in the region where the bulk \index{phase
separation} phase separates into Bi-rich and Ga-rich liquid
phases. The measurements reveal the evolution of the microscopic
structure of \index{nanoscale} nanometer-scale
\index{wetting!film} wetting films of the Bi-rich,
low-surface-tension phase along different paths in the bulk
\index{phase diagram} phase diagram. A balance between the surface
potential preferring the Bi-rich phase and the gravitational
potential which favors the Ga-rich phase at the surface pins the
interface of the two demixed liquid metallic phases close to the
free surface. This enables us to resolve it on an $\AA$ngstr\"om
level and to apply a  \index{mean-field square gradient model}
mean-field, square gradient model extended by thermally activated
capillary waves as dominant thermal fluctuations. The sole free
\index{fitting!parameters} parameter of the gradient model, i.e.
the so-called influence parameter, $\kappa$, is determined from
our measurements. Relying on a calculation of the liquid/liquid
interfacial tension that makes it possible to distinguish between
intrinsic and capillary wave contributions to the interfacial
structure we estimate that fluctuations affect the observed
short-range, \textit{complete} wetting phenomena only marginally.
A \textit{critical} wetting transition that should be sensitive to
thermal fluctuations seems to be absent in this binary metallic
alloy.

\section{Introduction}
The concept of a wetting transition that was introduced
independently by  \index{Cahn, John} Cahn \cite{Cahn77} and Ebner
and Saam \cite{Ebner77} in 1977, has stimulated a substantial
amount of theoretical and experimental work\cite{DeGennes85,
Sullivan86, Dietrich88, Law01, Bonn01}. Due to its critical
character it is not only important for a huge variety of
technological processes ranging from alloying to the flow of
liquids, but has the character of an extraordinarily versatile and
universal physical concept, which can be used to probe fundamental
predictions of statistical physics. For example, as shall be
depicted in more detail later on, it can be employed in the case
of wetting dominated by short-range interactions (SRW) in order to
test the predictions for the break-down of  \index{mean-field
square gradient model} mean-field behavior and the necessary
''transition'' to a \index{renormalization group} renormalization
group regime, where this surface phenomenon is significantly
affected by thermal fluctuations.

A wetting transition occurs for two fluid phases in or near
equilibrium in contact with a third inert phase, e.g., the
container wall or the liquid-vapor interface. On approaching the
coexistence critical point the fluid phase that is energetically
favored at the interface forms a nano scale \index{wetting!film}
wetting film that intrudes between the inert phase and the other
fluid phase. In general, this surface phenomenon is a delicate
function of both the macroscopic thermodynamics of the bulk phases
and the microscopic interactions.

Whereas one of the seminal theoretical works\cite{Ebner77} on the
wetting transition gave a microscopic view on this phenomenon,
experimental results at this same level of detail were only
recently obtained through application of X-ray and \index{neutron
scattering} neutron reflection and diffraction
techniques\cite{Plech00, Tostmann00, Schlossman02}. Moreover,
almost all of these experimental studies dealt with systems, like
methanol-cyclohexane and other organic materials, that are
dominated by long-range \index{interactions!Van der Waals}
van-der-Waals interactions\cite{Bonn01}. The principal exceptions
to this are the studies of the binary metallic systems
gallium-lead (Ga-Pb) \cite{Chatain96}, gallium-thallium (Ga-Tl)
\cite {Shim01}, and gallium-bismuth (Ga-Bi) \cite{Nattland95}, for
which the dominant interactions are short-range.

\begin{figure}[tp]
\centering
\includegraphics[width=0.6\columnwidth]{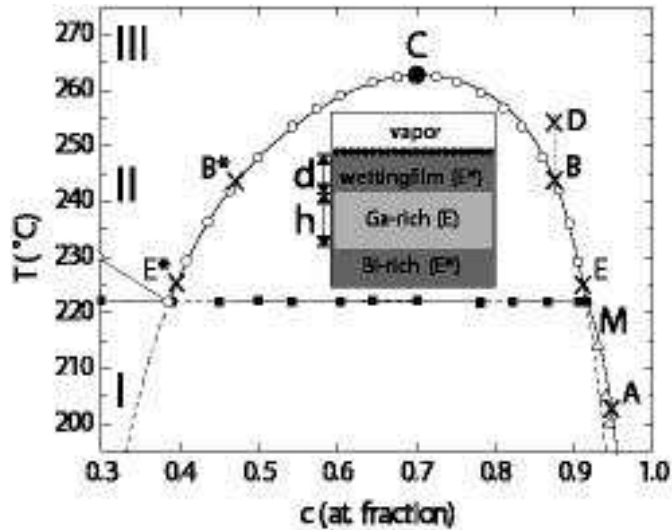}
\caption{\label{fig:GaBiPDcT}The atomic fraction
($c$)-temperature($T$) bulk \index{phase diagram} phase diagram of
Ga-Bi. The symbols indicate coexistence lines of Predel's phase
diagram. The lines show the phase boundaries calculated from
thermodynamic data. The dashed lines represent the metastable
extension of the (l/l) coexistence line below $T_M$. The points
are: C-bulk critical point, \index{monotectic point} M-monotectic
point, A,B,D,E-points on the experimental path. The insets
illustrate the surface and bulk phases. In region II the
\index{wetting!film} wetting film is 50~$\AA$ thick and the
Ga-rich fluid is 5 mm thick. The bold circles in the insets
symbolize the Bi-monolayer. }
\end{figure}

Here, we will present an x-ray reflectivity study of wetting
phenomena that occur at the free surface in the binary metallic
liquid Ga-Bi when the bulk demixes in two liquid phases, i.e. a
Bi-rich and a Ga-rich liquid phase. The fact that the wetting
geometry pins the liquid/liquid (l/l) interface that divides the
two liquids to the surface allows measurements that show the
compositional profile of the \index{nanoscale}  nanoscale
\index{wetting!film} wetting film at the $\AA$ngstr\"om level. We
will show results for the structure of the film as it evolves
towards the gravity-thinned film. Through a combination of this
structural information and the bulk thermodynamics of the system,
it was possible to extract detailed information on the dominant
interaction parameters governing the surface phenomenology. We
will show that a square gradient theory that is combined with the
effects of thermally excited capillary waves provides a reasonable
description of that interface. In fact, we will be able to extract
the sole free parameter of that model, i.e. the influence
parameter $\kappa$. This, in turn, will allow us to distinguish
between intrinsic (mean-field) and fluctuation (capillary waves)
contributions to the interfacial structure. On the basis of this
distinction we will estimate the influence of fluctuations on the
observed SRW at the free surface.

Moreover, the knowledge of the energetics of the (l/l) interface
will allow us to discuss from a more general perspective the rich
wetting phenomenology which has been reported here and in former
works on this system, ranging from \index{nearly free electron}
\index{surface!freezing} surface freezing to
\index{wetting!tetra-point} tetra point wetting.

This paper is structured as follows: In the first section, we
introduce the bulk \index{phase diagram} phase diagram of Ga-Bi
and relate its topology to the wetting transitions observable at
the free surface of this binary alloy. X-ray reflectivity
measurements on the \index{wetting!film} wetting films along
different paths in the bulk phase diagram will be discussed in the
second section. The third section will focus on the thermodynamics
and structure of the liquid-liquid interface. In this section we
will develop a \index{square gradient theory} square gradient
theory in order to model the concentration profile at the
liquid/liquid interface. Finally, in the last section we will
provide a more general discussion of the rich wetting
phenomenology at the free surface of this binary liquid metal.

\section{Bulk \& Surface Thermodynamics}
The bulk phase diagram of Ga-Bi, cp. Figure ~\ref{fig:GaBiPDcT},
was measured by Predel with differential thermal
analysis\cite{Predel60}. It is dominated by a \index{miscibility
gap} miscibility gap with consolute point C (critical temperature
$T_C=262.8^{\circ}C$, critical atomic fraction of Ga, $c_{
crit}=0.7$) and a \index{monotectic point} monotectic temperature,
$T_M=222^{\circ}C$. In the part of region I with $T<T_{ M}$, solid
Bi coexists with a Ga-rich liquid. At $T_{ M}$, the boundary
between region I and region II, a first order transition takes
place in the bulk due to the liquidisation of pure Bi. For $T_{
M}<T<T_{ C}$ (region II), the bulk separates into two
\index{immiscible phase} immiscible phases, a high density Bi-rich
liquid and a low density Ga-rich liquid. The heavier Bi-rich phase
is macroscopically separated from the lighter Ga-rich phase due to
gravity. In region III, beyond the miscibility gap, a homogeneous
liquid is found.

 Additionally, the plot depicts the extension of the l/l
coexistence line for $T<T_{ M}$. This illustrates \index{Dietrich,
Siegfried} Dietrich and \index{Schick, Michael} Schick's
observation that the path leading to coexistence on heating from A
to M is dictated by the topology of the \index{phase diagram}
phase diagram and corresponds to a path which probes
\index{wetting!complete} "complete wetting". Moreover, Fig
incorporates the thermodynamic path towards D
 the costarting at the critical point C running to the \index{monotectic point} monotectic
point M.

\begin{figure}[tp]
\centering
\includegraphics[width=1.0\columnwidth]{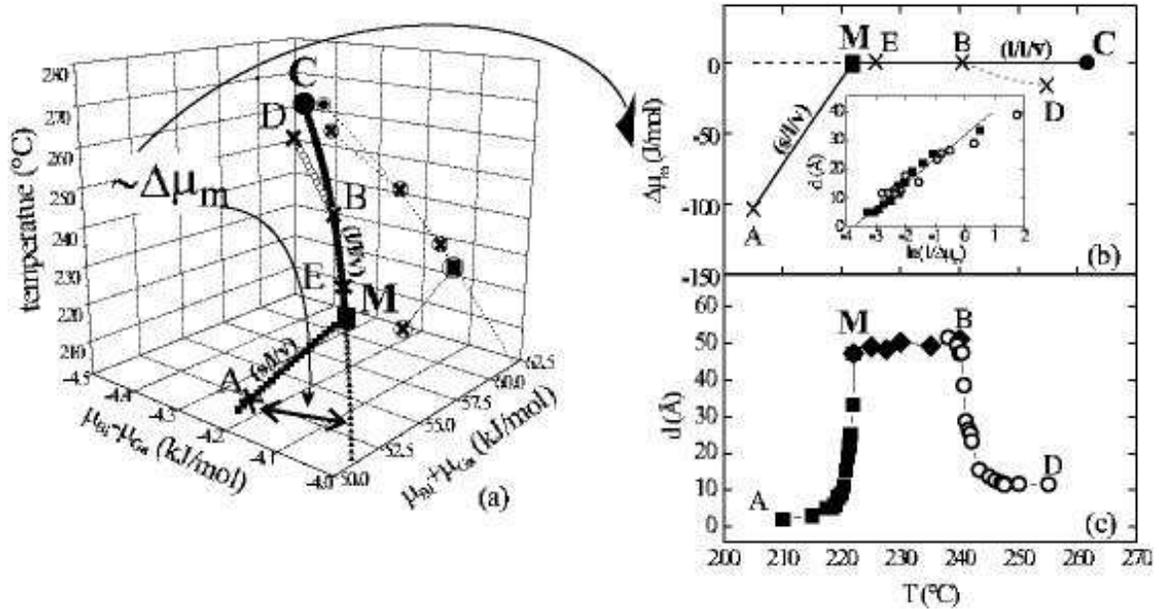}
\caption{\label{fig:GaBiPDmuT}\textbf{(a)} The \index{chemical
potential} chemical potential ($\mu$)-temperature ($T$) bulk
\index{phase diagram}  phase diagram of Ga-Bi. The axes are
temperature, $T$, the difference $\left ( \mu_{ Bi}-\mu_{ Ga}
\right) $ of the chemical potentials of the single species, and
their sum $\left ( \mu_{ Bi}+\mu_{ Ga} \right )$. The solid
symbols indicate the following coexistence lines: (solid circles)
liquid/liquid/vapor (l/l/v) triple line, (solid squares)
solid/liquid/vapor (s/l/v) triple line, (small triangles)
metastable extension of the liquid/liquid/vapor (l/l/v) triple
line below $T_{ M}$, (open circles) experimental path B-D probing
\index{wetting!complete} complete wetting. The points are: C-bulk
critical point, M-monotectic point, A,B,D,E-points on the
experimental path. To illustrate the 3-dimensional structure of
the phase diagram, a projection of the phase boundary lines on the
($\mu_{ Bi}-\mu_{ Ga}$,$T$) plane is drawn in the plot. The
symbols surrounded by a circle show where the aforementioned
characteristic points fall on this projection.
\textbf{(b)}($\Delta \mu_{ m}$,$T$) phase diagram: (A-M) is the
(s/l/v), and (M-C) is the (l/l/v) coexistence line. The path B-D
is in the single phase region, and M and C are the monotectic and
critical points. Inset: effective wetting layer thickness $d$ on
A$\rightarrow $M (squares) and B$\rightarrow $D (open circles).
The solid line is a fit to the A$\rightarrow $M $d$-values.
\textbf{(c)} The measured $d$ along the experimental path. }
\end{figure}

Following the observation of Perepezko\cite{Perepezko82} that on
cooling fine Ga-rich droplets are coated by a Bi-rich solid phase,
\index{Nattland, Detlef} Nattland et. al. studied the liquid-vapor
interface in region II using \index{ellipsometry} ellipsometry.
They found that a thin Bi-rich film intrudes between the vapor and
the Ga-rich subphase in defiance of gravity \cite{Nattland95}, as
implied by the illustration in Figure \ref{fig:GaBiPDcT}. This was
a clear example of the critical point wetting in binary systems
that was described by \index{Cahn, John} Cahn\cite{Cahn77}. On
approaching to point C, the Bi-rich phase necessarily becomes
energetically favored at the free surface and as a result it forms
the \index{wetting!film} wetting film that intrudes between the
Ga-rich subphase and the surface. In fact the situation is
slightly more complicated since x-ray studies indicate that
throughout region III the free surface is coated by a monolayer of
pure Bi\cite{Lei96}. More recently we showed that on heating the
binary liquid a thicker wetting layer of Bi rich liquid forms
between the Bi monolayer and the bulk Ga rich liquid. This appears
to be an unusual example of complete wetting that is pinned to the
monotectic temperature $T_M$. This phenomenon was first discussed
by \index{Dietrich, Siegfried} Dietrich and \index{Schick,
Michael} Schick in order to explain an analogous finding in the
binary metallic alloy Ga-Pb \cite{Chatain96, Dietrich97}. The
nature of this apparent coincidence of a surface transition with
the first order transition in the bulk at $T_M$ can most easily be
illustrated by a transformation of the ($c,T$)-diagram to the
appropriate \index{chemical potential} chemical
potential-temperature ($\mu$, $T$)-diagram that is depicted in
Figure  \ref{fig:GaBiPDmuT}(a) for Ga-Bi. The axis are
temperature, $T$, the difference $\left ( \mu_{ Bi}-\mu_{ Ga}
\right) $ of the \index{chemical potential} chemical potentials of
the single species, and their sum $\left ( \mu_{ Bi}+\mu_{ Ga}
\right )$. In this plot, the \index{miscibility gap}
(l/l)-miscibility gap of Figure \ref{fig:GaBiPDcT}, which strictly
spoken is a liquid/liquid/vapor (l/l/v) coexistence boundary,
transforms into a (l/l/v)-triple line extending from $M$ to $C$
(solid circles). At M this triple line intersects with another
triple line, i.e. the solid/liquid/vapor (s/l/v)-coexistence line
(solid squares), rendering M a \index{wetting!tetra-point}
\textit{tetra point} of four phase
coexistence\footnote{Additionally, the (s/l/v) triple line due to
the coexistence of a Bi-rich liquid, a pure Bi solid, and the
vapor phase - cp. the left side of the ($c$,$T$) \index{phase
diagram} phase diagram (Figure  \ref{fig:GaBiPDcT}) starts at M.
Since it is not relevant for the considerations of the wetting
thermodynamics at the surface, it is not shown in Figure
\ref{fig:GaBiPDmuT}(a),(b) for the sake of simplicity.}. At M a
solid Bi, a Ga-rich , a Ga-poor phase, and the vapor coexist.

The thermodynamics by which the surface transition at M is pinned
by the bulk phase transition is obvious if one considers the
topology of the ($\mu$,T)-plot in the proximity of M. The wetting
of the free surface by the Bi-rich phase as well as the bulk
transition are driven by the excess free energy, $\Delta\mu_{m}$,
of the Bi-rich phase over that of the Ga-rich liquid phase
\cite{Dietrich88}. This quantity is proportional to the distance
between the (l/l/v)-triple line and any other line leading off
(l/l/v)-coexistence, e.g. the (s/l/v)-triple line
(A$\rightarrow$M) or the line B$\rightarrow$D in Figure
\ref{fig:GaBiPDmuT}(a). The wetting thermodynamics is displayed in
a slightly simpler way by the plot $\Delta \mu_{ m}$ vs. $T$ in
Figure  \ref{fig:GaBiPDmuT}(b). In this figure for $T>T_{M}$, the
(l/l/v) coexistence line transforms into a horizontal straight
line that extends from M to C. For $T<T_{ M}$ the horizontal
dashed line indicates the metastable (l/l/v) extension of the
coexistence and that is above the solid-Bi/Ga-rich/vapor (s/l/v)
coexistence line that goes from M to A. This illustrates the
observation by \index{Dietrich, Siegfried} Dietrich and
\index{Schick, Michael} Schick\cite{Dietrich97} that the path
A$\rightarrow$ M leads to coexistence, and thus
\index{wetting!complete} complete wetting is dictated by the
topology of the phase diagram.

A more quantitative understanding of the surface wetting phenomena
can be developed by analyzing the grand canonical potential,
$\Omega _{ S}$ per unit area A of the surface\cite{Dietrich88}:
$\Omega _{ S}/A=d\, \Delta\mu+\gamma_0\, e^{-d/\xi}$. Here, $d$ is
the wetting film thickness, $\xi$ is the decay length of a
short-range, exponential decaying potential, $\gamma_0$ its
amplitude and $A$ an arbitrary surface area. The quantity
$\Delta\mu$ comprises all energies that are responsible for a
shift off true bulk (l/l/v) coexistence, i.e. the aforementioned
quantity $\Delta\mu_m$. The formation of the heavier Bi-rich
wetting layer at some height, $h$, above its bulk reservoir costs
an extra gravitational energy $\Delta\mu_g$=$g\Delta\rho_m h$
where $\Delta\rho_{ m}$ is the mass density difference between the
two phases. Minimization of $\Omega_{ S}$ in respect to $d$ then
yields the equilibrium \index{wetting!film} wetting film thickness
of the Bi-rich phase $d=\xi \ln(\gamma_0/\Delta\mu)$. In fact the
gravitational energy is only significant in comparison with the
other terms for very small values $\Delta\mu_m$ and for most of
the data shown in Figure \ref{fig:GaBiPDmuT} the gravitational
term can be neglected, allowing  of $\Delta\mu=\Delta\mu_m$. Thus,
one expects a logarithmic increase of the wetting film thickness
upon approaching M from A that is given by $d=\xi
\ln(\gamma_0/\Delta\mu_m)$, in agreement with the experimental
finding that was presented in our recent Physical Review Letter on
this phenomenon - cp. inset in Figure  \ref{fig:GaBiPDmuT}(b)
\cite{Huber02}. The slight deviations for small values of
$\Delta\mu_m$, as well as the finite value of $d\sim50~{\AA}$
along the coexistence line M$\rightarrow$B are due to the
gravitational term. Since this approach of the (l/l/v)-coexistence
line ends in M with its four phase coexistence, the phenomenon is
properly described as \index{wetting!tetra-point} \textit{tetra
point wetting}. As demonstrated here, the occurrence of this
complete wetting phenomenon at the surface is an intrinsic feature
of the bulk phase diagram.

In this paper we shall present additional x-ray reflectivity
measurements that show the evolution of the wetting film on
approaching coexistence from point D in regime III, and hence
along a path probing complete wetting that is not dictated by the
topology intrinsic to the bulk \index{phase diagram} phase
diagram, but rather by the experimentator's choice of the overall
atomic fraction of Ga in the sample $c_{ nom}$, i.e. the path
B$\rightarrow$D in Figure \ref{fig:GaBiPDcT} and Figure
\ref{fig:GaBiPDmuT}(a,b). Furthermore, we will probe the wetting
film structure along an on-(l/l/v) coexistence path B$\rightarrow$
E.

\section{X-ray reflectivity measurements}
\subsection{Experiment}

\begin{figure}[tp]
\centering
\includegraphics[width=1.0\columnwidth]{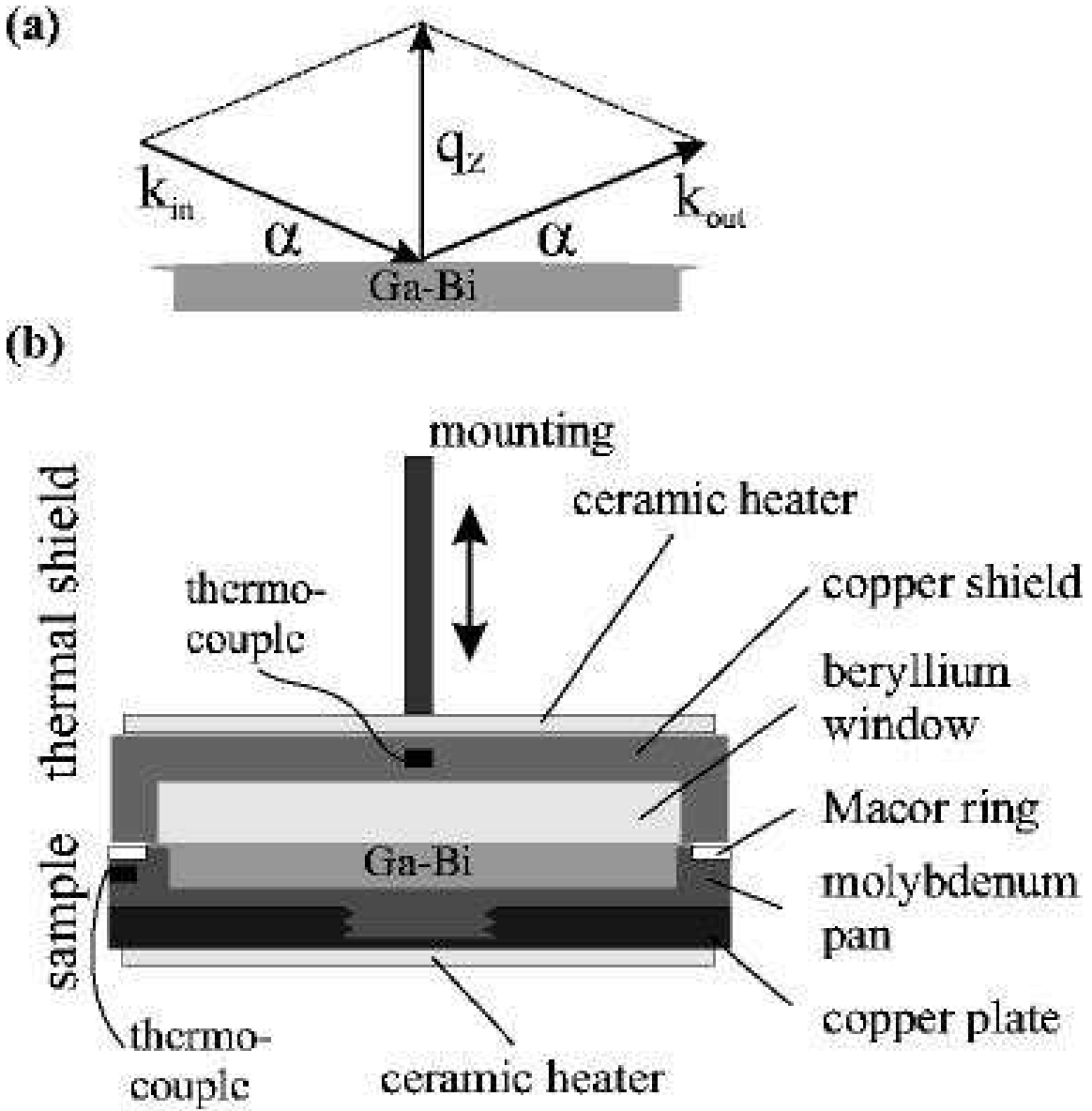}
\caption{\label{fig:ExpSetup} (a) scattering geometry (b) Sketch
of the experimental setup. The arrow close to the mounting
indicates the variable position of the thermal shield in respect
to the sample surface.}
\end{figure}

\subsubsection{Sample Preparation \& Sample Environment}
The Ga-Bi alloy was prepared in an inert-gas box using
$>99.9999\%$ pure metals. A solid Bi pellet was covered by an
amount of liquid Ga required for a nominal concentration $c_{
nom}=88$ at\% Ga. It was then transferred in air into an ultrahigh
vacuum chamber. A 24-hour period of bake-out yielded a pressure of
10$^{-10}$ torr. The residual surface oxide on the liquid's
surface was removed by sputtering with Ar$^{+}$ ions. Using
thermocouple sensors and an active temperature control on both
sample pan and its adjacent thermal shield a temperature stability
and uniformity of $\pm 0.05^{\circ }C$ was achieved. A sketch of
the experimental setup, sample environment resp. can be found in
Figure  \ref{fig:ExpSetup}(b).

A challenge in all x-ray reflectivity measurements of liquid
metals is the high surface tension of these systems, e.g. pure Ga
has a surface tension of $\approx$700mN/m. It often prevents
liquid metal from wetting non-reactive container walls or
substrates. This also leads to large \index{curvature} curvatures
of the liquid surface hampering x-ray reflectivity
measurements\cite{Regan97}. Remnant oxide layers that exists at
the liquid/container, liquid/substrate interface resp. even
enhance this non-wetting effect; therefore, we removed these oxide
layers from the Mo sample pan by sputtering with Ar$^{+}$ ions, at
a sputter current of 25mA and a sputter voltage of 2kV, and could
reach wetting of the Mo crucible by the liquid metal. This
resulted in small \index{curvature} curvatures of the free surface
as was judged by eye and later on by x-ray reflectivity
measurements. The resulting flat surfaces facilitated the
accumulation of reliable x-ray reflectivity data sets,
particularly for small incident angles, $\alpha$. On the other
hand, the wetting of the container wall by the liquid alloy
promoted some spilling of the liquid during the sample movements
that were necessary to track the sample position during the
reflectivity scan. We circumvented this problem by installing a
ceramic ring that surrounded the sample pan. The ceramics is not
wetted by the liquid metal and, therefore, inhibits any liquid
flow at the outermost circumference of the sample pan.

Another experimental challenge is related to x-ray reflectivity
measurements from liquids in general. The surface of a liquid is
sensitive to any kind of vibrations or acoustic noise. In order to
obtain a stable reflected signal from the liquid surface, we
mounted our \index{UHV!chamber} UHV chamber on an active
\index{vibration isolation} vibration isolation\cite{Regan97}.

\subsubsection{X-Ray Reflectivity}
X-ray reflectivity measurements were carried out using the liquid
surface reflectometer at beamline X22B at National
\index{synchrotron!source} Synchrotron Light Source with an x-ray
wavelength $\lambda =1.54~$\AA . Background and bulk scattering
were subtracted from the specular signal by displacing the
detector out of the reflection plane by $0.3 ^{\circ}$. The
scattering geometry can be found as inset in Figure
\ref{fig:ExpSetup}. The intensity $R(q_{ z})$, reflected from the
surface, is measured as a function of the normal component $q_{ z}
=4\pi / \lambda \sin(\alpha)$ of the momentum transfer and yields
information on the surface-normal structure of the electron
density $\rho (z)$ as described by the so-called master
formula\cite{Pershan84}. This formula relates the
Fourier-transformed electron density gradient with the
experimentally obtained $R/R_{ F}$. The symbol $R_{ F}$ denotes
the Fresnel reflectivity that is expected from an ideally flat and
sharp surface having the electron density of the Ga-rich liquid.
The standard procedure for determining the electron density
profile $\rho(z)$ from the measured reflectivity $R(q_{ z})$ is to
construct a simple and physically meaningful density model and to
fit its Fourier transform to experimentally obtained data sets of
$R/R_{ F}$\cite{Pershan94}. We employ a three-box model
\cite{Tostmann00}, where the upper box represents the
\index{Gibbs!adsorption} Gibbs-adsorbed Bi monolayer the second
represents the Bi-rich \index{wetting!film} wetting film and the
lower box represents the bulk liquid. The relative electron
densities of these boxes correspond to pure Bi (top box) and to
the electron density of the Bi-rich wetting film, $\rho$. The
quantity $\rho_{ sub}$ denotes the electron density of the Ga-rich
subphase. In the simplest approximation the electron density
profiles of the interfaces between the different phases can be
described analytically by error-functions. The relevant model for
this problem includes three error-functions (erf)-diffusenesses
for the following interfaces: vapor/Bi monolayer, Bi monolayer/Bi
rich film, Bi rich film/Ga rich bulk phase. The first two
interfaces describing the monolayer feature were remained
unchanged in the presence of the Bi-rich film, while only the
diffuseness of the Bi rich film/Ga rich bulk phase interface,
$\sigma_{ obs}$, was a variable parameter as a function of $T$.
The model also included size parameters that describe the
thickness of the two upper boxes. The box that describes the bulk
sub-phase extends to infinity. During the \index{fitting!function}
fitting the thickness of the monolayer box was kept constant while
the thickness of the Bi-rich film was allowed to vary.

\begin{figure}[tp]
\centering
\includegraphics[width=1.0\columnwidth]{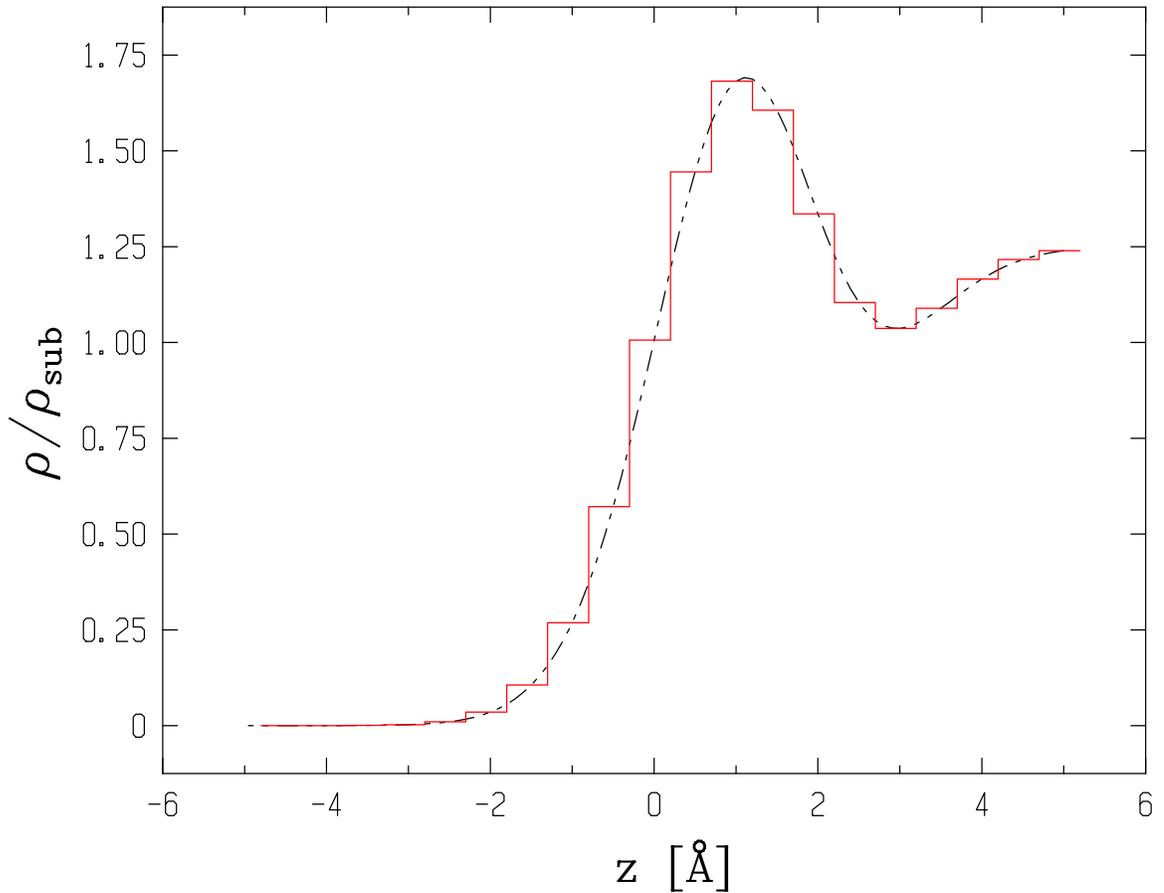}
\caption{\label{fig:Parratt} Illustration of the approximation of
the analytic electron density profile (dashed line) by thin slabs
in order to apply the \index{Parratt formalism} Parratt formalism.
Here, only the region close to the monolayer feature is depicted.}
\end{figure}

The use of the master formula tacitly assumes the validity of the
\index{Born approximation} Born approximation. This assumption
holds true for wave vectors, $q_{ z}$, much larger than the value
of the \index{critical!wavevector} critical wavevector $q_{ c}$ of
the Ga-rich subphase, where multi-scattering effects can be
neglected\cite{Tolan99}. Here, however, we will be interested in
features corresponding to 30-60~$\AA$ thick \index{wetting!film}
wetting films, which when translated to momentum-space corresponds
to features in $q_{ z}$ around 0.05-0.1~${\AA}^{-1}$ that are
comparable to the value of the critical wavevector $q_{ c} \approx
0.05~{\AA}^{-1}$ (The q$_c$-values change only slightly depending
on $T$, cp. Table \ref{tab:FitsAB}). In order to deal with this we
resort to the recursive \index{Parratt formalism} Parratt
formalism \cite{Parratt54} for $q_{ z}$-values close to $q_{ c}$.
In this formalism one develops a 2x2 matrix that relates the
amplitudes and phases for the incoming and outgoing waves on both
sides of a slab of arbitrary dielectric constant. For the present
problem we approximate the above mentioned three-box profile by a
large number of thinner slabs whose amplitudes are chosen to
follow the envelope of the analytic profile - cp. Figure
\ref{fig:Parratt}. Typically the number of slabs was of the order
of 300.

\subsection{Wetting film structures approaching (l/l/v)-coexistence}

\begin{figure}[tp]
\centering
\includegraphics[width=0.8\columnwidth]{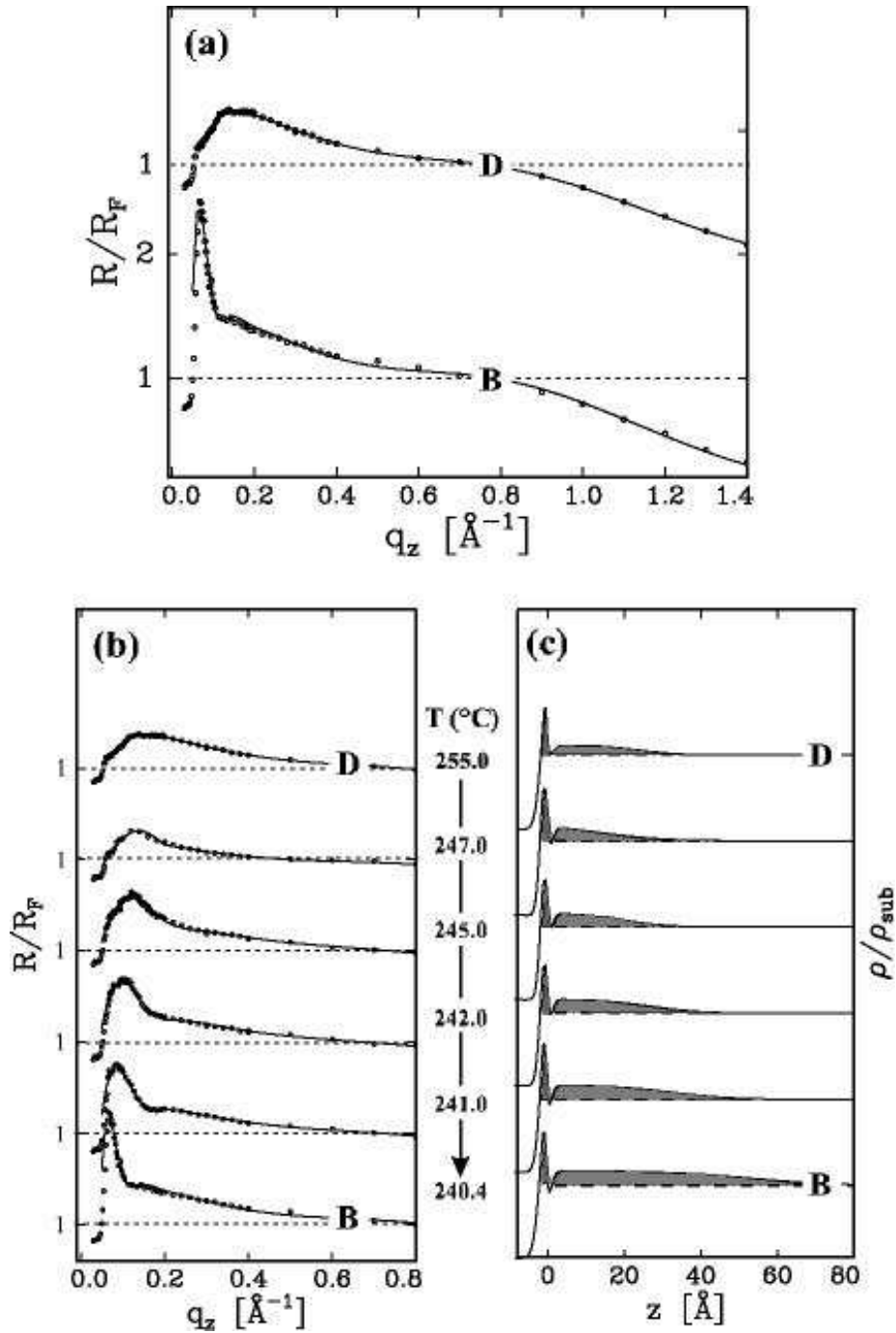}
\caption{\label{fig:RRFDensPathBD}\textbf{(a)} Normalized
reflectivities $R/R_{ F}$ for selected points D and B
corresponding to $T_{ D}$=255.0$^{\circ}$C, $T_{
B}=240.4^{\circ}$C. Dashed lines indicate $R/R_{ F}=1$.
\textbf{(b)} $T$-dependent normalized reflectivities $R/R_{ F}$
while approaching coexistence on path D $\rightarrow$B along with
fits. Dashed lines indicate $R/R_{ F}=1$.\textbf{(c)} Electron
density profiles $\rho$/$\rho_{ sub}$. All regions with
$\rho/\rho_{ sub}>1$ indicating Bi-enrichment as compared with the
Ga-rich subphase are gray shaded.}
\end{figure}

Here we will present x-ray reflectivity measurements from the free
surface of Ga-Bi along a path approaching (l/l/v) coexistence from
the homogeneous part of the bulk phase diagram (regime III), i.e.
the path D$\rightarrow$B . This path, particularly the position of
its characteristic point B, is solely determined by the overall
atomic fraction of Ga in the sample pan, $c_{ nom}$. While cooling
this path intersects the (l/l/v) triple line (cp. Figure
\ref{fig:GaBiPDmuT}) at the point B, corresponding to temperature
$T_{ B}$ of the miscibility boundary (cp. Figure
\ref{fig:GaBiPDcT}). For $T<T_{ B}$ the heretofore homogeneous
bulk liquid \index{phase separation} phase separates into the
heavier Bi rich liquid that settles towards the bottom of the pan
and the lighter Ga rich liquid on top. The actual position in the
($c$,$T$)-plane was chosen from consideration of the vanishing
electron density contrast between Ga-rich and Bi-rich phase upon
approaching C on the one hand or the (l/l/v) triple line (cp.
Figure \ref{fig:GaBiPDmuT}(a)) enforcing the \index{spinodal
demixing} spinodal demixing of the afore homogeneous bulk liquid.
In our case, its position in the ($c$,$T$)-plane was chosen after
a consideration of the vanishing electron density contrast between
Ga-rich and Bi-rich phase upon approaching C, on the one hand, and
a reasonably ''long path'' on (l/l/v) coexistence on the other
hand. The compromise was a value of about $12\%$ for c$_{ nom}$,
which corresponds to $T_{ B}=240.4^{\circ}C$.

X-ray reflectivity $R(q_{ z})$ was measured at selected
temperatures on path D$\rightarrow$B. In order to avoid artifacts
associated with the fact that the evolution of the wetting film is
governed by slow diffusion processes \cite{Lipowsky86, Wu86,
Fenistein02} equilibration was monitored by continuous taking
repeated reflectivity scans following increments of small
temperature steps of $0.5^{\circ}C$. Typically the measured
reflectivity fluctuated wildly for a couple of hours, after which
it slowly evolved to a stable equilibrium. The fits (lines) to
these ''equilibrium'' $R/R_{ F}$ (points) are shown in Figure
\ref{fig:RRFDensPathBD}(a,b), and the corresponding $\rho (z)$
profiles - in Figure ~\ref{fig:RRFDensPathBD}(c).  At point D
[T$_{ D}=255^{\circ}C$], typical of region III, $R/R_{ F}$
exhibits a broad peak at low $q_{ z}$ as well as an increased
intensity around $q_{ z}=0.8~\AA^{-1}$. The corresponding electron
density profile $\rho(z)$ obtained from the fit indicates a thin,
inhomogeneous film of increased electron density (as compared with
$\rho_{ sub}$) close to the surface along with a peak right at the
surface. It is consistent with a segregated monolayer of pure Bi
along with a thin layer of a Bi-enriched phase. As the temperature
is decreased towards B, the peak is gradually shifting to lower
$q_{ z}$ and its width decreases. This behavior manifests the
continuous growth in thickness of the wetting layer upon
approaching $T_{ B}$ and is in agreement with the thermodynamic
path probing \index{wetting!complete} complete wetting:
$\Delta\mu_{ m}\rightarrow0$ on path D$\rightarrow$B.

Our experimental data clearly indicate film structures dominated
by sizeable gradients in the electron density, which contrasts
with frequently used "homogeneous slab" models, but is in
agreement with theoretical calculations. These range from density
functional calculations via square gradient approximations to
Monte Carlo simulations for wetting transitions at hard walls
\cite{Ebner77, Davis96, Schmid01}. Inhomogeneous profiles have
also been observed experimentally in microscopically resolved
wetting transitions for systems dominated by long-range
\index{interactions!Van der Waals} van-der-Waals interactions
\cite{Plech00, Plech01}. Clearly detailed interpretation of the
inhomogeneity for the GaBi nano-scale \index{wetting!film} wetting
films will require either a density functional analysis, or some
other equivalent approach. Nevertheless, even a simple model
approximating the wetting layer by a slab of thickness $d$ allows
a reliable determination of the surface potential governing this
complete wetting transition. In order to do so, effective film
thicknesses $d$ have been extracted form the $\rho(z)$ profiles.
In Figure  \ref{fig:GaBiPDmuT}(c) we show plots of these
$d$-values versus both $T$ and $\Delta\mu_{ m}$. The last plot
shows the expected logarithmic behavior and allows determination
of the values for the parameters of the short-range potential,
i.e. $\gamma_{ 0}=400mN/m$ and $\xi=5.4~\AA$ \cite{Huber02}.
Moreover, agreement between the new ($d$,$\mu_m$)-behavior along
$D \rightarrow B$ and the data for the path A$\rightarrow$M
provides an experimental prove that both paths have the same
thermodynamic character, i.e. they probe complete wetting:
$\Delta\mu_{ m}\rightarrow 0 $ on path A$\rightarrow$M and
D$\rightarrow$B.

\subsection{Wetting film structures: On (l/l/v)-coexistence}
\label{fig:RRFDensAB}
\begin{figure}[tp]
\centering
\includegraphics[width=1.0\columnwidth]{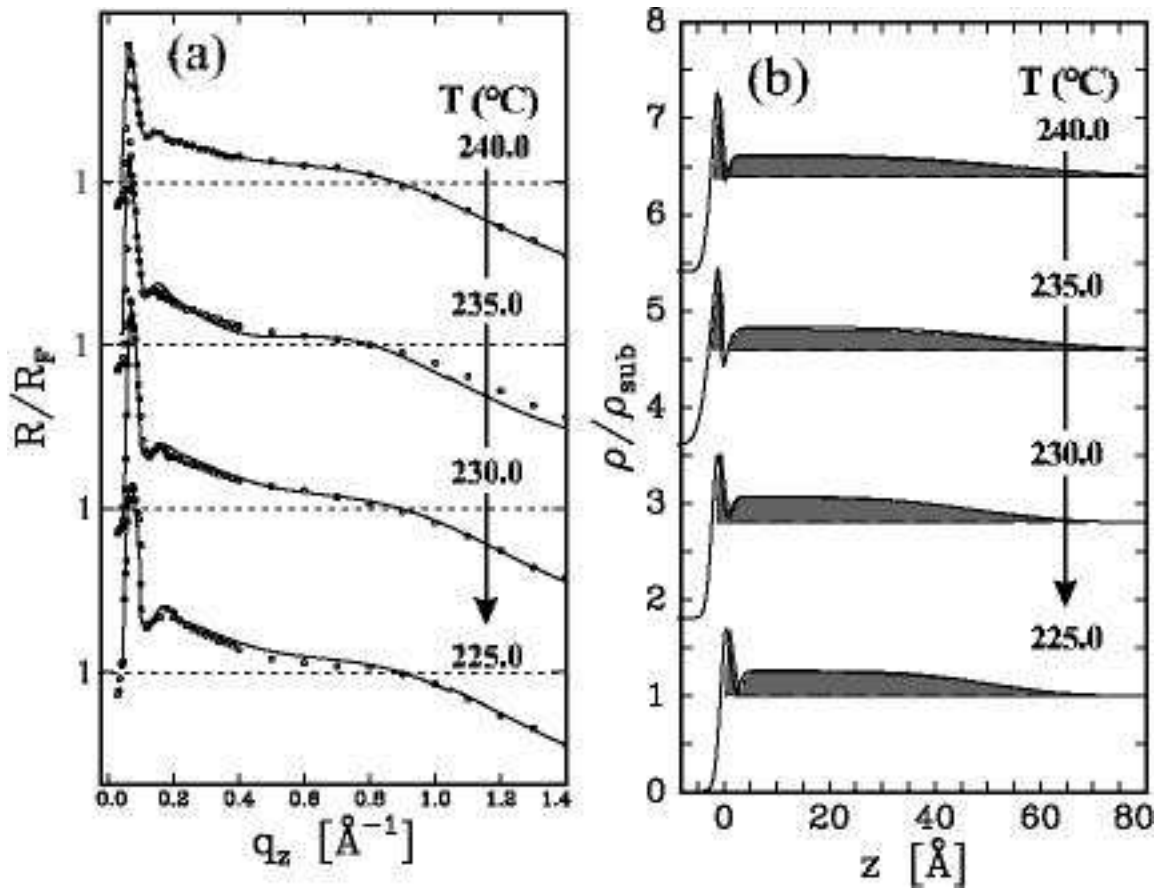}
\caption{\textbf{(a)} T-dependent normalized reflectivities R/R$_{
F}$ while leaving coexistence on path B $\rightarrow$D along with
fits. Dashed lines indicate $R/R_{ F}=1$. With increasing $T$,
each $R/R_F$ is shifted by $1.2$. \textbf{(b)} Electron density
profiles $\rho/\rho_{ sub}$. All regions with $\rho/\rho_{ sub}>1$
indicating Bi-enrichment as compared with the Ga-rich subphase are
gray shaded.}
\end{figure}

As discussed before, on cooling from the one phase region (III) we
reach the (l/l/v)-coexistence at B. The data for the measured
dependence of $R/R_{ F}$ on $q_{ z}$ along coexistence between
points B and M are shown in Figure  \ref{fig:RRFDensAB}(a). The
existence of two peaks at low $q_{ z}$ indicates the presence of a
fully formed thick film ($\sim 50~\AA$). The solid lines indicate
the best fit results corresponding to the real space profiles that
are shown in Figure  \ref{fig:RRFDensAB}(b). The best fit value
for maximum density of the thick film $\rho /\rho _{ sub}=1.20$
agrees reasonably well with the value of 1.21 that is calculated
form the \index{phase diagram} phase diagram at point B.

These results are also reasonably consistent with what is expected
for a gravity limited slab of uniform density. Using
$\xi=5.4~\AA$, $\sigma_0=400mN/M$ and the known material constants
that make up $\mu_m$ (see table \ref{tab:FitsAB} below) the
calculated value for $d=d_{ g}=\xi \ln (\sigma_{ 0}/\Delta\mu_{
g})=15.6\xi=85~\AA$. In view of the fact that this estimate does
not take into account the excess energy associated with associated
with concentration gradients across the interfaces some
overestimation of $d_{ g}$ is not too surprising. Nevertheless,
this rough calculation does show that the \index{wetting!film}
wetting film thickness is expected to be on a mesoscopic rather
than on the macroscopic length scale that has been observed for
similar wetting geometries in systems governed by long-range,
\index{dispersion} dispersion forces\cite{Law01}.

Upon cooling from point B to M, the intensity of the first peak of
$R/R_{ F}$ increases as expected due to the gradual increase of
the electron density contrast $\rho/\rho_{ sub}$ that follows from
the increase in the Bi concentration of the bulk phase. Similarly,
there is a small shift in the position of the peak indicating that
the thickness varies slightly from $53~\AA$ near B to $50~\AA$
near E that follows from the $T$-dependence of the density
contrast $\Delta\rho_{ m}(T)$ as estimated from the bulk phase
$\Delta\rho_{ m}(T)$ - cp. Table \ref{tab:FitsAB}. The
on-coexistence path B$\rightarrow$E is too far away from C to
expect more pronounced effects on the wetting layer thickness due
to a vanishing density contrast or the increase of the influence
of the criticality on the interaction potentials \cite{Dietrich88,
Bonn00, Bonn01}.

Moreover the gravitational thinning of the on-coexistence wetting
film thickness to a length scale comparable to the range of the
exponentially decaying short-range interactions prevents testing
the influence of long-range, \index{interactions!Van der Waals}
van-der-Waals like atomic interactions on the wetting behavior
that is present regardless of whether the system is insulating or
metallic\cite{Ashcroft91, Pluis89}.

Finally, we would like to highlight the unique wetting geometry
encountered here: The subtle balance between the surface potential
that favors the Bi-rich liquid phase at the surface and the
gravitational potential that favors the Ga-rich liquid phase above
the denser Bi-rich phase, pins the (l/l) interface between the two
coexisting phases close to the free surface. It is this property
that makes it possible to study the structure of these
\index{wetting!film} wetting films. For example, the width of the
interface between that film and bulk, $\sigma_{ obs}$, is the case
of the decay of the Kiessig fringes at low $q_{ z}$. This will be
discussed further in the following section.

\begin{table}
\begin{tabular}{ccccccccccc}
  $T (^{\circ} C)$ &$c^{ I}$ &$c^{ II}$ &$\Delta \rho$ $_{ m}$ &$\rho/\rho_{ sub}$ &$\rho/\rho_{ sub} $&q$_c$ &d$_g$ &$d$ &$\sigma_{ obs}$ &$\sigma_{ calc}$ \\
  \hline
  & & & & calc & observ & & & & & \\
  \hline
  225.0 &0.39& 0.91& 2.30 &1.24 &1.25 &0.0498 & 84.7 &50 &11.78&12.0 \\
  \hline
  230.0 &0.41& 0.90& 2.20 &1.24 &1.25 &0.0499 & 84.9 &51 &14.2 &13.3 \\
  \hline
  235.0 &0.43& 0.89& 2.09 &1.22 &1.22 &0.0500 & 85.2 &52 &15.0 &14.9 \\
  \hline
  240.0 &0.45& 0.88& 1.95 &1.21 &1.21 &0.0502 & 85.5 &53 &17.5 &17.25 \\
\end{tabular}
\caption{\label{tab:FitsAB}Material parameters of the coexisting
liquid phases as calculated from the bulk phase diagram as well as
electron density profiles as obtained from our fits to $R/R_{ F}$
for selected temperatures $T$ along the on-coexistence path
B$\rightarrow$E. The atomic fraction of Ga in the coexisting
Ga-rich, Bi-rich liquid phase are $c^{ I}$, $c^{ I}$ resp. The
symbols $\rho/\rho_{ sub}(calc)$, $\rho/\rho_{ sub}(obs)$ denote
the calculated, observed electron density ratios resp.}
\end{table}

\section{Liquid/Liquid Interface} In this section, we will discuss the microscopic structure
of the (l/l)-interface separating the gravitationally thinned film
of the Bi-rich phase from the Ga-rich subphase along the
on-coexistence path B$\rightarrow$E. We shall first describe a
simple square gradient theory for this interface and then extend
it by the consideration of thermally excited capillary waves. We
will use this theory in order to extract the (l/l)-interfacial
profile and the (l/l)-interfacial tension from our measurements of
the average surface structure.

\subsection{Square Gradient Theory}
At the (l/l)-interface between the two phases, the concentration
varies continuously from the bulk concentration of the homogeneous
Bi-rich phase, $c^{ I}$, to the bulk concentration of the
homogeneous Ga-rich phase, $c^{ II}$. Assuming that the variation
in concentration within the interface is gradual in comparison
with the intermolecular distance, the excess free energy for the
inhomogeneous region can be expanded in terms of local variables
$c(\overrightarrow{r})$ and $\bigtriangledown
c(\overrightarrow{r})$\cite{Davis96}. The \index{Gibbs!free
energy} Gibbs free energy density cost within the interface can
then be expressed in terms of a combination of a local function
$g(c(\overrightarrow{r},T)$ and a power series in
$\bigtriangledown c(\overrightarrow{r})$ \cite{VanderWaals94,
Cahn58, Safran94,Davis96}:
\begin{eqnarray}
\widetilde{G}=N\int_{ V}\left[ g(c(\overrightarrow{r}),T)+
\frac{1}{2}\kappa \left(\nabla c(\overrightarrow{r})
\right)^{2}+... \right] dV \label{eq:GTaylor}
\end{eqnarray}
where $N$ is the number of molecules per unit volume. The function
$g(c(\overrightarrow{r},T)$ is the \index{Gibbs!free energy} Gibbs
free energy density that a volume would have in a homogeneous
solution. The next term is the leading term in the power series
expansion. The coefficient $\kappa$ referred to as the influence
parameter characterizes the effect of concentration gradients on
the free energy. From first principles, it is related to the
second moment of the \index{Ornstein-Zernike function}
Ornstein-Zernike direct correlation function and can be replaced
by the pair potential weighted mean square range of intermolecular
interactions of the system\cite{Davis96}. Equation
(\ref{eq:GTaylor}) can be considered the \index{Landau-Ginzburg
functional} Landau-Ginzburg functional for this problem.

Applying (\ref{eq:GTaylor}) to the one-dimensional composition
change $c(z)$ across the interface and neglecting terms in
derivatives higher than the second gives for the total
\index{Gibbs!free energy} Gibbs free Energy $\widetilde{G}$ of the
system:
\begin{eqnarray}
\widetilde{G}=NA_{ 0}\int_{-\infty}^{+\infty} \left[ g(c(z)) +
\frac{1}{2}\kappa \left( \frac{dc(z)}{dz} \right)^2 \right]dz
\label{eq:GradientApprox}
\end{eqnarray}
The variable $z$ represents the direction perpendicular to the
interface. The symbol $A_{ 0}$ denotes an arbitrary surface area.
The interfacial tension $\gamma_{ ll}$ is defined as the excess
energy of this inhomogeneous configuration in respect to
the\index{Gibbs!free energy} Gibbs free energy $G$ of the
homogeneous liquid of one of the coexisting phases. The reference
system we have chosen is the homogeneous Bi-rich liquid with
concentration $c^{ I}$:
\begin{eqnarray}
\gamma_{ ll}=\frac{1}{A_{ 0}}\left[ \widetilde{G}(c(z)-G(c^{ I})
\right] \label{eq:SigmaDef}
\end{eqnarray}
Combining the gradient approximation (\ref{eq:GradientApprox})
with the thermodynamic definition of the interfacial tension
(\ref{eq:SigmaDef}) yields:
\begin{eqnarray}
\gamma_{ ll} = N \int_{-\infty} ^{+\infty} \left[ \Delta
g(c(z))+\frac{1}{2}\kappa \left(\frac{dc(z)}{dz}\right)^{2}\right]
dz
\end{eqnarray}
where the grand thermodynamic potential $\Delta g(c)$ is given by
\begin{eqnarray}
\Delta g(c(z))=g(c(z))-g(c^{ I})
\end{eqnarray}
It follows then from (\ref{eq:SigmaDef}) that the interfacial
tension is a functional of the concentration profile $c(z)$ at the
interface. The equilibrium surface tension is obtained by
minimizing this functional leading to the following Euler-Lagrange
equation:
\begin{eqnarray}
\Delta g(c)=\kappa\;\frac{d^2c(z)}{dz^2} \label{eq:DiffProfile}
\end{eqnarray}
This differential equation along with the boundary conditions that
the concentration has to vary from $c^{ I}$ on the one side of the
interface to $c^{ II}$ on the other side determines $c(z)$
unambiguously. In fact, the simplicity of (\ref{eq:DiffProfile})
allows its direct integration, yielding an expression for $z(c)$:
\begin{eqnarray}
z(c)=z_0+ \int_{c^{ I}}^{c^{{II}}} \sqrt{\frac{\kappa}{\Delta
g(c)}}\; dc \label{eq:DiffProfileQuadratur}
\end{eqnarray}
where $z_{ 0}$ and $c_{ 0}$ represent an arbitrary chosen origin
and composition. It should be noted, that from a renormalization
of the integrand of equation (\ref{eq:DiffProfileQuadratur}), one
can conclude that any characteristic length scale, such as the
intrinsic width of the interfacial profile, $\sigma_{ intr}$, has
to scale as $\sqrt{\kappa}$.

An attractive feature of this gradient theory is that it relates
the profile to $\kappa$ and $g(c(\overrightarrow{r},T)$ without
regard for the theoretical basis by which
$g(c(\overrightarrow{r},T)$ is derived. In this paper we will use
the extended regular solution model that will be presented in the
appendix to model the homogeneous Free Energy of the binary liquid
metal, particularly the one used to model its \index{miscibility
gap} miscibility gap. The sole quantity required then in order to
calculate the (l/l) interfacial profile and tension is an
expression for the influence parameter $\kappa$, that precedes the
square gradient\cite{Enders98}. It follows that the influence
parameter can be extracted from the measurement of the microscopic
structure of the interface. For example, from an appropriate
choice for $\kappa$, solving of (\ref{eq:DiffProfile}) or
(\ref{eq:DiffProfileQuadratur}) will yield a profile characterized
by the intrinsic width $\sigma_{intr}$ that can be compared with
$\sigma_{obs}$.

\subsection{Capillary Wave Excitations on the (l/l) Interface}
The formalism presented so far relies on a simple
\index{mean-field square gradient model} mean-field picture that
neglects any thermal fluctuations on the (l/l) interface. By
contrast, the real experiment is sensitive to not only the
mean-field diffuseness of the interface, but is also sensitive to
the fact that thermal fluctuations contribute additional
broadening to the fluid interface. An intuitive,
semiphenomenological approach to handle the interplay of these two
quantities was initiated by Buff, Lovett, and Stillinger
\cite{Buff65, Rowlinson82}. Essentially, they imagined the
interface as if it were a membrane in a state of tension
characterized by the bare \index{interfacial energy} interfacial
energy $\gamma_{ll}$, as calculated in the former paragraph; it
sustains a spectrum of thermally activated capillary waves modes
whose average energy is determined by equipartition theorem to be
$k_{B}T/2$. By integration over the spectrum of capillary waves,
one finds the average mean square displacement of the interface,
$\sigma_{cap}$, as
\begin{eqnarray}
\sigma_{cap}^2=\frac{k_{ B}T}{4 \pi \gamma_{ ll}(\kappa)} \ln
\frac{q_{ min}}{q_{ max}} \label{eq:WidthCapillary}
\end{eqnarray}
Here $q_{ max}$ is the largest capillary  wave vector that can be
sustained by the interface. For bulk liquids this is typically of
the order of $\pi$/molecular diameter; however, it does not seem
realistic to describe excitations with wavelengths that are
smaller than the intrinsic interfacial width as surface capillary
waves{\cite{Wadewitz99, Evans81}. Thus $\sigma_{ intr}$ describes
the intrinsic interfacial width $q_{ max}=\pi/\sigma_{ intr}$. The
value of the quantity $q_{min}$ is slightly more complicated since
it depends on whether the resolution is high enough to detect the
long wavelength limit at which an external potential $v(z)$, such
as either gravity or the \index{interactions!Van der Waals} van
der Waals interaction with the substrate,  quenches the capillary
wave spectrum.  If the resolution is sufficiently high then $q_{
min}=\pi/L_{ v}$ where $L_{ v}^2=\gamma_{ ll}^{
-1}\frac{\partial^2 v(z)}{\partial z^2}$. If it is not then the
value of $\sigma_{ cap}^{ 2}$ is limited by the resolution limited
average length scale $L_{ res}$ over which the observed
fluctuations can be resolved. In this case $q_{ min}=\Delta q_{
res}=\pi/L_{ res}$ where $\Delta q_{res}$ is the projection of the
detector resolution on the plane of the interface. In our case
$q_{ res}$ is significantly larger than $q_{ v}$ therefore, $q_{
min}=q_{ res}=0.04~\AA^{-1}$.

Using the fact that capillary waves obey a \index{gaussian}
gaussian statistics which manifests itself as an
\index{erf-function} erf-function type profile for the average
interfacial roughness\cite{Daillant99} together with the
observation that the intrinsic profiles $c(z)$ are also well
described by erf-function profiles, the two contributions to the
interfacial width can be added in \index{gaussian} gaussian
quadrature\cite{Daillant99}. Thus we write the total observed
interfacial width, $\sigma_{ calc}$, as
\begin{eqnarray}
  \sigma_{ calc}^{2}=\sigma_{ intr}(\kappa)^2+ \sigma_{ cap}(\gamma_{ ll}(\kappa))^2
\label{eq:WidthQuadrature}
\end{eqnarray}

Comparison of (\ref{eq:WidthCapillary}) and (\ref{eq:DiffProfile})
reveals that both, $\sigma_{ intr}$ and $\sigma_{ cap}$ depend on
$\kappa$. Thus, the determination of $\kappa$ leads to the problem
of finding the self-consistent value that yields a value for
$\sigma_{ calc}$ that agrees with the experimentally determined
roughness $\sigma_{ obs}$.

\subsection{Intrinsic Profile and (l/l) interfacial tension of
Ga-Bi}

We performed a self consistent calculation of $\kappa$ for the
observed (l/l) interface at point E, where $\sigma_{
obs}(E)\approx12~\AA$. In order to do so, we solved the
Euler-Lagrange Equation \index{Maple}
(\ref{eq:DiffProfile})\footnote{We used the standard procedures
for solving differential equations included in the \protect
\textit{Maple 7} software package (Waterloo Maple Inc.) as well as
the \protect \textit {"Shooting Technique for the solution of
two-point boundary value problems"} as implemented for this
software package by D.B. Maede\cite {Maede96}.} which yields the
intrinsic $c(z)$ profile as depicted and compared with an
erf-profile in Figure \ref{fig:LiqLiqIntrins}(a). From that we
calculated the corresponding erf-diffuseness $\sigma_{ intr}(E)$
and via numerical integration of (\ref{eq:SigmaDef}) the
corresponding interfacial tension $\gamma(E)$ as well as the
resultant capillary induced roughness and the overall average
diffuseness, $\sigma_{ cap}$ and $\sigma_{ intr}$ resp. After
starting with a reasonable value of $\kappa$, as judged by a
scaling analysis of (\ref{eq:DiffProfile}), this procedure was
iterated until $\sigma_{ calc}(E)=\sigma_{ obs}(E)$. The value we
obtained was $\kappa_{ opt}=5.02\cdot10 ^{-13}Nm^3/mol$. The other
quantities of interest at E are: $\sigma_{ cap}(E)=10.32~\AA$ ,
$\sigma_{ intr}(E)=6.35~\AA$ and $\gamma_{ ll}(E)=3.3~mN/m$.

It is interesting to observe that (\ref{eq:DiffProfile}) implies
$\sigma_{ intr}\propto \frac {1}{\kappa}$, whereas from
(\ref{eq:WidthCapillary}) one obtains $\sigma_{ cap} \propto
\sqrt{\kappa}$. Hence, the function of $\sigma_{ calc}(\kappa)$
exhibits a minimum for $\sigma_{ calc}$ at a characteristic value,
$\kappa_{ min}$. It is interesting to note, that the $\kappa_{
opt}$, determined from our measurements, corresponds to this
peculiar value $\kappa_{ min}$ meaning that the system selects
from all possible $\sigma_{ intr}$, $\sigma_{ cap}$ combinations,
the one with the minimum overall width for the (l/l)-interface.

Assuming a $T$-independent influence parameter along with the
available thermochemical data sets that are available over a wide
range of the \index{phase diagram} phase diagram allows
calculation of the $T$-dependent interfacial profiles and (l/l)
interfacial tension from very low up to very high $T$, close to
$T_C$. In Figure \ref{fig:LiqLiqIntrins}(b) the intrinsic (l/l)
interfacial profiles are plotted for selected $T$'s versus the
reduced temperature, $t=|T-T_{ C}|/T_{ C}$ (Here $T$ denotes the
absolute temperature in Kelvin units). The increasing width of the
interface while approaching $T_C$ ($t\rightarrow0$) is
well-demonstrated. In Figure  \ref{fig:gammaGaBi} the
$t$-dependent interfacial tension, $\gamma_{ll}(T)$, is depicted.
As expected, it vanishes as $t\rightarrow0$.
\begin{figure}[tp]
\centering
\includegraphics[width=0.7\columnwidth]{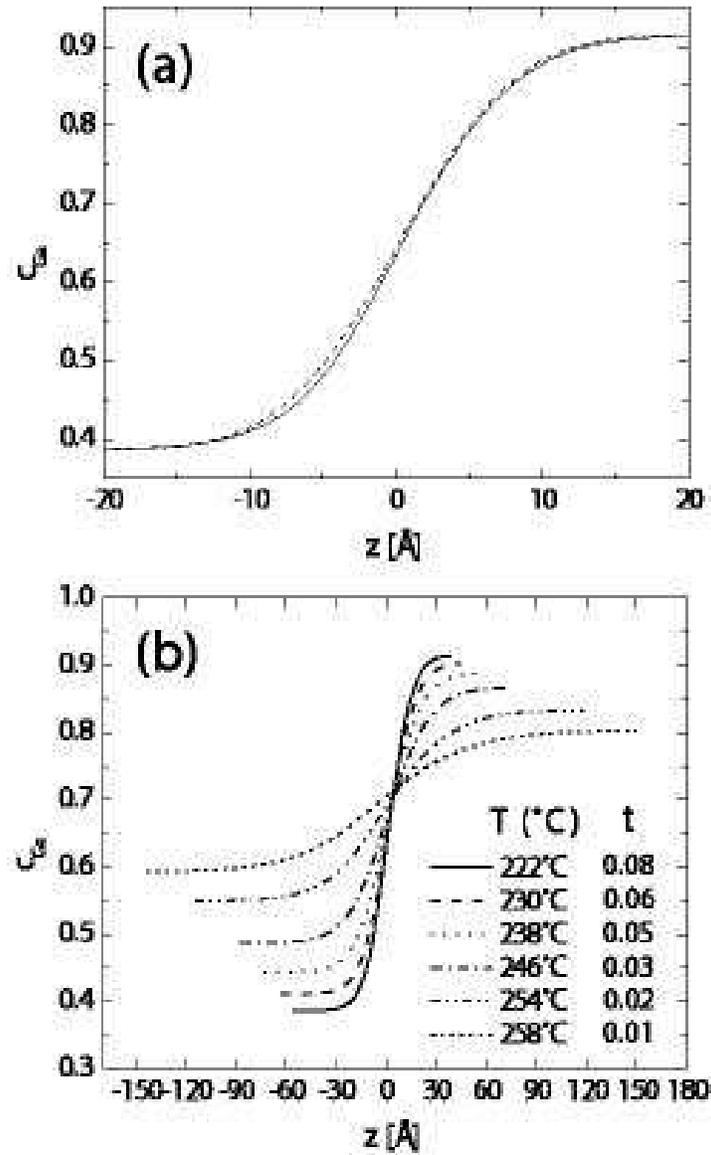}
\caption{\label{fig:LiqLiqIntrins} \textbf{(a)} Calculated
intrinsic profile $c(z)$ ($-$) for $T=T_E=225^{\circ}C$, $t=0.07$
in comparison with the \index{erf-function}  erf-function profile
(dashed line): $c(z)=\frac{c^{ I}+c^{ II}}{2}+\frac {(c^{ II}-c^{
I})}{2}$ erf $\left ( \frac{z}{\sqrt{2} \sigma_{ intr}(E)}
\right)$ with $\sigma_{ intr}(E)=6.35~\AA$. \textbf{(b)}
Calculated intrinsic concentration profiles $c(z)$ for selected
temperatures $T$, reduced temperatures $t$ resp. as indicated in
the figure.}
\end{figure}

\begin{figure}[tp]
\centering
\includegraphics[width=1.0\columnwidth]{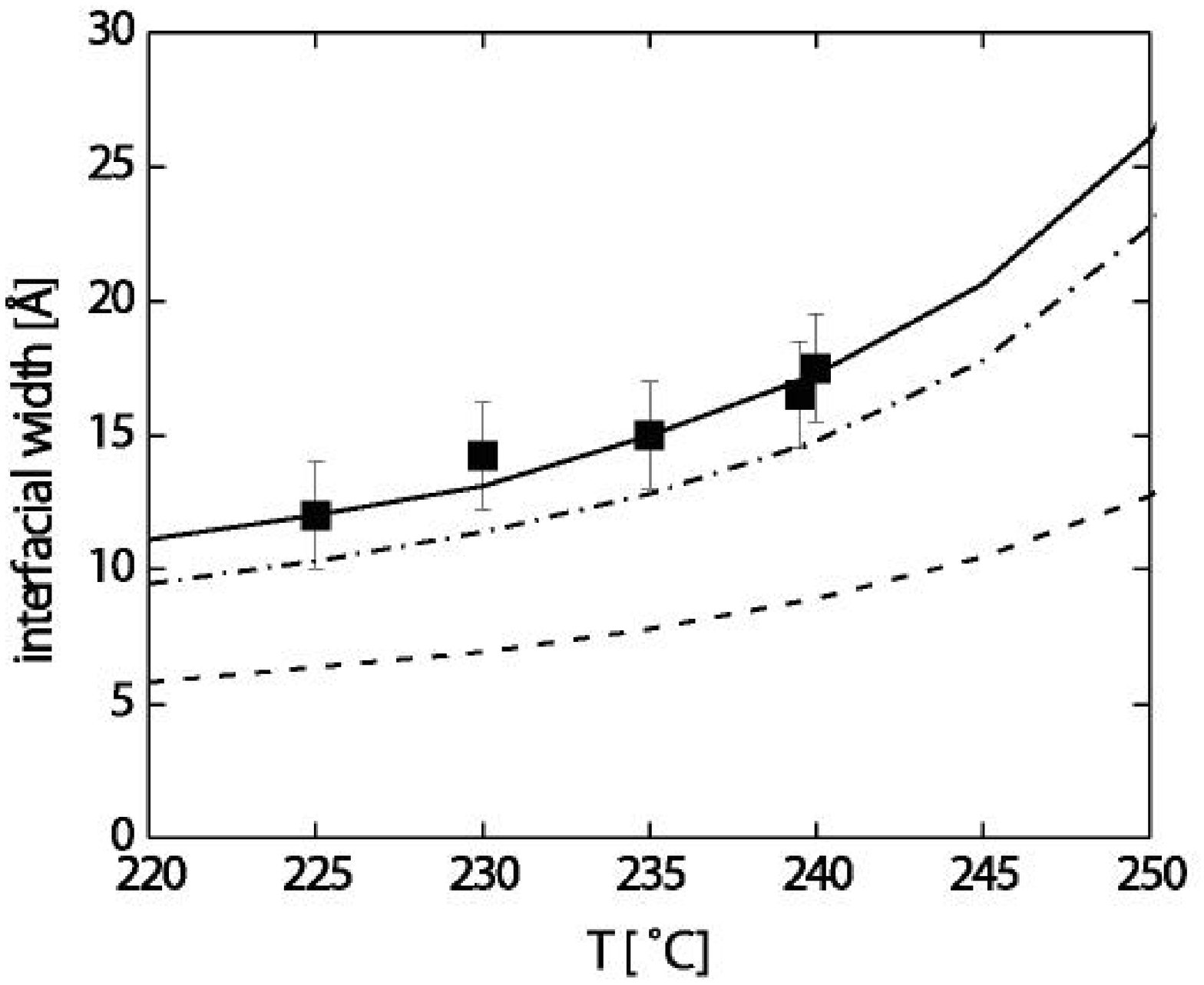}
\caption{\label{fig:CpCMW}Comparison of temperature dependent,
measured interfacial width $\sigma $ ($\blacksquare$) with
calculated interfacial widths (lines): $\sigma _{ calc}$ =
$\sqrt{\sigma_{intr}^{2}+\sigma_{cap}^{2}}$ ($-$) , intrinsic
width $\sigma_{intr}$ ($\cdot-$), and capillary width
$\sigma_{cap}$ ($--$).}
\end{figure}


Comparison of $\gamma_{ ll}(T)$ with other measurements would be a
useful test of our measurements and interpretation. Unfortunately,
measurements are rather difficult and we are not aware of any such
measurements for Ga-Bi. Fortunately, it is still possible to
estimate $\gamma_{ ll}(T)$ using other available experimental data
sets on this alloy. For example, Predel\cite{Predel60} noted that
in the vicinity of C the (l/l)-coexistence curve of Ga-Bi can be
represented by a function of the form $|\phi_{ c}-\phi|=K(T_{
c}-T)^{ \beta}$ with $\beta\cong0.33$, where $\phi$ is the volume
fraction of Ga, Bi resp. and $K$ a constant \cite{Predel60}. This
behavior suggests that the \index{demixing transition} demixing
transition in Ga-Bi belongs to the same universality class as
binary nonmetallic liquid mixtures, i.e. \index{Ising model} Ising
lattice model with dimensionality D=3. Moreover, it allows for an
estimation of the interfacial tension based on a scaling relation
for these quantities, i.e. the hypothesis of
\index{two-scale-factor universality} two-scale-factor
universality (TSFU) \cite{Stauffer72} should apply. Using this
concept along with available $T$-dependent measurements of the
specific heat, Kreuser and Woermann extracted an expression for
the $T$-dependent interfacial tension in Ga-Bi \cite{Kreuser93},
$\gamma_{ ts} = \gamma_{ ts0}~t^{ \mu}$, where $\mu$ is a critical
exponent with universal character for a bulk \index{demixing
transition}demixing transition, $\mu \cong 1.26$ and $\gamma_{
ts0}=66\pm15mN/m$. The corresponding $t$-dependent $\gamma_{ ts}$
($-$) is plotted in Figure \ref{fig:gammaGaBi}. A good agreement
with the $t$-dependence of $\gamma_{ ll}$ calculated by our square
gradient theory is found, e.g. at $t_{ E}=0.07, T_{
E}=225^{\circ}C$: $\gamma(E)= 3.31mN/m$, $\gamma_{ts}(E)=5.7mN/m$
\footnote{In the strict sense, the TSFU hypothesis is only valid
in a $T$-range, where mapping of the \index{demixing transition}
demixing process on a second order phase transition is justified,
which is typically the case for reduced temperatures, $t<10^{
-1}$. Despite of this restriction, it often fits well experimental
results\cite{Merkwitz98} and obviously also our results of the
gradient theory in a much wider temperature range.}


\begin{figure}[tp]
\centering
\includegraphics[width=1.0\columnwidth]{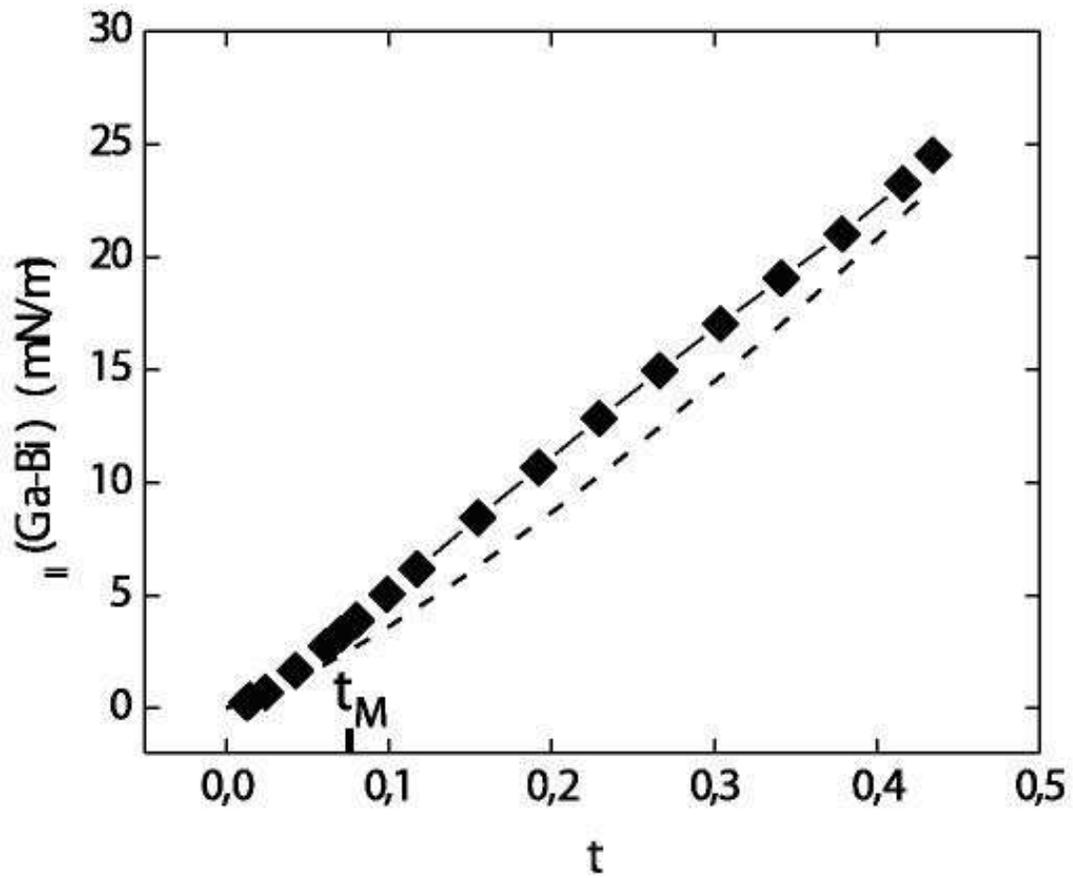}
\caption{\label{fig:gammaGaBi} Calculated (l/l) interfacial
tension $\gamma_{ ll}$ of Ga-Bi as a function of the reduced
temperature $t$. The dashed line represents the TSFU prediction
for Ga-Bi. The symbol $t_{ M}$ indicates the \index{monotectic
point} monotectic temperature at $t_{ M}$(Ga-Bi)$=0.08$.}
\end{figure}

\subsection{(l/l) interfacial tension of Ga-Pb}

Aside from the Ga-Bi system the only other metallic system for
which detailed temperature dependent measurements\cite{Merkwitz98}
and estimations \cite{Wynblatt98} of the (l/l) interface are
reported, is Ga-Pb . Given the close relationship between Ga-Bi
and Ga-Pb (similar constituents, identical phase diagram topology
with consolute point C and monotectic point M), we felt encouraged
to also apply our gradient \index{square gradient theory} theory
to this binary system. Moreover, we assumed that the influence
parameter, $\kappa$, as determined by the measurements on Ga-Bi,
also provides a reasonable value for this metallic system.
Available thermochemical data sets for the \index{miscibility gap}
miscibility gap of Ga-Pb\cite{Ansara91} allowed us to calculate
the $t$-dependent (l/l) interfacial tension. As can be seen in
Figure \ref{fig:gammaGaPb}, where our calculated values for
$\gamma_{ ll}$ are shown in comparison with the experimental
values, an excellent agreement is found.

\begin{figure}[tp]
\centering
\includegraphics[width=1.0\columnwidth]{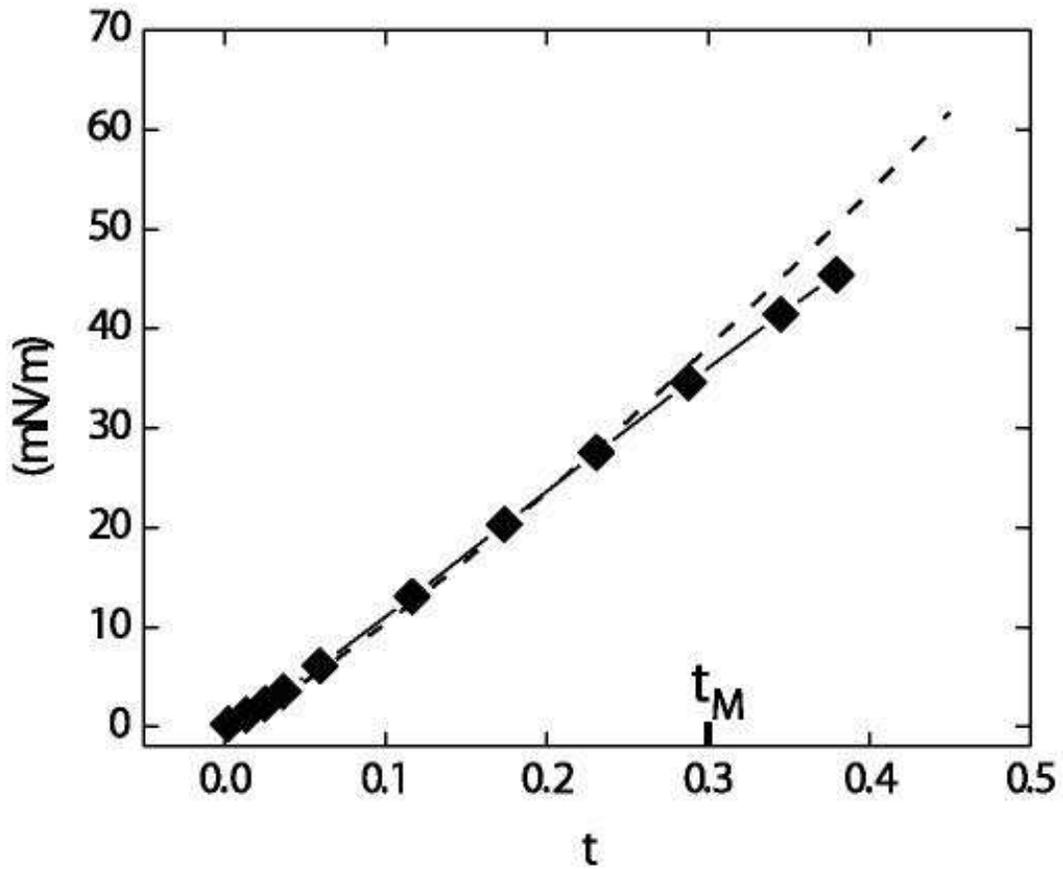}
\caption{\label{fig:gammaGaPb} Calculated (l/l) interfacial
tension $\gamma_{ ll}$ of Ga-Pb as a function of the reduced
temperature $t$. The solid points are our calculated values. The
dashed line represents a fit to the measurements by Merkwitz et.
al.. The symbol $t_{ M}$ indicates the monotectic temperature at
$t_{ M}$(Ga-Pb)$=0.3$.}
\end{figure}

Overall, our results suggest that our model of an intrinsic,
\index{mean-field square gradient model}  mean-field interface
broadened by capillary waves describes the (l/l) interface
reasonably well. One must, however, bear in mind the approximation
implicit in the presented analysis. For instance, we have assumed
that the free surface of the liquid alloy behaves like a rigid
wall, whereas it will have its own spectrum of capillary waves
that could be coupled to the waves at the interface. However,
since the interfacial tension of the free surface is about two
orders of magnitude larger than the (l/l) \index{interfacial
energy} interfacial energy in this metallic system, we believe
that the rigid wall assumption is a reasonable starting point.
Furthermore, the calculation presented above has the shortcoming
of a $T$-independent influence parameter $\kappa$.
It should be mentioned, though, that theory predicts a divergence
of $\kappa$ close to $T_{ C}$ for real fluids,
$\kappa\approx(T-T_{C})^{ -0.02}$ for $T\rightarrow T_{ C}$. In
fact, it can even be shown that this divergence is necessary to
obtain the correct scaling behavior of $\gamma_{ ll}(T)$ near C
\cite{Cornelisse97, Fisk69, Sengers89}.

\section{Wetting Phenomenology at the Free Surface}
\subsubsection{Effects of Fluctuations on \index{wetting!short-range} Short-Range Wetting}
Our analysis in section 1 of the \index{wetting!complete} complete
wetting transition at B was based on a  \index{mean-field square
gradient model} mean-field (MF) model for a SRW transition. Such a
model accurately predicts the critical behavior (i.e. the critical
exponents) of systems close to a phase transition, provided
fluctuations can be neglected. For SRW transitions, however, it
can be shown that the upper critical dimensionality
$D_u=3$\cite{Schick90}. The value $D_u$ is the dimension beyond
which MF theory can be applied successfully. If the dimension is
smaller than $D_u$ fluctuations are important and one has to
resort to \index{renormalization group} renormalization group (RG)
methods to describe the critical behavior of a phase transition.
Because its upper critical dimensionality is exactly equal to 3,
the SRW transition has received a great deal of theoretical
attention, since it allows one to explore the regime where the MF
behavior breaks down due to fluctuations, and the RG approach
becomes applicable. This break-down depends on the so-called
fluctuation parameter, $\omega=k_{ B} \cdot T /(4\pi\gamma_{
ll}\xi_{ b}^2)$ (where $\xi_b$ is the bulk \index{correlation
length} correlation length, and $\gamma_{ ll}$ is the (l/l)
interfacial tension) which measures the magnitude of the dominant
thermal fluctuations, the thermally induced capillary waves at the
fluid, unbinding interface, in our case at the (l/l) interface of
the coexisting Bi- and Ga-rich liquids.

For a complete wetting transition RG analysis of the divergence of
the \index{wetting!film} wetting film thickness yields a changed
prefactor of the logarithmic divergence, $d_{ RG}\sim\xi_{
RG}\,(1+\omega /2)\, \ln\, (1 /\Delta \mu)$\cite{Dietrich88,
Brezin83, Kroll85}. An estimation of $\omega$ using the bulk
\index{correlation length} correlation length, $\xi_b$, estimated
from the aforementioned TSFU prediction, yields for the wetting
transitions at M, $\omega_{ M}=0.3\pm0.2$ and for the observed
wetting complete wetting transition at B, $\omega_{ B}=0.4\pm0.2$.
Thus, the prefactors of the logarithmic increase of the wetting
film thickness with $\Delta\mu$ is changed by only $\approx$ 10\%,
20\% resp. in comparison with the mean-field prediction. This is
well within our experimental error of about 30\% for the
determination of $\xi$. (The large error bar on $\xi$ is due to
its extraction from a slab thickness analysis whereas the measured
wetting films have significant inhomogeneous structures). A clear
distinction between RG and MF behavior cannot be drawn in our
case.

A \textit{critical} wetting transition characterized by the
\index{wetting!film} wetting film formation
on-(l/l/v)-coexistence, while changing $T$, with a similar value
of $\omega$ should show more pronounced deviations from MF
behavior. Thus, we shall estimate the characteristic temperature
$T_{ W}$ of that transition in the following.

\subsubsection{The \index{wetting!critical} Critical Wetting Transition in Ga-Bi}
It follows from our measurements that the critical wetting
transition has to be hidden in the metastable range of the (l/l/v)
coexistence line somewhere below $T_{ M}$. An estimation of $T_{
W}$ is possible by a consideration of the Free energy at the
surface for the wet (Bi-rich wetting at surface) and non-wet
(Ga-rich phase at the surface) situation. In order to do so, we
introduce the spreading energy, $\Theta(T)$. It estimates the
preference of the Bi-rich phase in respect to the Ga-rich phase at
the free surface at any temperature and is given by the following
energetic contributions. In the wet-situation, the energy is given
by the sum of the free surface tension of the Bi-rich phase and
the tension of the (l/l) interface between the Bi-rich phase and
the Ga-rich subphase. The non-wet situation is characterized by
the surface energy of the bare Ga-rich liquid phase,
$\gamma(Ga-rich)_{ lv}$. We define $\Theta(T)$ therefore in the
following way:
\begin{eqnarray}
   \Theta(T) &=&\gamma(Ga-rich)_{ lv}(T)-\nonumber \\
             & &\left ( \gamma(Bi-rich)_{ lv}(T)+\gamma_{ ll}(T) \right )\nonumber \\
             &=&\left ( \gamma(Ga-rich)_{ lv}(T)-\gamma(Bi-rich)_{ lv}(T) \right )\nonumber \\
             & &-\gamma_{ ll}(T)\nonumber \\
          &=&\Gamma(T)-\gamma_{ ll}(T)
\label{eq:SpreadingEnergy}
\end{eqnarray}
In equation (\ref{eq:SpreadingEnergy}) the quantity $\Gamma(T)$
denotes the difference in the bare liquid-vapor surface tensions
of the coexisting phases.

From the definition of $\Theta(T)$ it follows that establishment
of the wetting film is energetically preferred if $\Theta>0$
(wet-situation), whereas for $\Theta<0$ it is unfavorable (non-wet
situation). The condition $\Theta(T)=0$ marks, therefore, the
transition temperature $T_{ W}$, where the system switches
on-coexistence from a non-wet to a wet-situation or vice versa.

In order to estimate $\Theta(T)$, in addition to $\gamma_{
ll}(T)$, as extracted in the previous section we need measurements
or estimations of $\gamma(Ga-rich)_{ lv}(T)$ and
$\gamma(Bi-rich)_{ lv}(T)$. Measurements of these quantities are
reported in the literature\cite{Khokonov74, Ayyad02, Ayyad02b},
the one by Ayyad and \index{Freyland, Werner} Freyland that
employed the noninvasive method of capillary wave spectroscopy is
the most recent one. From these measurements, we get a
conservative $T$-independent estimate for $\Gamma$ of the order of
100mN/m. Our calculation of $\gamma_{ ll}(T)$, particularly its
extrapolation towards very low temperature $t\rightarrow1$ which
yields $\gamma_{ ll}(0K)=75mN/m$, indicates that it is always
significant smaller than $\Gamma$. Our analysis suggests,
therefore, that $\Theta(T)>0$ for all $T$ meaning that \textit{no}
\index{wetting!critical} critical wetting transition occurs on the
metastable extension of the (l/l/v) coexistence line.

This remarkable conclusion is, on the one hand, in contrast to the
findings for the few examples of binary metallic wetting systems
studied so far: In Ga-Pb and Ga-Tl, \index{Wynblatt, Paul}
Wynblatt and \index{Chatain, Dominique} Chatain et al. estimated a
value for $T_{ W}$ significantly below the corresponding
\index{monotectic point} monotectic temperatures T$_{ M}$, but
still at final values of the order of $0.3 \cdot T_C(K)$, $0.2
\cdot T_C(K)$ for Ga-Pb, Ga-Tl resp. ($T_{ C}(K)$ denotes the
critical temperature of bulk demixing on the Kelvin temperature
scale.) On the other hand, it reflects the general finding that in
binary metallic systems with short-range interactions the critical
wetting transition occurs at significantly lower temperatures than
this is the case in organic liquid systems, where $T_{ W}$ is
found to lie above $0.5 \cdot T_{ C}(K)$. This tendency seems to
be driven to its extreme in the case of Ga-Bi: $T_W=0 \cdot T_{
C}(K)=0K$.

A semiquantitative argument for the absence of a critical wetting
transition in Ga-Bi as compared to the two other metallic systems
investigated so far might relate to the fact that in comparison
with the others the liquid/liquid coexistence region for GaBi
system is significantly shifted towards the Ga-rich, high surface
tension phase and, in addition the miscibility gab is much
narrower.  As a result the change in concentration across the l/l
interface is smaller for the GaBi system than for the others. In
view of the fact that one can express
\begin{eqnarray}
   \gamma_{ ll} = \kappa \int dz \frac{dc(z)}{dz}^{ 2}\approx \sqrt{\kappa}\left ( \Delta c \right )^2
\label{eq:GammaApprox}
\end{eqnarray}
this suggests that $\gamma_{ll}$ could be expected to be smaller
for GaBi than for the others. If this effect is important, and if
it is not compensated by an accompanying reduction in the value of
$\Gamma$, then it is possible that
$\Theta(T)=\Gamma-\gamma_{ll}>0$ for all temperatures, thereby
precluding a wetting transition.

\subsubsection{Tetra Point Wetting $\Leftrightarrow$ \index{surface!freezing} Surface Freezing}
\index{wetting!tetra-point} Finally, we would like to comment on
the fact that the observed wetting by the Bi rich liquid at the
GaBi monotectic point is not necessarily the only phenomena that
can occur. In principle one expects at this tetra point a similar
or even more diverse wetting phenomenology as has been predicted
\cite{Pandit83} and found \cite{Zimmerli92} in the proximity of a
\index{wetting!triple point} triple point for one component
systems. For example, another possibility is that the free surface
could be wet by solid Bi. In the bulk both solid Bi and the Bi
rich liquid are stable for $T>T_M$; however,  for $T<T_{ M}$ the
free energy of the Bi rich liquid is larger than that of solid Bi.
Consequently, if the solid Bi surface phase had a favorable
spreading energy it would certainly wet the free surface for
$T<T_{ M}$. In fact, on the basis of optical and surface energy
measurements \index{Turchanin, Andrei} Turchanin et al.
\cite{Turchanin01, Turchanin02} proposed a surface \index{phase
diagram} phase diagram for which a surface frozen" phase of solid
Bi forms at the free surface of the GaBi liquid for temperatures
below $T_M$. Unfortunately, we have a problem with this. The first
may be purely semantic, but  the idea of a "surface frozen phase"
originates in the experiments by Earnshaw et al. and
\index{Deutsch, Moshe} Deutsch, \index{Gang, Oleg} Gang,
\index{Ocko, Benjamin} Ocko et al \cite{Earnshaw92, Wu93} on the
appearance of solid surface phases at temperatures above that of
bulk freezing for \index{alkanes} alkanes and related compounds.
For the GaBi system the crystalline phase of bulk Bi is stable at
temperatures well above those of at which \index{Turchanin,
Andrei} Turchanin et al made their observations. We believe that
if their observations do correspond to  thermal equilibrium
phenomena, the effect would be more appropriately described as
wetting of the free surface by solid Bi. Unfortunately, the second
problem we have is that we do not believe that surface wetting by
solid Bi is favored by the spreading energy. If wetting by solid
Bi were favored, as implied by \index{Turchanin, Andrei} Turchanin
et al, then the wetting layers of Bi rich liquid that are the
subject of this paper would have to be metastable, existing only
because of a kinetic barrier to the nucleation of the solid
wetting film. In fact we have regularly observed   that on cooling
below $T_{ M}$ the original fluid, smooth, highly reflecting
surfaces became both rough and rigid. The effect is exacerbated by
the presence of temperature gradients and by rapid cooling. We
interpreted this as the formation of bulk solid Bi, rather than
wetting,  since we never saw any direct evidence for the formation
of films, rather than bulk Bi.  We argued  that if there not
enough time for the excess Bi in the bulk liquid phase below the
surface to diffuse, and precipitate as bulk solid Bi at the bottom
of the liquid pan, the bulk liquid would simply undergo bulk
\index{spinodal demixing} spinodal \index{phase separation} phase
separation. This is consistent with our observations.
Unfortunately it is very difficult to absolutely prove that  our
films are stable and this issue can't be  resolved from the
existing evidence.  Nevertheless we find it difficult to
understand why solid Bi would wet the surface for $T<T_{ M}$ and
not wet the surface as $T$ approaches $T_{ M}$ from above.

\section{Summary}
In this paper we reported x-ray reflectivity measurements of the
temperature dependence of wetting phenomena that occur at the free
surface of the binary metallic alloy Ga-Bi when the bulk demixes
into two liquid phases, i.e. a Bi-rich and a Ga-rich liquid phase.
We characterized the temperature dependence of the thickness and
interfacial profile, with $\AA$ngstr\"om resolution,  of the
Bi-rich wetting films that form at the free surface of the liquid
on approaching the (l/l/v) coexistence triple line,  i.e. along a
path of \index{wetting!complete} complete wetting .  The results,
which are characterized by large concentration gradients, agree
with density functional calculations for such transitions at hard
walls. On the basis of this it was possible to determine the
short-range surface potential that favors the Bi-rich phase at the
free surface and hence is governing the wetting phenomenology. The
short-range \textit{complete wetting} transition turns out to be
only marginally affected by thermal fluctuations.  According to
\index{renormalization group} renormalization group analysis the
\index{wetting!critical} \textit{critical wetting} transition
should be more sensitive to critical fluctuations; however,
according to our analysis the wetting layer of Bi rich liquid
should be present for all temperatures. This is in contrast to the
results for the other binary liquid metal systems for which the
\index{wetting!transition} wetting transitions exists but since
the metastable l/l/v line it falls below the liquidus line it is
hidden.

The largest thickness of the gravitationally limited Bi rich
wetting layers that formed at the free surface at
(l/l/v)-coexistence were typically of the order of $\approx
50~\AA$. As a result of the fortuitous balance between the surface
potential that favors the Bi-rich Phase and the gravitational
potential which favors the lighter Ga-rich phase the interface
dividing the two coexisting phases is sufficiently close to the
free surface that x-ray reflectivity was sensitive to its
microscopic structure.  We interpreted these measurements in terms
a \index{Landau, Lev} Landau type of \index{square gradient
theory} square gradient phenomenological theory from which it was
possible to extract the sole free parameter of that model, i.e.
the influence parameter $\kappa$. To the best of our knowledge
this is the first time that such a parameter has been directly
determined from measurement. Furthermore, it was possible to make
a distinction between the intrinsic width of the interface and the
additional contributions to the width from thermal capillary wave
features fluctuations. On making using of available bulk
thermodynamic data, along with the extracted influence parameter,
we were able to calculate the experimentally difficult to get
(l/l) interfacial tension for a wide temperature range. As a test
of our methods we performed the same calculation of the (l/l)
interfacial tension for Ga-Pb with its analogous miscibility. We
assumed that the influence parameter, $\kappa$, as determined from
our measurements on the (l/l) interfacial structure in Ga-Bi, was
also applicable to Ga-Pb and, using the thermochemical data sets
available for Ga-Pb, we obtained good agreement between our
calculated values and macroscopic measurements of the (l/l)
interfacial tension. This suggests that the value of the influence
parameter extracted in this study might provide a reasonable value
for the larger class binary metallic alloys. If this proves to be
true, then the value determined here can be used, along with the
surface potential, to predict wetting properties of this larger
class of systems.

\section{Appendix: Bulk thermodynamics} Here\footnote{Note unless it is is explicitly indicated otherwise, in this paper ''$T$'' denotes the absolute temperature in Kelvin units.}, we
focus on the part of the \index{phase diagram} phase diagram
dominated by the \index{miscibility gap} miscibility gap and the
\index{monotectic point} monotectic point M. In a liquid-liquid
(l/l) equilibrium of binary systems, consisting of two components
(Ga and Bi resp.), at atmospheric pressure p, temperature T, the
compositions $c_{ I}$, $c_{ II}$ of two coexisting bulk phases (I
and II) are given by the thermodynamic conditions:
\begin{eqnarray}
\mu_{ Ga}^{ I}(T)=\mu_{ Ga}^{ II}(T)~~~~ \mu_{Bi}^{ I}(T)=\mu_{
Bi}^{ II}(T) \label{eq:muequilib}
\end{eqnarray}
where $\mu$ is the \index{chemical potential} chemical potential.
The experimental conditions of the experiments that we are going
to present are constant temperature, pressure and a fixed number
of moles of each component, therefore we have to resort to the
\index{Gibbs!free energy} Gibbs free energy $G$ in order to
describe the thermodynamic equilibrium:
\begin{eqnarray}
G&=&U-TS+PV \\
G&=&n\;g=n_{ Ga}\;\mu_{ Ga}\,+\,n_{ Bi}\;\mu_{ Bi}
\end{eqnarray}
Since the total number of particles in mol $n=n_{ Ga}+n_{ Bi}$ is
constant and only two components have to be considered, the system
can be described with a molar Gibbs Free Energy g(c,T) that
depends on only on the molar fraction of one component:
\begin{eqnarray}
G\,&=&\, n\;(c\, \mu_{ Ga}\,+\,(1-c)\,\mu_{ Bi}\,)\\
\Rightarrow g(c,T)\,&=&\,c\,\mu_{ Ga}+(1-c)\,\mu_{ Bi}
\end{eqnarray}
With this notation the phase equilibrium conditions (equation
(\ref{eq:muequilib}) transform into the following two equations:
\begin{eqnarray}
\label{commontangent}
 \frac{\partial g}{\partial c} |c^{ I}
&=&\frac{\partial g}{\partial c} |c^{ II}\\
 \frac{\partial g}{\partial c}|c^{ I}&=&\frac
 {g(c^{I})-g(c^{II})}{c^{I}-c^{II}}
\label{eq:commontangent}
\end{eqnarray}
Having a model for the free energy of a binary system, it is
possible, then, to calculate $c_{ I}$ and $c_{ II}$ of the
coexisting phases by solving of equation (\ref{eq:commontangent}).
The standard approach to describe a miscibility gap in a binary
demixing system is the regular solution model for the
\index{Gibbs!free energy} Gibbs free energy \cite{Atkins01}. The
resulting miscibility gap has a symmetric shape and is centered
around the consolute point $c_{ crit} = 0.5$ in the
($c$,$T$)-plane, quite in contrast to the miscibility gap of Ga-Bi
with its asymmetric shape, centered around $c_{ crit}$=0.7 in the
($c$,$T$)-plane. Therefore, it is necessary to resort to an
\textit{extended} regular solution model. Relying on data sets for
Ga-Bi from the \index{Calphad} Calphad initiative\cite{Kaufman01},
we use such a model based on \index{Redlich-Kister} Redlich-Kister
polynoms $L_{ \nu}(T)$\cite{Kattner97, Lukas82} and express
$g(c,T)$ in the following way:
\begin{center}
\begin{eqnarray}
g(c,T)&=&c\cdot g_{ 0}(Ga)(c,T)+(1-c)\cdot g_{ 0}(Bi)\nonumber\\
      & &+R\cdot T \left [\,c\;ln(c)\;+\;(1-c)\,ln(1-c)\right ] \label{eq:gERSM}\\
      & &+\;\Delta g_{ mix}(c,T)\nonumber\\
 \Delta g_{
mix}(c,T)~&=&~c(c-1)~\sum_{ \nu=1}^{5}L_{
\nu}(T)(1-2c)^{\nu}\nonumber
\end{eqnarray}
\end{center}
The first two terms correspond to the \index{Gibbs!free energy}
Gibbs energy of a mechanical mixture of the constituents; the
second term corresponds to the entropy of mixing for an ideal
solution, and the third term, $\Delta g_{ xs}$ is the so-called
excess term. Here, it is represented by an extension of the usual
regular solution expression, i.e. a sum of \index{Redlich-Kister}
Redlich-Kister polynomials, $L_{ \nu}(T)$, as listed in Table
\ref{tab:RedKisterPolys}.

\begin{table}
\begin{tabular}{c|c}
  $\nu$ & Redlich-Kister polynomial L$_{ \nu}(T)$\\
  \hline
  0 & 80000-3389+T \\
  \hline
  1 & -4868-2.4342$\cdot T$ \\
  \hline
  2 & -10375-14.127$\cdot T$ \\
  \hline
  3 & -4339.3 \\
  \hline
  4 & 2653-9.41$\cdot$T\\
  \hline
  5 & -2364\\
\end{tabular}
\caption{\label{tab:RedKisterPolys} \index{Redlich-Kister}
Redlich-Kister polynomials used to model the (l/l) miscibility gap
of Ga-Bi.}
\end{table}

By solving the nonlinear equations (\ref{eq:commontangent})
applied on $g(c,T)$ for the temperature range $T_{ M} < T <
262^{\circ}C$ we could calculate the binodal coexistence line as
plotted in Figure  \ref{fig:GaBiPDcT}. The agreement of the
calculated phase boundaries with Predel's measured phase
boundaries is excellent. Particularly, the measured consolute
point C ($T_C=262^{\circ}C$, $c_{ crit}=0.7$ is nicely reproduced
by the calculated critical values $T_{ C}=262.8°C$ and $c=0.701$.
Furthermore, the knowledge of $g(c,T)$ allows us to extrapolate
the binodal lines below $T_{ M}$ into the region of metastable
(l/l) coexistence, and, hence, to get information on the
energetics of the metastable Ga-rich, Bi-rich phase resp. Similar
to the procedure presented above, we also calculated the phase
boundaries below $T_{ M}$. Here, we used data sets for the Gibbs
free energy of pure solid Bi, which coexists for $T<T_{ M}$ with a
Ga-rich liquid.

\bibliographystyle{unsrt}


%% file: Thesis_all.bbl
\begin{thebibliography}{1}
\bibitem{resolution}
    The reflectivity measured for
    $q_z \stackrel{>}{\approx} 1.2$~\AA$^{-1}$
    varies by a few percent depending on the choice of vertical
    detector resolution, due to diffuse scattering
    near the specular condition which arises from surface fluctuations.
    We have not incorporated this small effect into our
    analysis of the reflectivity.  A detailed discussion is given
    in Ref.~5.
\bibitem{note}  In this paper, we consistently use $n=0$ to refer to the
surface layer. This produces discrepancies with the equations
given in Ref.~3.
\end{thebibliography}
\begin{thebibliography}{1}
\bibitem{comment1} The microscopic definition of the local density profile $\tilde{\rho}(z-h(\vec{r}_{xy}))$
requires averaging over density fluctuations with wavelengths shorter than $\xi$. It follows
that fluctuations with surface wavevectors $q_{xy} \leq \pi/\xi$ are contained in the definiton
of $\tilde{\rho}(z-h(\vec{r}_{xy}))$.
\bibitem{comment2} This cutoff is selected such that the number of capillary modes is
equal to the number of surface atoms. Choice of a larger cutoff length would imply that
the local density profile includes contributions from averaging over the capillary waves
with the shortest wavelengths.
First principle arguments that might support such a larger cutoff length are beyond the scope of this
paper. A smaller cutoff length on the other hand would imply capillary waves with wavelengths
smaller than the particle size which is not physically reasonable.
\end{thebibliography}
\begin{thebibliography}{1}
\bibitem{peak} The position of  the layering peak
is expected to be close to that of
the first peak of the bulk $S(q)$.
\end{thebibliography}
\begin{thebibliography}{10}
\bibitem{oops} Because the sample surface
was curved, these measurements required an integration over a
range of $\alpha$ at each $q_z$ point.
\end{thebibliography}
\begin{thebibliography}{1}
\bibitem{error1} The expression in Eq.~(3)
    is given incorrectly in  Eq.~(12) of Ref.~15.)
\bibitem{error2} This reduced form factor is given incorrectly in Ref.~12.
\bibitem{gid} Grazing incidence diffraction measurements, which probe
    in-plane correlations of the first 50~\AA\ at the surface,  were
    performed for the Hg 0.13\% Au alloy at room temperature and at
    $-20^\circ$C.  No sharp peaks were observed; the data contained
    only features characteristic of the bulk Hg structure factor.
\end{thebibliography}
\begin{thebibliography}{1}
\bibitem{born} The kinematic limit, or Born approximation, is valid where
$R(q_z) << 1$.  In this work, data are fit only for $q_z > 0.2$, where
$R(q_z) < 3 \times 10^{-4}$, and the structure is determined principally
by data with $q_z > 0.4$, where $R(q_z)$ is in the range
$10^{-5}$--$10^{-9}$.
\end{thebibliography}
\begin{thebibliography}{10}
\bibitem{dip} The dip in density between the outermost Bi adlayer and the thick wetting layer
can be modeled with an atomic   model similar to the one used for Bi segregated on top of Ga
with subsurface layering (see Ref.\,10).
Including subsurface layering  provides a slightly improved
fit but does not substantially alter the parameters of the wetting layer.
\end{thebibliography}
\begin{thebibliography}{10}
\bibitem{Kaufman01}
The calculations of the phase boundaries rely on data sets for
Ga-Bi from the
  Calphad initiative (Larry Kaufman (MIT)), which have been refined in order to
  more accurately reproduce predel's phase diagram.
\end{thebibliography}
